\documentclass[twocolumn]{aastex63}
\usepackage{graphicx}
\usepackage[countmax]{subfloat}
\usepackage{booktabs}
\usepackage{epsfig}
\usepackage{float}
\usepackage{placeins}
\usepackage{mwe}
\usepackage{amsmath}
\newcommand{\RNum}[1]{\uppercase\expandafter{\romannumeral #1\relax}}
\newcommand{\bl}{BL\,Lacertae}

\begin{document}

\title[Intra-night variability of \bl]{Analysis of the intra-night variability of \bl\ during its August 2020 flare}

\correspondingauthor{B. Mihov}
\email{bmihov@astro.bas.bg}

\author[0000-0003-4682-5166]{A. Agarwal}
\affiliation{Raman Research Institute, C. V. Raman Avenue, Sadashivanagar, Bengaluru~-- 560 080, India}

\author[0000-0002-1567-9904]{B. Mihov}
\affiliation{Institute of Astronomy and NAO, Bulgarian Academy of Sciences, 72 Tsarigradsko Chaussee Blvd., 1784 Sofia, Bulgaria}

\author{V. Agrawal}
\affiliation{Twilio, RMZ Ecoworld, Bellandur, Bangalore, India}

\author{S. Zola}
\affiliation{Astronomical Observatory, Jagiellonian University, ul. Orla 171, 30-244 Krakow, Poland}

\author{Aykut \"Ozd\"onmez}
\affiliation{Ataturk University, Faculty of Science,  Department of Astronomy and Space Science, 25240, Yakutiye, Erzurum, Turkey}

\author{Erg\"un Ege}
\affiliation{Istanbul University, Faculty of Science, Department of Astronomy and Space Sciences, 34116, Beyazıt, Istanbul, Turkey}

\author{L. Slavcheva-Mihova}
\affiliation{Institute of Astronomy and NAO, Bulgarian Academy of Sciences, 72 Tsarigradsko Chaussee Blvd., 1784 Sofia, Bulgaria}

\author{D. E. Reichart}
\affiliation{University of North Carolina at Chapel Hill, Chapel Hill, NC 27599, USA}

\author{D. B. Caton}
\affiliation{Dark Sky Observatory, Dept. of Physics and Astronomy, Appalachian State
University, Boone, NC 28608, USA}

\author{Avik Kumar Das}
\affiliation{Raman Research Institute, C. V. Raman Avenue, Sadashivanagar, Bengaluru~-- 560 080, India}


\begin{abstract}
We present an analysis of the $BVRI$ photometry of the blazar \bl\ on diverse timescales from mid-July to mid-September 2020. We have used 11 different optical telescopes around the world and have collected data over 84 observational nights. 
The observations cover the onset of a new activity phase of \bl\ started in August 2020 (termed as the August 2020 flare by us), and the analysis is focused on the intra-night variability. On short-term timescales, (i) flux varied with $\sim$2.2\,mag in $R$ band, (ii) the spectral index was found to be weakly dependent on the flux (i.e., the variations could be considered mildly chromatic) and (iii) no periodicity was detected. 
On intra-night timescales, \bl\ was found to show bluer-when-brighter chromatism predominantly. We also found two cases of significant inter-band time lags of the order of a few minutes. The duty cycle of the blazar during the August 2020 flare was estimated to be quite high ($\sim$90\% or higher).
We decomposed the intra-night light curves into individual flares and determined their characteristics. On the basis of our analysis and assuming the turbulent jet model, we determined some characteristics of the emitting regions: Doppler factor, magnetic field strength, electron Lorentz factor, and radius. The radii determined were discussed in the framework of the Kolmogorov theory of turbulence.
We also estimated the weighted mean structure function slope on intra-night timescales, related it to the slope of the power spectral density, and discussed it with regard to the origin of intra-night variability.
\end{abstract}

\keywords{galaxies: general~-- galaxies: active~-- \bl\ objects: general~-- \bl\ objects: individual: \bl}

\section{Introduction}
\label{sec:intro}

Blazars are a subclass of radio-loud active galactic nuclei whose relativistic jets are closely aligned with the line of sight \citep{1995PASP..107..803U}.
Blazars display peculiar characteristics across the entire electromagnetic spectrum, including non-thermal continuum emission variables on timescales ranging from a few minutes to years
\citep[e.g.][]{1995ARA&A..33..163W,2008AJ....136.2359G,2015MNRAS.452.2004M,2020ApJ...891..120B,2021A&A...645A.137A}, strong optical linear polarization, and superluminal motions \citep{2019ApJ...874...43L}.
Blazars are divided into two categories, namely \bl\ objects (BL\,Lacs) and flat-spectrum radio quasars, based on their optical spectra and compact radio morphology.
Flat-spectrum radio quasars show strong emission lines, while BL Lacs display very weak or no emission lines in their optical spectra.

The observed spectral energy distribution (SED) of blazars shows two broad humps: the first one extends from 10$^{12}$\,Hz to 10$^{17}$\,Hz, while the second one is peaking between 10$^{21}$\,Hz and 10$^{26}$\,Hz \citep[e.g.][]{2010ApJ...716...30A}. The low-frequency hump is attributed to the synchrotron radiation of the relativistic electrons in the magnetic field of Doppler-boosted jets. On the other hand, the high-energy hump is generally associated with the
inverse Compton scattering of the infrared/optical/ultraviolet photons by the jet electrons \citep{2009ApJ...704...38S}. The seed photons for the inverse Compton scattering could be originating from the synchrotron emission within the jet, commonly known as synchrotron self-Compton \citep{2002ApJ...581..143B}, or from the external photon fields such as accretion disk, broad emission line region, and dusty torus and named as external Compton \citep{1994ApJ...421..153S}.
Blazars are further classified based on the location of their synchrotron peak as follows \citep{2010ApJ...716...30A}: high synchrotron peaked ($\nu_{\rm peak} \ge 10^{15}\,{\rm Hz}$), intermediate synchrotron peaked ($10^{14}\,{\rm Hz} \le \nu_{\rm peak} \le 10^{15}\,{\rm Hz}$), and low synchrotron peaked ($\nu_{\rm peak} \le 10^{14}\,{\rm Hz}$).

\bl\ is the prototype of the BL\,Lac class of blazars and has a redshift of $z=0.0686 \pm 0.0004$ \citep{1995ApJ...452L...5V}. It is classified as a low-synchrotron-peaked blazar \citep{2018A&A...620A.185N}. 
\bl\ has been of great interest for numerous intense multi-wavelength (MWL) campaigns \citep[e.g.][]{2002A&A...390..407V,2003ASPC..299..221V,2003ApJ...596..847B,2010A&A...524A..43R,2015A&A...573A..69W,2017MNRAS.469..813A,2019A&A...623A.175M,2020ApJ...900..137W,2022Natur.609..265J,2023ApJ...943..135K,2023MNRAS.519.3798S}; in particular, \bl\ is one of the favorite targets of the campaigns organized by the Whole Earth Blazar Telescope collaboration.
More-than-century-long observations of \bl\ reveal intense variability on diverse timescales ranging from a few minutes \citep[e.g.][]{2002A&A...390..407V,2015MNRAS.452.4263G,2017MNRAS.469.3588M,2022ApJ...926...91F} to years \citep{1992AJ....104...15C,2004A&A...421..103V,2004A&A...424..497V,2009A&A...501..455V,2013MNRAS.436.1530R}. As an example of yearly variability, \citet{1992AJ....104...15C} detected an erratic behavior of the source with a $V$ band magnitude ranging from 14 to 16 over about 17 years of observations.
\bl\ shows outbursts of a few magnitudes, which is typical for blazars; for example, \citet{2004A&A...421..103V} reported a brightness excursion of about 3\,mag in all bands during the 1997 outburst \citep[see also][]{2018BlgAJ..28...22B}.

\bl\ generally shows a bluer-when-brighter (BWB) chromatism, whose strength was found to be related to the timescale considered: \citet{2002A&A...390..407V} reported strongly BWB chromatic, fast flares on intra-night timescales and mildly chromatic variations on longer timescales \citep[see also][]{2004A&A...421..103V,2018Galax...6....2B,2019MNRAS.484.5633G}. The mildly chromatic component was explained as arising because of the Doppler factor change, while the strongly chromatic flares were assumed to be of synchrotron origin.

Previous studies of \bl\ in optical bands show both the lack \citep[e.g.][]{1998A&A...332L...1N,2021RAA....21..259L} and the presence of inter-band time lags, $\tau$: \citet{2003A&A...397..565P} found a time lag of $\tau=13.8^{+11.4}_{-9.0}$\,min between $B$ and $I$ bands ($B$ band leads), \citet{2006MNRAS.373..209H} found a lag of 11.6\,min between $e$ and $m$ bands ($e$ band leads), \citet{2017MNRAS.469.3588M} found a lag of 11.8\,min between $R$ and $V$ bands ($R$ band leads), and \citet{2022ApJ...926...91F} found a lag of $\sim$16\,min between $B$ and $V$ bands ($B$ band leads) and a lag of $\sim$18\,min between $B$ and $R$ bands ($B$ band leads).
Therefore, the so-called soft lag~-- that is, the lower-frequency/softer energy emission variations are lagging~-- dominates the inter-band time lags observed in \bl.

The Doppler factor, $\delta$, is an important jet characteristic, and for \bl\, it was determined by a number of authors using various approaches. 
\citet{2017ApJ...846...98J} used the observed variability timescale and the angular size of the six moving knots, observed by the Very Long Baseline Array, to get Doppler factors of $6.2\pm1.5$, $11.0\pm5.6$, $5.6\pm3.3$, $8.4\pm1.7$, $8.6\pm2.6$, and $7.1\pm4.3$.
\citet{2017MNRAS.466.4625L} and \citet{2018ApJ...866..137L} compared observed and intrinsic brightness temperatures and got the following variability Doppler factors $6.1 \pm 0.8$ and $12.17^{+3.44}_{-2.81}$, respectively, while
\citet{2018ApJS..235...39C} used broadband SED to derive $\delta=3.8$.
\citet{2020ApJ...897...10Z} proposed a new method to estimate the Doppler factor for a source of known $\gamma$-rays and broad emission line luminosities; the authors got $\delta=8.13$ for \bl.
\citet{2021PASJ...73..775Y} used the relation between the core and extended radio luminosities to estimate $\delta=14.22$ for a continuous jet and $\delta=6.66$ for a moving blob; to get these values, the authors assumed a spectral index $\alpha=0.5$ ($F_{\nu} \propto \nu^{-\alpha}$, where $F_{\nu}$ is the monochromatic flux density). Generally, the different methods result in different Doppler factors because of the different assumptions made.

During the summer of 2020, a new phase\footnote{During this long-lasting activity phase \bl\, reached its historical maximum of $R=11.271 \pm 0.003\,\rm mag$ at JD 2459426.4930 \citep[Jul 30, 2021,][]{2021ATel14820....1K}} of the \bl\ activity began, which continued throughout 2021. The source was reported as flaring during August 2020 in the optical \citep{2020ATel13930....1G,2020ATel13956....1J,2020ATel13958....1S} and high-energy $\gamma$-rays \citep{2020ATel13933....1C,2020ATel13964....1O}. The MAGIC system of Cherenkov telescopes detected very high energy $\gamma$-rays during the night of Aug 19 \citep{2020ATel13963....1B}; the next peak of the very high energy $\gamma$-rays was detected on Sep 19 \citep{2020ATel14032....1B}. A significant optical intra-night variability (INV) was also observed \citep{2020ATel13956....1J}. 

In this paper, we report the results from our observations of \bl\ on intra-night timescales during its August 2020 flare; the mid-August to mid-September \bl\ activity will be termed by us as an August 2020 flare throughout the paper.
In particular, we focus on the analysis of the individual intra-night light curves (INLCs) recorded in the course of our monitoring.

The paper is organized as follows. In Section~\ref{sec:obs} we describe our observations and data reduction. In Section~\ref{sec:tech} the analysis techniques used by us are described in detail. In Section~\ref{sec:res} we present the results obtained, and in Section~\ref{sec:disc} we discuss them.

\section{Observations and Data Reductions}
\label{sec:obs}

To understand the source behavior in the optical regime, we carried out optical observations of \bl\ from July to September 2020 using 11 different optical telescopes around the globe over 84 observational nights and gathering $\sim$12\,800 frames in $BVRI$ bands.
The telescopes used are as follows: 50\,cm OAUJ-CDK500 (Corrected Dall-Kirkham Astrograph, telescope A) of the Astronomical Observatory operated by the Jagiellonian University, Krakow, Poland; Kirkham astrograph telescope (KRK, telescope B) of the Jagiellonian University, Krakow, Poland; 40\,cm PROMPT-USASK telescope of Sleaford Observatory (PSASK, Telescope C); 60\,cm Rapid Response Robotic Telescope (RRRT, telescope D) of the Fan Mountain Observatory, SUH (telescope E); 50/70\,cm Schmidt telescope at the Rozhen National Astronomical Observatory, Bulgaria \citep[telescope F,][]{2010gfss.conf..137K}; 2.01\,m RC Himalayan Chandra Telescope (HCT, telescope G) at Indian Astronomical Observatory, Hanle, India; 40\,cm telescope of the Dark Sky Observatory (DSO, telescope H); 40\,cm telescope of the Montana Learning Center (MLC-COS16, telescope I); 60\,cm RC robotic telescope, Turkey (telescope J); and 1.0\,m RC telescope, Turkey (telescope K).
Telescopes F and G are described in \citet{2019MNRAS.488.4093A}, and telescopes J and K are described in \citet{2021A&A...645A.137A}. The technical details about the rest of the telescopes are given in Table~\ref{tab:telescopes}. Telescopes A, C, D, H, and I work in the robotic mode under the Skynet Robotic Telescope Network software \citep{2021RMxAC..53..206Z}.
The complete log of our observations is presented in Table~\ref{tab:obs_log1}.

The data reduction procedure includes bias/dark subtraction, flat-fielding, and cosmic-ray treatment which was performed using the standard IRAF\footnote{IRAF is distributed by the National Optical Astronomy Observatories, which are operated
by the Association of Universities for Research in Astronomy, Inc., under a cooperative agreement with the
National Science Foundation.} tasks. This was followed by the extraction of the instrumental magnitudes of the source and standard stars in the field using the Dominion Astronomical
Observatory Photometry ({\tt DAOPHOT II}) software \citep{S1987PASP, S1992ASPC}.
To perform differential photometry, we finally chose stars B and C from the source
field\footnote{https://www.lsw.uni-heidelberg.de/projects/extragalactic/charts/2200+420.html} that are
in close proximity to the target and with magnitudes similar to the blazar.
A more detailed data reduction procedure is discussed in \citet{2019MNRAS.488.4093A}.

To get the optimum aperture for each night, we performed aperture photometry for different radii: 1.0, 1.2, 1.4, 1.6, 1.8, 2.0, 2.5, and 3.0 times the full width at the half-maximum (FWHM) of the field stars. For background subtraction, we selected the sky annulus to approximately 5$\times$FWHM. We finally selected the aperture with the best signal-to-noise ratio and minimum standard deviation of the difference between instrumental magnitudes of standard stars.
The above procedure was applied on all the $BVRI$ frames, and the calibrated magnitudes of the source were derived.

The calibrated $BVRI$ magnitudes of the blazar were dereddened by subtracting the Galactic extinction values from the NASA/IPAC Extragalactic Database: $A_B=0.43$\,mag, $A_V=0.54$\,mag, $A_R=0.64$\,mag,
and $A_I=0.80$\,mag. The flux from the nucleus of the source is contaminated by its elliptical host galaxy. Hence, to perform host galaxy subtraction, we converted extinction-corrected magnitudes to fluxes using the zero point values from \citet{1998A&A...333..231B}.
Thereafter using the measurements from \citet{2007A&A...475..199N}, we estimated the host galaxy emission in the $R$ band. This $R$ band value is further used
to obtain the corresponding contributions for the $BVI$ bands by using the galaxy colors \citep{1995PASP..107..945F} as $B-V = 0.96$\,mag, $V-R = 0.61$\,mag, and $R-I = 0.70$\,mag. 

\begin{rotatetable}
\begin{deluxetable*}{@{}cccccccc@{}}
\centerwidetable
\tablecaption{Details about the telescopes and instruments used \label{tab:telescopes}}
\tablewidth{0pt}
\tablehead{
\colhead{Telescope} & \colhead{A} & \colhead{B} & \colhead{C} & \colhead{D} & \colhead{E} & \colhead{H} & \colhead{I} \\
\colhead{} & \colhead{50\,cm Cassegrain} & \colhead{50\,cm Cassegrain} & \colhead{40\,cm RC} & \colhead{60\,cm RC} & \colhead{60\,cm RC} & \colhead{40\,cm RC} & \colhead{40\,cm RC}
}
\tabletypesize{\small}
\startdata
 CCD model & U47 & Alta U47 & FLI & SBIG STX-16803 & CG47 & U47 & SBIG STXL-11002 \\
 & F42 & U42/F42 & & & & & \\
 Chip size & 1024$\times$1024\,px & 1024$\times$1024\,px & 2014$\times$2014\,px & 4096$\times$4096\,px & 1024$\times$1024\,px & 1024$\times$1024\,px & 4008$\times$2672\,px \\
 & 2048$\times$2048\,px & 2048$\times$2048\,px & & & & & \\
 Scale & 0\farcs81/px & 0\farcs399/px & 0\farcs12/px & 1\farcs13/px & 1\farcs12/px & 1\farcs23/px & 1\farcs53/px \\
 & 0\farcs81/px & 0\farcs399/px & & & & & \\
 Field & 13\farcm3$\times$13\farcm3 & 6\farcm8$\times$6\farcm8 & 10\farcm1$\times$10\farcm1 & 25\farcm8$\times$25\farcm8 & 19\farcm1$\times$19\farcm1 & 10\farcm5$\times$10\farcm5 & 34\farcm1$\times$22\farcm7 \\
 & 27\farcm7$\times$27\farcm7 & 13\farcm6$\times$13\farcm6 & & & & & \\
 Gain & 1.41\,$\rm e^-$/ADU & 1.41\,$\rm e^-$/ADU & 1.21\,$\rm e^-$/ADU &  1.27\,$\rm e^-$/ADU & 1.61\,$\rm e^-$/ADU & 1.29\,$\rm e^-$/ADU &  1.74\,$\rm e^-$/ADU \\
 & 1.69\,$\rm e^-$/ADU & 1.69\,$\rm e^-$/ADU & & & & & \\
 Read-out noise & 8.3\,$\rm e^-$\,rms & 8.3\,$\rm e^-$\,rms & 17.0\,$\rm e^-$\,rms & 16.96\,$\rm e^-$\,rms & 12.81\,$\rm e^-$\,rms & 8.74\,$\rm e^-$\,rms  & 11.13\,$\rm e^-$\,rms \\
 &  10.1\,$\rm e^-$\,rms & 10.1\,$\rm e^-$\,rms  & & & & & \\
 Binning used & 1$\times$1 & 2$\times$2 & 1$\times$1 & 2$\times$2 & 3$\times$3 & 1$\times$1  & 1$\times$1 \\
 Typical seeing & 2\farcs5 to 3\farcs5 & $\sim$3\farcs5 & 1$\arcsec$ to 3$\arcsec$ & 1$\arcsec$ to 3$\arcsec$ & 1\farcs5 to 2.5$\arcsec$ & 2\farcs5 to 3\farcs5 & 1\farcs5 to 2\farcs5 \\
\noalign{\smallskip}
\enddata
\end{deluxetable*}
\end{rotatetable}

\clearpage
\startlongtable
\begin{deluxetable*}{cccccccccccc}
\tablecaption{Log of photometric observations for the blazar \bl \label{tab:obs_log1}}
\tablewidth{0pt}
\tablehead{
\colhead{Date} & \colhead{Telescope} & \multicolumn{4}{c}{Number of data points} & \colhead{Date} & \colhead{Telescope} & \multicolumn{4}{c}{Number of data points} \\
\cline{3-6}
\cline{9-12}
\colhead{(yyyy mm dd)} & \colhead{} & \colhead{$B$} & \colhead{$V$} & \colhead{$R$} & \colhead{$I$} & \colhead{(yyyy mm dd)} & \colhead{} & \colhead{$B$} & \colhead{$V$} & \colhead{$R$} & \colhead{$I$} 
}
\tabletypesize{\small}
\startdata
2020 07 13 & J & 1 & 1 & 1 & 1        & 2020 08 27 & J & 0 & 2 & 1 & 1 \\
2020 07 14 & J & 1 & 1 & 1 & 1        & 2020 08 28 & B & 0 & 0 & 152 & 0 \\    
2020 07 15 & J & 1 & 1 & 1 & 1        & 2020 08 28 & C & 0 & 0 & 10 & 0 \\     
2020 07 16 & J & 1 & 1 & 1 & 1        & 2020 08 28 & F & 328 & 18 & 18 & 348 \\
2020 07 17 & J & 1 & 1 & 1 & 1        & 2020 08 28 & J & 0 & 2 & 1 & 1 \\     
2020 07 19 & J & 1 & 1 & 1 & 1        & 2020 08 29 & D & 0 & 0 & 23 & 0 \\     
2020 07 20 & J & 1 & 1 & 1 & 1        & 2020 08 29 & J & 0 & 2 & 1 & 1 \\      
2020 07 21 & J & 0 & 1 & 0 & 1        & 2020 08 30 & D & 0 & 0 & 9 & 4 \\      
2020 07 22 & J & 0 & 1 & 0 & 1        & 2020 08 30 & J & 0 & 2 & 0 & 1 \\      
2020 07 23 & J & 0 & 0 & 1 & 1        & 2020 08 30 & K & 2 & 2 & 518 & 2 \\    
2020 07 24 & J & 1 & 0 & 0 & 1        & 2020 08 31 & K & 3 & 77 & 75 & 73 \\   
2020 07 25 & J & 1 & 1 & 1 & 1        & 2020 09 01 & J & 0 & 2 & 1 & 1 \\      
2020 07 26 & J & 1 & 1 & 1 & 1        & 2020 09 02 & B & 30 & 30 & 1000 & 30 \\
2020 07 28 & J & 0 & 3 & 1 & 1        & 2020 09 02 & C & 0 & 0 & 9 & 0 \\      
2020 07 29 & J & 0 & 3 & 0 & 1        & 2020 09 02 & J & 1 & 2 & 1 & 1 \\      
2020 07 30 & J & 1 & 3 & 0 & 1        & 2020 09 03 & A & 45 & 39 & 27 & 28 \\  
2020 07 31 & K & 3 & 2 & 92 & 3       & 2020 09 03 & B & 13 & 17 & 553 & 15 \\ 
2020 07 31 & J & 0 & 2 & 0 & 0        & 2020 09 04 & G & 7 & 1 & 7 & 7 \\      
2020 08 01 & J & 0 & 1 & 1 & 0        & 2020 09 04 & J & 0 & 0 & 1 & 1 \\      
2020 08 02 & J & 0 & 1 & 1 & 1        & 2020 09 05 & D & 25 & 25 & 46 & 0 \\   
2020 08 04 & J & 1 & 2 & 0 & 1        & 2020 09 06 & D & 0 & 0 & 87 & 0 \\     
2020 08 05 & J & 1 & 2 & 0 & 1        & 2020 09 06 & G & 1 & 1 & 1 & 1 \\      
2020 08 06 & J & 1 & 1 & 1 & 1        & 2020 09 06 & J & 1 & 2 & 1 & 1 \\      
2020 08 07 & J & 1 & 0 & 1 & 1        & 2020 09 07 & A & 30 & 29 & 24 & 30 \\  
2020 08 12 & J & 1 & 2 & 0 & 1        & 2020 09 07 & D & 0 & 0 & 34 & 0 \\     
2020 08 13 & J & 1 & 2 & 0 & 1        & 2020 09 07 & J & 1 & 2 & 1 & 1 \\      
2020 08 14 & J & 1 & 2 & 1 & 0        & 2020 09 08 & A & 32 & 33 & 32 & 30 \\  
2020 08 16 & J & 1 & 2 & 0 & 1        & 2020 09 08 & B & 0 & 0 & 236 & 0 \\    
2020 08 18 & J & 1 & 2 & 1 & 1        & 2020 09 08 & J & 0 & 2 & 1 & 1 \\      
2020 08 19 & G & 0 & 2 & 0 & 0        & 2020 09 09 & A & 43 & 45 & 44 & 45 \\  
2020 08 20 & A & 0 & 0 & 370 & 0      & 2020 09 09 & B & 0 & 0 & 508 & 0 \\    
2020 08 20 & B & 95 & 118 & 118 & 108 & 2020 09 09 & J & 1 & 2 & 1 & 1 \\      
2020 08 20 & C & 0 & 0 & 6 & 0        & 2020 09 10 & A & 49 & 50 & 217 & 54 \\ 
2020 08 20 & G & 0 & 1 & 0 & 0        & 2020 09 10 & B & 0 & 0 & 187 & 0 \\    
2020 08 21 & A & 0 & 0 & 327 & 0      & 2020 09 10 & J & 1 & 2 & 1 & 1 \\      
2020 08 21 & C & 0 & 0 & 6 & 0        & 2020 09 10 & K & 2 & 2 & 206 & 3 \\    
2020 08 21 & G & 2 & 1 & 144 & 2      & 2020 09 11 & A & 242 & 59 & 62 & 61\\  
2020 08 22 & C & 0 & 0 & 25 & 0       & 2020 09 11 & B & 0 & 0 & 573 & 0 \\    
2020 08 22 & G & 2 & 1 & 2 & 2        & 2020 09 11 & C & 0 & 0 & 9 & 0 \\      
2020 08 23 & A & 7 & 7 & 7 & 7        & 2020 09 11 & J & 1 & 2 & 0 & 1 \\      
2020 08 23 & C & 0 & 0 & 20 & 0       & 2020 09 12 & A & 41 & 12 & 370 & 11 \\ 
2020 08 23 & E & 38 & 36 & 37 & 40    & 2020 09 12 & B & 0 & 0 & 56 & 0 \\     
2020 08 23 & G & 3 & 0 & 0 & 3        & 2020 09 12 & C & 0 & 0 & 17 & 0 \\    
2020 08 23 & J & 0 & 0 & 1 & 0        & 2020 09 12 & I & 0 & 0 & 10 & 0 \\     
2020 08 24 & A & 6 & 7 & 6 & 6        & 2020 09 12 & J & 1 & 2 & 1 & 1 \\      
2020 08 24 & B & 0 & 0 & 18 & 0       & 2020 09 12 & K & 1 & 2 & 241 & 1 \\    
2020 08 24 & C & 0 & 0 & 23 & 0       & 2020 09 13 & A & 14 & 232 & 18 & 221 \\
2020 08 24 & G & 2 & 0 & 0 & 2        & 2020 09 13 & B & 0 & 0 & 376 & 0 \\    
2020 08 24 & J & 1 & 2 & 1 & 1        & 2020 09 13 & C & 0 & 0 & 6 & 0 \\      
2020 08 25 & B & 54 & 54 & 949 & 54   & 2020 09 13 & I & 0 & 0 & 2 & 0 \\     
2020 08 25 & C & 0 & 0 & 12 & 0       & 2020 09 13 & J & 1 & 2 & 1 & 1 \\      
2020 08 25 & J & 1 & 2 & 1 & 1        & 2020 09 13 & K & 2 & 2 & 233 & 2 \\   
2020 08 26 & C & 0 & 0 & 16 & 0       & 2020 09 14 & A & 18 & 224 & 18 & 223 \\
2020 08 26 & F & 299 & 14 & 15 & 296  & 2020 09 14 & B & 6 & 6 & 710 & 6 \\   
2020 08 26 & J & 1 & 2 & 1 & 1        & 2020 09 14 & C & 0 & 0 & 13 & 0 \\     
2020 08 27 & C & 0 & 0 & 9 & 0        & 2020 09 14 & D & 0 & 0 & 26 & 0 \\     
2020 08 27 & F & 332 & 18 & 18 & 331  & 2020 09 14 & J & 1 & 0 & 1 & 1 \\      
\noalign{\smallskip}
\enddata
\end{deluxetable*}

\section{Analysis Techniques}
\label{sec:tech}

Having obtained the light curves (LCs) in flux units, we
\begin{enumerate}
    \item combined the LCs in the case in which multi-telescope data are available and cleaned the combined LCs of the outliers if any; and
    \item corrected the combined LCs for the smooth flux variation in the case in which the LCs show two variability components.
\end{enumerate}
The corrected LCs were further
\begin{enumerate}
    \item decomposed into individual flares; and
    \item used to build the structure functions (SFs).
\end{enumerate}
In addition, the corrected MWL LCs were
\begin{enumerate}
    \item used to build the color-magnitude diagrams (CMDs); and
    \item used to search for inter-band time lags.
\end{enumerate}
Below we shall describe in detail the analysis techniques used in each of the above steps.

\subsection{Variability Detection and Amplitude}
\label{sec:test}

We quantified the flux variability of \bl\ using $C$-, $F$-, and $\chi^{2}$-tests and the percentage amplitude variation, $A$. A brief introduction to these methods is given below.

\subsubsection{C-test}

The most frequently used variability detection criterion is the $C$-test \citep{1999A&AS..135..477R}, which is defined as

\begin{equation}
C_{1} = {\rm \frac {\sigma(BL-S_{B})}{\sigma(S_{B}-S_{C})}}, \quad
C_{2} = {\rm \frac {\sigma(BL-S_{C})}{\sigma(S_{B}-S_{C})}},
\end{equation}
where BL$-$S$_{\rm B}$, BL$-$S$_{\rm C}$, and S$_{\rm B}$$-$S$_{\rm C}$ are the differential instrumental LCs of the blazar (BL) against the standard star B (S$_{\rm B}$), BL against the standard star C (S$_{\rm C}$), and S$_{\rm B}$ against S$_{\rm C}$, respectively, while $\sigma$(BL$-$S$_{\rm B}$), $\sigma$(BL$-$S$_{\rm C}$), and $\sigma$(S$_{\rm B}$$-$S$_{\rm C}$) are the standard deviations of the respective LCs. 
If $C \geq 2.576$, then we marked the LC as a variable at a confidence level of $99.5$\% or greater; otherwise, we call it a non-variable (here $C$ is a mean over $C_1$ and $C_2$). As pointed out by \citet{2017MNRAS.467..340Z}, through their study of INV in active galactic nuclei using various statistical methods, the $C$-test could be considered a suitable test to detect variability with more reliable results as compared to the $F$-test.
 
\subsubsection{F-test}

The $F$-test \citep{2017MNRAS.467..340Z} is a powerful tool to quantify variability at diverse timescales and is defined as

\begin{equation}
 \label{eq.ftest}
 F_1=\frac{\sigma^2(\mathrm{BL-S_B)}}{\sigma^2(\mathrm{S_B-S_C})}, \quad F_2=\frac{\sigma^2(\mathrm{BL-S_C})}{\sigma^2(\mathrm{S_B-S_C})},
\end{equation}
where BL$-$S$_{\rm B}$, BL$-$S$_{\rm C}$, and S$_{\rm B}$$-$S$_{\rm C}$ are the differential instrumental LCs of BL against S$_{\rm B}$, BL against S$_{\rm C}$, and S$_{\rm B}$ against S$_{\rm C}$, respectively, while $\sigma^2$(BL$-$S$_{\rm B}$), $\sigma^2$(BL$-$S$_{\rm C}$), and $\sigma^2$(S$_{\rm B}$$-$S$_{\rm C}$) are the variances of the respective LCs. 
Averaging $F_1$ and $F_2$ gives the mean observational $F$ value, which is then compared with the critical value, $F_{\rm c}=F^{(\alpha)}_{\nu_{\rm BL},\nu_{\rm S}}$,
where $\nu_{\rm BL}$ and $\nu_{\rm S}$ give the number of degrees of freedom for the blazar and star LCs, respectively, estimated as the number of measurements, $N_{\rm data}$, minus 1 ($\nu = N_{\rm data} - 1$). The significance level, $\alpha$, is set as 0.1\% and 1\% (i.e. $3 \sigma$ and $2.6 \sigma$) for this work.
If the mean $F$ value is more than the critical value, the null hypothesis (i.e. no variability) is rejected and the LC is marked as variable. 

\subsubsection{$\chi^{2}$-test}
Further, to detect the genuine variability in our source, we also used the $\chi^{2}$-test, which is interpreted as:

\begin{equation}
\chi^2 = \sum_{i=1}^N \frac{(V_i - \overline{V})^2}{e_i^2},
\end{equation}
where $\overline{V}$ is the mean magnitude and $V_i$ the magnitude corresponding to the $i$-th observation
with a respective uncertainty $e_i$. Estimating the exact values of uncertainties is unattainable in the IRAF package used for data reduction,
whereas, the theoretical uncertainties have been found to be smaller by 1.3--1.75 \citep{2003ApJ...586L..25G}. For our data, the factor is around 1.6, on average. Therefore, for a better calculation of photometric uncertainties, we should multiply the uncertainties obtained from data analysis by the above factor.
The obtained $\chi^{2}$ value is then compared with a critical value $\chi_{\alpha,\nu}^2$ where $\alpha$ is the significance level and $\nu = N_{\rm data}-1$ is the degree of freedom. When $\chi^2 > \chi_{\alpha,\nu}^2$, it indicated the presence of variability.

Depending on the sampling of the individual INLCs and on the monitoring duration, there could happen the blazar to have both variable and non-variable status for one and the same night; the nights of Sep 2 and Sep 7 are examples in this context.
In such cases, we adopted the status obtained by testing the better LCs in terms of sampling and/or duration.

\subsubsection{Percentage amplitude variation}

To estimate the percentage amplitude change in our LCs, we calculated the variability amplitude parameter $A$ \citep{1996A&A...305...42H}:
\begin{eqnarray}
A = 100\times \sqrt{{(m_\mathrm{max}-m_\mathrm{min}})^2 - 2\langle e^2\rangle} ~ [\%],
\end{eqnarray}
where $m_\mathrm{max}$ and $m_\mathrm{min}$ are the maximum and minimum magnitudes attained by the blazar and $\langle e^2\rangle$ the mean squared uncertainty of the measurements.

\subsection{Combination of the Light Curves}
\label{sec:comb}

The INLCs obtained with two or more telescopes were combined in order to get a single LC for the given night and band.
If the individual LCs have overlapping parts, then, before the combination, the LCs were adjusted such that (i) a single band LC was adjusted to match the corresponding LC from a MWL data set (in order to avoid the systematic uncertainties when the LCs are used to build CMDs) and (ii) a poorly sampled LC was adjusted to match the densely sampled one (if it does not contradict the first condition). 
Technically, the adjustment was made as follows: we interpolated the first LC over the second one in their overlapping parts, computed the median offset and its standard uncertainty, and applied the so-obtained offset according to the above conditions. If the LCs have no overlapping parts, then the LCs were combined without adjustment. Finally, the observations consisting of a few data points per band were (adjusted and) combined with the so-built composite INLCs (an exception were the telescope J data, see below). 
During the combination of the LCs, a few outlying measurements were identified and cleaned.

The so-combined INLCs were merged with the rest of the data to build the short-term\footnote{Variability on timescales from days to weeks/months is usually termed as the short-term variability \citep[STV,][]{2020AN....341..713S}.} variability LCs (STLCs) of \bl\ for each band. To these STLCs, the telescope J STLCs were adjusted (actually, the adjustment was needed only for the $BV$ bands) and combined. 

We are interested in the analysis of the INV, and so the above procedure is optimized for the accurate combination of the individual INLCs, but not for the STLCs of the individual telescopes. This would result in increased night-to-night scatter in the STLCs, but this is not an issue for the presented research.

\subsection{Correction for the Smooth Flux Variation}
\label{sec:dopp}

Generally, the INLCs obtained in the course of our study could be described as flares superimposed onto a smooth flux variation; that is, the LCs show two variable components. The flare timescales are much shorter than the smooth component timescale. The latter timescale is usually longer than several hours, which is longer than the typical duration of a single-telescope intra-night monitoring session.

We are interested in the analysis of the flaring activity of \bl\, and so a correction has to be done in order to minimize the contribution of the smooth variability component.
For example, to make flares more
evident, \citet{1997A&A...327...61G} divided their LC by a curve interpolated through the local minima of the same LC. 

The correction of the LCs for the smooth flux variation (or detrending for short) was done following an approach closely related to that of \citet{2004A&A...421..103V}; see also \citet{2020ApJS..247...49X} and \citet{2021MNRAS.501.1100R}.
Firstly, we selected the regions of the LC that are free of flares~-- they were assumed to be related to the smooth component we want to correct for. Secondly, we fitted to these regions a low-degree polynomial. For more complicated LCs, the fitting was done by splitting the LC into segments and fitting a polynomial to each segment. The polynomials could be of different degrees for different segments, or, for some of the segments, the polynomial could be replaced by another fitting function (e.g. cubic spline or Gaussian).
Upon completion of the fit, care was taken to ensure the individual fitting functions were joined smoothly.
If MWL data are available for a given night, then the fitted regions and the fitting functions are one and the same for all bands.
Finally, we rescaled each data point of the LC by dividing the corresponding flux value by the scaling factor $C_k(t)=F_{k\rm ,fit}(t)/F_{k\rm ,min}$ (here $k$ represents the $BVRI$ bands), which is the ratio between the value of the (composite) fitting function at the corresponding time and the fitting function minimum value. That minimum value served as the base level in the LC decompositions.

\subsection{Decomposition of the INLCs}
\label{sec:decompo}

The INLCs that show flaring activity were decomposed using the following double exponential function \citep[DE,][]{2010ApJ...722..520A}:
\begin{multline}
    F(\Delta t) = F_{\rm base} + \\
    F_0 \left[\exp\left(\frac{\Delta t_0-\Delta t}{\mathcal{T}_{\rm r}}\right)+\exp\left(\frac{\Delta t-\Delta t_0}{\mathcal{T}_{\rm d}}\right)\right]^{-1},
\end{multline}
where $F_{\rm base}$ is the constant base level, $F_0$ twice the flare amplitude (with respect to the base level), $\Delta t_0$ the approximate position in the time of the flare peak, and $\{\mathcal{T}_{\rm r},\mathcal{T}_{\rm d}\}$ the rise and decay timescales.
If the LC has been detrended, then the base level was set to the minimal value of the function, used to fit the smooth component, and was held fixed during the decomposition. If no detrending has been done, then the base level is left free (we, however, have no such LCs).
The time variable, $\Delta t = t-t_0$, we used represents the time since the earliest observation (taken at $t_0$) among the available data sets for the given night; the JD of the earliest observation is indicated in the LC plots. 

The characteristics of the DE function can be summarized as follows.
The actual position in the time of the flare maximum is
\begin{equation}
    \Delta t_{\rm max}=\Delta t_0+\frac{\mathcal{T}_{\rm r}\mathcal{T}_{\rm d}}{\mathcal{T}_{\rm r}+\mathcal{T}_{\rm d}}\ln\left(\frac{\mathcal{T}_{\rm d}}{\mathcal{T}_{\rm r}}\right)
\end{equation}
and it is equal to $\Delta t_0$ in the case of symmetric flares, $\mathcal{T}_{\rm r}=\mathcal{T}_{\rm d}$.
An estimate of the total duration of the flare could be found as $\Delta\mathcal{T}\simeq2\,(\mathcal{T}_{\rm r}+\mathcal{T}_{\rm d})$. The asymmetry parameter is defined as
\begin{equation}
    \xi=\frac{\mathcal{T}_{\rm d}-\mathcal{T}_{\rm r}}{\mathcal{T}_{\rm d}+\mathcal{T}_{\rm r}} \qquad 
      \begin{cases}
    \xi \in [-1,1];\\
    \xi = 0 \implies \text{symmetric flare}.
  \end{cases}
\end{equation}
Finally, the doubling and halving timescales are equal to $\ln(2)\mathcal{T}_{\rm r}$ and $\ln(2)\mathcal{T}_{\rm d}$, respectively \citep{2007ApJ...669..862A}.

\subsection{Structure Function}
\label{sec:sf}

The SF was introduced by \citet{1985ApJ...296...46S} and is particularly useful for analyzing unevenly sampled astronomical data \citep[e.g.][]{2018Galax...6....2B}.
Various aspects of the SF application are thoroughly discussed by \citet{2010MNRAS.404..931E} and \citet{2016ApJ...826..118K}.

For a time separation $\delta t$ and a bin of size ${\text d}t$, we calculated the first-order SF as
\begin{equation}
    D^1(\delta t,{\text d}t)=\frac{1}{N(\delta t,{\text d}t)}\sum_{i>j}[F(t_i)-F(t_j)]^2,
\end{equation}
where $N(\delta t,{\text d}t)$ is the number of pairs $(t_i,t_j)$ for which $\delta t<t_i-t_j< \delta t+{\rm d}t$.
The choice of bin size depends on the LC sampling.
The uncertainties of the SF were calculated simply as the standard uncertainty of the mean in the bins \citep[see][for discussion about the SF uncertainties]{2020MNRAS.491.5035S}. The value of $\delta t$ in each bin was set to the middle of the bin.

Ideally, the SF has two plateaus connected by a curve, whose slope depends on the nature of the observed flux variation \citep[shot noise, flicker noise, etc.; see][]{1992ApJ...396..469H,2020MNRAS.491.5035S}. Let us assume that the LC can be represented by the sum ($s+n$), where $s$ is the signal and $n$ is the noise, both having Gaussian distribution. Then, the first plateau (at $\delta t \rightarrow 0$) equals $2\sigma^2_n$ and the second one equals to $2\sigma^2_s$, where $\sigma^2$ represents the corresponding variances. 
These plateaus bracket the time separations over which the flux variations are correlated. The upward-sloping curve between the plateaus is usually characterized by its logarithmic slope ${\rm d}[\log(D^1)]/{\rm d}[\log(\delta t)]$. The time separation at which this curve flattens could be considered as a robust characteristic variability timescale; if the second plateau is not reached, then the timescale is longer than the observation span.
Next, the SF could be used to study the time asymmetry of the LCs \citep{1998ApJ...504..671K,2017MNRAS.471.2216B,2021BlgAJ..34...79B}.
Finally, if the LC shows periodicity, then the SF has a dip at the time separation equal to the corresponding period.

It is common practice for the measurement uncertainties to be subtracted off during the SF build, and there are various ways to do that \citep[see][for discussion on this topic]{2016ApJ...826..118K}.
If the measurement uncertainties, $e$, are assumed to follow a Gaussian distribution, then $\sigma^2_n$ could be approximated as $\sigma^2_n\simeq\langle e^2\rangle$ and, therefore, $D^1(\delta t)-2\langle e^2\rangle$ is the noise-free SF estimate we wanted. 
The problem here is that any incorrectness in the measurement uncertainty estimation affects the slope of the SF. Hence, we prefer to add $2\sigma^2_n$ as a free parameter during the SF fitting rather than subtracting $2\langle e^2\rangle$ from the SF. In particular, in this way, we could obtain an independent estimate of the mean measurement uncertainty.

In the case of no noise subtraction, we fitted the SF using a single power-law (SPL) model plus a noise term to determine the SF slope:
\begin{equation}
D^1(\delta t) = 2\sigma^2_n+D^1_0\left(\frac{\delta t}{\delta t_0}\right)^{\varrho},
\end{equation}
where $D^1_0$ is the variability amplitude at the fixed timescale $\delta t_0$ (we arbitrarily choose $\delta t_0=1$\,min), $\varrho$ the power-law index, and $\sigma^2_n$ the variance of the measurement noise.
The fitting was done up to the turnover point, $\delta t_{\rm to}$, at which the SF changes its slope and starts to flatten. After that point, the SPL overestimates the SF.

It is worth mentioning two issues that affect the SF fitting, namely the lack of statistical independence and Gaussianity; that is, the individual SF estimates are not independent of each other and the distribution of the SF estimates within the individual bins is not Gaussian \citep{2010MNRAS.404..931E}. The latter problem could be solved particularly by fitting not $D^1(\delta t)$, but $\log[D^1(\delta t)]$ as we actually did; see \citet{2010MNRAS.404..931E} and \citet{2015MNRAS.451.4328K} for details about these issues.

There is an approximate relation between the slopes of the power spectral density (PSD), $\varkappa$, and SF, namely $\varkappa \simeq \varrho+1$
\citep[the equality is obtained under special conditions, see][for details]{2010MNRAS.404..931E}.

\subsection{Color-magnitude Diagram}
\label{sec:cmd}

Given the \bl\ fluxes $F_{\nu}$, we built the following CMDs: $F_{\nu_1}/F_{\nu_3}$ vs $F_{\nu_2}$ if three- or four-band data are available ($\nu_1>\nu_2>\nu_3$, where $\nu_i$ is the frequency corresponding to the $i$-th band) and $F_{\nu_1}/F_{\nu_2}$ vs $(F_{\nu_1}\!+\!F_{\nu_2})/2$ if two-band data are available.
The CMD forms were chosen to minimize the possibility of introducing spurious effects if we are correlating the flux ratio with one of the fluxes used to build the ratio itself \citep{1996A&A...312..810M,2007A&A...470..857P}. The flux ratios we used are representative for the two-point spectral index, $\alpha_{\nu_1\nu_2} \propto -\log(F_{\nu_1}/F_{\nu_2})$, under the assumption that $F_{\nu} \propto \nu^{-\alpha}$.

The CMDs were built by selecting the data points from the corresponding LCs closest to each other. In addition, we required the time intervals among the data points used to get a single CMD data point to be smaller than a predefined threshold, which depends on the sampling of the LCs used and was typically set to a few minutes.

The CMDs were fitted by the power-law model $F_{\nu_1}/F_{\nu_2} \propto X^{\varpi}$, where $\varpi$ is the power-law index\footnote{The power-law index corresponds to the slope of the CMD in magnitude units \citep[e.g.][]{2007A&A...470..857P}.} and $X$ either the flux or the mean flux depending on the CMD form used. Further analysis of the CMDs was done after taking a logarithm of both sides of the above equation.

To consider a CMD trend significant at 99\% confidence level, we required (i) the linear Pearson correlation coefficient to be $|r| \ge 0.5$ and (ii) the probability to get such a correlation coefficient by chance to be $p \le 0.01$ \citep[e.g.][]{2016MNRAS.458.1127G,2021A&A...645A.137A}.
For the nights for which we have $BVRI$ band data, we used the following CMD forms: $F_i/F_I$ vs $F_R$ ($i=B,V$). 
To assign a significant BWB or redder-when-brighter CMD trend for these nights, we further required both CMDs to show a significant correlation.

\subsection{Cross-correlation Analysis}
\label{sec:dcf}

To search for inter-band time lags, we used a Python implementation {\tt pyDCF}\footnote{https://github.com/astronomerdamo/pydcf} \citep{2015MNRAS.453.3455R} of the discrete cross-correlation function \citep[DCF,][]{1988ApJ...333..646E}, which is suitable to cross-correlate unevenly sampled time series. 
In our runs (i) the measurement uncertainties were not taken into account in the build of the DCF following \citet{1994PASP..106..879W} and (ii) the Gaussian weighting scheme was applied in order to assign higher importance to the values closer to the bin center.  

The estimation of the time lag and its uncertainty was done utilizing the flux randomization/random subset selection method \citep[FR/RSS,][]{1998PASP..110..660P,2004ApJ...613..682P} based on Monte Carlo simulations.
During the RSS process, the data points counted more than once were rejected. At the end of each FR/RSS run, the time lag was found as the centroid of the DCF, defined as the DCF-weighted mean lag. 
The centroid was calculated using DCF points above a predefined threshold, which was set to the DCF peak value of less than one to three times its uncertainty~-- we varied the threshold value so as to ensure at least ten data points for the centroid calculation.
We ran a total of 2500 cycles, and the resulting time lags were used to build the cross-correlation centroid distribution (CCCD). Given the CCCD, the time lag is estimated as the 50th percentile (or the median) of the CCCD, while the 16th and 84th percentiles serve as the $1\sigma$ uncertainties. 

The significance of the cross-correlation results was estimated by means of Monte Carlo simulation following the approach of \citet{2014MNRAS.445..437M}. 
To generate the LCs, we used a Python implementation {\tt DELCgen}\footnote{https://github.com/samconnolly/DELightcurveSimulation} \citep{2015arXiv150306676C} of the method of \citet{2013MNRAS.433..907E}, which accounts for the flux probability density function (PDF) and PSD of the observed LC; the alternative LC generation method of \citet{1995A&A...300..707T} produces LCs having a Gaussian flux PDF. 
To produce evenly sampled LCs needed for the PSD build, we used interpolation onto a regular grid having a time interval of 2\,min. We fitted the PSD by a single-slope power law, ${\rm PSD}\propto f^{-\varkappa}$ \citep[here $f$ is the temporal frequency;][]{2005A&A...431..391V,2010MNRAS.402..307V,2012A&A...544A..80G}. The PDF was approximated either with a Gaussian or with a sum of Gaussians. 
Each simulated LC has the same statistical properties and sampling as the observed one. In addition, the noise was added to each simulated LC according to the mean observational uncertainty. We generated a total of 2500 LCs for each of the bands involved in the cross-correlation.
Then, we cross-correlated the simulated LCs in the same way as we have done for the observed ones. Finally, the distribution of the simulated cross-correlation coefficients for each time lag bin was used to estimate the significance levels of the observed coefficients. 

The LCs produced during a typical intra-night monitoring session are of good sampling, so it is worth trying the interpolated cross-correlation function (ICF) for the time lag search. We used a Python implementation {\tt PyCCF}\footnote{https://bitbucket.org/cgrier/python\_ccf\_code} \citep{2018ascl.soft05032S} of the method of \citet{1998PASP..110..660P}. To estimate the lag and its uncertainty, we used the cross-correlation peak distribution (CCPD) because there is no need for additional free parameters, namely the bin size and threshold. 

\begin{figure*}[t!]
\centering
\includegraphics[width=\linewidth,clip=true]{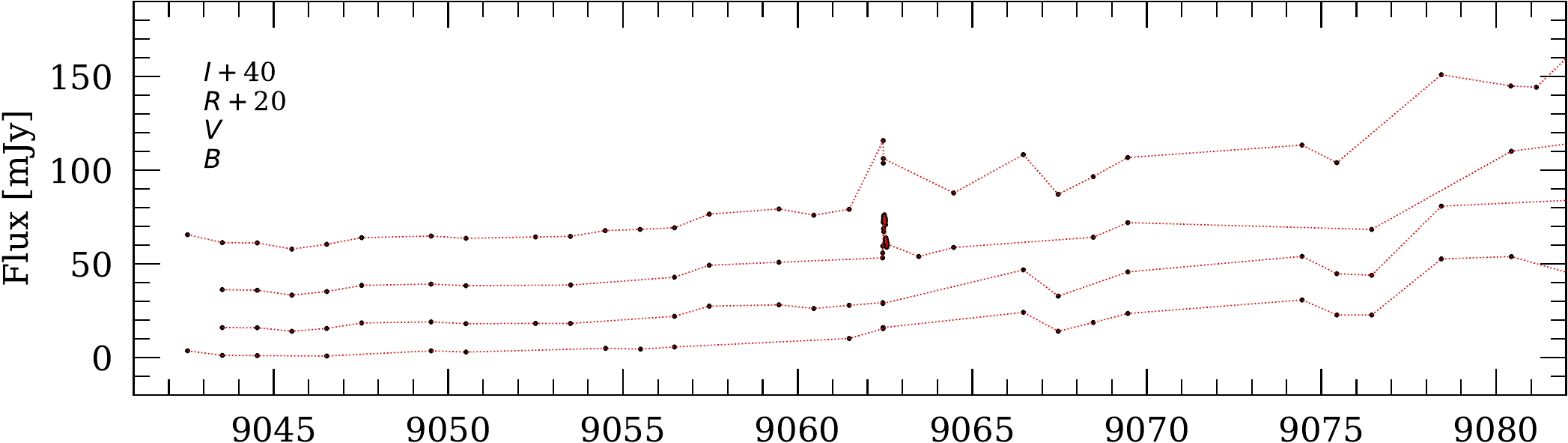}\\
\vspace{.25cm}
\includegraphics[width=\linewidth,clip=true]{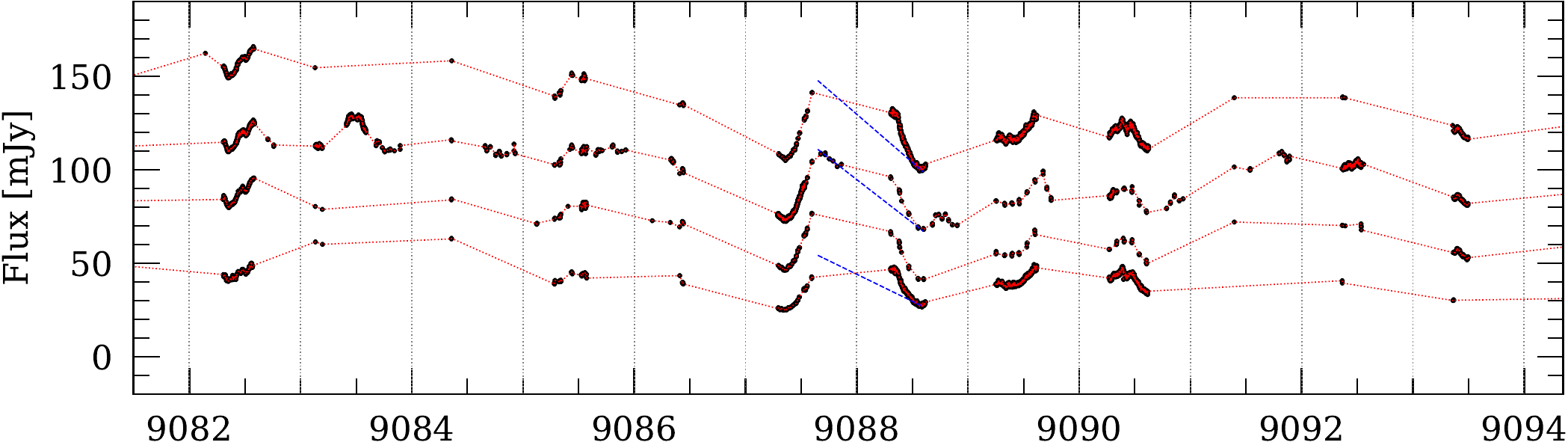}\\
\vspace{.25cm}
\includegraphics[width=\linewidth,clip=true]{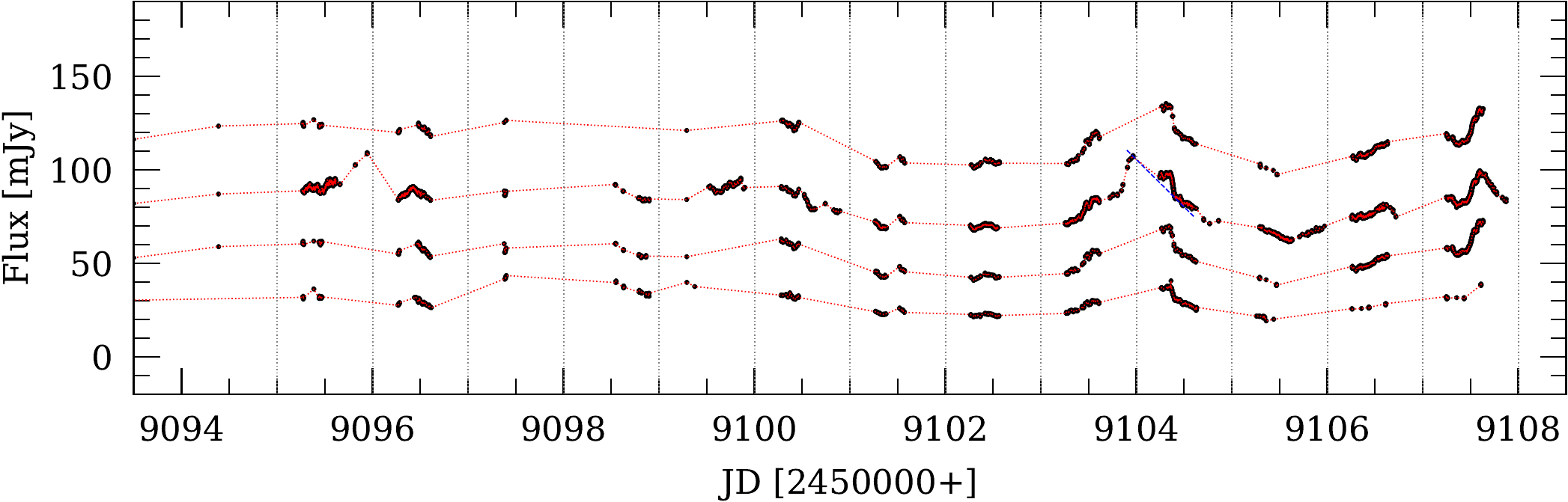} 
\caption{Light curves in $BVRI$ bands from Jul 11 to Sep 14, 2020. The LCs are ordered as indicated in the top panel; $RI$ band LCs are shifted by the corresponding offsets for display purposes. The blue dashed lines are the fits used to determine the shape of the smooth component for the corresponding nights~-- see Section~\ref{sec:res:inv} for the description of the $BRI$ band LCs around JD = 2459088 and of the $R$ band LC around JD = 2459104.}
\label{fig:stv:lc}
\end{figure*}
 
\begin{figure}[t!]
\centering
\includegraphics[width=\linewidth,clip=true]{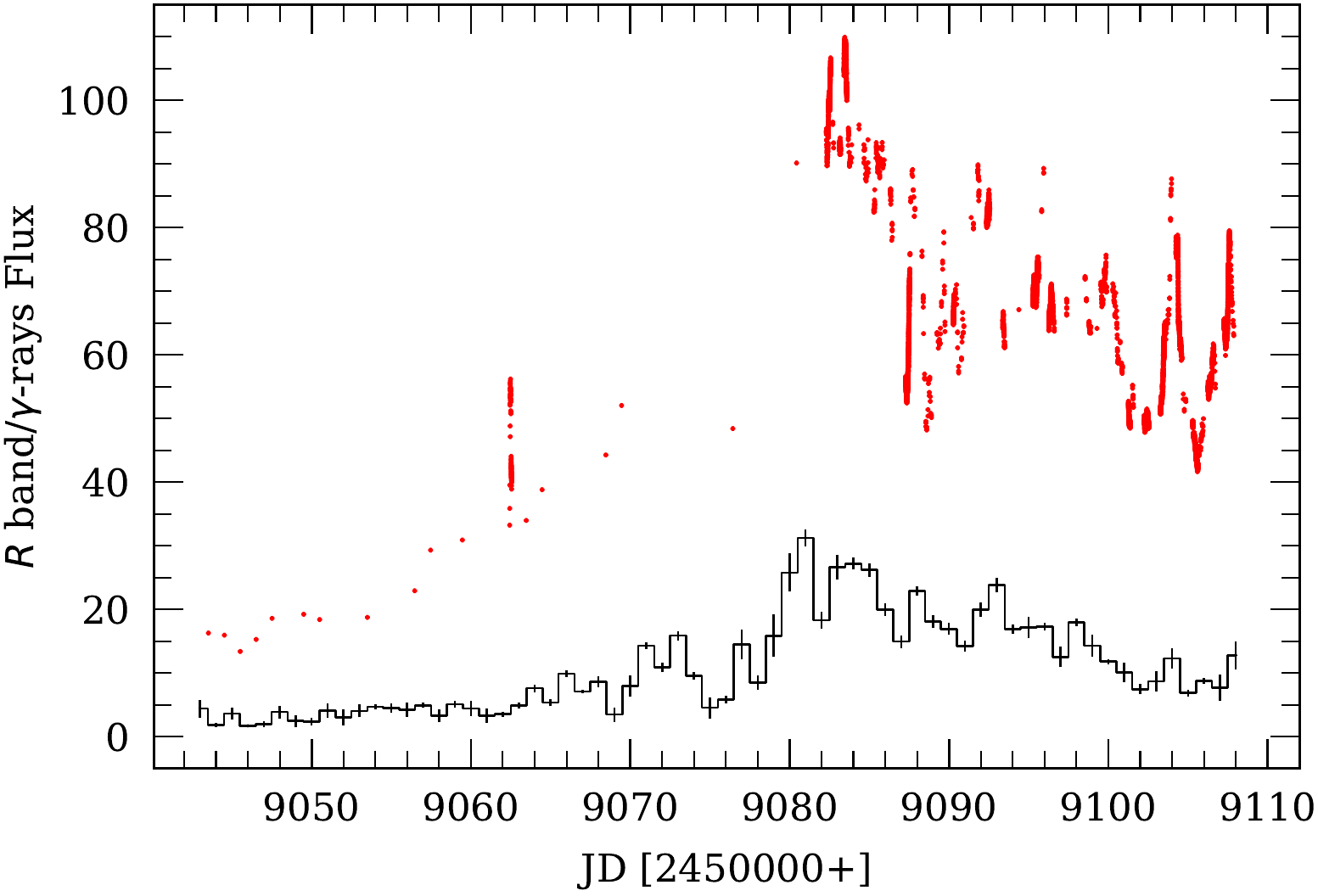} 
\caption{Optical ($R$ band, red circles) and the ``quick-look'' $\gamma$-rays (0.1--300\,GeV band, black stepped curve) LCs from Jul 11 to Sep 14, 2020. The $R$ band flux is in units of mJy, while the $\gamma$-rays flux is in units of $10^{-7}\,\rm photons~s^{-1}\,cm^{-2}$.}
\label{fig:stv:lc:opt-g}
\end{figure}

\begin{figure}[t!]
\centering
\includegraphics[width=\linewidth,clip=true]{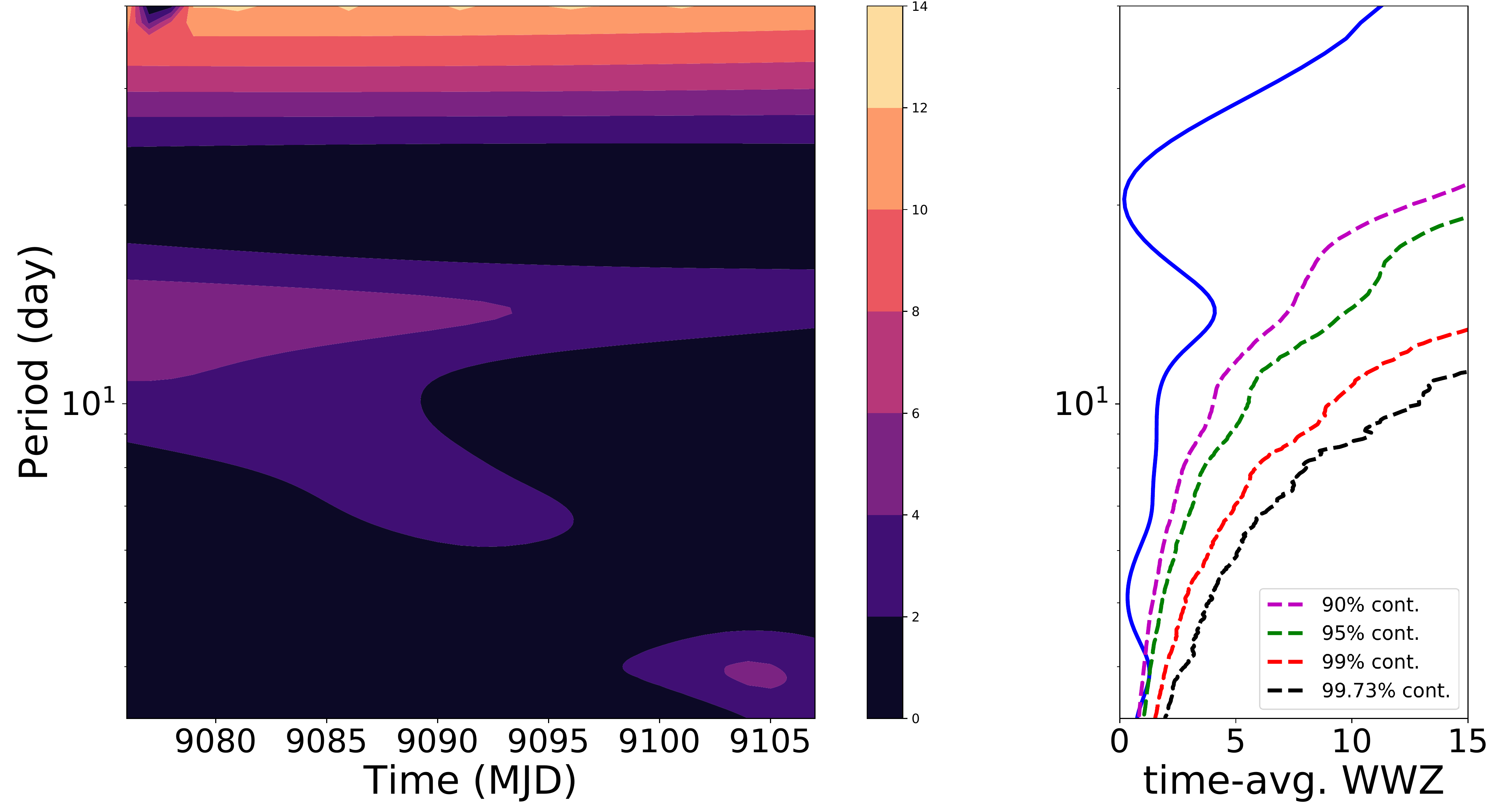} 
\caption{Weighted wavelet $Z$-transform of the nightly binned and cut $R$ band LC (see text). {\em Left panel}: the colored WWZ power in the time-period plane. {\em Right panel}: the time-averaged WWZ power as a function of the period. The colored dashed curves represent the corresponding local significance contours.}
\label{fig:wwz}
\end{figure}

\section{Results}
\label{sec:res}

\subsection{Short-term Variability}
\label{sec:res:stv}

The LCs from Jul 11 to Sep 14, 2020 (built as described in Section~\ref{sec:comb}) are shown in Figure~\ref{fig:stv:lc}. In Figure~\ref{fig:stv:lc:opt-g}, we show the $R$ band LC along with the $\gamma$-rays\footnote{The $\gamma$-rays LC is derived at the Large Area Telescope Instrument Science Operations center in a ``quick-look'' analysis. These preliminary flux estimates should be used with caution, so we shall use them only for illustrative purposes.} LC in the 0.1--300\,GeV band for inter-band comparison. The comparison reveals a good correlation between the optical and $\gamma$-rays LCs. In general, the LCs could be split visually into two parts~-- a pre-flare and a flare \citep[see also][]{2023MNRAS.519.3798S}.

The pre-flare LCs (untill the end of July 2020 = JD 2459062, the top panel of Figure~\ref{fig:stv:lc}) are characterized by a smooth and gradual flux increase. Since the beginning of August 2020, the flux increase has continued, but it is not as smooth as in July.
During the pre-flare period, we recorded the minimal $R$ band flux of 13.37\,mJy (or calibrated magnitude of 14.0545 $\pm$ 0.0016, telescope J) at JD 2459045.52063. The pre-flare is followed by a period of flaring activity, namely the August 2020 flare, which starts in the first decade of August and continues beyond the end of the time interval considered in this paper. 
The maximal $R$ band flux of 109.88\,mJy (or calibrated magnitude of 11.8190 $\pm$ 0.0033, telescope A) for the monitoring period was reached at JD 2459083.45823~-- that is, soon after the August 2020 flare onset.
Unfortunately, the period between the flare onset and the flare peak is very sparsely covered by data points, so we cannot study the shape of the rising part of the August 2020 flare. According to the preliminary $\gamma$-rays LC, it seems that the flux rise is steeper than the flux decay; that is, there is an asymmetry.
We also have no information about the optical intra-night activity of \bl\ at that period~-- we have detected only a non-well-sampled flare on Jul 31.
We cannot, however, rule out the presence of other flares during the rising phase of the August 2020 flare because of the sparse sampling and the lack of intra-night monitoring sessions. On the other hand, the decaying phase of the August 2020 flare shows the high activity of \bl\ on intra-night timescales. That activity will be our focus from now on: 
in what follows, we shall not consider the pre-flare, and all analysis will be related to the August 2020 flare.

\subsubsection{Searching for Periodicity}
\label{sec:res:stv:period}

To search for periodicity in the STLCs of \bl, we used the Lomb-Scargle periodogram \citep[][]{1976Ap&SS..39..447L,1982ApJ...263..835S} and weighted wavelet $Z$-transform  \citep[WWZ,][]{1996AJ....112.1709F} techniques.
Before the periodicity search, we performed nightly binning of our data following the approach of \citet{2021A&A...645A.137A}~-- in this way, we removed the influence of the different number of data points per night on the search results. We also cut out the weakly variable part of the LCs (namely before JD~= 2459075).
Given our data, we found no signs of periodicity for all bands using both techniques (Figure~\ref{fig:wwz}). Recently, \citet{2022Natur.609..265J} reported a detection of a transient periodicity of 0.55 days in the $R$ band LC generated by the Whole Earth Blazar Telescope; their WWZ time interval encompasses ours.

\begin{figure}[t!]
\centering
\includegraphics[width=\linewidth,clip=true]{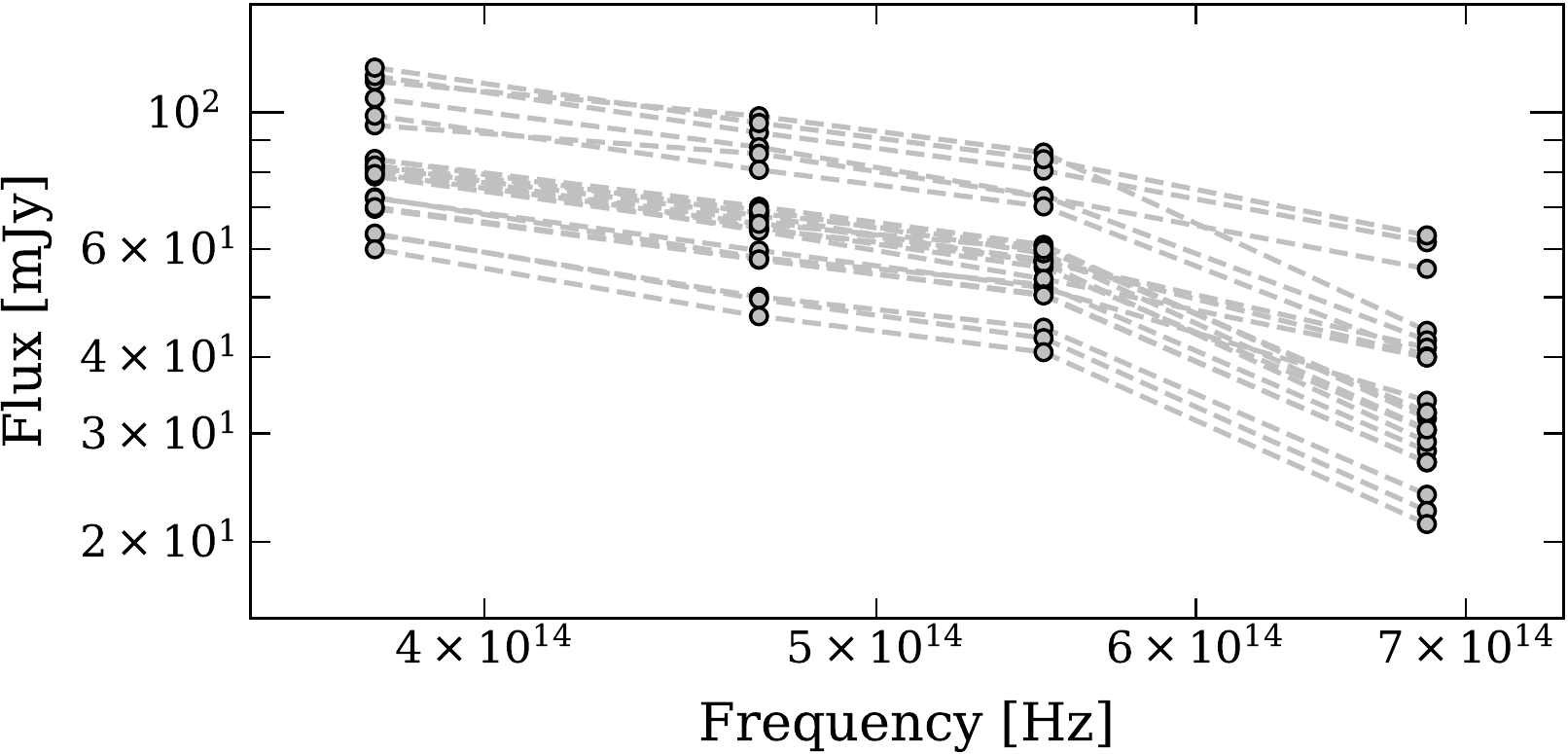} 
\caption{Spectral energy distribution for the individual nights. Note the scatter in the $B$ band fluxes (see text).}
\label{fig:sed}
\end{figure}

\begin{figure}[t!]
\centering
\includegraphics[width=\linewidth,clip=true]{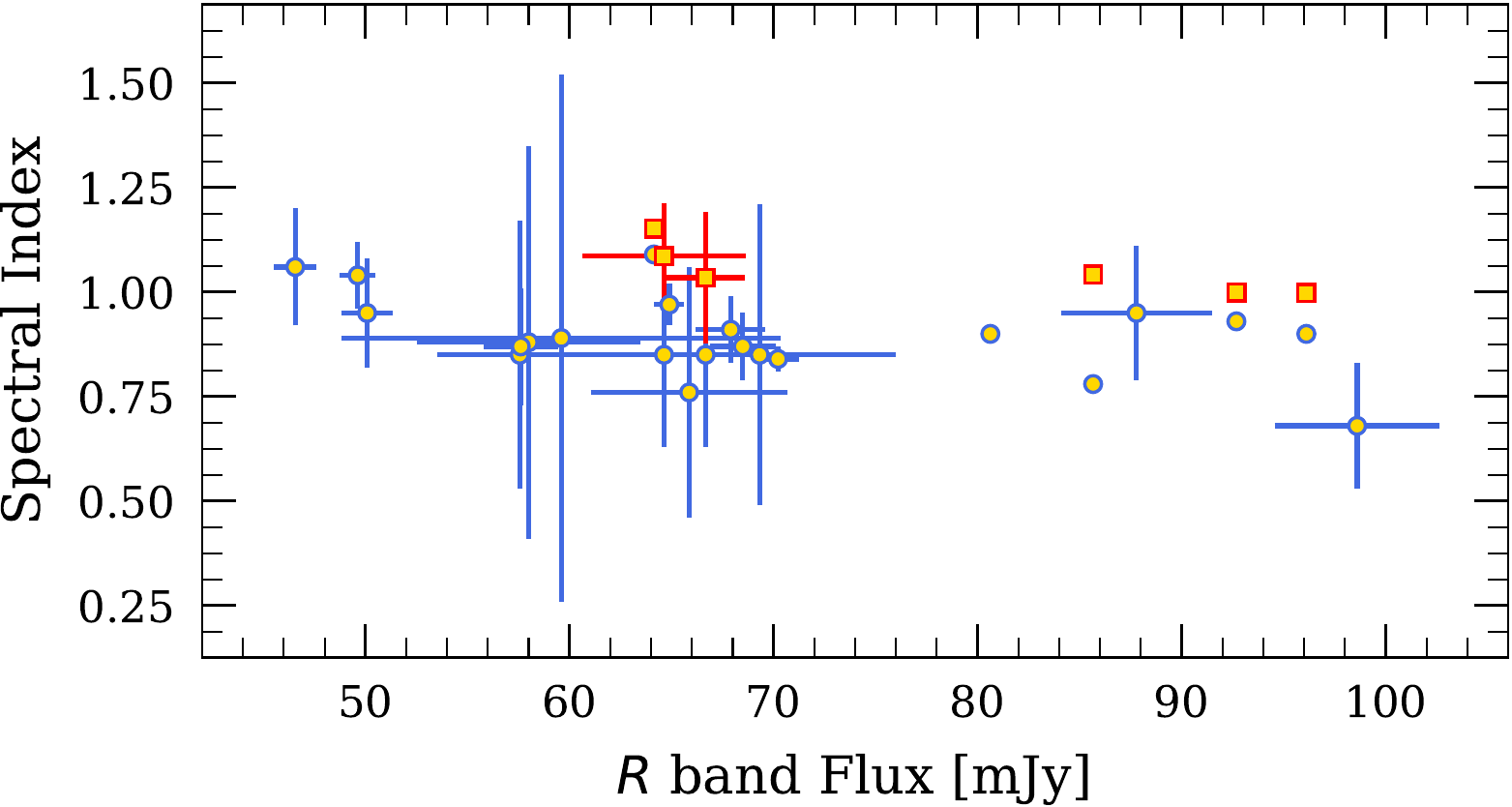} 
\caption{Dependence of the spectral index on the $R$ band flux. The blue circles are the spectral indices calculated using the $VRI$ bands, while the red squares are the spectral indices calculated using the $BVRI$ ones. The error bars reflect the variability amplitude in the cases when the intra-night monitoring data are included in the spectral index calculation (see text).}
\label{fig:flux:alpha}
\end{figure}

\subsubsection{Spectral Energy Distribution}
\label{sec:sed:res}

For the nights of $BVRI$ observations, we built the SEDs as follows.
If a single measurement is available for the given night, then we use the corresponding flux directly.
If repeating observations were performed during the given night, then we calculated the weighted mean fluxes for the corresponding bands. The averaging was done over the same time interval for the corresponding bands to avoid the influence of the different duration of the INLCs on the mean value obtained. 
This time interval was taken to be the duration of the shortest LC for the given night. The uncertainty of the mean flux was taken to be the larger between (i) the weighted standard deviation and (ii) the standard uncertainty of the weighted mean. 
The effective wavelengths for the $BVRI$ bands were taken from \citet{1998A&A...333..231B}. The so-derived SEDs have been plotted in Figure~\ref{fig:sed} for all nights jointly.

To estimate the spectral index, we fitted a linear polynomial of the form $\log(F_{\nu}) = -\alpha \log(\nu) + \rm const$ to each SED. We used only $VRI$ bands in the fitting because of the large scatter of the $B$ band fluxes: for most of the nights the $B$ band flux is below the power-law model expectation. 
The similar behavior of the $B$ band measurements was discussed by \citet{2020ApJ...900..137W}. They attributed this behavior to the combination of the wide $B$ filter band and the spectral shape of \bl.

We show in Figure~\ref{fig:flux:alpha} the relation between the spectral index and the $R$ band flux. There is a hint of steepening of the spectral index as the flux decreases. However, the overall spectral index behavior of \bl\ on short-term timescales could be considered mildly chromatic~-- the dependence of $\alpha$ on the flux level is weak.
The median spectral index over the August 2020 flare was found to be $\langle \alpha_{VRI} \rangle_{\rm med}=0.885\pm0.020$ (a standard deviation of 0.096).

For six nights, we were able to calculate the spectral index using the $BVRI$ bands~-- for these nights, the $B$ band flux behaves not so unusually (see above). The corresponding median spectral index was calculated to be $\langle \alpha_{BVRI} \rangle_{\rm med}=1.038\pm0.025$ (a standard deviation of 0.061). In any case, the inclusion of the $B$ band leads to slightly steeper indices (Figure~\ref{fig:flux:alpha}).

\begin{figure*}[t!]
\gridline{\fig{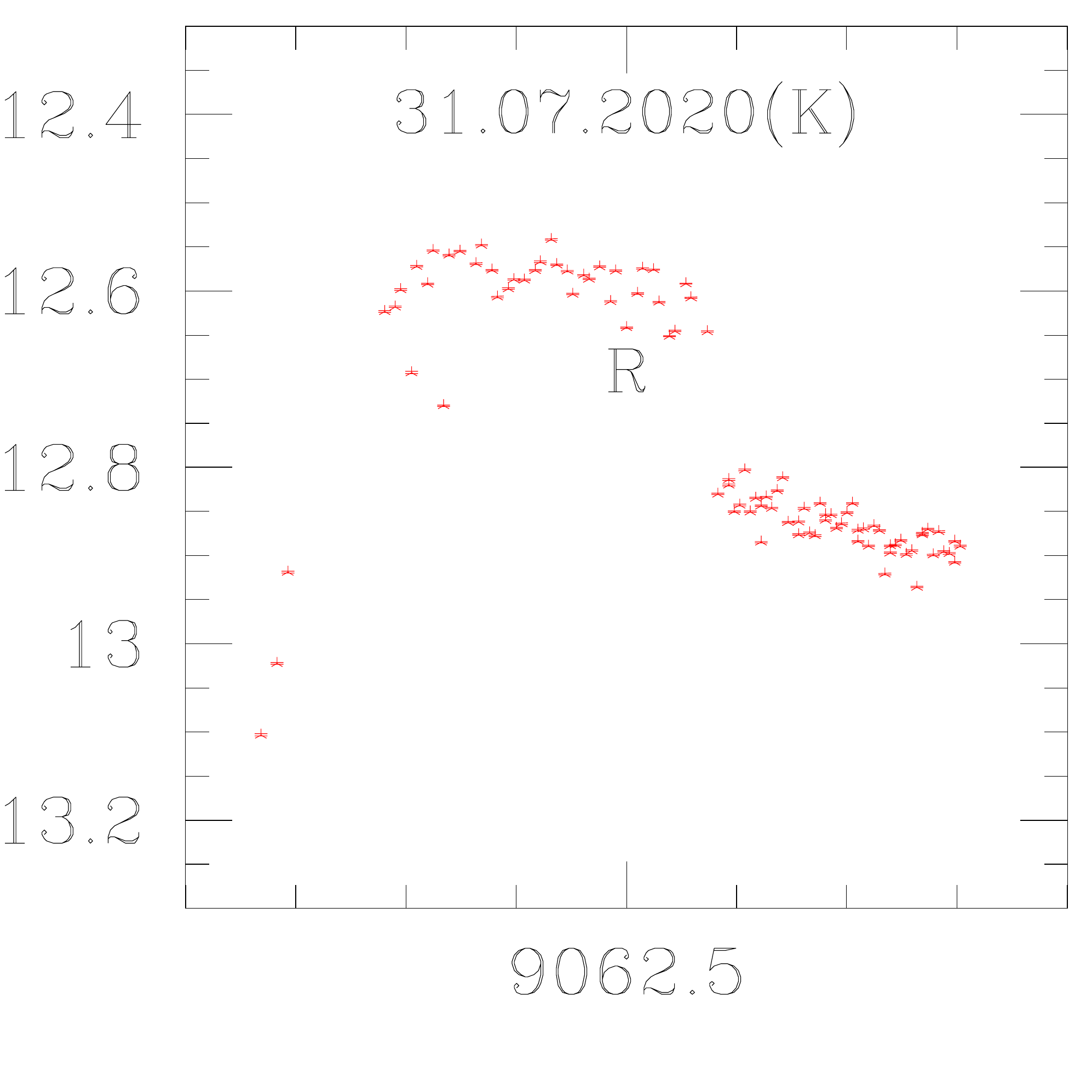}{0.25\textwidth}{}
          \fig{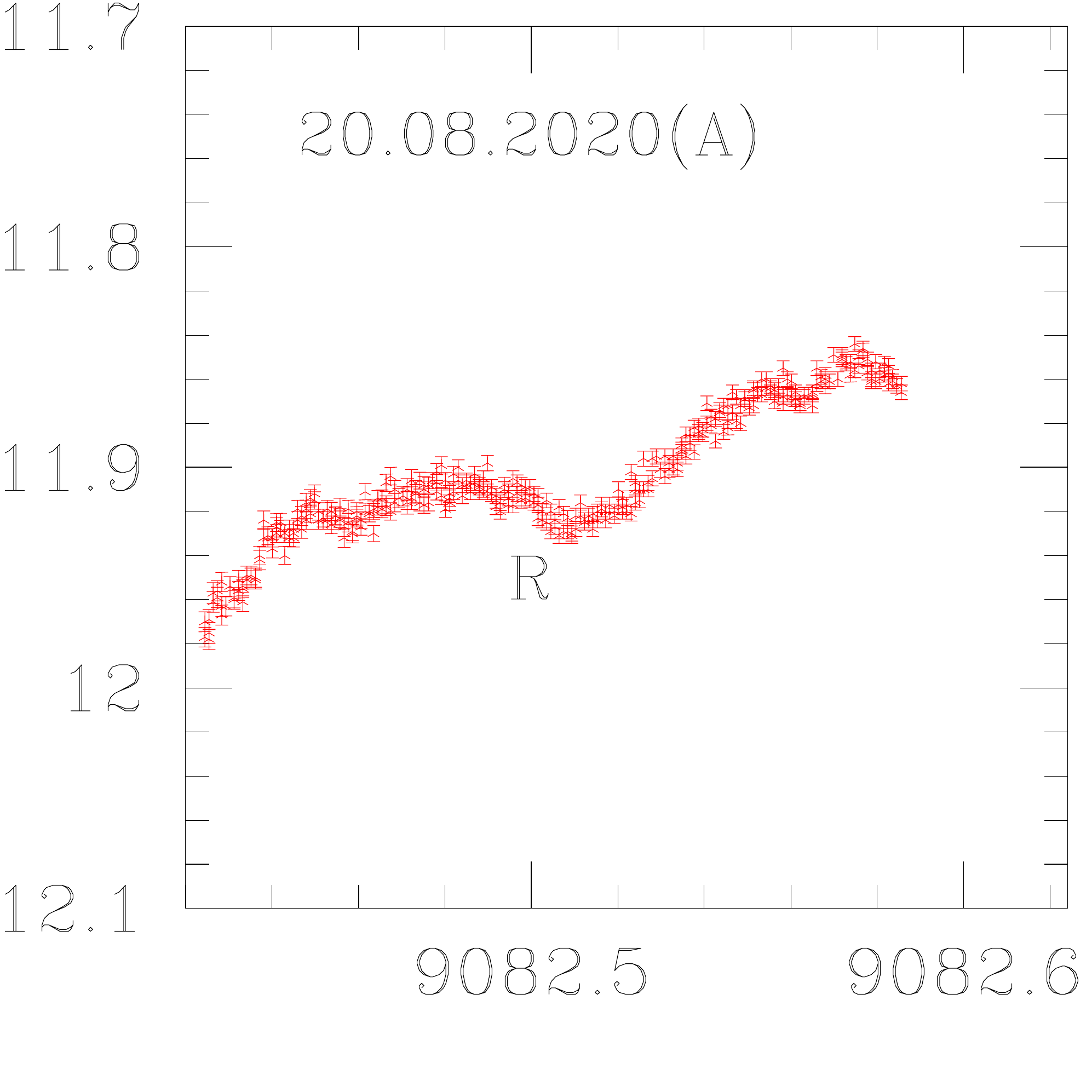}{0.25\textwidth}{}
          \fig{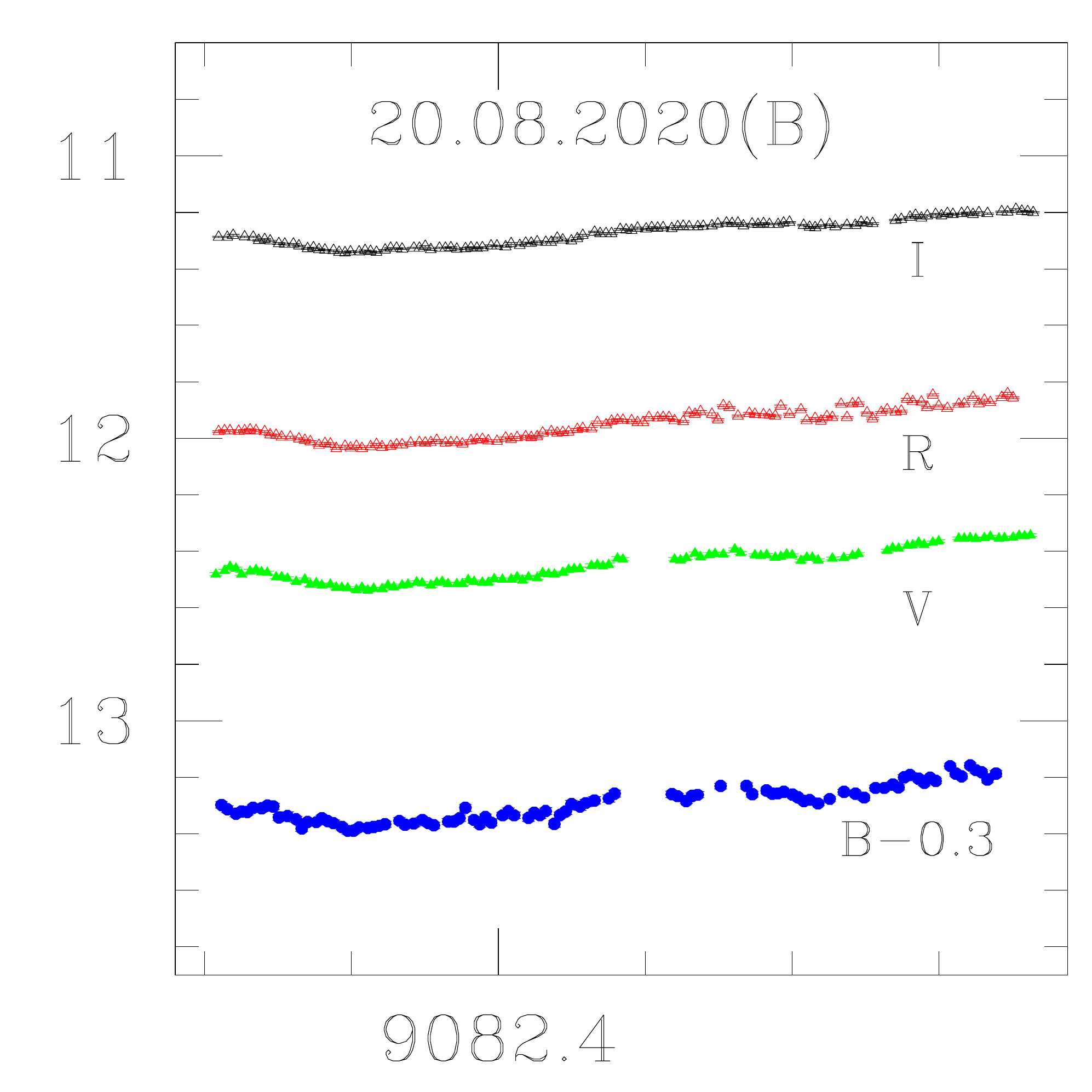}{0.25\textwidth}{}}
\gridline{\fig{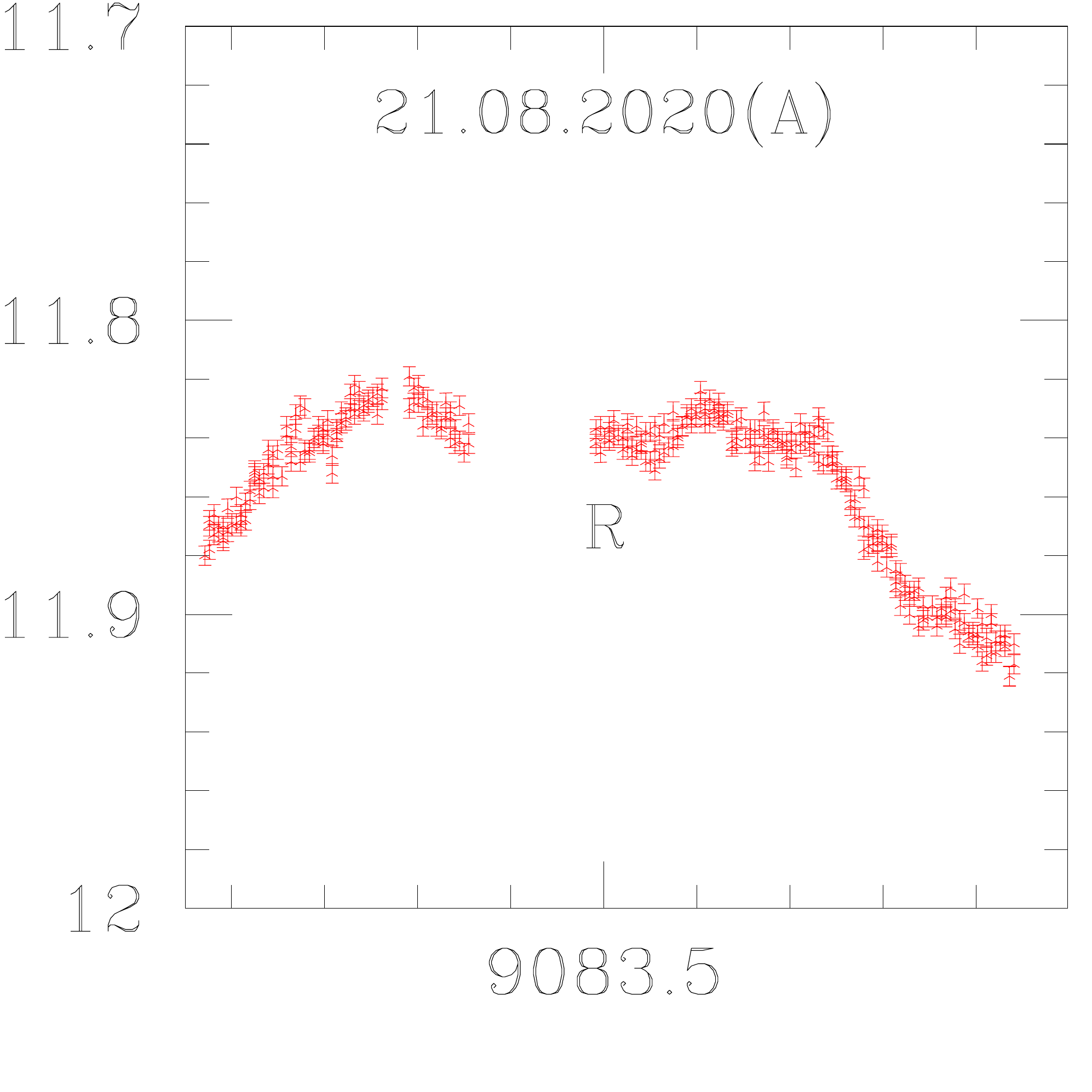}{0.25\textwidth}{}
          \fig{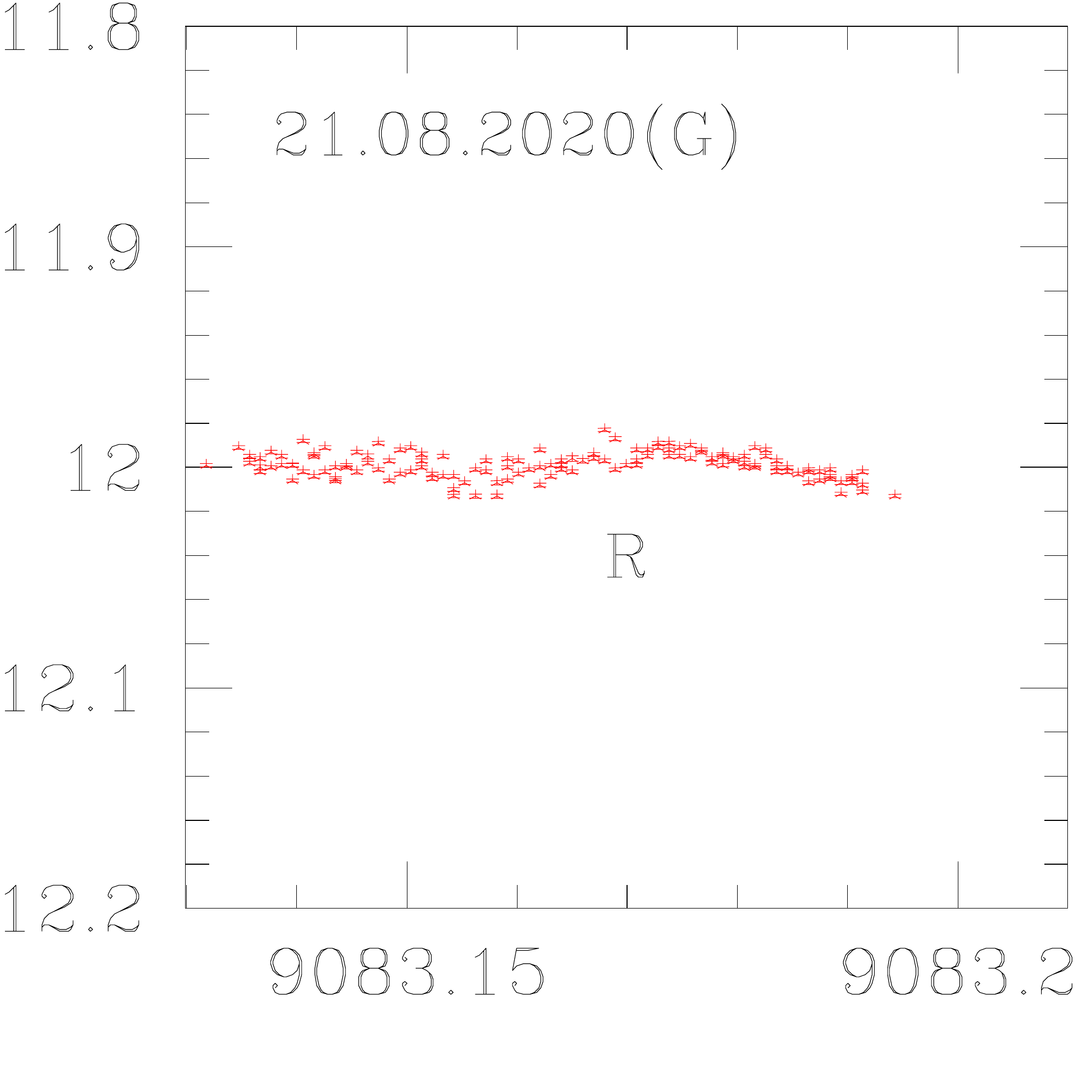}{0.25\textwidth}{}
          \fig{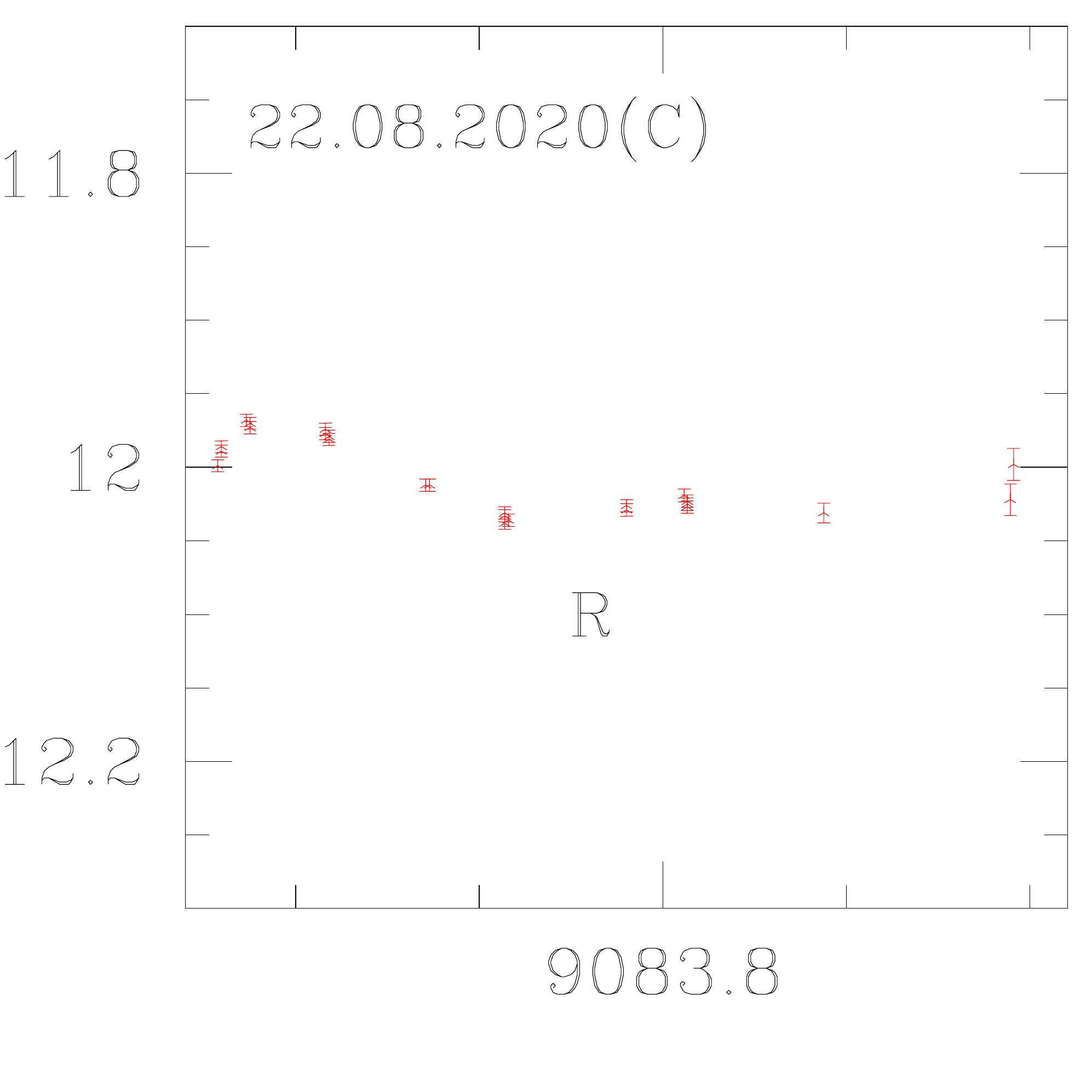}{0.25\textwidth}{}}
\gridline{\fig{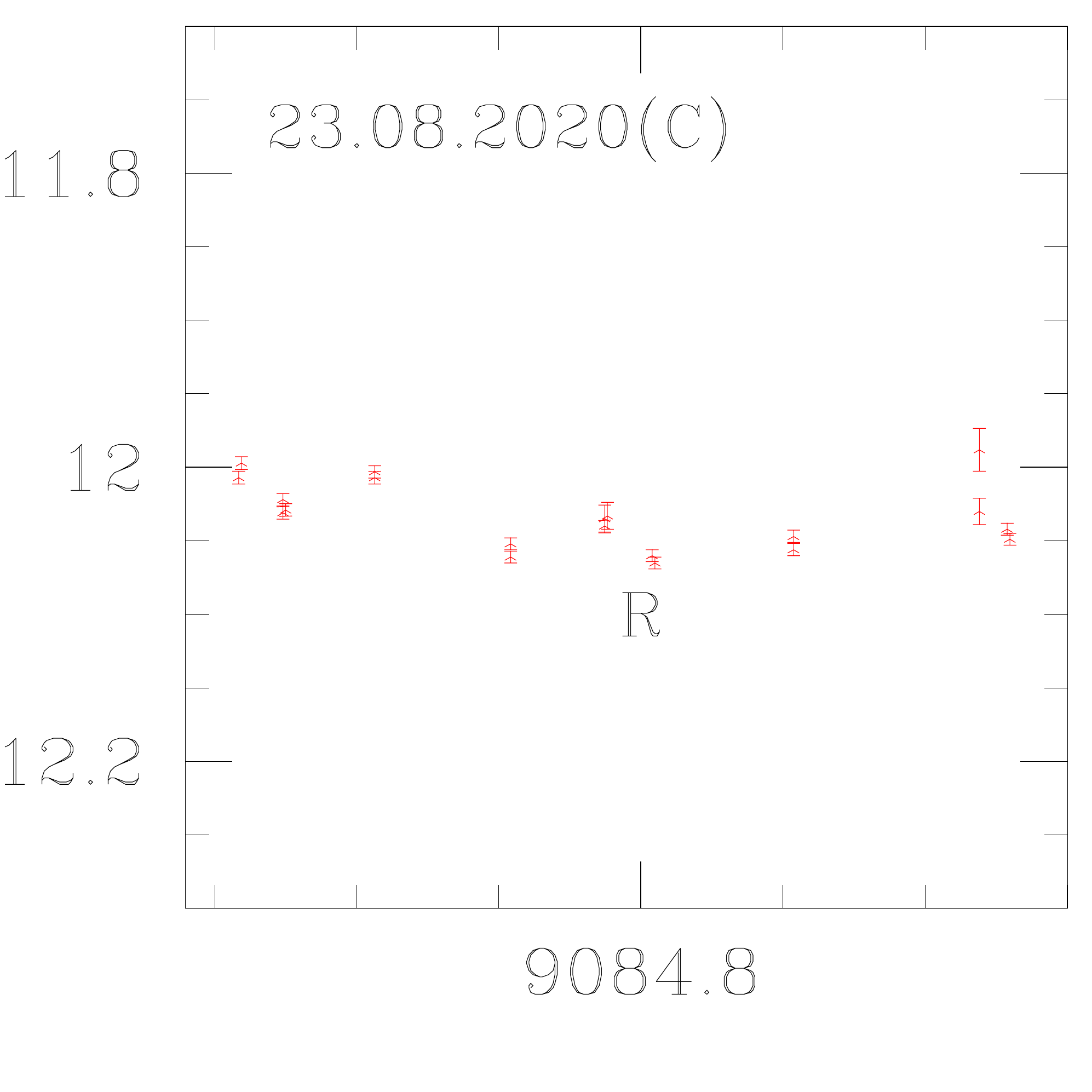}{0.25\textwidth}{}
          \fig{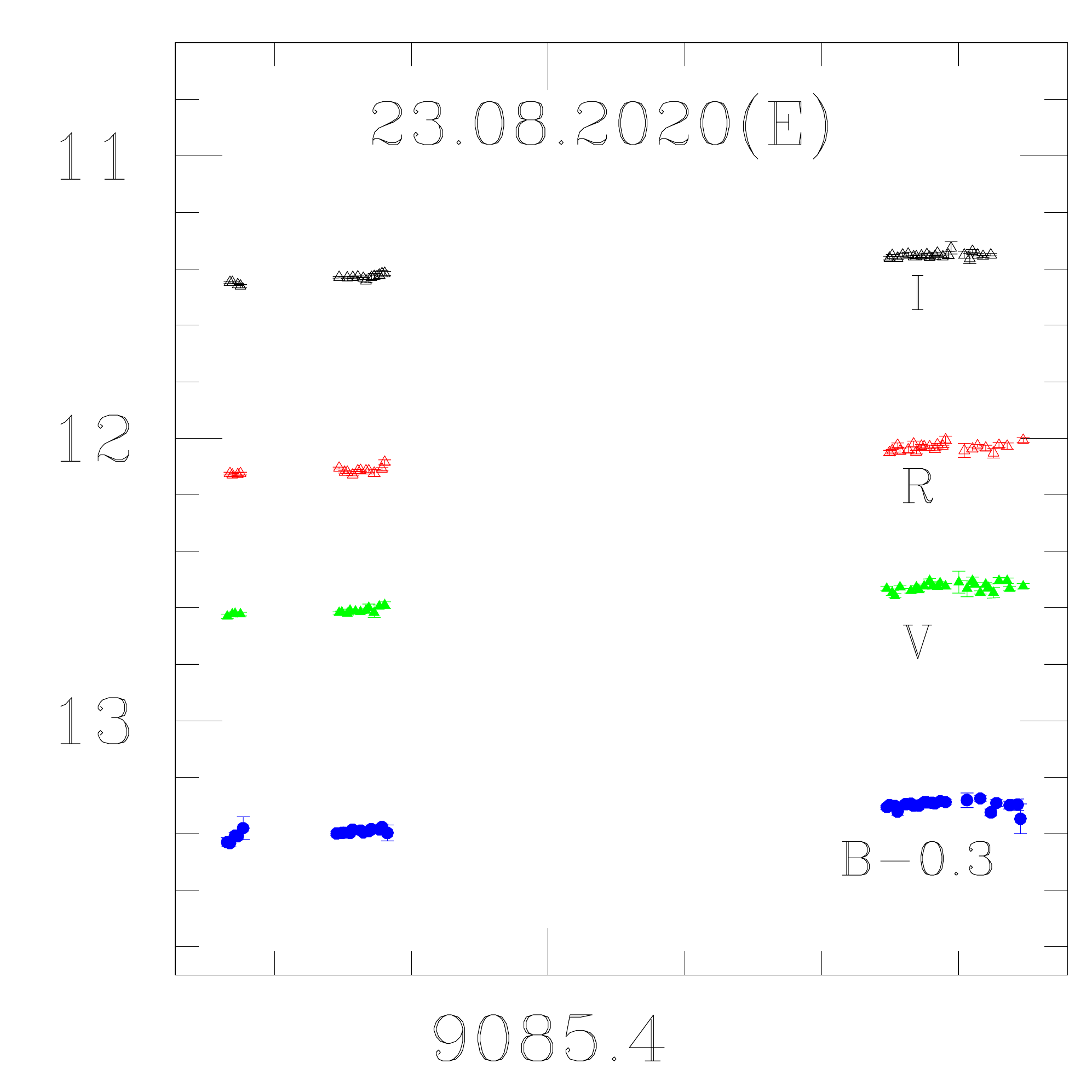}{0.25\textwidth}{}
          \fig{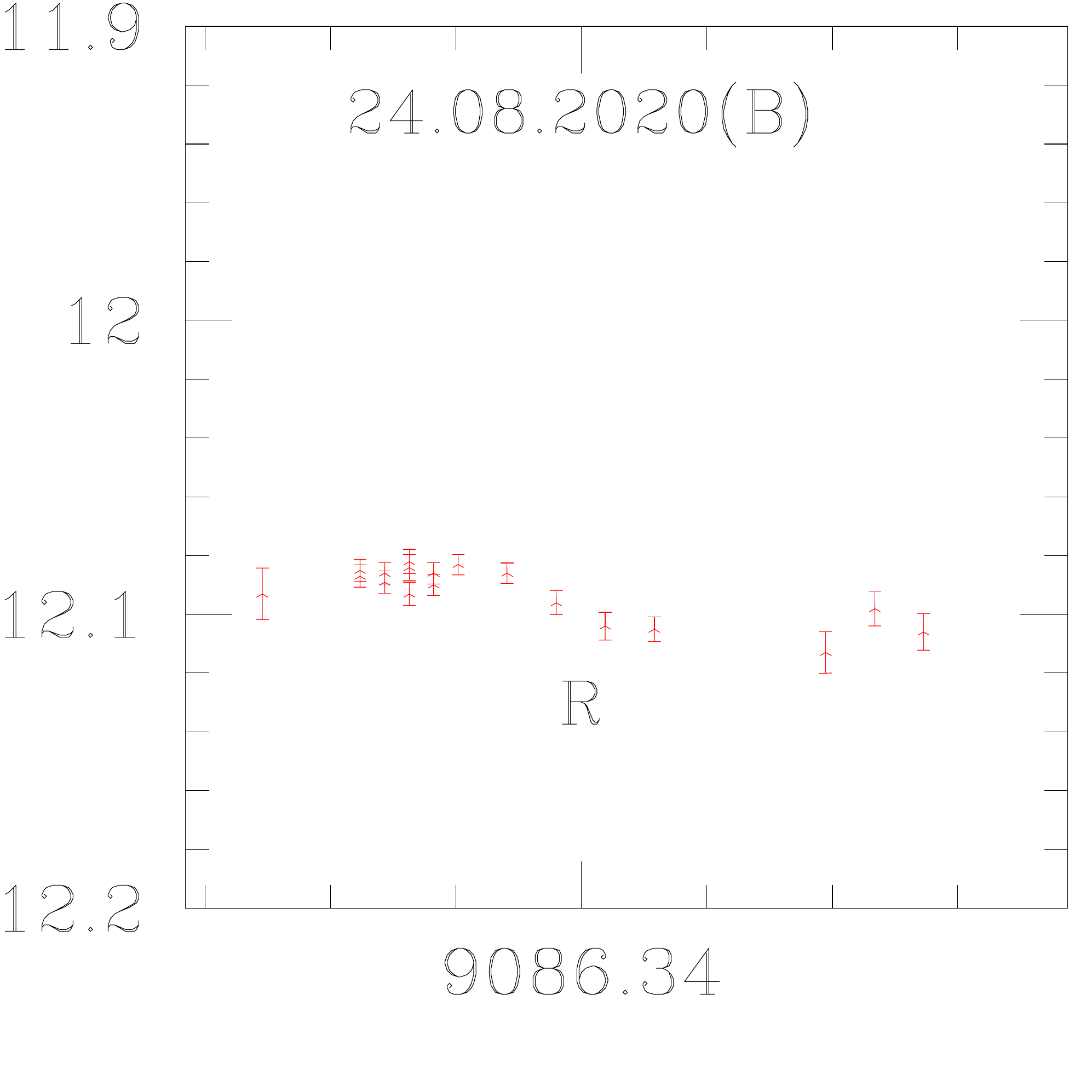}{0.25\textwidth}{}}
\gridline{\fig{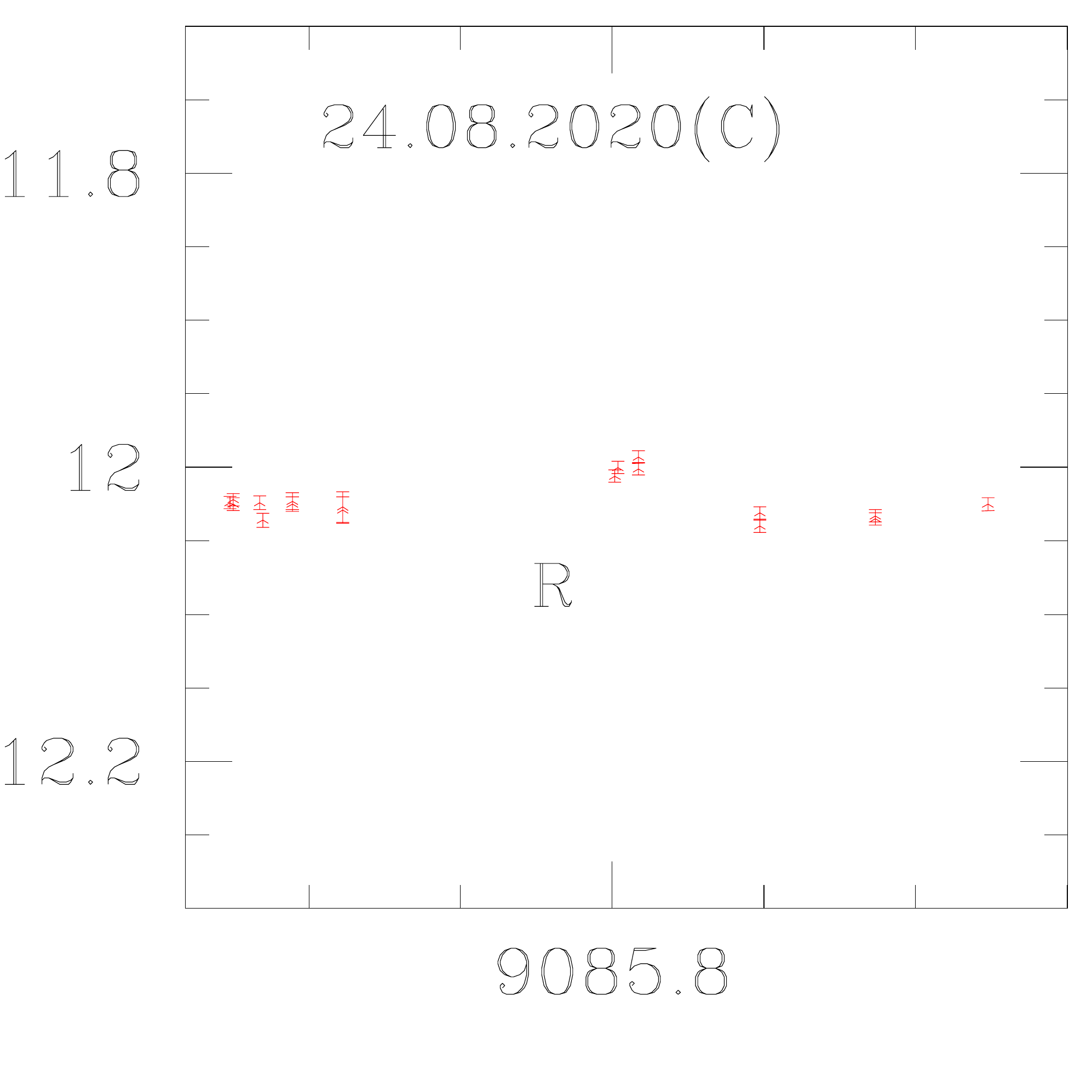}{0.25\textwidth}{}
          \fig{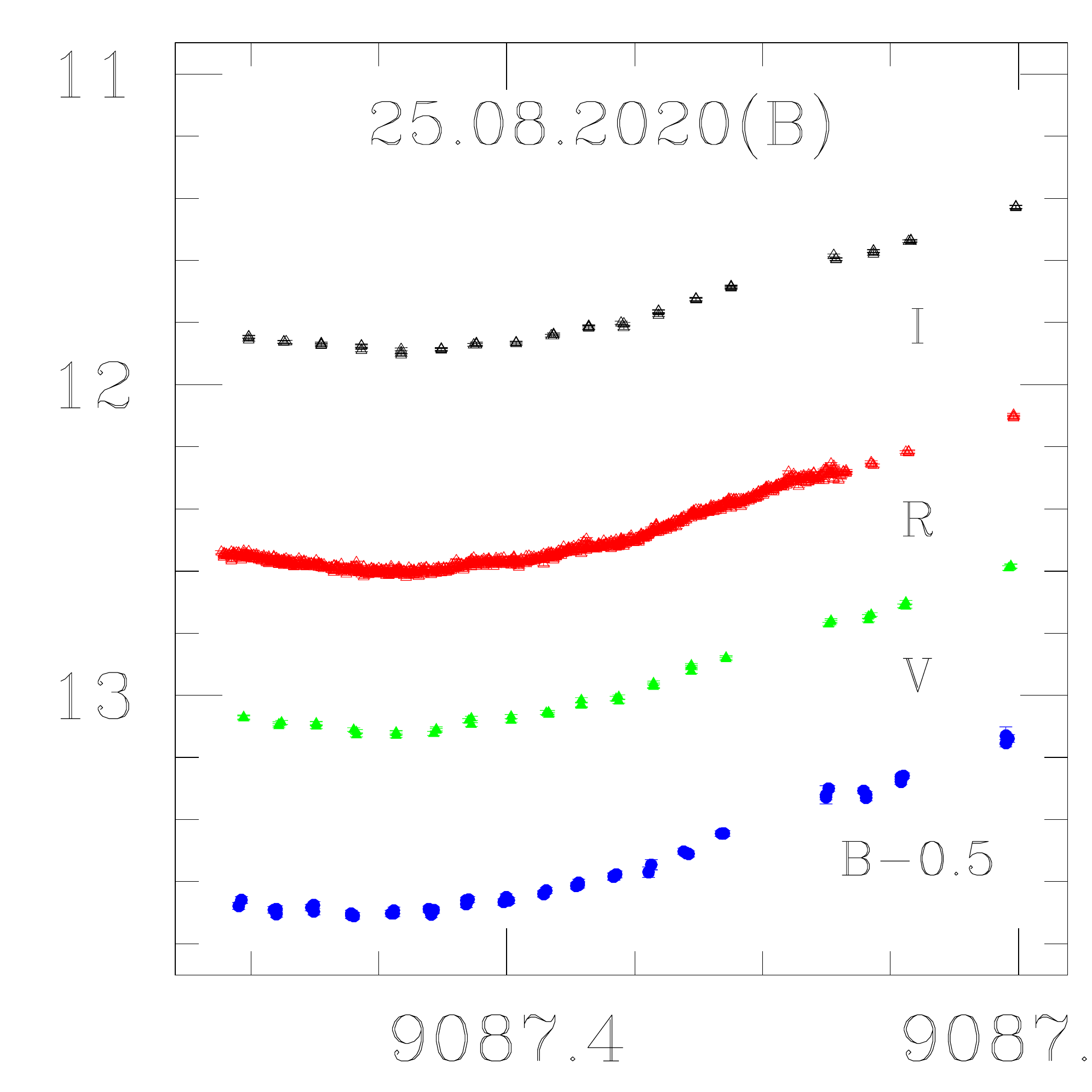}{0.25\textwidth}{}
          \fig{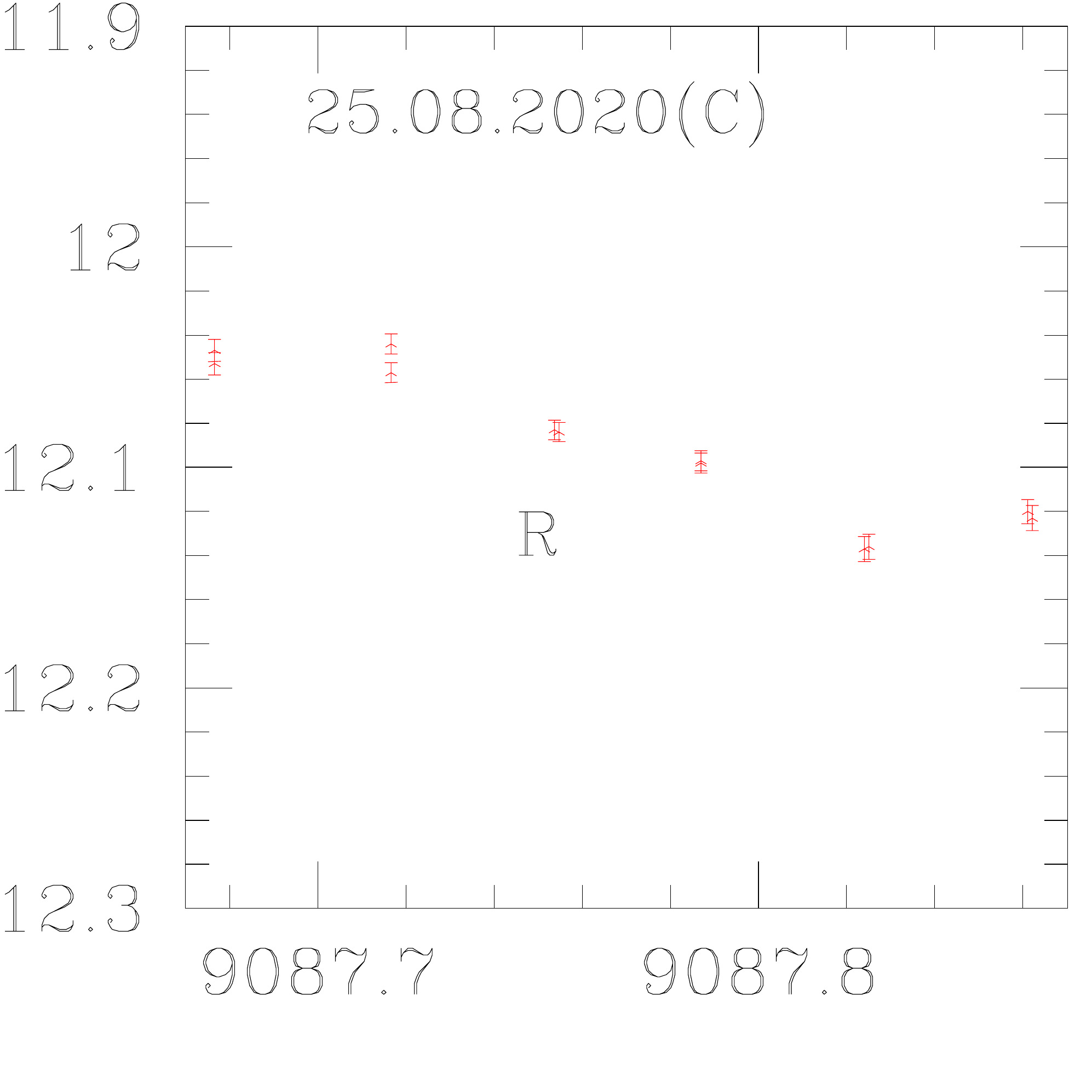}{0.25\textwidth}{}}
\caption{Intra-night LCs of \bl. The blue, green, red, and black colored data points code $BVRI$ bands, respectively; the $B$ band offsets are indicated. In each plot, the JDs are along the $x$-axis and the \bl\ brightness in magnitudes is along the $y$-axis.
The observation date and the telescope used are indicated in each plot.}
\label{fig:inlc:mag}
\end{figure*}

\setcounter{figure}{5}
\begin{figure*}[t!]
\gridline{\fig{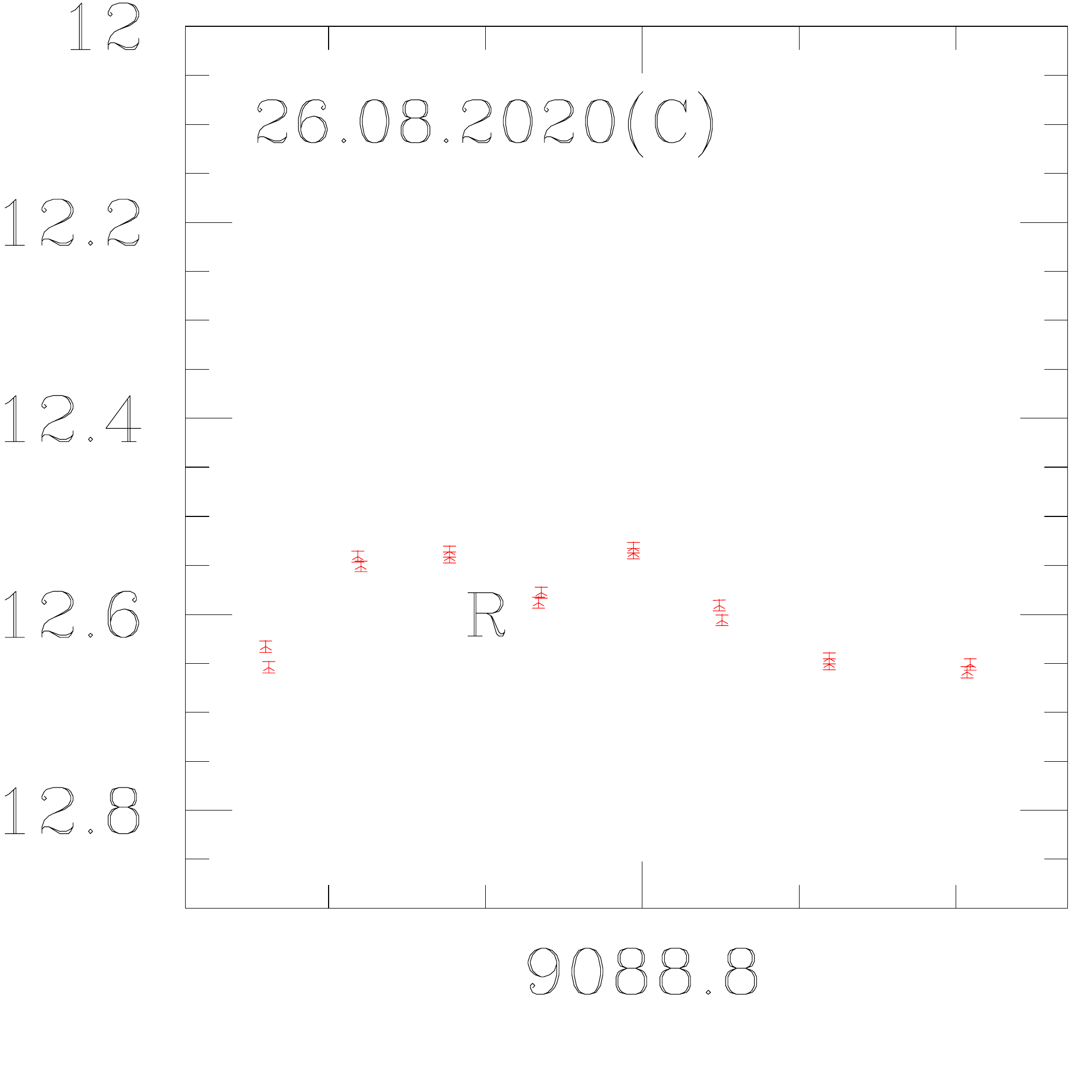}{0.25\textwidth}{}
          \fig{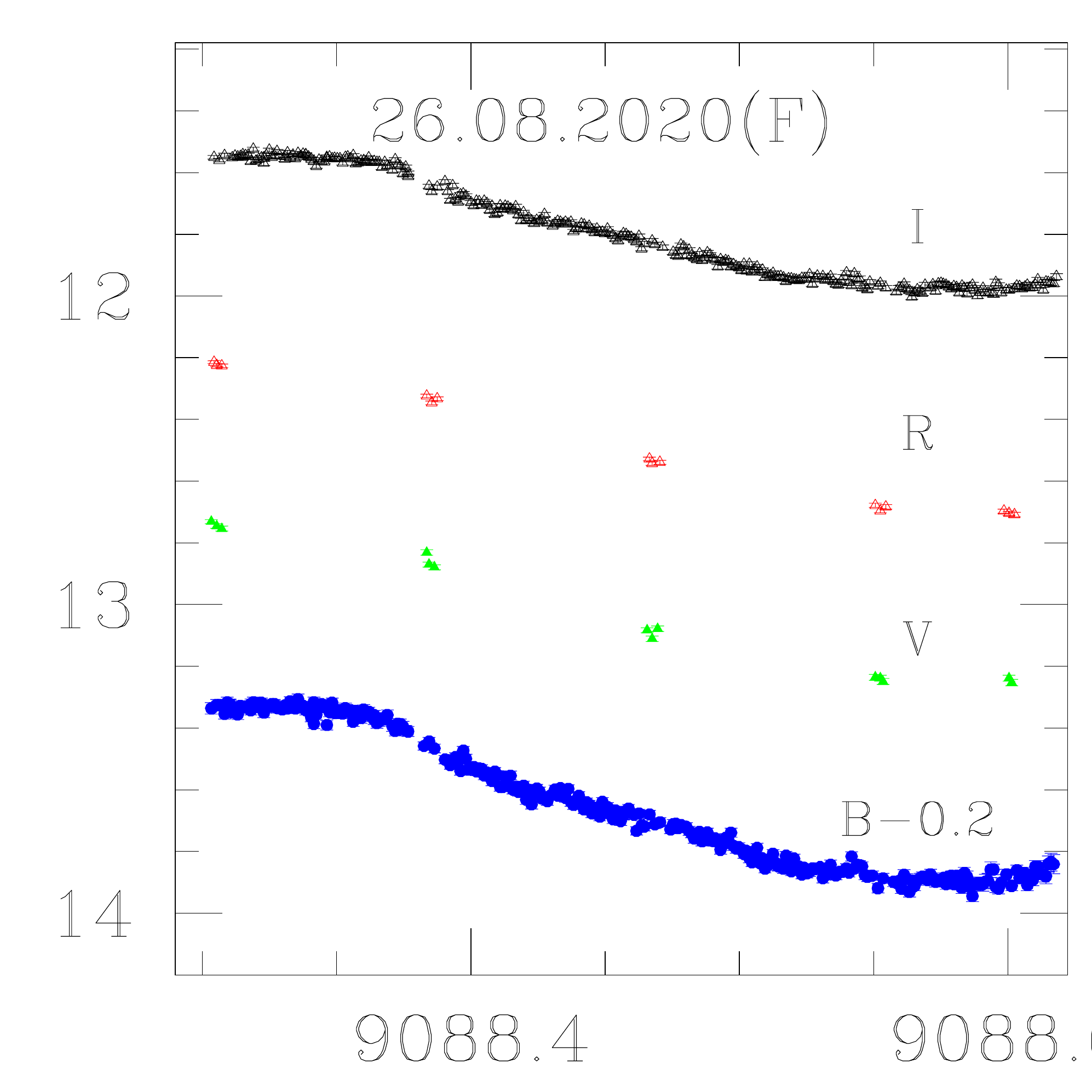}{0.25\textwidth}{}
          \fig{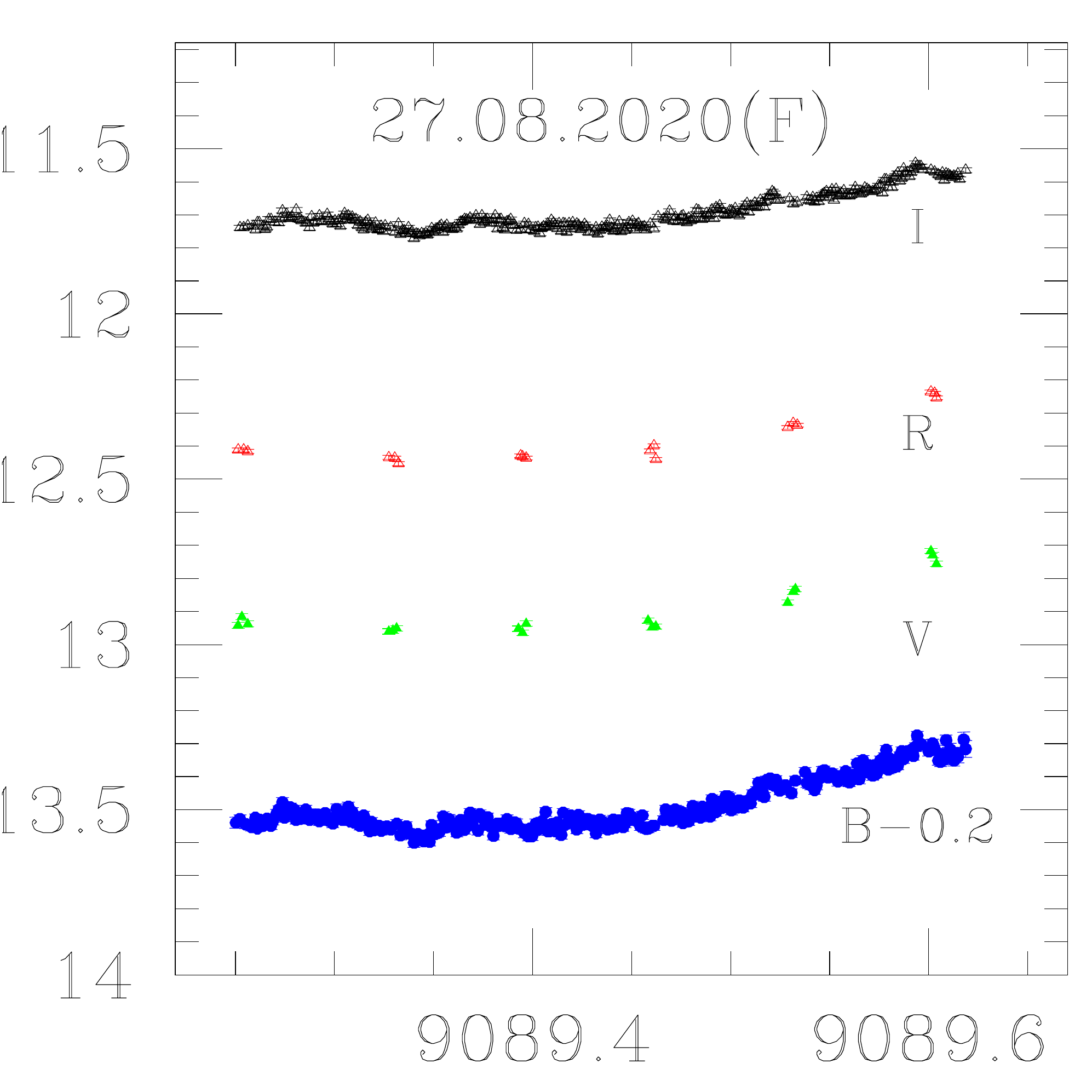}{0.25\textwidth}{}}
\gridline{\fig{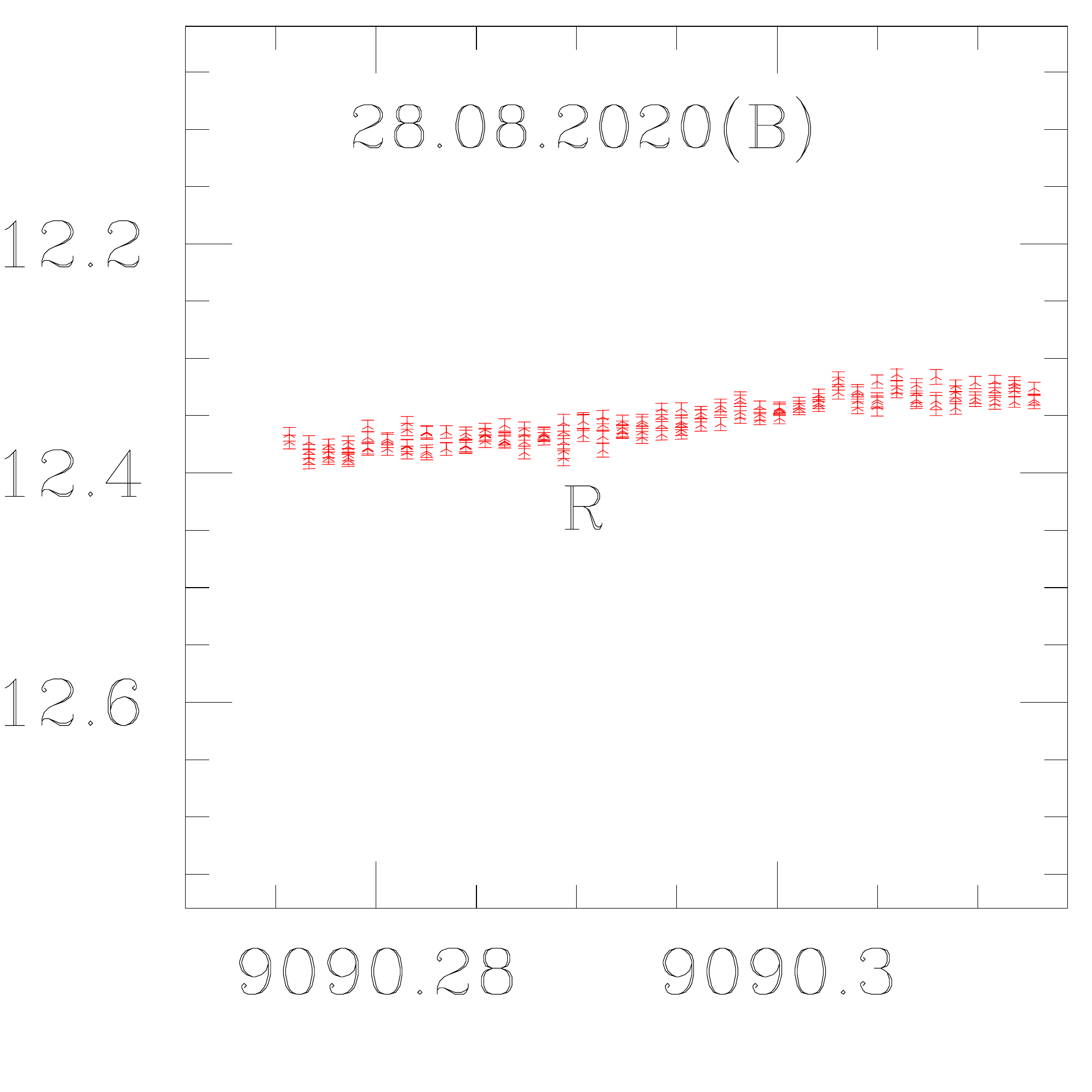}{0.25\textwidth}{}
          \fig{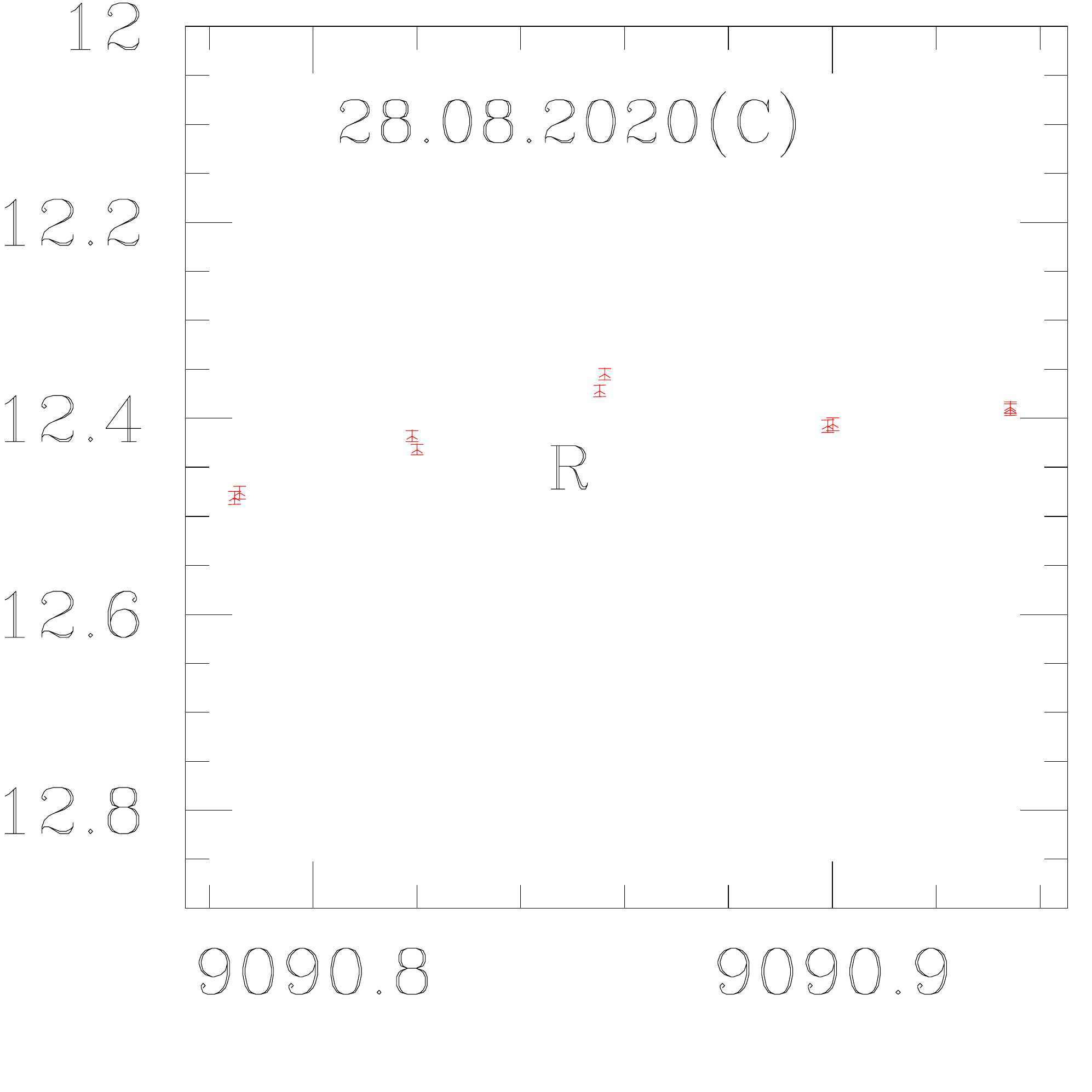}{0.25\textwidth}{}
          \fig{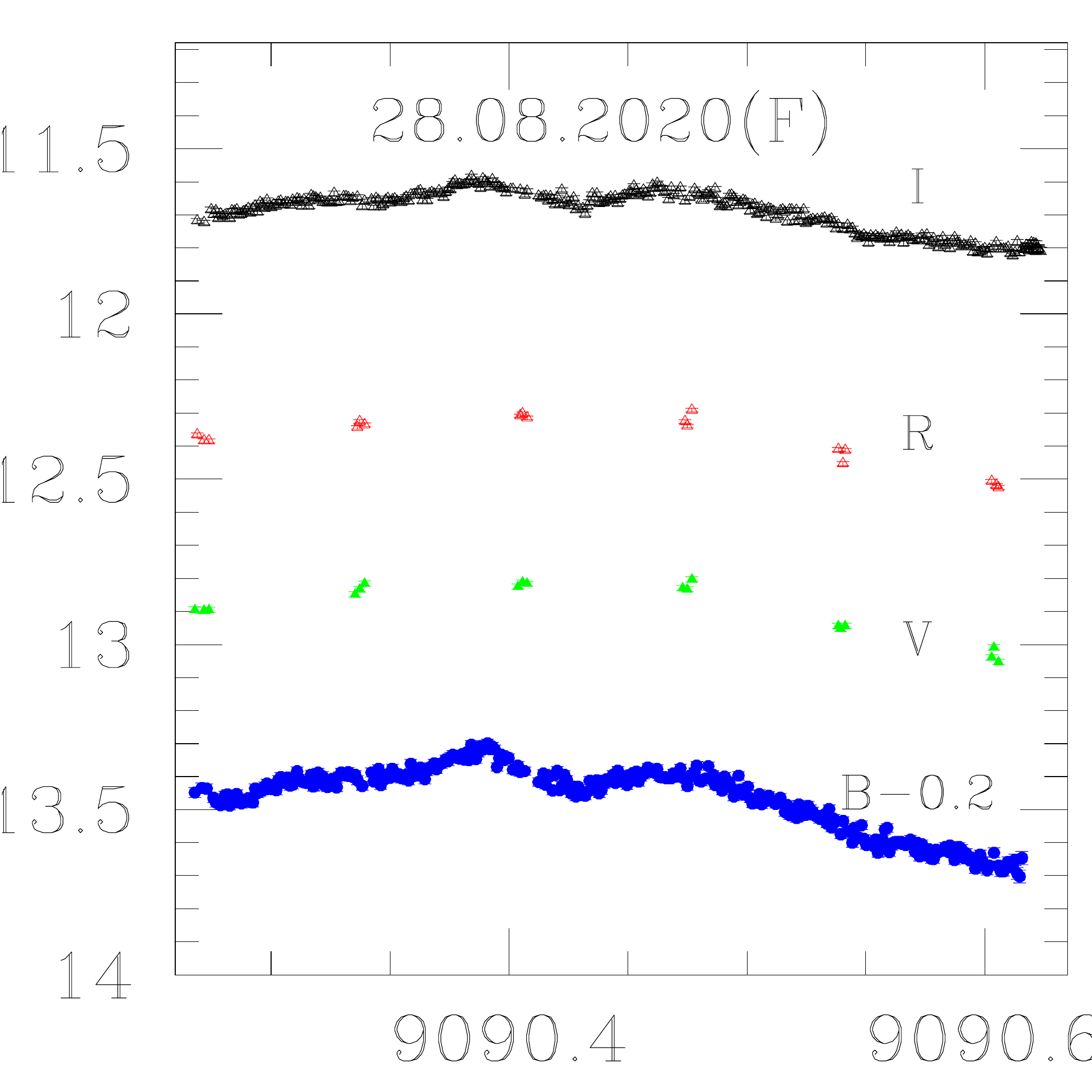}{0.25\textwidth}{}}
\gridline{\fig{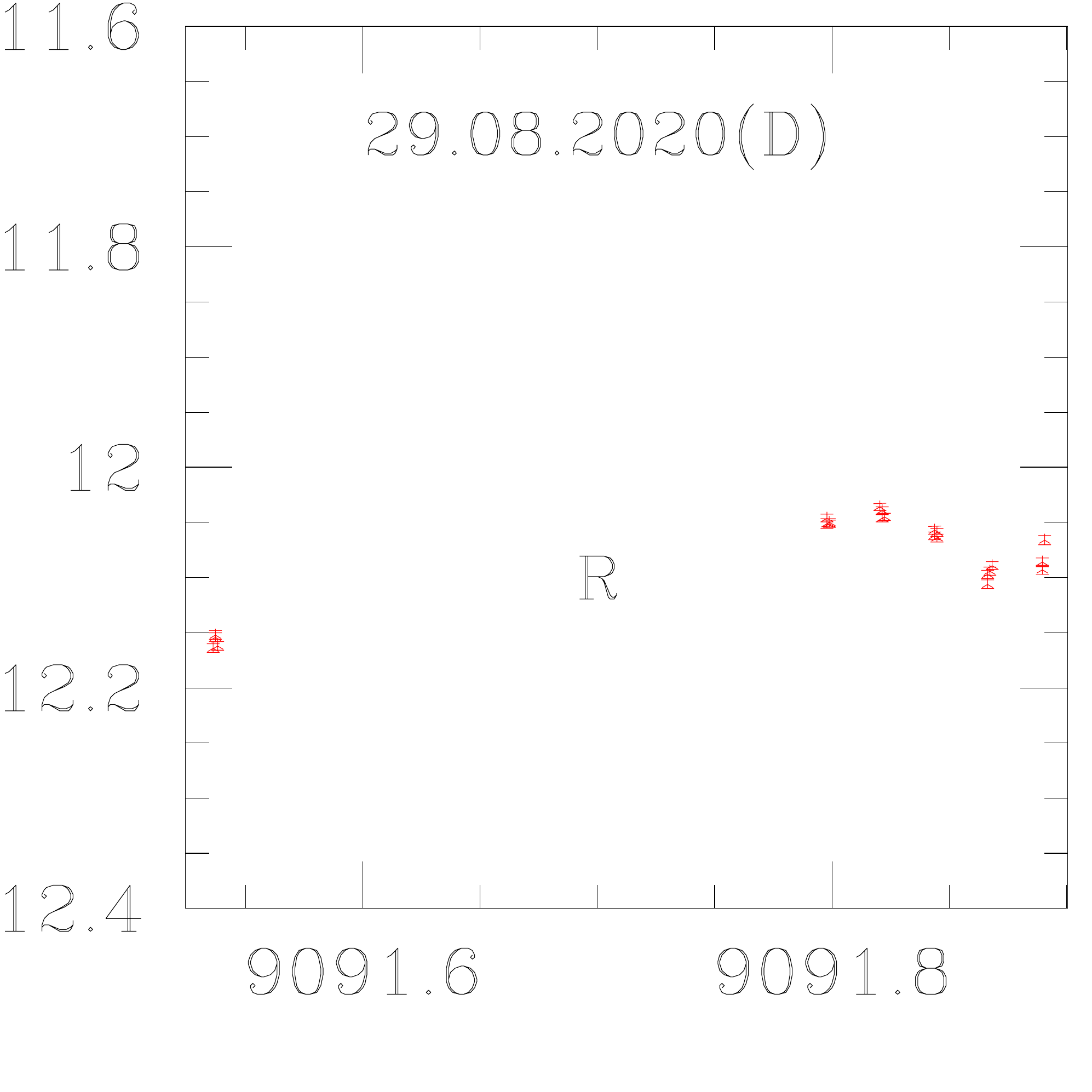}{0.25\textwidth}{}
          \fig{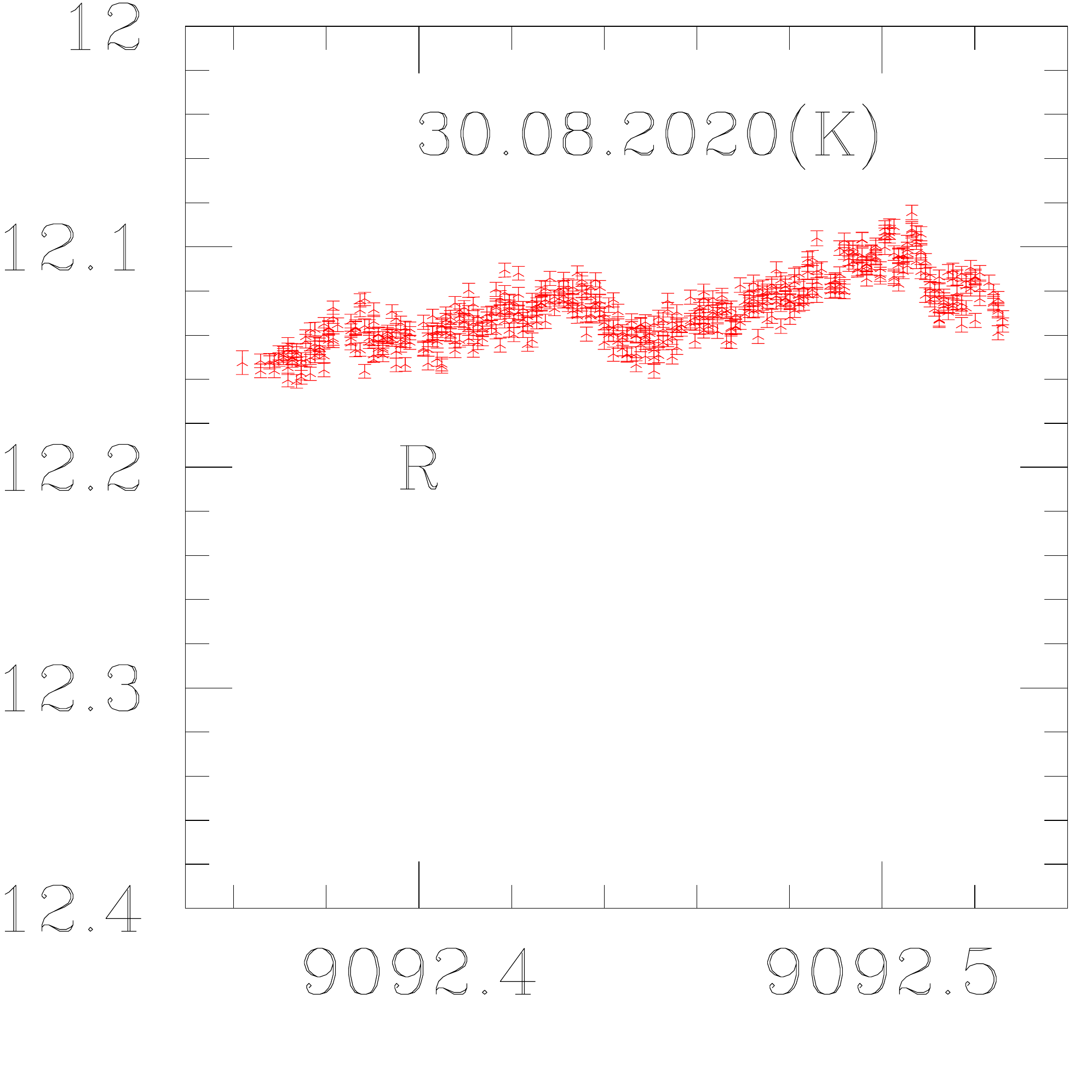}{0.25\textwidth}{}
          \fig{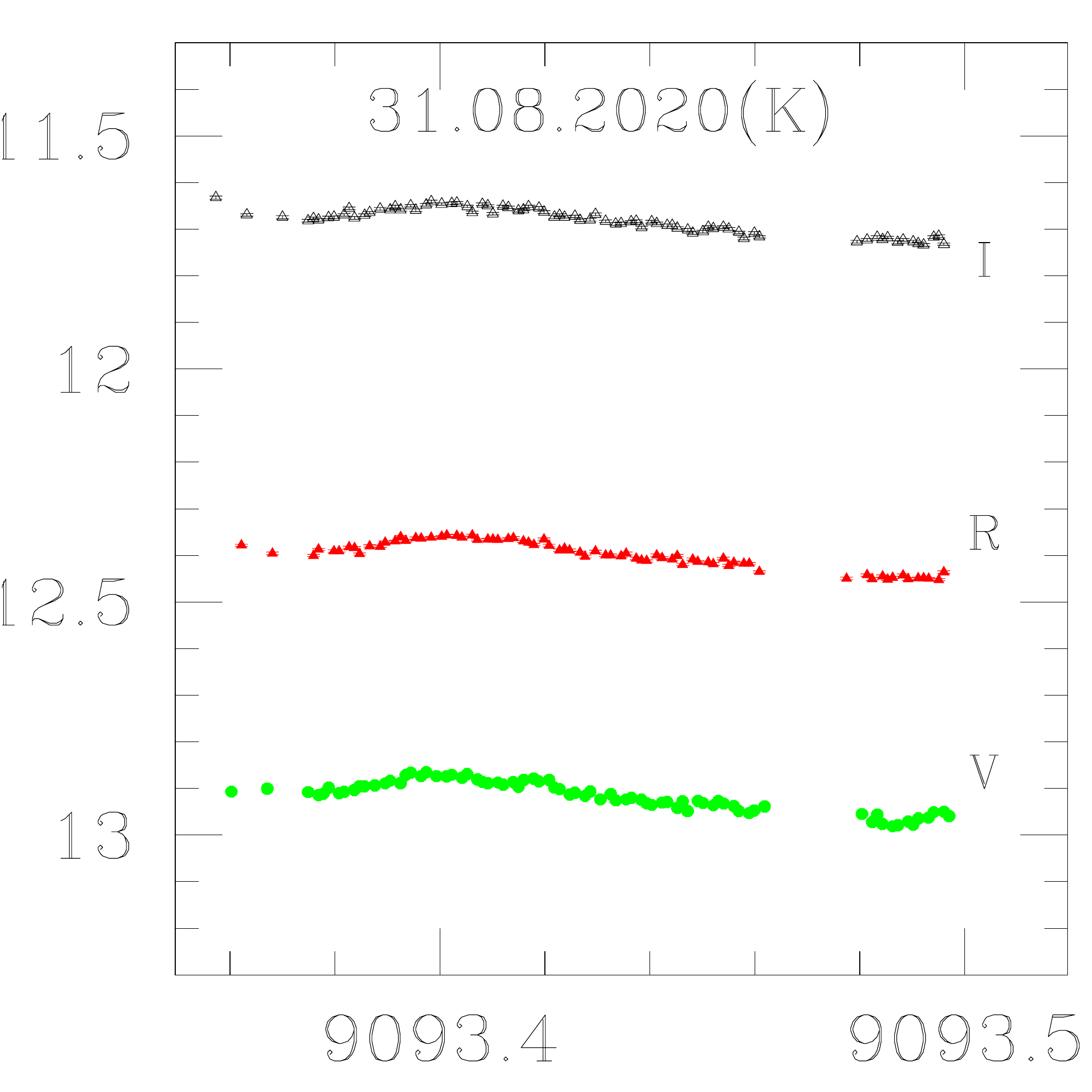}{0.25\textwidth}{}}
\gridline{\fig{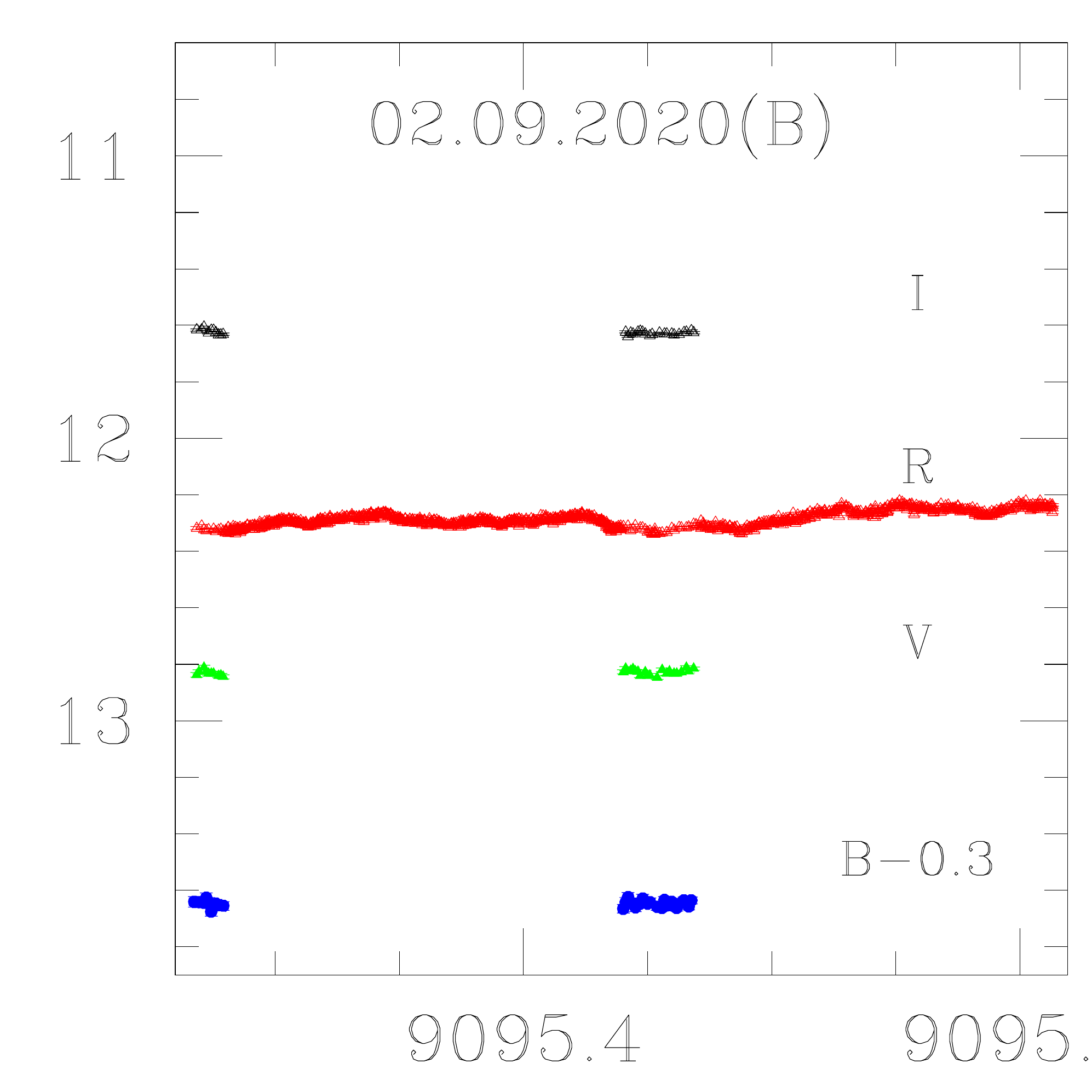}{0.25\textwidth}{}
          \fig{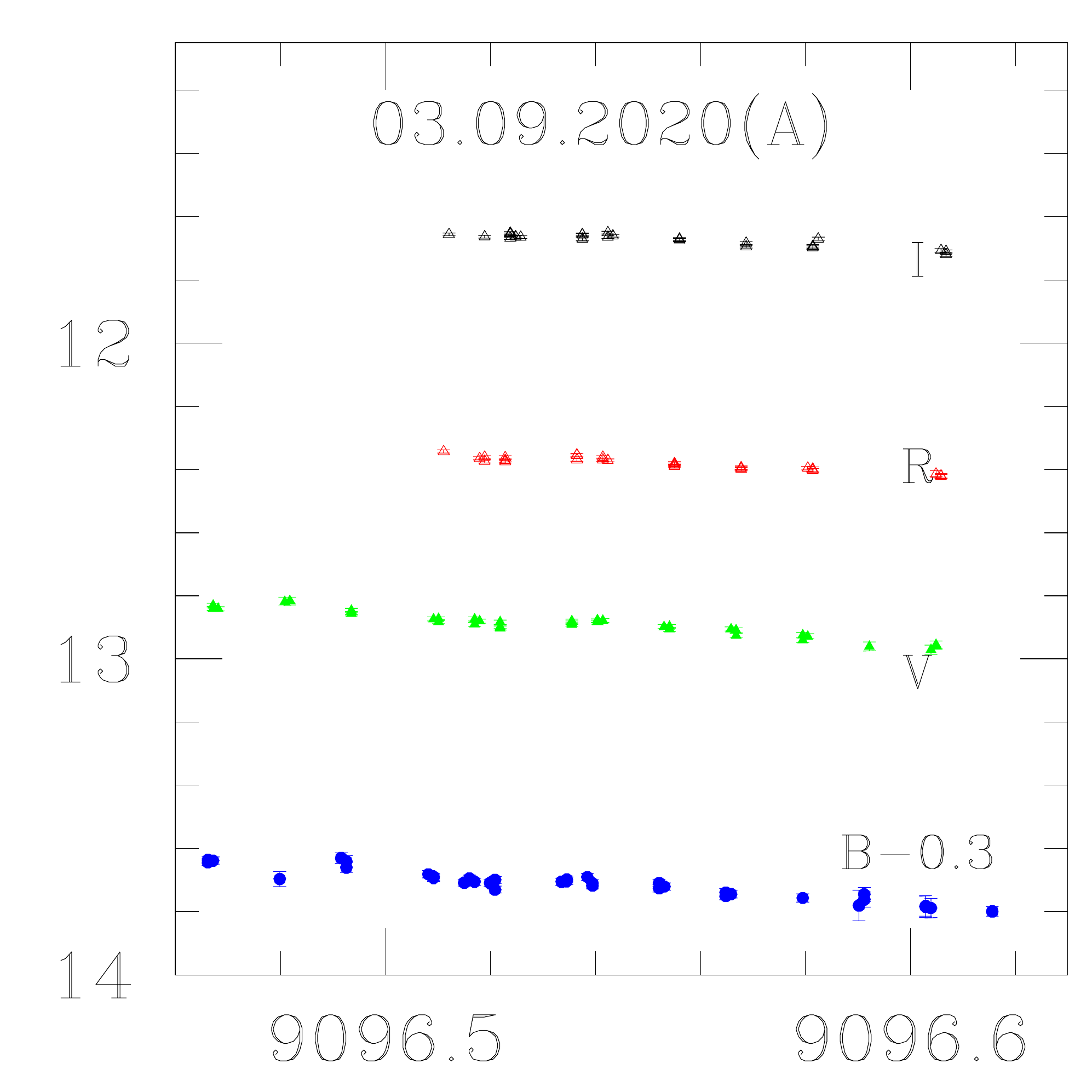}{0.25\textwidth}{}
          \fig{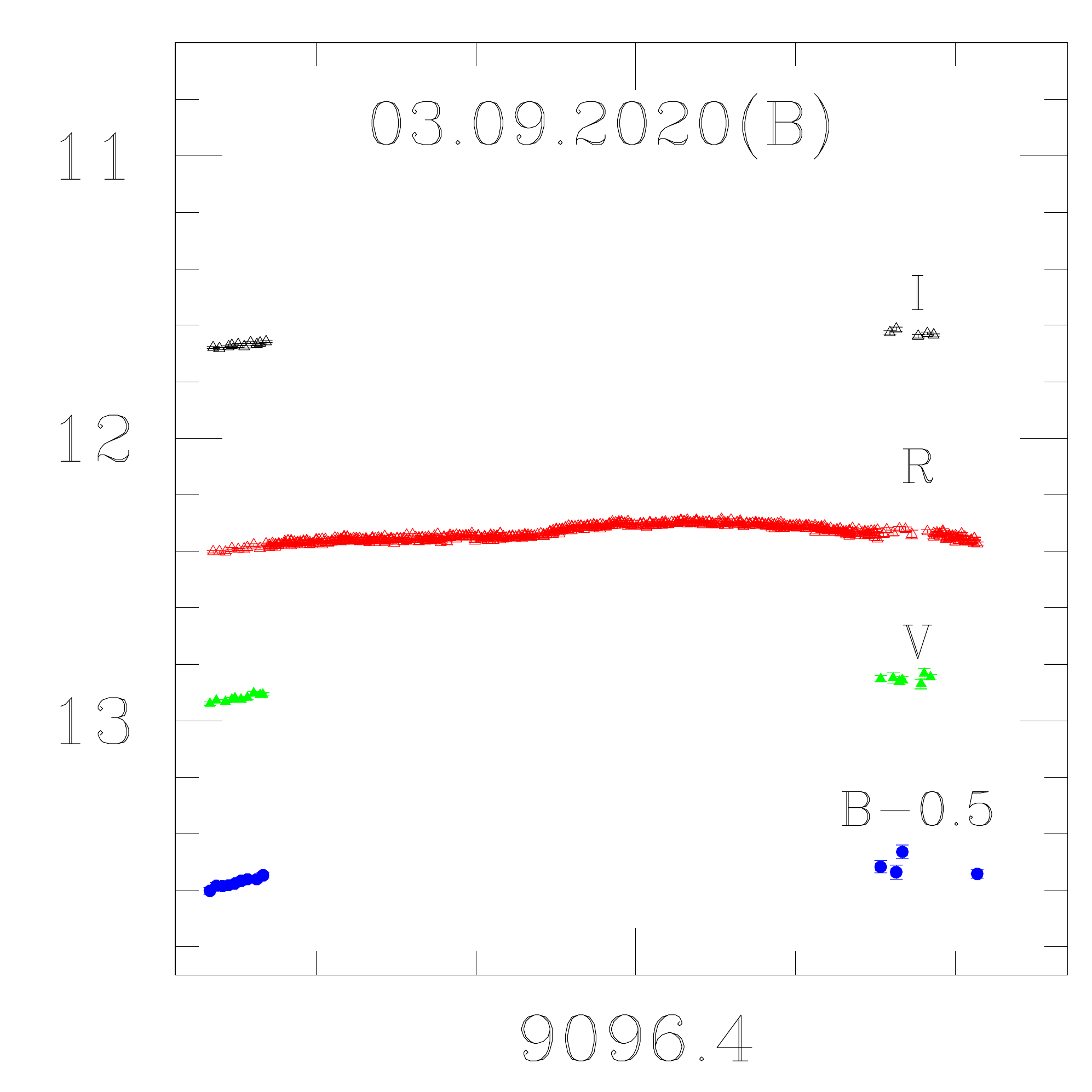}{0.25\textwidth}{}}
\caption{Continued.}
\end{figure*}

\setcounter{figure}{5}
\begin{figure*}[t!]
\gridline{\fig{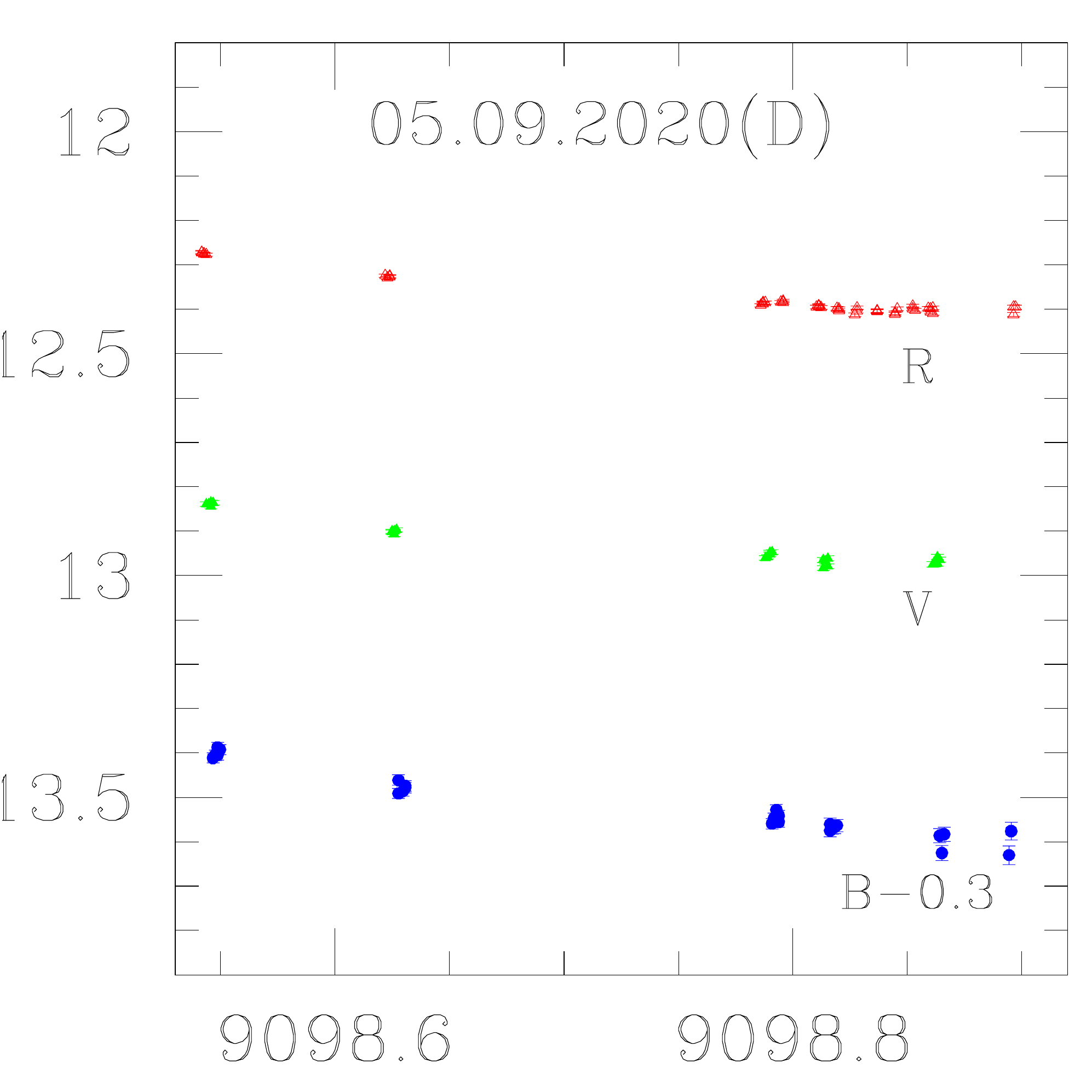}{0.25\textwidth}{}
          \fig{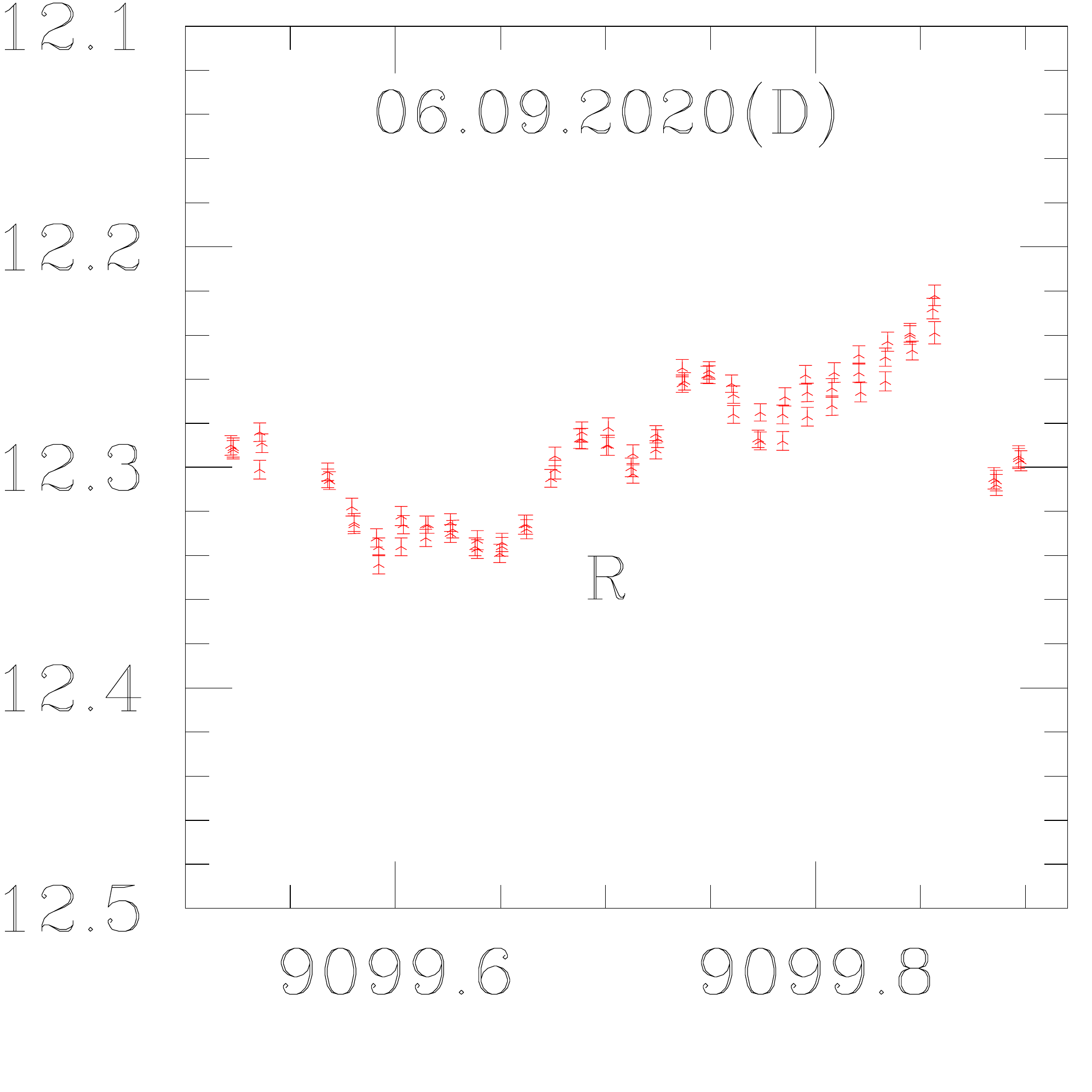}{0.25\textwidth}{}
          \fig{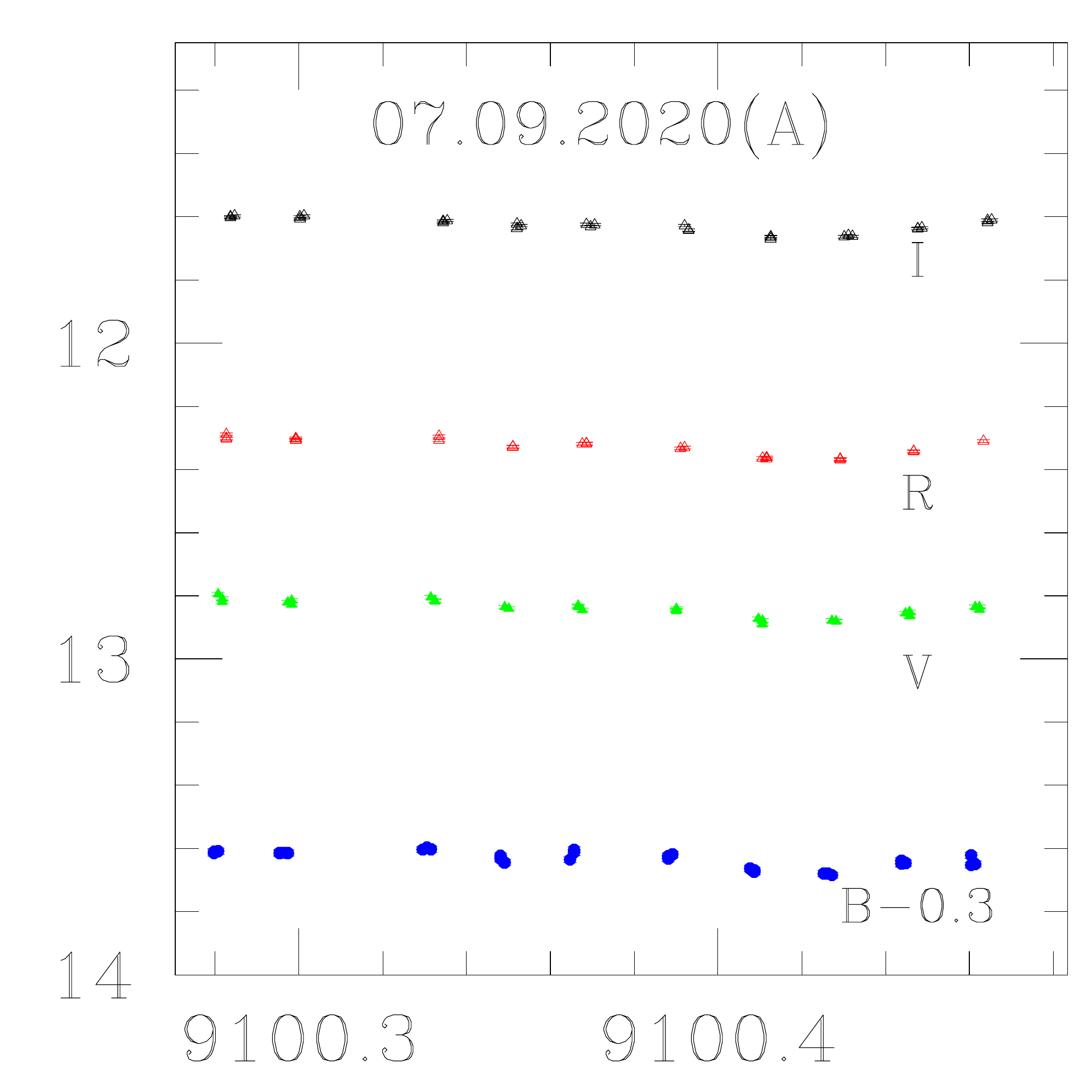}{0.25\textwidth}{}}
\gridline{\fig{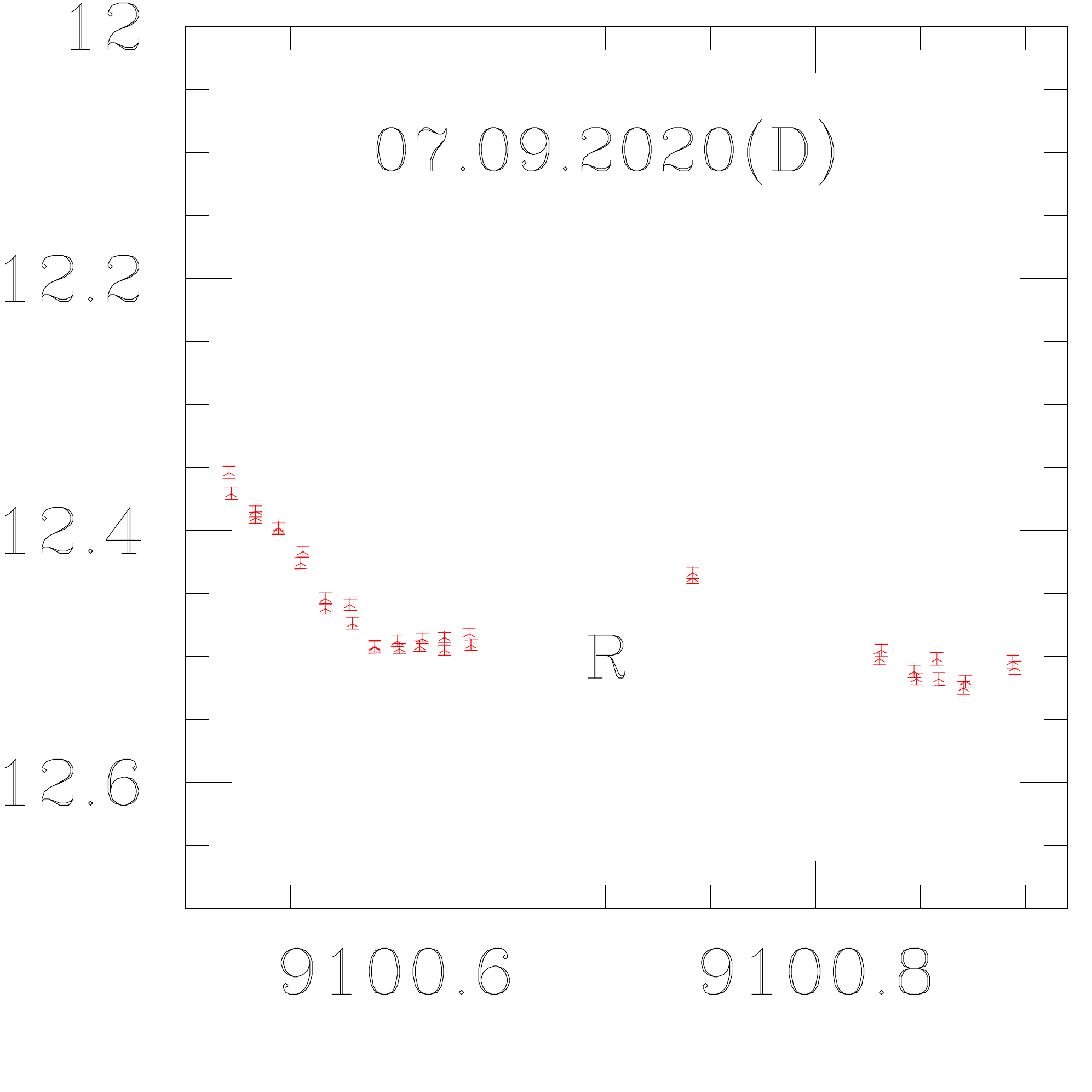}{0.25\textwidth}{}
          \fig{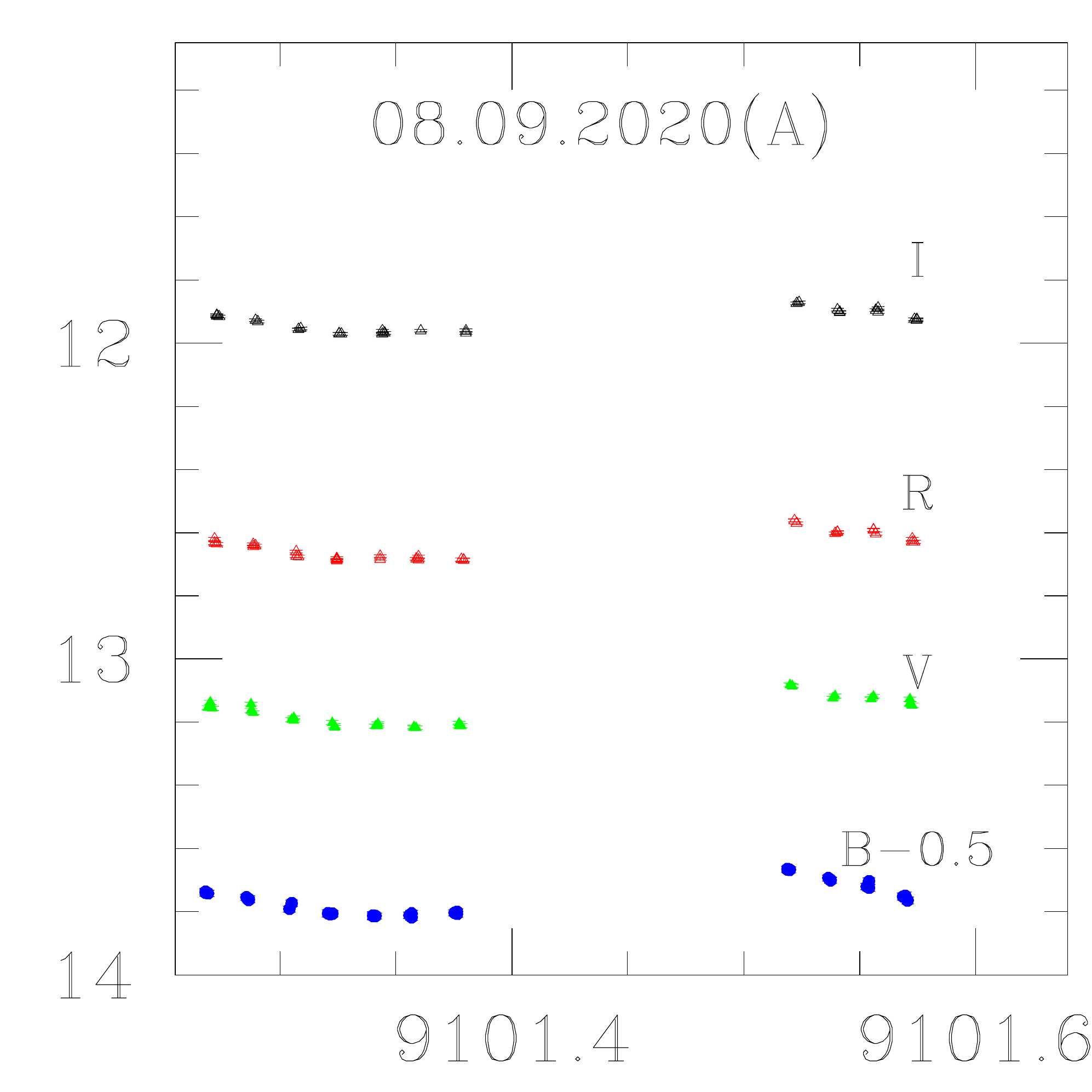}{0.25\textwidth}{}
          \fig{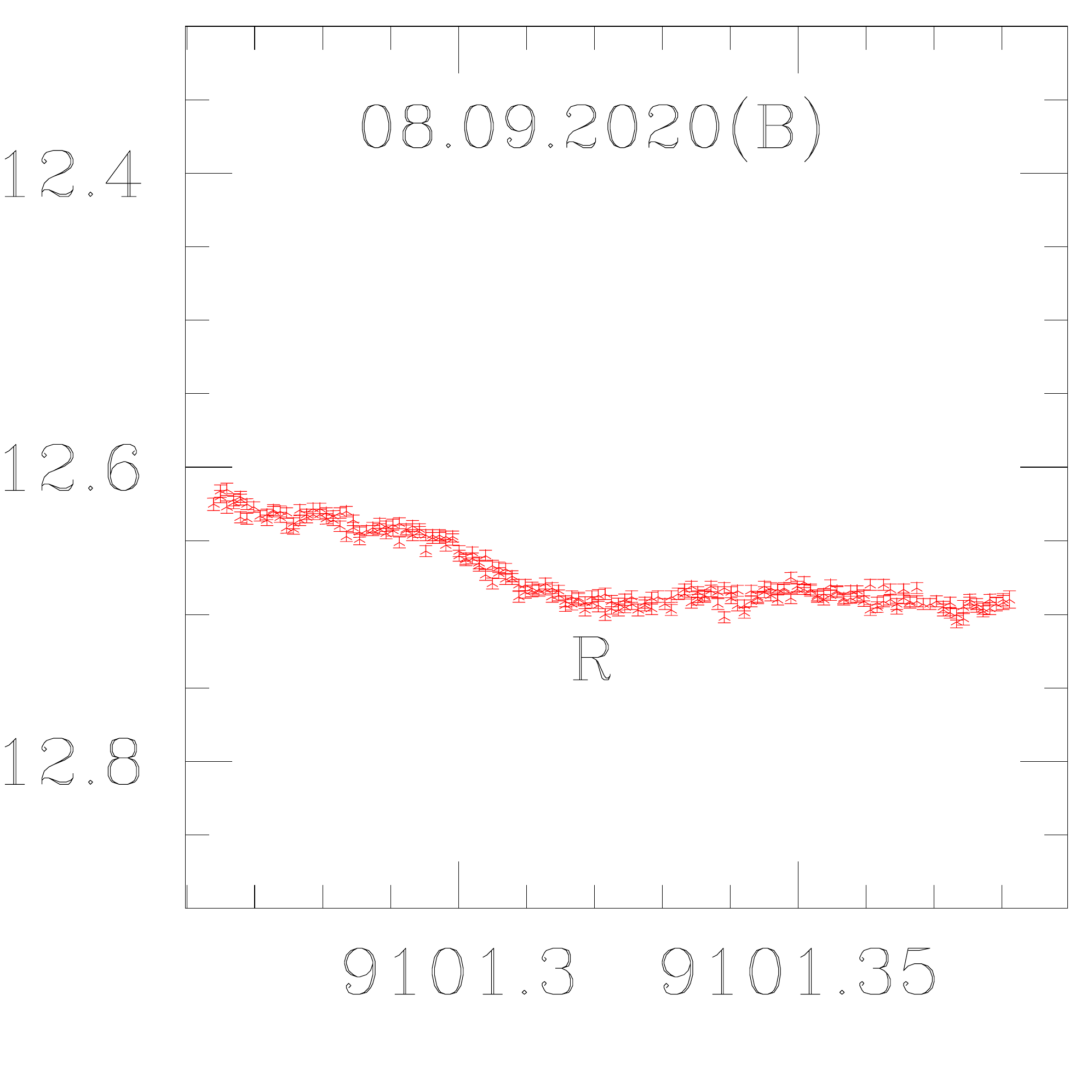}{0.25\textwidth}{}}
\gridline{\fig{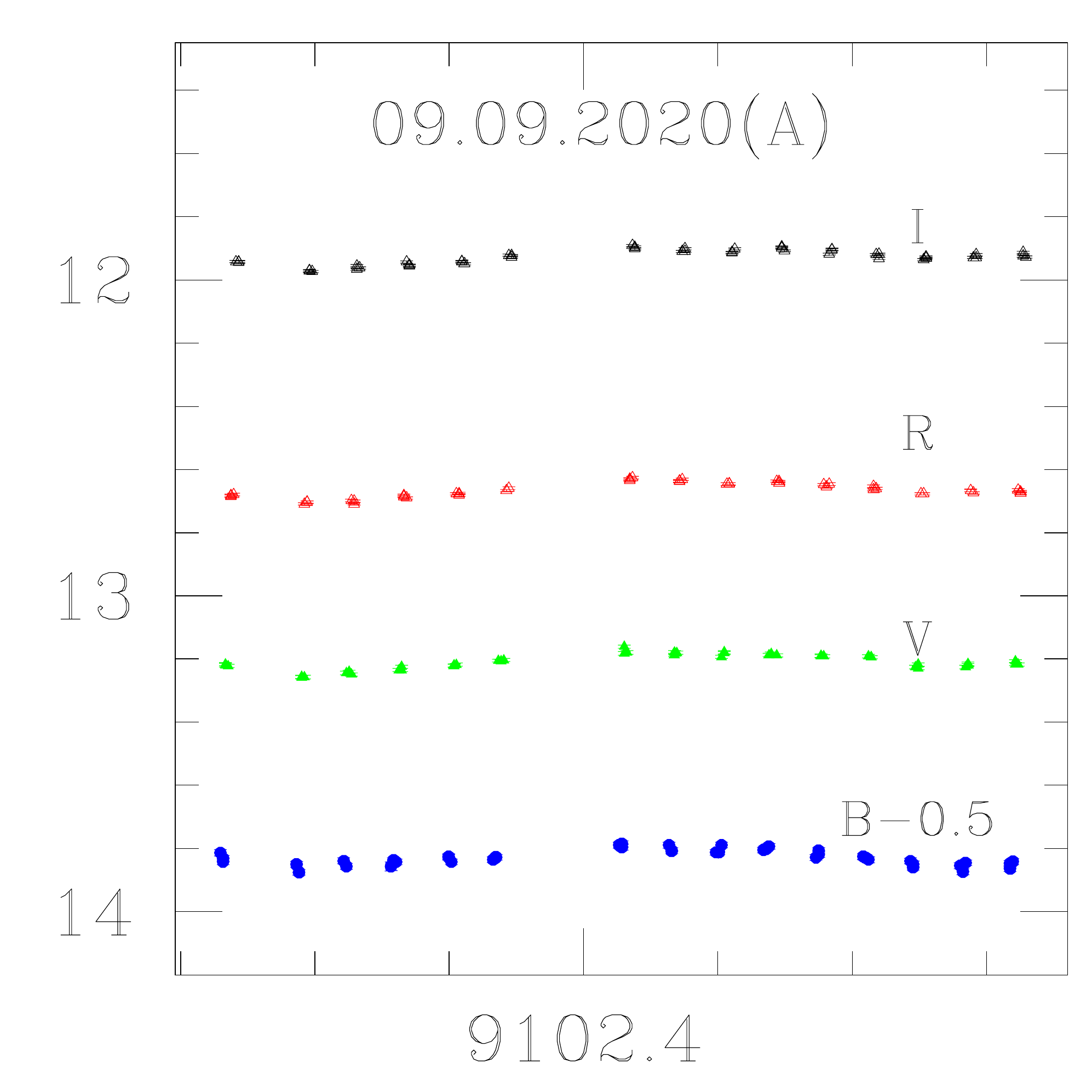}{0.25\textwidth}{}
          \fig{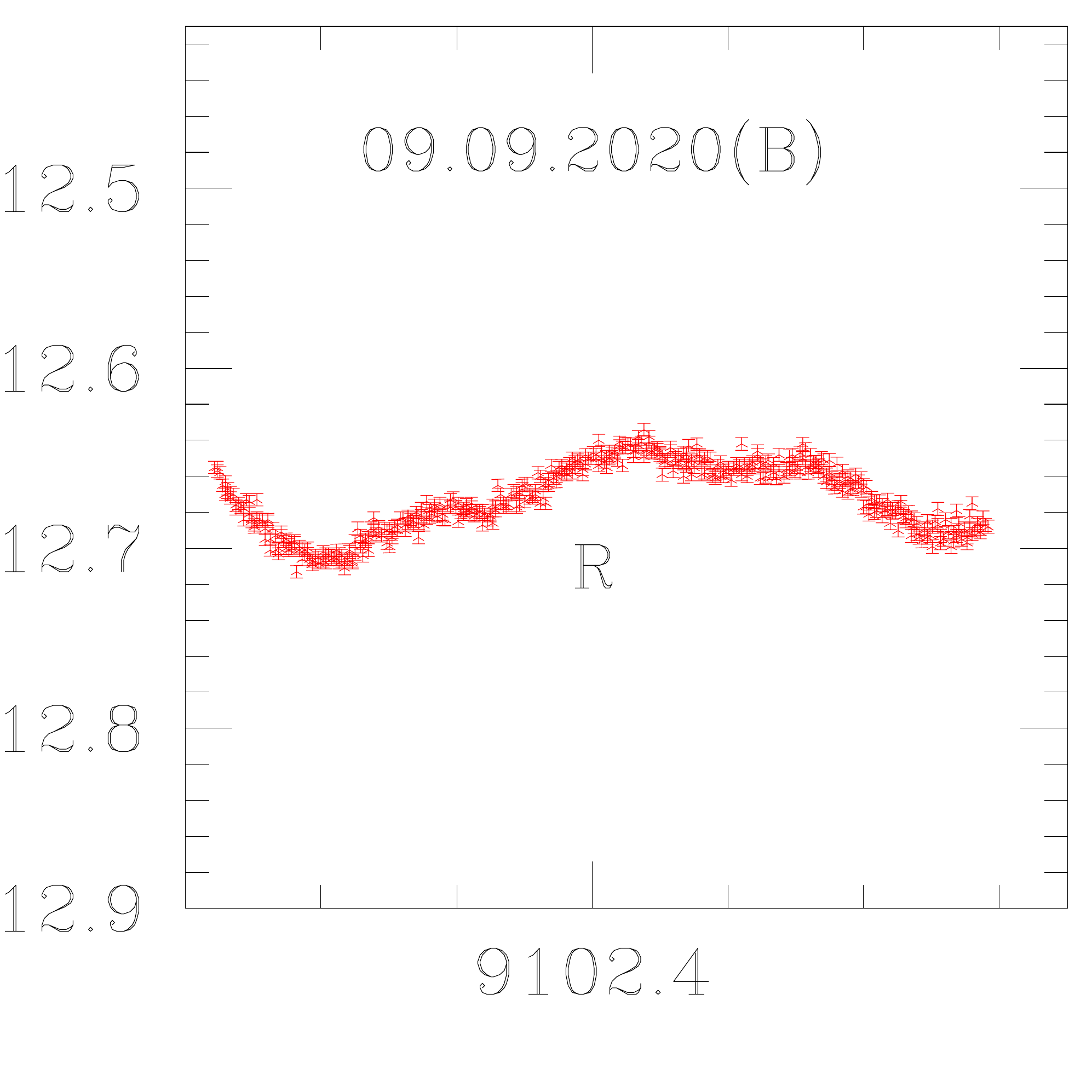}{0.25\textwidth}{}
          \fig{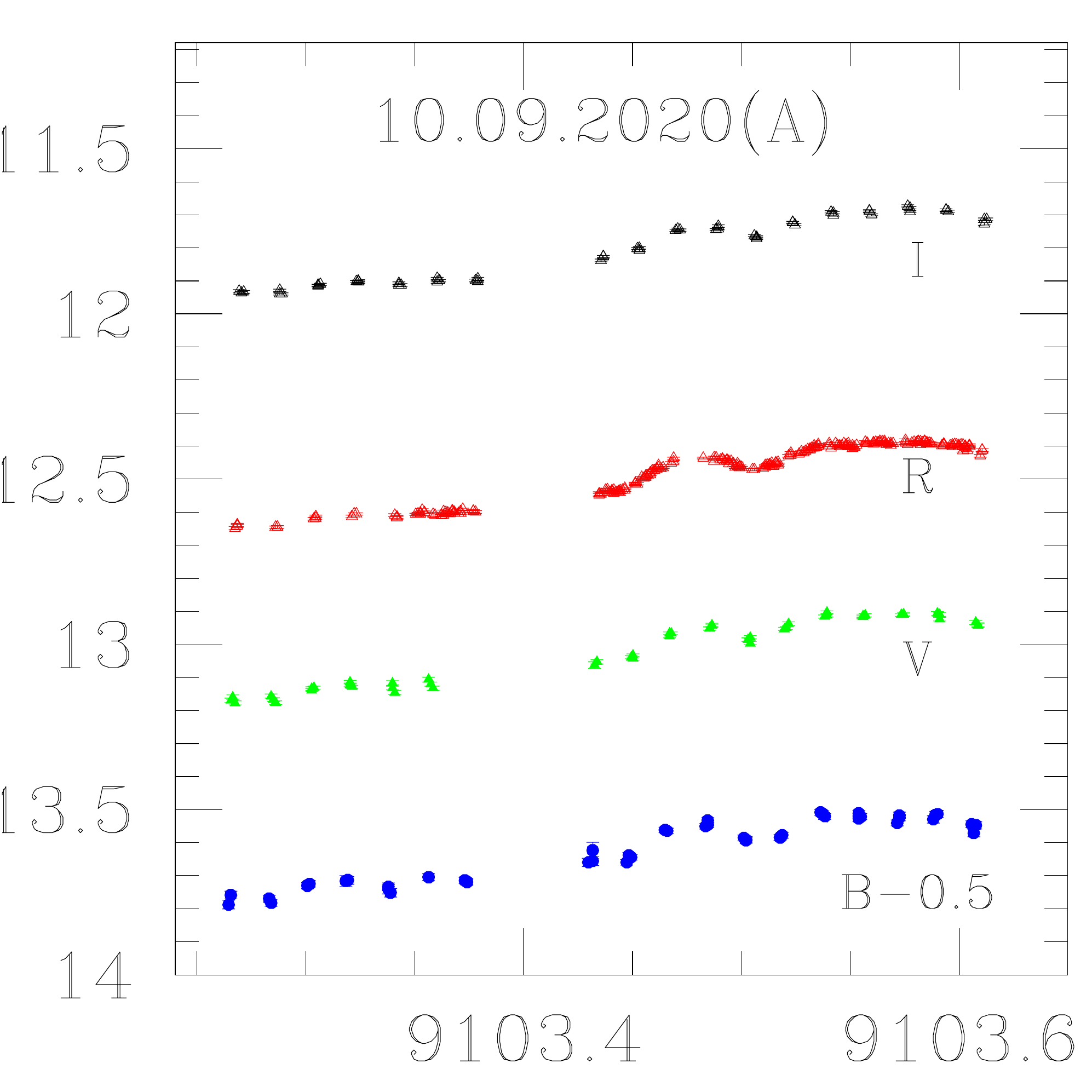}{0.25\textwidth}{}}
\gridline{\fig{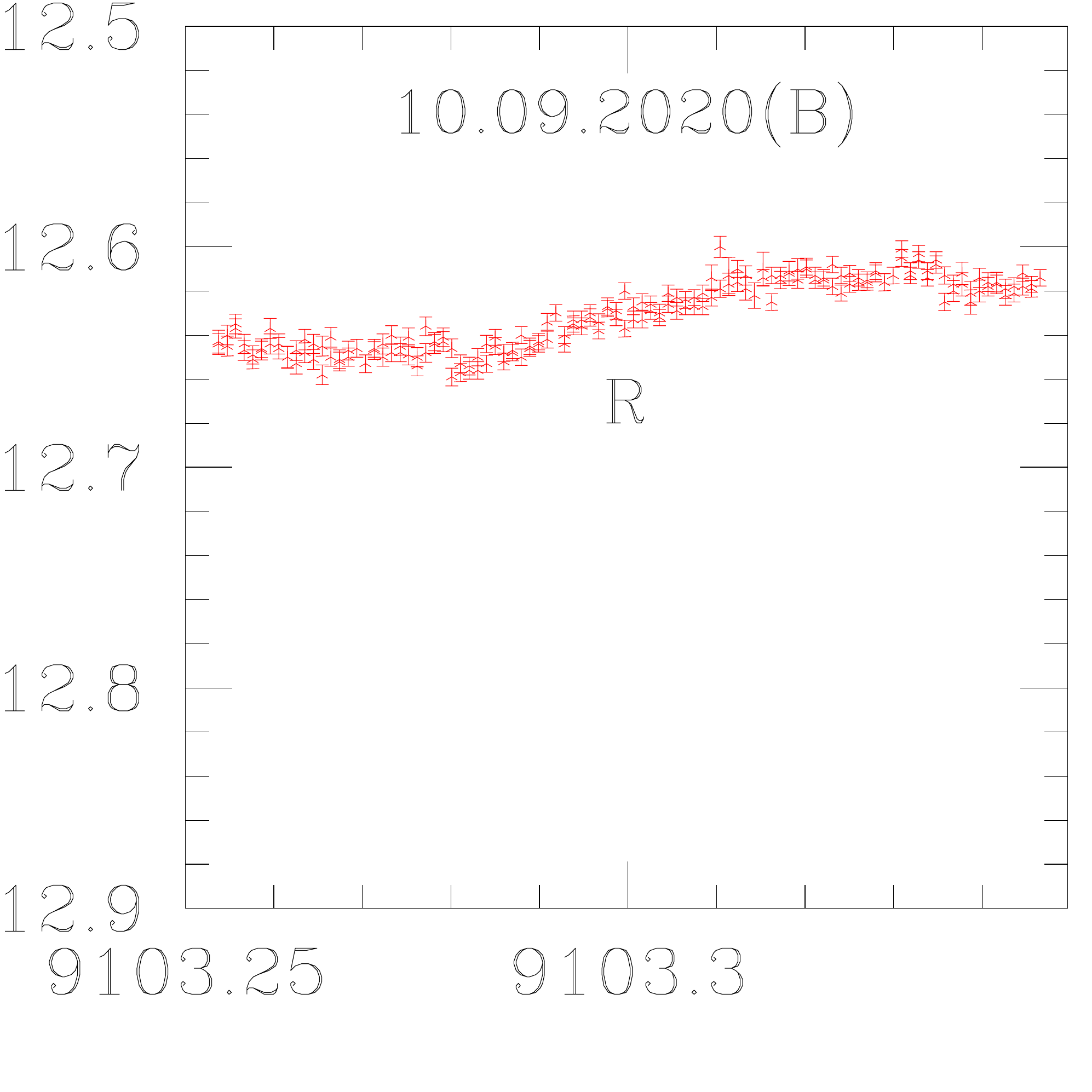}{0.25\textwidth}{}
          \fig{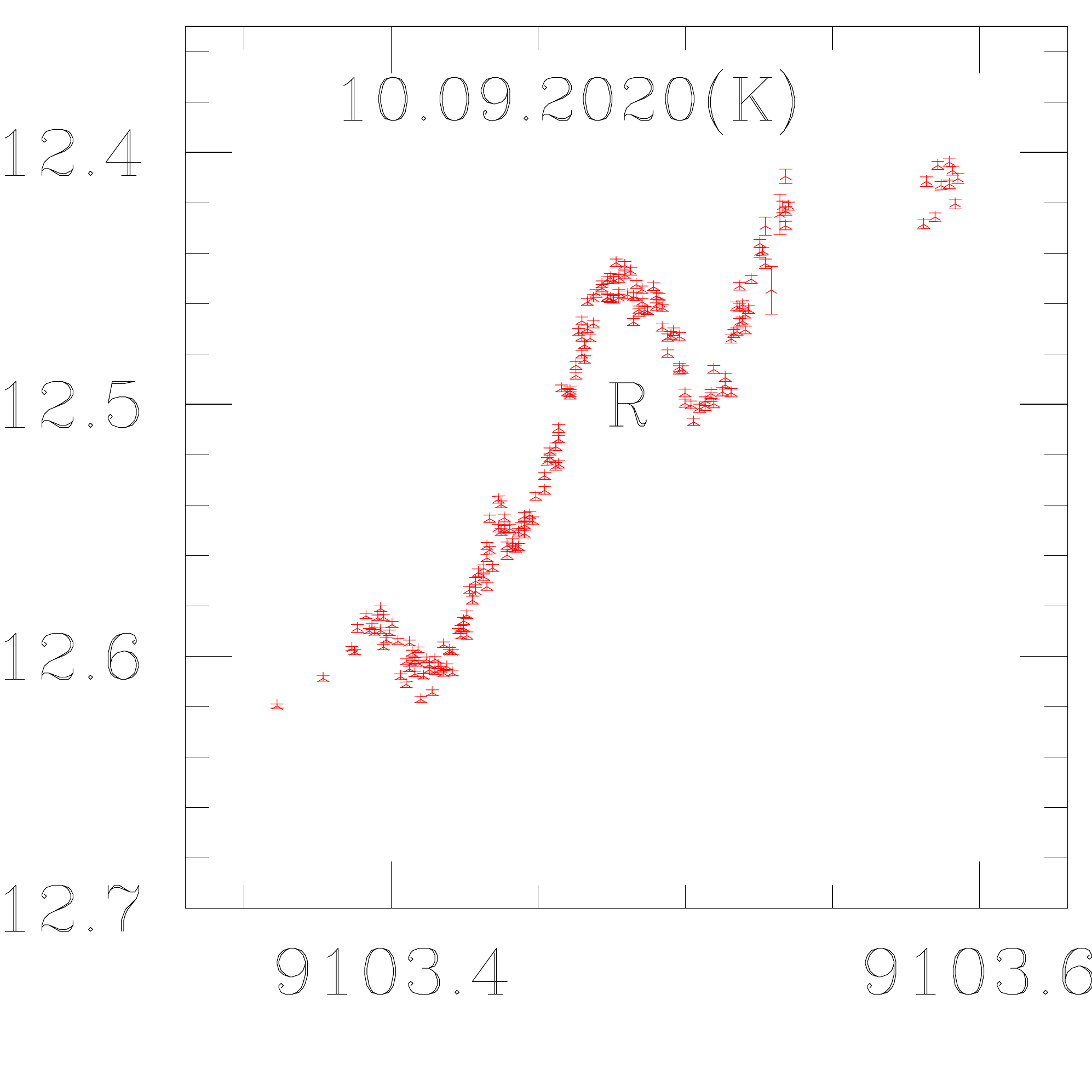}{0.25\textwidth}{}
          \fig{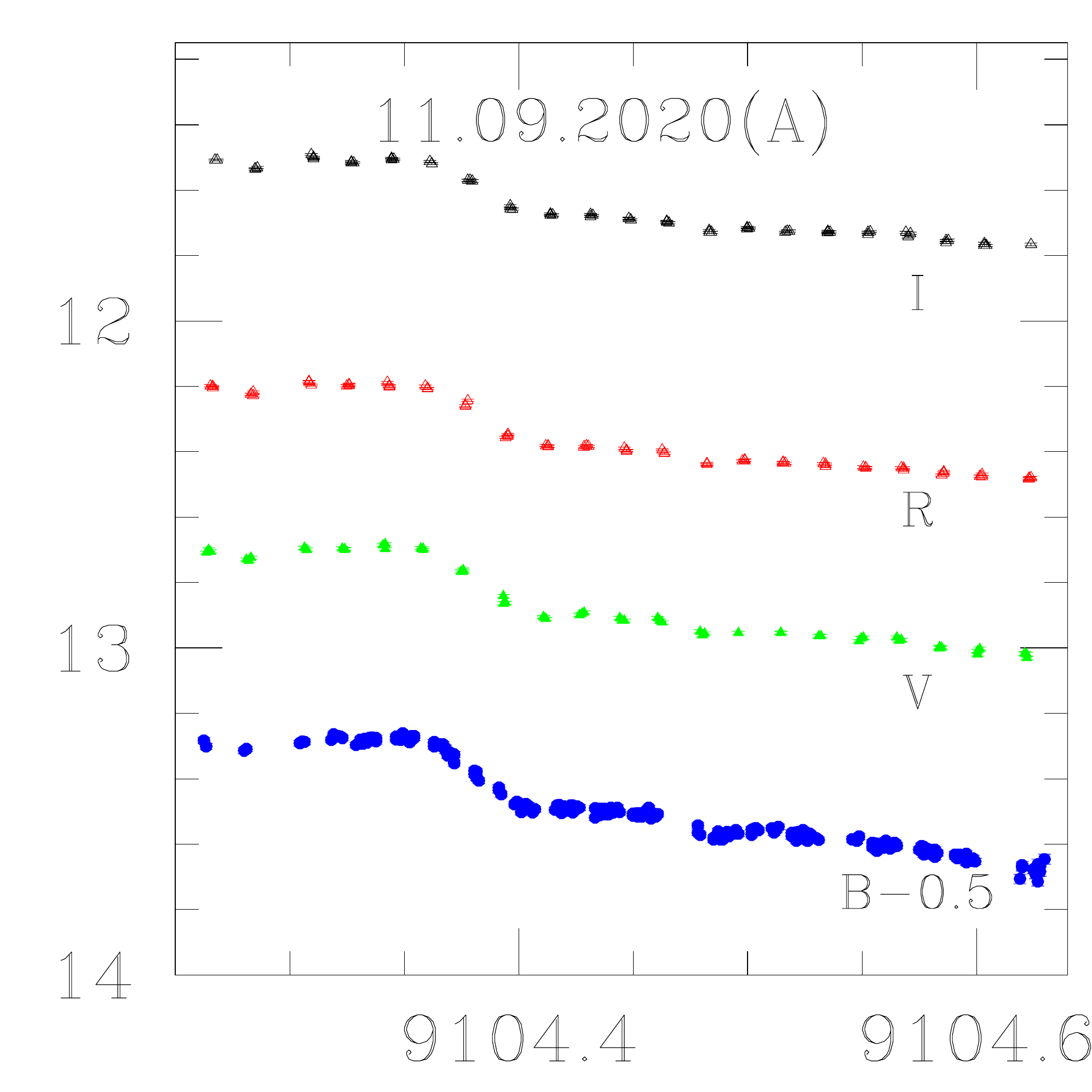}{0.25\textwidth}{}}
\caption{Continued.}
\end{figure*}

\setcounter{figure}{5}
\begin{figure*}[t!]
\gridline{\fig{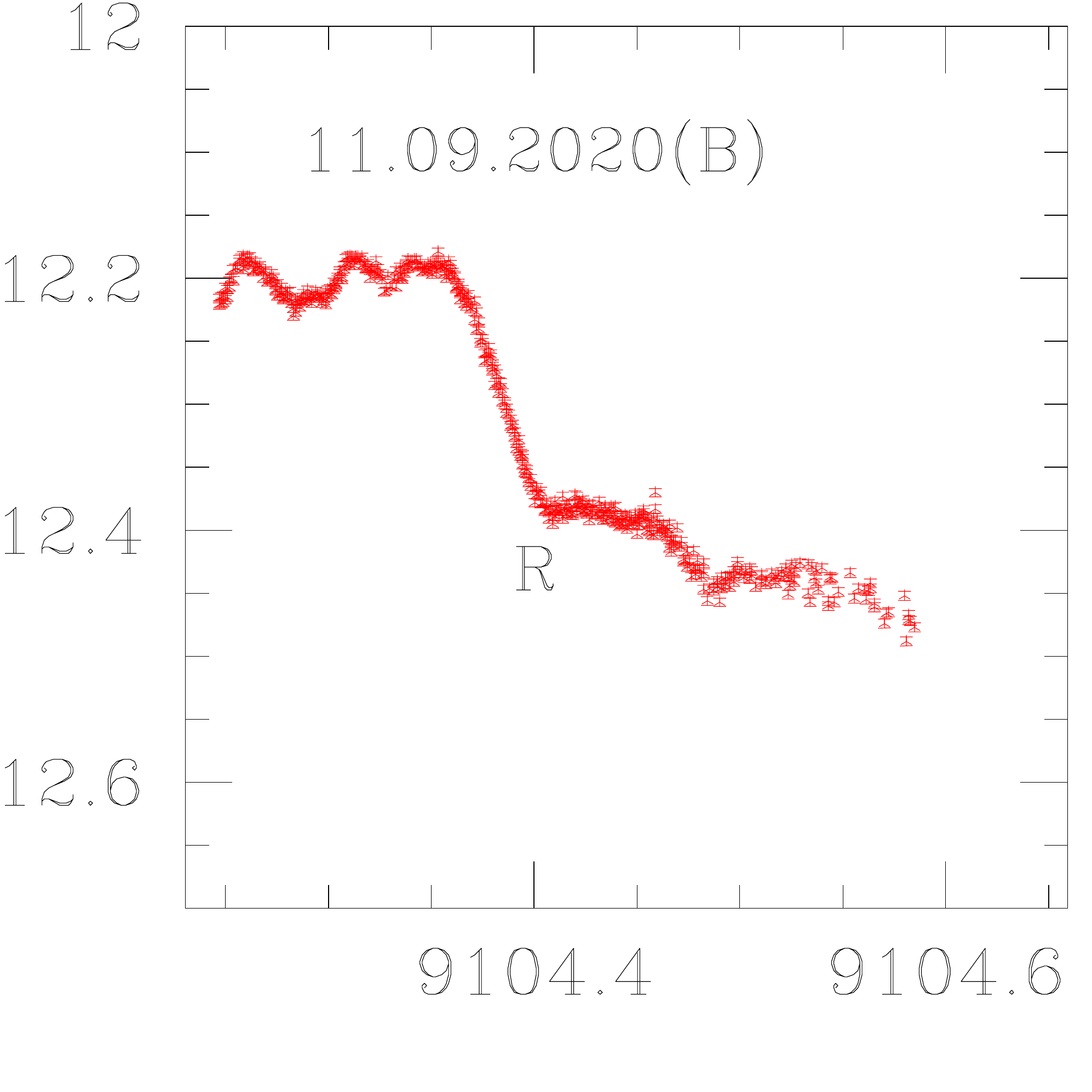}{0.25\textwidth}{}
          \fig{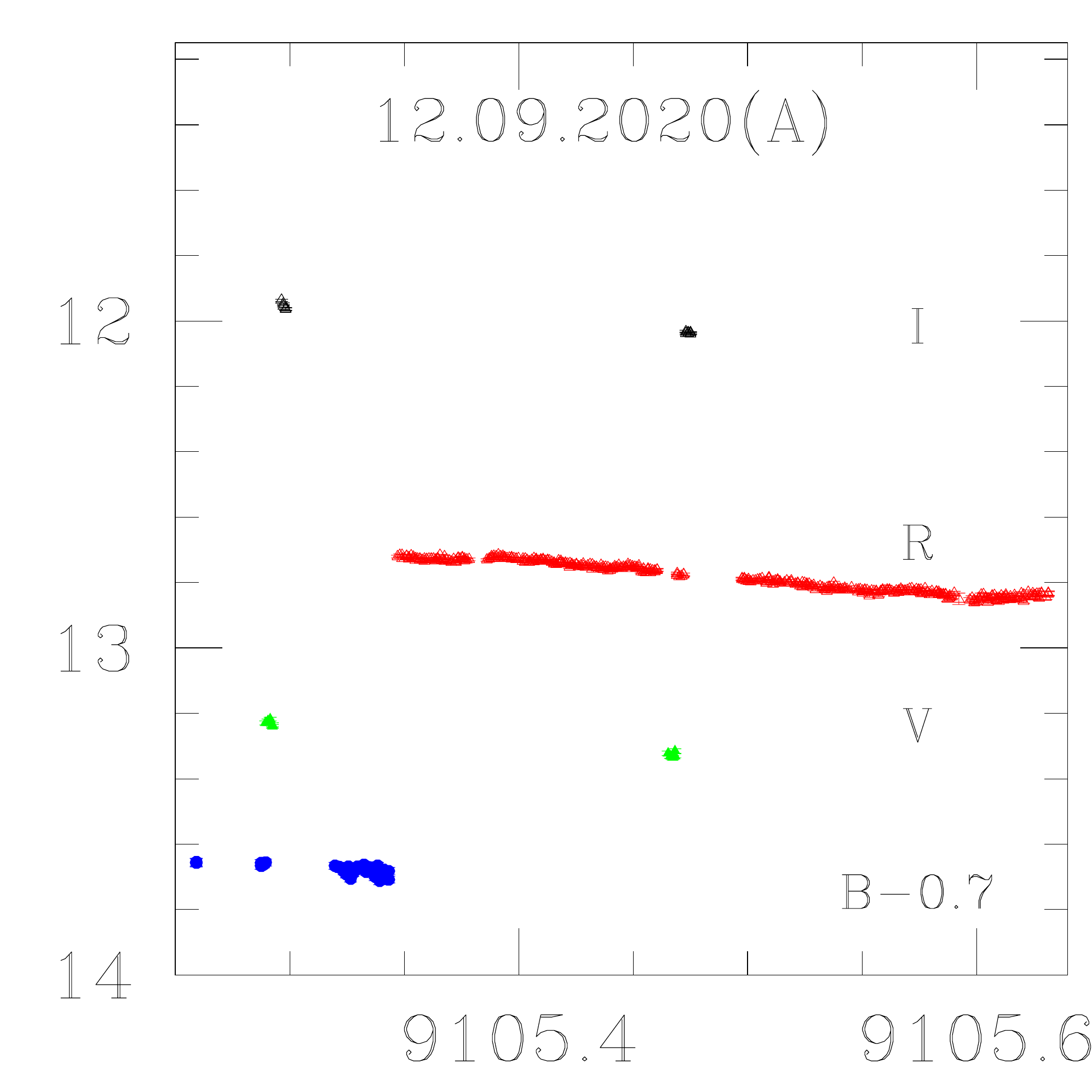}{0.25\textwidth}{}
          \fig{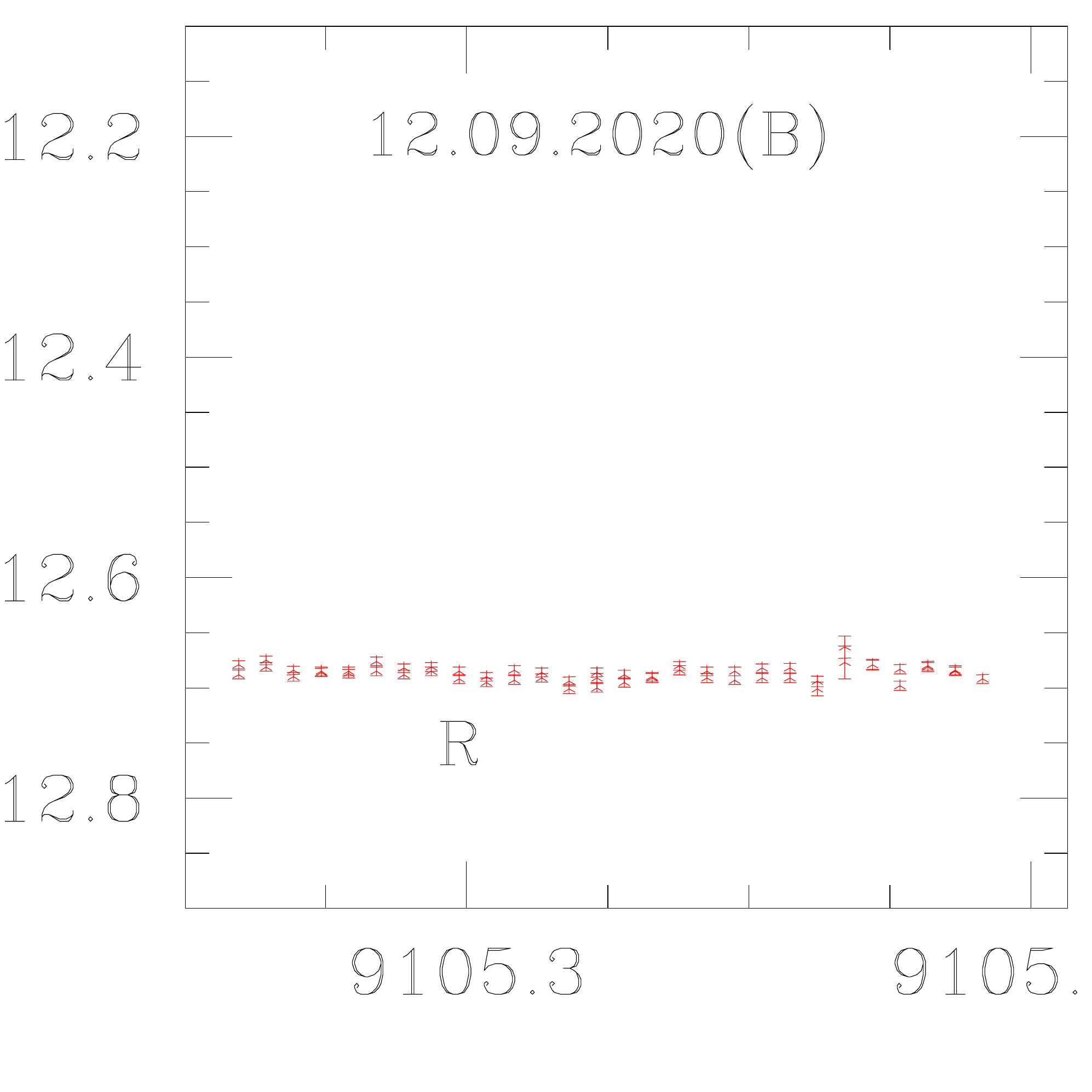}{0.25\textwidth}{}}
\gridline{\fig{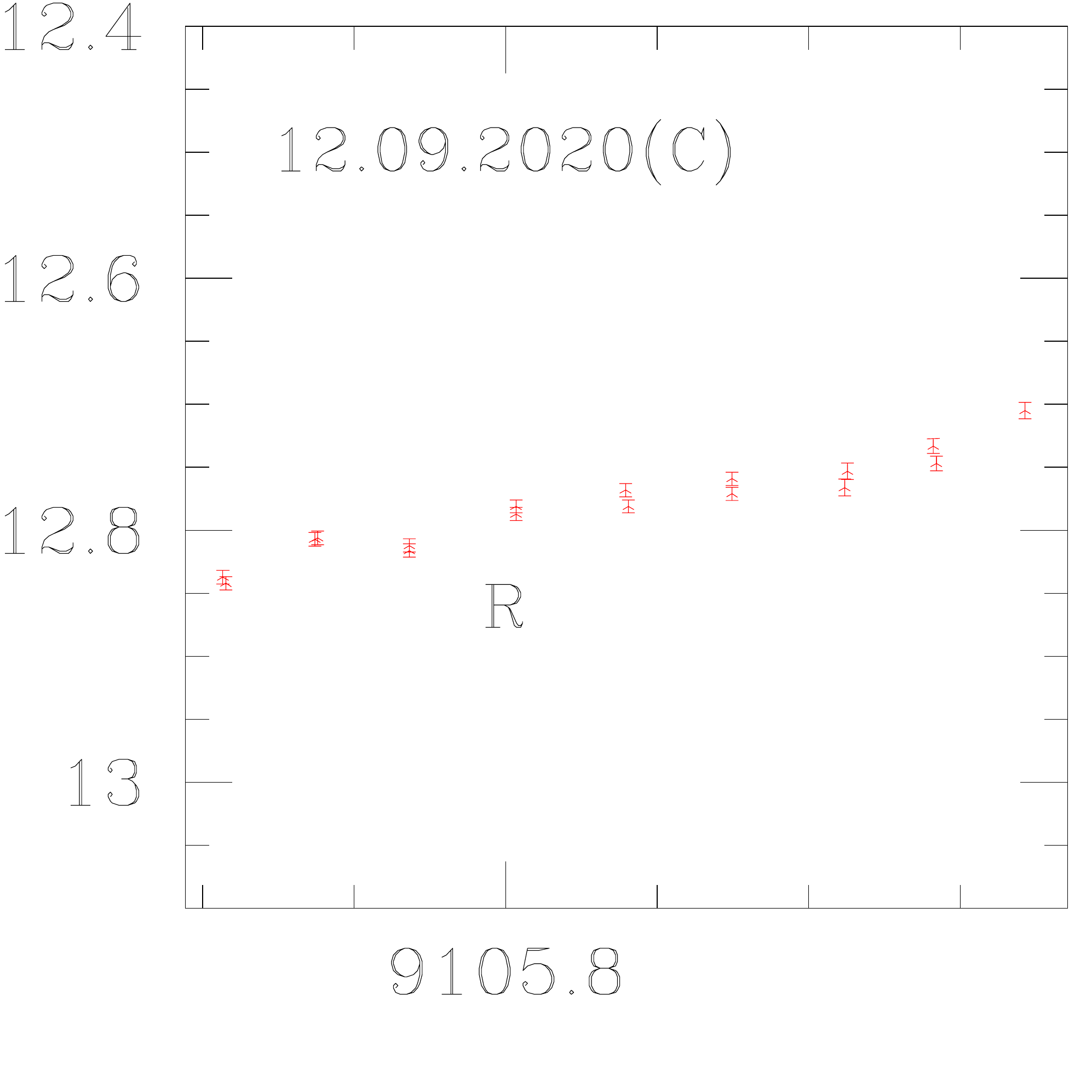}{0.25\textwidth}{}
          \fig{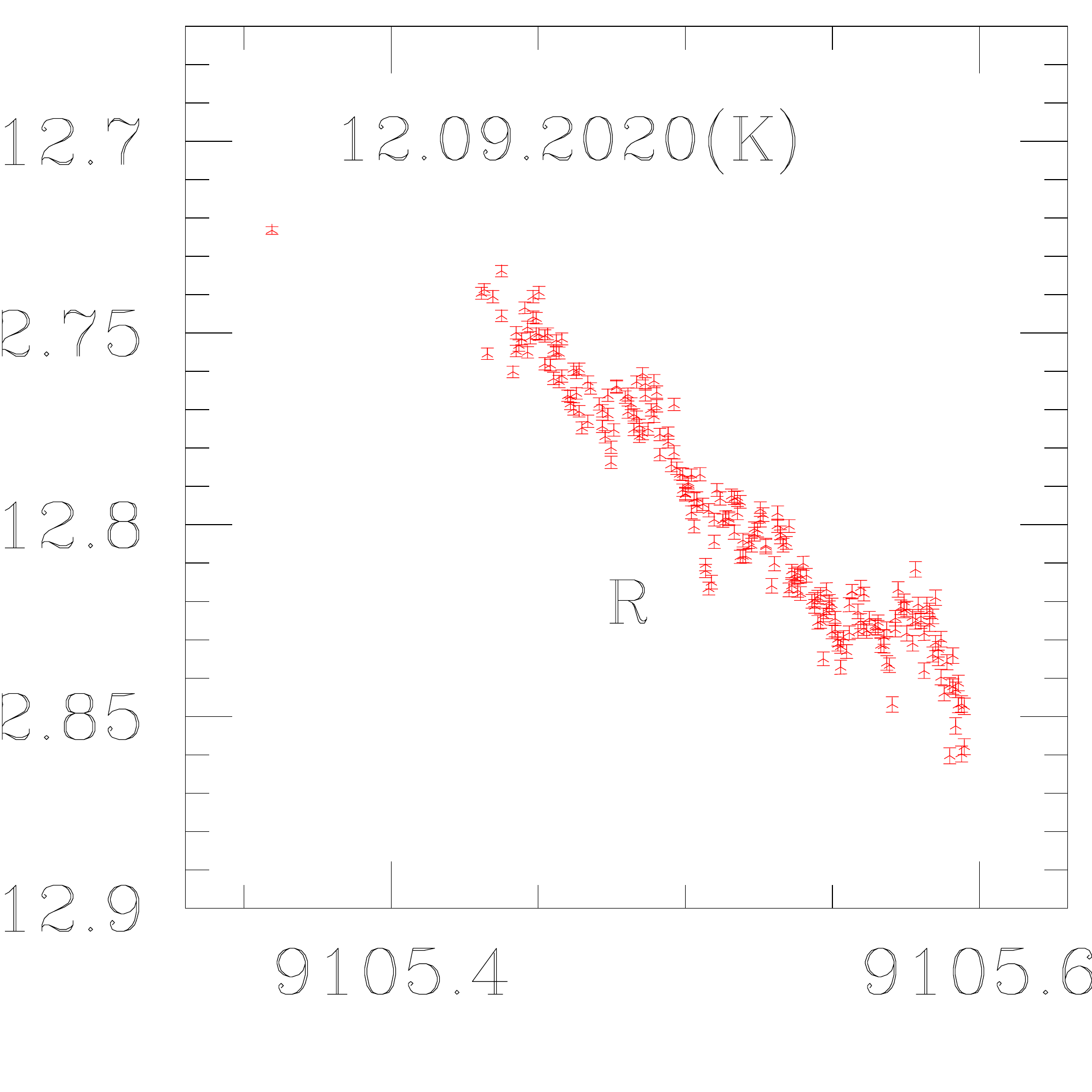}{0.25\textwidth}{}
          \fig{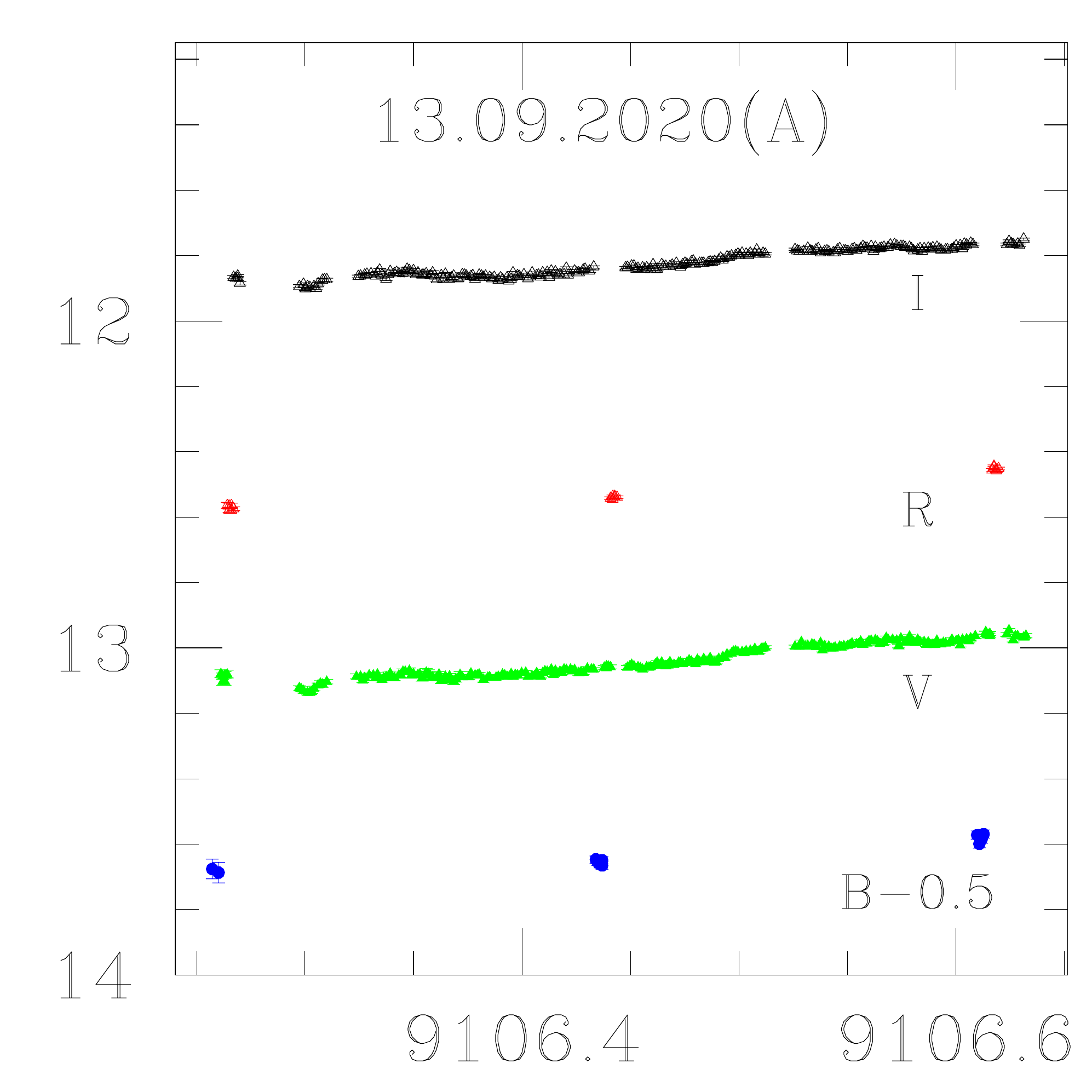}{0.25\textwidth}{}}
\gridline{\fig{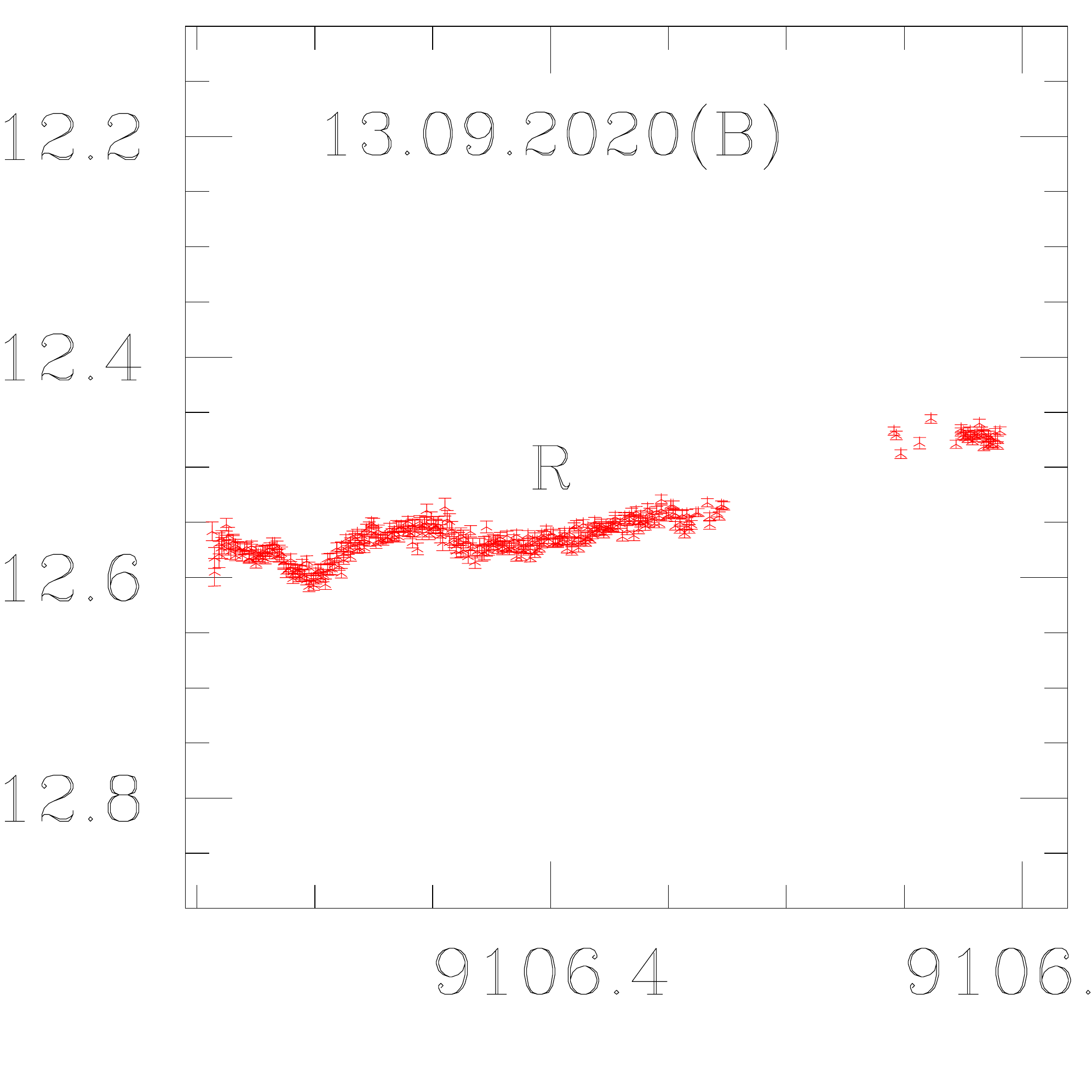}{0.25\textwidth}{}
          \fig{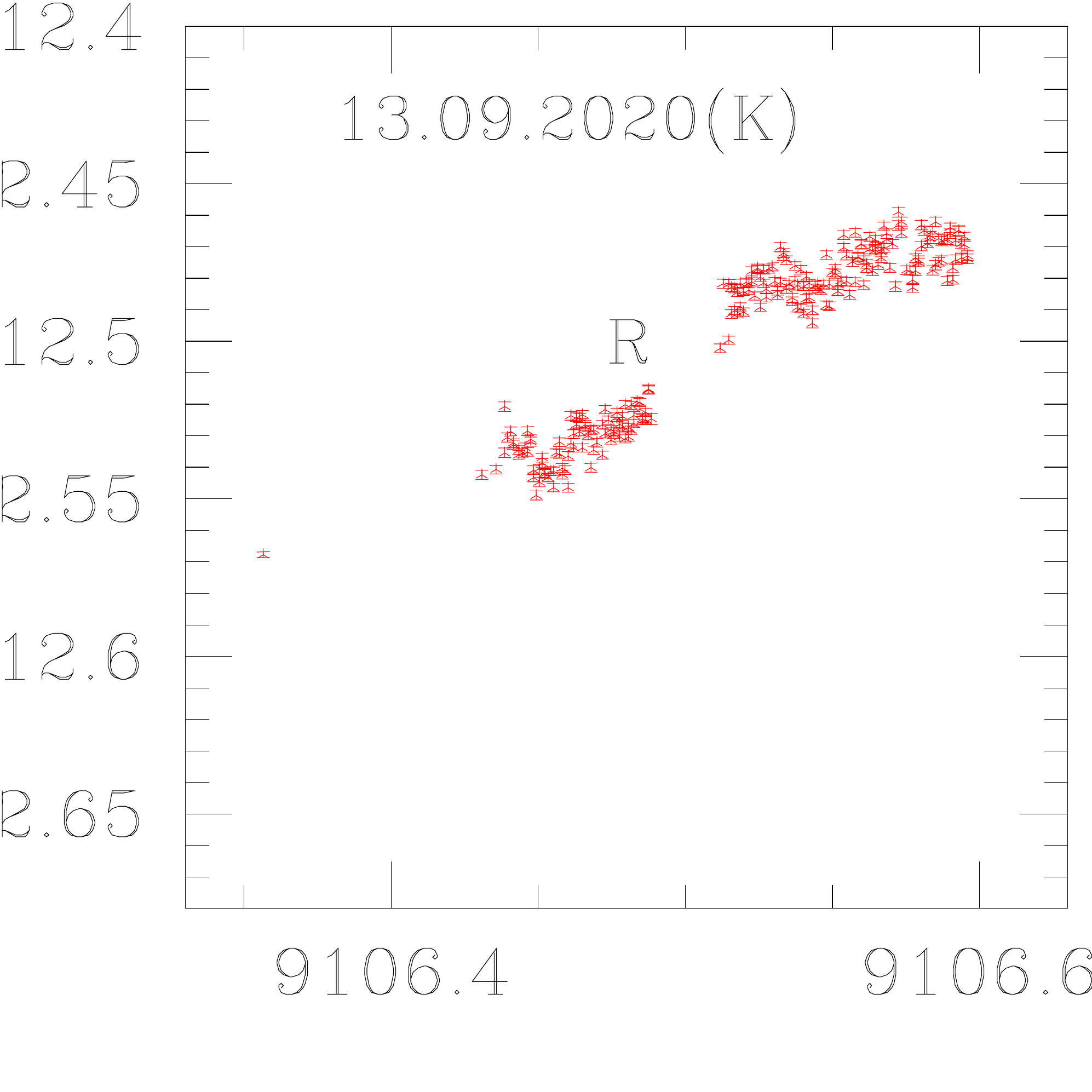}{0.25\textwidth}{}
          \fig{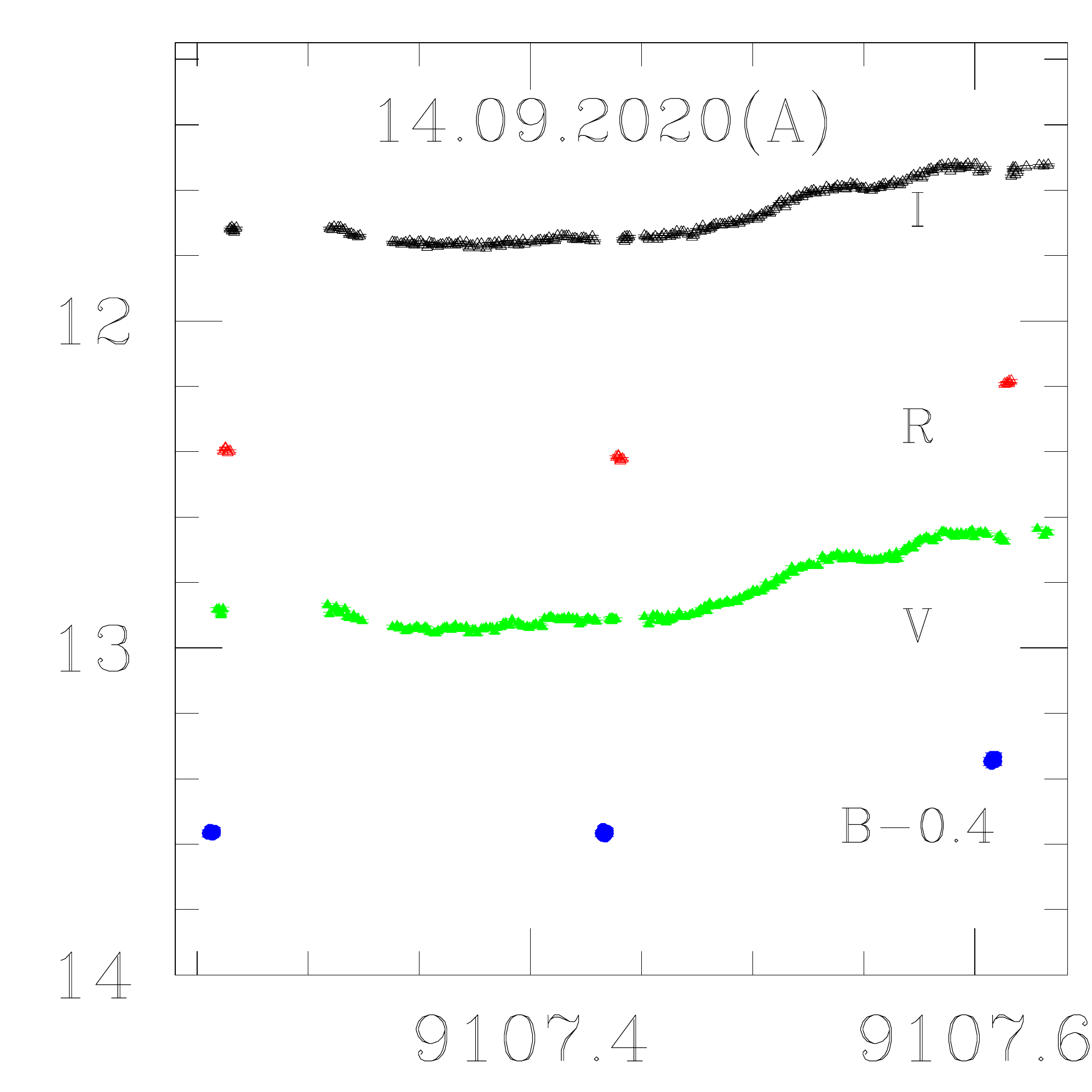}{0.25\textwidth}{}}
\gridline{\fig{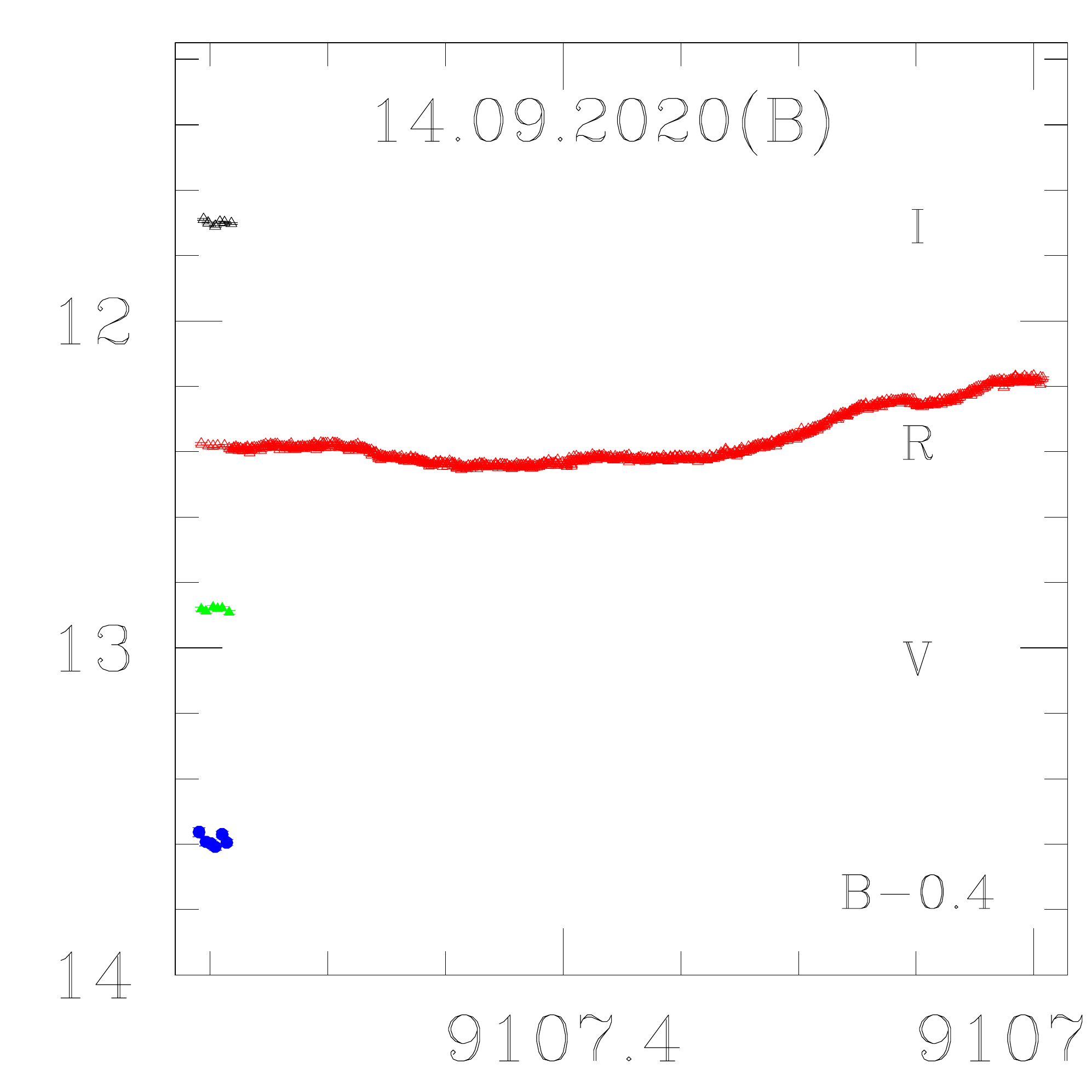}{0.25\textwidth}{}
          \fig{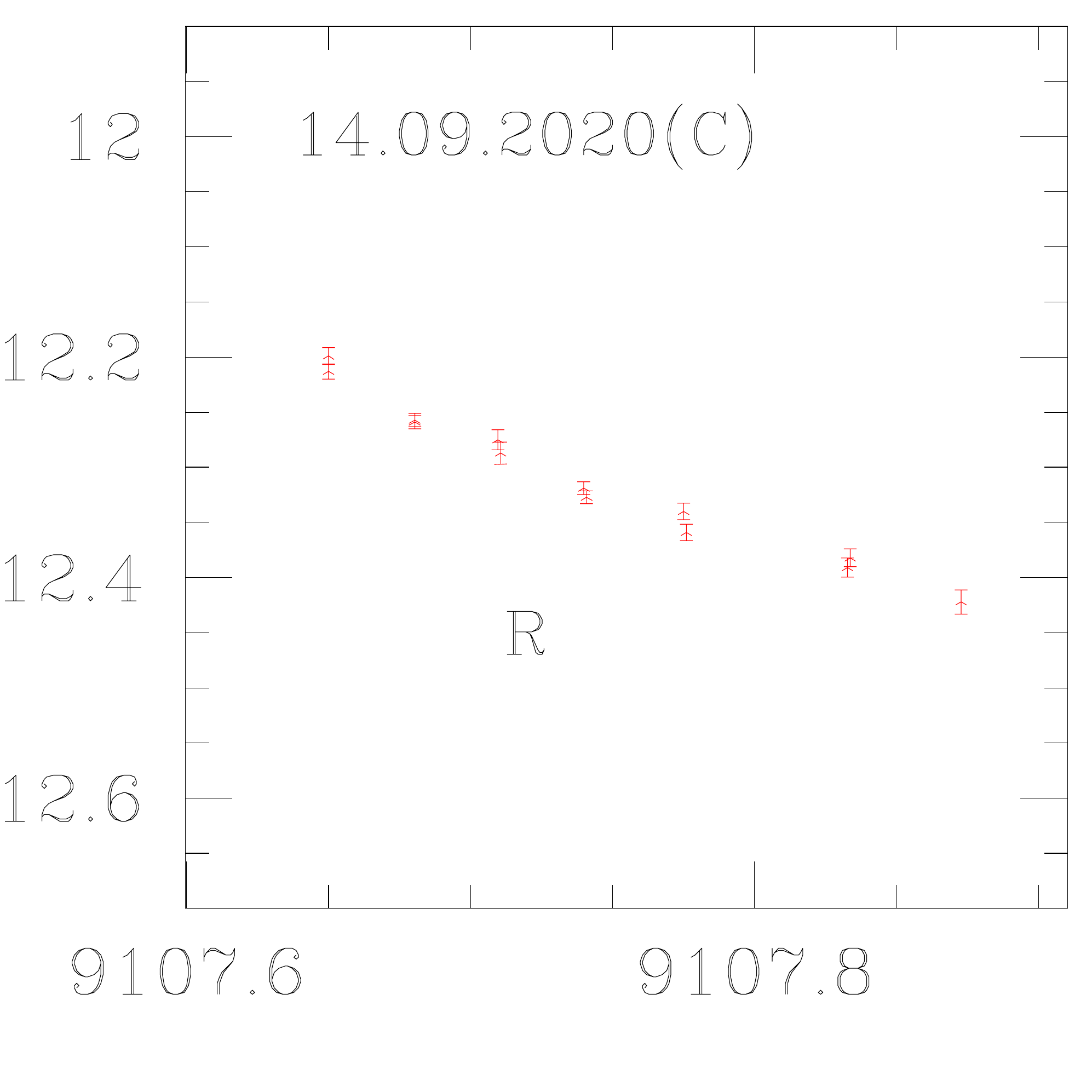}{0.25\textwidth}{}
          \fig{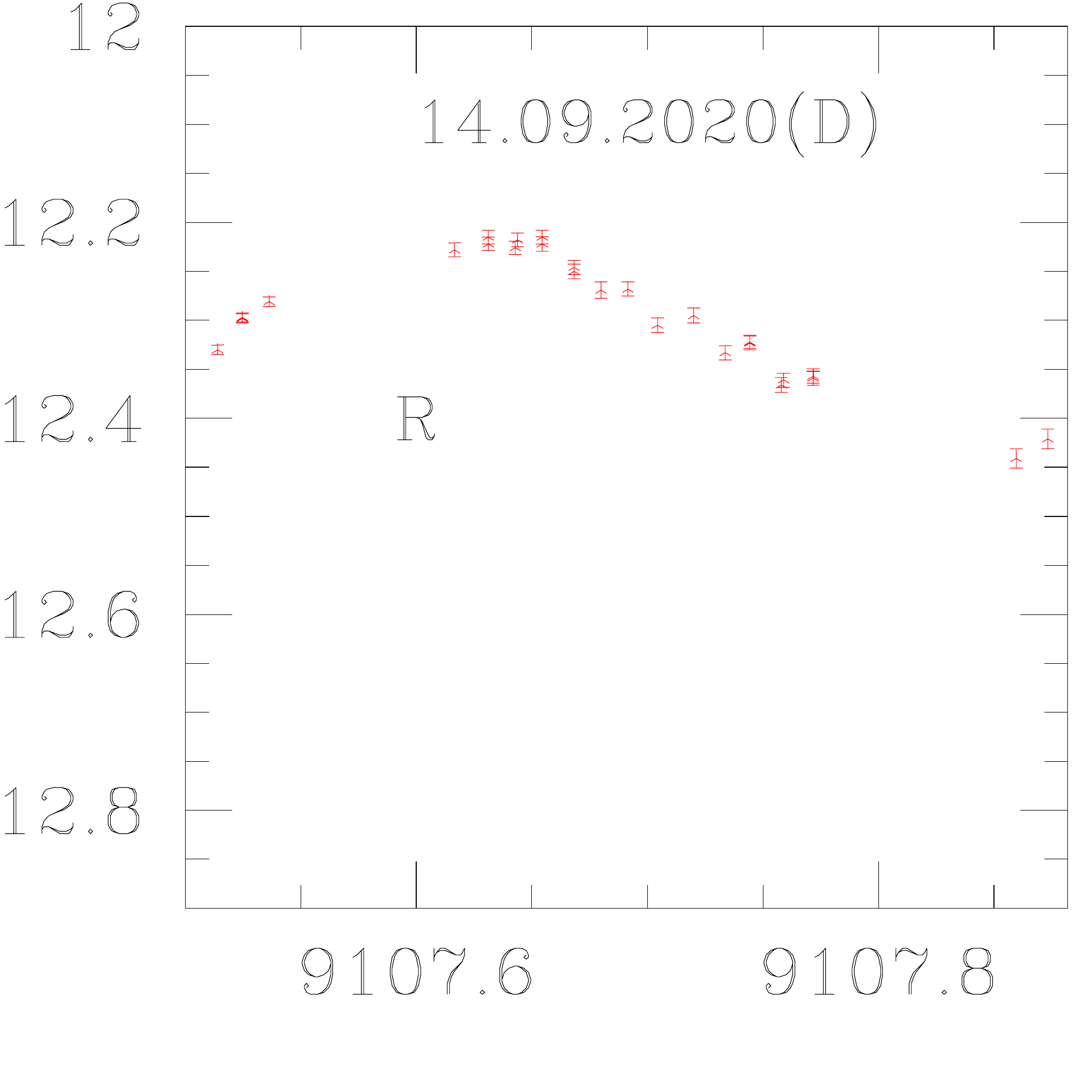}{0.25\textwidth}{}}
\caption{Continued.}
\end{figure*}

\clearpage
\startlongtable
\begin{longrotatetable}
\begin{deluxetable*}{ccccccccccc}
\centerwidetable
\tablecaption{Results from the INV tests of \bl \label{tab:tests}}
\tablewidth{0pt}
\tablehead{
\colhead{Date(Telecope)} & \colhead{Band} & \colhead{$N$} & \colhead{$C$-test} & \colhead{$F$-test} & \colhead{$\chi^{2}$-test} & \colhead{Status} & \colhead{$A$} \\
\colhead{(yyyy mm dd)} & \colhead{} & \colhead{} & \colhead{$C_{1},C_{2},C$} & \colhead{$F_{1},F_{2},F,F_{\rm c}(0.99),F_{\rm c}(0.999)$} & \colhead{$\chi^{2}_{1},\chi^{2}_{2},\chi^{2}_{\rm av},\chi^{2}_{0.99},\chi^{2}_{0.999}$} & \colhead{} & \colhead{(\%)}
}
\tabletypesize{\small}
\startdata
 2020.07.31(K)   & $R$     & 92  & 6.32, 6.62, 6.47  & 40.00, 43.89, 41.94, 1.63, 1.93  & 4955.9, 5553.5, 5254.5, 125.29, 138.44   & V &18.10 \\
 2020.08.20(A)   & $R$     & 370  & 3.61, 3.33, 3.47  & 13.14, 11.11, 12.06, 1.27, 1.38  & 18064.0, 4148.8, 11106.4, 435.12, 458.68   & V &13.39 \\
 2020.08.20(B)   & $B$     & 95  & 2.48, 2.59, 2.53  & 6.14, 6.72, 6.43, 1.62, 1.90  & 2182.1, 847.15, 1514.62, 128.80, 142.12        & V &23.15 \\
              & $V$     & 108  & 4.24, 4.31, 4.28  & 18.09, 18.62, 18.36, 1.57, 1.83  & 9901.5, 2365.8, 6133.65, 143.94, 157.95    & V &19.49 \\
              & $R$     & 118  & 3.90, 3.78, 3.84  & 15.25, 14.29, 14.77, 1.54, 1.78  & 8487.3, 1791.4, 5139.35, 155.50, 170.01    & V &19.39 \\
              & $I$     & 118  & 4.68, 5.05, 4.87  & 21.92, 25.50, 23.71, 1.54, 1.78  & 18721.0, 31986.0, 25353.5, 155.50, 170.01  & V &15.30 \\
\noalign{\smallskip}
 2020.08.21(A)   & $R$     & 327  & 2.70, 3.10, 2.90  & 7.30, 9.60, 8.45, 1.29, 1.41  & 8576.2, 3202.1, 5889.15, 388.32, 410.64    & V &10.19 \\   
 \noalign{\smallskip}
 2020.08.21(G)   & $R$     & 144  &  0.87, 1.17, 1.02 & 0.75, 1.36, 1.06, 1.48, 1.63 & 353.6, 181.7, 267.7, 185.25, 201.00   & NV & -- \\  
\noalign{\smallskip}
 2020.08.22(C)   & $R$     & 25  & 1.99, 2.24, 2.11  & 3.96, 5.01, 4.49, 2.66, 3.73 & 466.93, 159.95, 313.44, 42.98, 51.18    & PV & 6.96   \\
\noalign{\smallskip}
 2020.08.23(C)   & $R$     & 20  & 1.36, 1.80, 1.58  & 1.85, 3.24, 2.54, 3.03, 4.47 & 206.00, 83.94, 144.97, 36.19, 43.82    & NV & -- \\
 2020.08.23(E)   & $B$     & 38  & 2.70, 2.60, 2.65  & 7.28, 6.77, 7.03, 2.18, 2.84  & 1363.2, 644.12, 1003.66, 59.89, 69.35        & V &15.64 \\
              & $V$     & 40  & 2.58, 2.69, 2.63  & 6.64, 7.23, 6.94, 2.13, 2.76  & 1991.2, 745.50, 1368.35, 62.43, 72.05    & V &12.50 \\
              & $R$     & 37  & 2.43, 2.70, 2.56  & 5.91, 7.27, 6.59, 2.20, 2.89 & 2093.8, 608.01, 1350.90, 58.62, 67.98    & V &12.44 \\
              & $I$     & 36  & 2.63, 2.70, 2.66  & 6.91, 7.28, 7.10, 2.23, 2.93  & 3843.2, 788.51, 2315.85, 57.34, 66.62  & V &13.46 \\
\noalign{\smallskip}
 2020.08.24(B)   & $R$     & 18  & 0.68, 1.09, 0.88  & 0.46, 1.19, 0.83, 3.24, 4.92  & 33.26, 23.78, 28.52, 33.41, 40.79    & NV & -- \\
 2020.08.24(C)   & $R$     & 23  & 0.81, 0.98, 0.89  & 0.65, 0.95, 0.80, 2.78, 3.98  & 59.27, 19.05, 39.16, 40.29, 48.27    & NV & -- \\  
\noalign{\smallskip}
 2020.08.25(B)   & $B$     & 54  & 8.18, 7.78, 7.98  & 66.88, 60.59, 63.73, 1.91, 2.38  & 8098.8, 4149.8, 6124.3, 79.84, 90.57      & V & 57.98 \\
              & $V$     & 54  & 10.57, 10.02, 10.29  & 111.74, 100.36, 106.05, 1.91, 2.38 & 14087.0, 4962.8, 9524.9, 79.84, 90.57  & V &54.59 \\
              & $R$     & 949  & 9.87, 9.99, 9.93  &  97.41, 99.82, 98.62, 1.16, 1.22 & 282320.00, 90054.20, 186187.10, 1052.23, 1088.28 & V & 52.09\\
              & $I$     & 54  & 9.37, 8.60, 8.98  &  87.70, 73.97, 80.84, 1.91, 2.38 & 16613.00, 3338.00, 9975.5 79.84, 90.57   & V & 47.70 \\
 2020.08.25(C)   & $R$     & 12  & 2.82, 3.33, 3.08  & 8.00, 11.09, 9.53, 4.46, 7.76  & 336.15, 105.48, 220.82, 24.72, 31.26    & V & 9.27 \\
\noalign{\smallskip}
 2020.08.26(C)   & $R$     & 16  & 2.72, 3.25, 2.99 & 7.40, 10.56, 8.98,  3.52, 5.53 & 335.33, 140.81, 238.07, 30.58, 37.70 & V & 12.68\\
 2020.08.26(F)   & $B$     & 299  & 9.34, 9.02, 9.18 & 87.31, 81.33, 84.32, 1.31, 1.43 & 46939.39, 28023.80, 37481.60, 357.72, 379.17 & V & 63.76 \\
              & $V$     & 14  &   & 14.13, 13.43, 13.78, 3.91, 6.41 & 6045.40, 2331.0, 4188.2, 27.69, 34.53  & V &52.09 \\
              & $R$     & 15  & 16.92, 16.73, 16.83 & 286.47, 280.05, 283.26, 3.70, 5.93 & 10953.42, 37940.38, 24446.90, 29.14, 36.12 & V & 49.40\\
              & $I$     & 296  & 10.72, 10.69, 10.70 & 114.81, 114.30, 114.56, 1.31, 1.44 & 130575.58, 31436.29, 40502.97, 354.43, 375.79 & V & 47.89 \\
\noalign{\smallskip}
 2020.08.27(F) & $B$     & 332  & 4.05, 4.28, 4.17 & 16.43, 18.29, 17.36, 1.29, 1.41 & 10057.95, 7286.32, 8672.13, 393.78, 416.24 & V & 32.35 \\
              & $V$     & 18  & 3.96, 4.00, 3.98  & 15.71, 15.96, 15.83, 3.23, 4.92 & 676.71, 269.64, 473.17, 33.41, 40.79 & V &24.77 \\
              & $R$     & 18  & 6.93, 7.23, 7.08 & 47.98, 52.25, 50.11, 3.23, 4.92  & 2306.1, 827.23, 1566.66, 33.41, 40.79 & V & 21.78 \\
              & $I$     & 331  & 3.70, 3.83, 3.76 & 13.71, 14.64, 14.17, 1.29, 1.41  & 17955.38, 4469.69, 11212.53, 392.69, 415.12 & V & 22.68 \\     
\noalign{\smallskip}
 2020.08.28(B)   & $R$   & 152 & 1.41, 1.24, 1.32 & 1.99, 1.53, 1.76, 1.46, 1.66 & 227.69, 940.44, 584.06, 194.34, 210.44 & NV & -- \\
 2020.08.28(C)   & $R$   & 10 & 4.42, 4.27, 4.35 & 19.52, 18.20, 18.86 5.35, 10.11 & 651.35, 144.66, 398.01, 21.67, 27.88 & V & 12.57 \\
 2020.08.28(F)   & $B$    & 328  & 4.69, 4.41, 4.55 & 21.97, 19.41, 20.69, 1.29,  1.41  & 13419.00, 7752.80, 10585.90, 389.42, 411.76 & V & 40.00 \\
              & $V$ & 18 & 8.05,  7.77,  7.91 &  64.76,  60.38,  62.57,  3.24,  4.92 & 25557.00,  10055.00,  17806.00,  33.41,  40.79 & V & 24.97 \\
              & $R$ & 18 & 7.57,  7.57,  7.57 &  57.34,   57.28,  57.31,  3.24,  4.92 & 2567.80,  907.91,  1737.86,  33.41,  40.79 & V &  23.39 \\
              & $I$ & 348 & 3.63,  3.83,  3.73 & 13.21,  14.64,  13.93,  1.28,  1.39 & 17354.00,  4375.70,  10864.85, 411.21,  434.14 & V & 23.98 \\
\noalign{\smallskip}
 2020.08.29(D)   & $R$ &  23 & 2.32,  2.90,  2.61 &  5.38,  8.38,  6.88,  2.78,  3.98 & 441.00,  168.93, 304.97,  40.29,  48.27 & V & 12.79 \\
\noalign{\smallskip}
 2020.08.30(K)   & $R$ &  518 & 1.16, 1.36, 1.26 & 1.35, 1.86, 1.60, 1.23, 1.31 & 5849.1, 935.11, 3392.1, 594.73, 622.09 & PV & 7.59 \\
\noalign{\smallskip}
 2020.08.31(K) & $V$ & 77 & 3.00, 3.20, 3.10 &  8.98, 10.27, 9.62, 1.71, 2.05 & 3441.5, 766.02, 2103.8, 107.58, 119.85 & V & 11.59\\
               & $R$ &  75 & 3.53, 3.53, 3.53 & 12.43, 12.43, 12.43, 1.73, 2.07 & 6268.8, 871.79, 3570.0, 105.20, 117.35 & V & 9.59 \\
              & $I$ &  73 & 2.38, 2.74, 2.56 &  5.65, 7.51, 6.58, 1.74, 2.09 & 4278.14, 517.97, 2398.05, 102.82, 114.83 & V & 10.09 \\
\noalign{\smallskip}
 2020.09.02(B)   & $B$ & 30 & 0.42,  0.92,  0.67 & 0.18,  0.86,  0.52,  2.42,  3.29 & 15.61,  26.43,  21.02,  49.59,  58.30 & NV &  -- \\
              & $V$  & 30 &  0.94,  1.13,  1.04 &  0.89,  1.27,  1.08, 2.42,  3.29 & 89.05,  34.49,  61.77,  49.59,  58.30 & NV & -- \\
              & $R$  & 1000 & 2.33,  2.70,  2.52 & 5.45,  7.28,  6.37,  1.16, 1.22 & 20214.00,  7039.00,  13626.50,  1105.92, 1142.85 & V & 11.88 \\
              & $I$ & 30 & 0.66,  1.01,  0.84 & 0.43,  1.02,  0.73,  2.42,  3.29 & 58.15,  28.05,  43.10,  49.59,  58.30 & NV & -- \\
\noalign{\smallskip}
 2020.09.03(A)   & $B$     & 45  & 2.87, 3.11, 2.99  & 8.22, 9.70, 8.96, 2.04, 2.60  & 1150.3, 617.43, 883.86, 68.71, 78.75    & V & 19.56 \\
              & $V$     & 39  & 2.67, 2.50, 2.59  & 7.16, 6.27, 6.71, 2.16, 2.80 & 601.69, 197.10, 399.39, 61.16, 70.70  & V & 15.36 \\
              & $R$     & 27  & 1.51, 1.49, 1.50  &  2.28, 2.23, 2.25, 2.55, 3.53 & 178.76, 50.45, 114.60, 45.64, 54.05  & PV & 7.86 \\
              & $I$     & 28  & 1.52, 1.96, 1.73   &  2.30, 3.82, 3.06, 2.51, 3.44 & 221.04, 82.88, 151.96, 46.96, 55.48   & PV & 6.87 \\
 2020.09.03(B)   & $B$  & 13 & 2.54,  3.15, 2.85 & 6.43, 9.91, 8.17, 4.15, 7.00 & 159.97, 119.12, 139.55, 26.22, 32.91 & V &  13.63 \\
              & $V$ & 17 & 3.75,  4.07,  3.91 &  14.09,  16.55,  15.32, 3.37,  5.21 & 1113.30,  413.97,  763.64,  32.00,  39.25 & V & 10.51 \\
              & $R$  & 553  & 2.85,  3.01,  2.93 & 8.13,  9.08,  8.61, 1.22,  1.30 & 1588.70,  5406.10,  3497.40,  632.22,  660.40 & V & 11.68 \\
              & $I$ & 15 & 2.40, 2.82, 2.61 & 5.74,  7.95,  6.85, 3.70,  5.93 & 377.34,  114.85,  246.10,  29.14,  36.12 & V &  6.77 \\
\noalign{\smallskip}
 2020.09.05(D)   & $B$     & 25 & 2.63, 2.73, 2.68 &  6.90, 7.43, 7.17, 2.60, 3.73 & 407.17, 260.16, 333.67, 42.98, 51.18 & V & 24.22\\
              & $V$     & 25 & 4.21, 4.91, 4.56 & 17.76, 24.15, 20.96, 2.66, 3.73 & 1145.60, 599.41, 872.51, 42.98, 51.18 & V & 14.78 \\
              & $R$     & 46  & 4.78, 5.22, 5.00  & 22.86, 27.28, 25.07, 2.02, 2.57 & 3506.20, 1310.00, 2408.10, 69.96, 80.08 & V & 14.09\\
\noalign{\smallskip}
 2020.09.06(D)   & $R$   & 87  & 3.34, 3.25, 3.30  & 11.18, 10.56, 10.87 1.66, 1.96  & 3189.50, 869.83, 2029.67, 119.41, 132.28  & V  & 12.19 \\
\noalign{\smallskip}
 2020.09.07(A)   & $B$     & 30 & 1.62, 2.04, 1.83 & 2.62, 4.16, 3.39, 2.42, 3.29 & 162.77, 144.31, 153.54, 49.59, 58.30 & PV & 8.60 \\
              & $V$     & 29 & 2.18, 2.73, 2.46 & 4.75, 7.43, 6.09, 2.46, 3.36 & 331.61, 207.57, 269.59, 48.28, 56.89 & PV & 9.47 \\
              & $R$     & 24 & 2.09, 2.82, 2.46 & 4.35, 7.94, 6.15, 2.72, 3.85 & 290.08, 170.85, 230.47, 41.64, 49.73 & PV & 8.08\\
              & $I$    & 30 & 1.89, 2.33, 2.11 & 3.56, 5.45, 4.51, 2.42, 3.29 & 396.10, 151.15, 273.63, 49.59, 58.30 & PV & 7.28 \\
 2020.09.07(D)    & $R$     & 34 & 5.22, 5.41, 5.32 & 27.22, 29.25, 28.24, 2.29, 3.04 & 2622.70, 814.59, 1718.65, 54.77, 63.87 & V & 17.09\\
\noalign{\smallskip}
 2020.09.08(A)   & $B$    & 32 & 2.85, 3.30, 3.08  &  8.14, 10.91, 9.53, 2.35, 3.15 &  504.54, 381.87, 443.21, 52.19, 61.10 & V & 15.33\\
              & $V$     & 33 & 4.01, 3.47, 3.74 &  16.09, 12.02, 14.06, 2.32, 3.09 & 1074.60, 361.48, 718.04, 53.48, 62.49 & V & 13.67 \\
              & $R$     & 32  & 2.88, 2.28, 2.58  & 8.27, 5.20, 6.73, 2.35, 3.15 & 615.64, 143.07, 379.36, 52.19, 61.10 & V & 12.78 \\
              & $I$     & 30 & 2.79, 2.30, 2.55 & 7.51, 4.46, 5.99, 2.42, 3.29 & 684.74, 117.63, 401.19, 49.59, 58.30 & V & 9.98 \\
 2020.09.08(B)   & $R$     &  236 & 3.58, 3.58, 3.58 & 12.78, 12.84, 12.81, 1.36, 1.50 & 6944.60, 2597.40, 4771.00, 288.35, 307.73 & V & 8.98\\
\noalign{\smallskip}
 2020.09.09(A)   & $B$     & 43  & 2.05, 1.68, 1.86  & 4.20, 2.78, 3.50, 2.07, 2.66  & 315.49, 127.60, 221.54, 66.21, 76.08  & PV & 8.90 \\
              & $V$     & 45  & 1.44, 1.06, 1.25  & 2.07, 1.13, 1.60, 2.04, 2.60 & 177.93, 44.32, 111.12, 68.71, 76.08  & NV & -- \\
              & $R$     & 44  & 1.71, 1.25, 1.48  &  2.94, 1.56, 2.25, 2.06, 2.63 & 284.56, 58.14, 171.35, 67.46, 77.42  & NV & -- \\
              & $I$     & 45  & 1.55, 1.18, 1.36  & 2.41, 1.38, 1.90,  2.04, 2.60  & 320.68, 53.79, 187.23 68.71, 76.08   & NV & -- \\
 2020.09.09(B)   & $R$     & 508 & 2.98, 3.02, 3.00 & 8.91, 9.13, 9.02, 1.23, 1.32 & 10540.00, 4086.20, 7313.10, 584.01, 611.13 & V & 7.88\\
\noalign{\smallskip}
 2020.09.10(A)   & $B$     & 49  & 5.11, 5.48, 5.29  & 26.09, 29.98, 28.03, 1.98, 2.49  & 2667.50, 1764.70, 2216.1, 73.68, 84.04 & V & 27.86 \\
              & $V$     & 50  & 10.12, 9.59, 9.85  & 102.41, 91.96, 97.18, 1.96, 2.46  & 11515.00, 4402.5, 7958.5, 74.92, 85.35  & V & 27.39 \\
              & $R$     & 217  & 11.18, 10.85, 11.02 & 125.05, 117.74, 121.40, 1.37, 1.52  & 65010.00, 22353.00, 43681.5, 267.27, 285.96  & V & 26.60 \\
              & $I$     & 54  & 6.75, 5.88, 6.31  & 45.57, 34.53, 40.05, 1.91, 2.38  & 7885.6, 1645.50, 4765.55, 79.84, 90.57  & V & 26.60 \\
 2020.09.10(B)   & $R$     & 187 & 2.00, 1.90, 1.95  & 4.0, 3.62, 3.81, 1.41, 1.58 &    1979.90, 620.30, 1300.1, 251.34, 233.79 & PV & 5.87\\
 2020.09.10(K)   & $R$     & 206 & 6.39, 6.54, 6.52  & 40.83, 44.14, 42.49, 1.39, 1.55 &    48841.00, 11330.00, 30086.0, 252.79, 271.00 & V & 21.60\\
\noalign{\smallskip}
 2020.09.11(A)   & $B$     & 242  & 7.91, 7.53, 7.72  & 62.57, 56.68, 59.62, 1.35, 1.49  & 31244.00, 16181.00, 23712.5, 294.99, 314.58  & V & 45.28 \\
              & $V$     & 59  & 14.87, 15.37, 15.12  & 221.04, 236.16, 228.60, 1.86, 2.18 & 30647.00, 13225.00, 21936, 85.95, 97.04  & V & 34.79 \\
              & $R$     & 62  & 9.22, 10.05, 9.64  &  85.02, 101.09, 93.05, 1.83, 2.24 & 14464.00, 5626.00, 10045.0, 89.59, 100.89   & V & 29.80 \\
              & $I$     & 61  & 7.93, 8.82, 8.38  & 62.96, 77.73, 70.35, 1.84, 2.25  & 13822.0, 4266.9, 9044.45, 88.38, 99.61   & V & 27.80 \\
 2020.09.11(B)   & $R$     & 573 & 13.72, 13.58, 13.65  & 188.32, 184.30, 86.31, 1.21, 1.29 &  338540, 100030, 219285.0, 653.61, 682.24 & V & 30.90 \\
\noalign{\smallskip}
 2020.09.12(A)   & $B$     & 41  & 0.99, 1.26, 1.13  & 0.99, 1.60, 1.29, 2.11, 2.73  & 65.51, 68.46, 66.98, 63.69, 73.40  & NV & -- \\
              & $V$     & 12  & 4.48, 5.23, 4.86  & 20.12, 27.41, 23.76, 4.46, 7.76  & 416.35, 265.35, 340.85, 24.72, 31.26  & V & 11.77 \\
              & $R$     & 370  & 5.10, 3.37, 5.24  &  26.06, 28.81, 27.44, 1.27, 1.38  & 22405.00, 9697.10, 16051.05, 435.12, 458.68  & V & 14.58 \\
              & $I$     & 11  & 3.82, 4.57, 4.20  & 14.59, 20.93, 17.76, 4.85, 8.75  & 421.66, 182.39, 302.02, 23.21, 29.59   & V & 10.19 \\  
 2020.09.12(B) & $B$    & 56  & 0.64, 1.04, 0.84  & 0.41, 1.09, 0.75, 1.89, 2.34  & 47.20, 47.71, 47.45, 82.29, 93.17  & NV & -- \\
 2020.09.12(C) & $R$    & 17  & 2.99, 3.38, 3.18  & 8.92, 11.44, 10.18, 3.37, 5.21 & 432.23, 153.43, 292.83, 32.00, 39.25  & V & 13.68 \\
 2020.09.12(K) & $R$    & 241  & 4.00, 3.64, 3.82  & 16.01, 13.24, 14.63, 1.35, 1.50 & 2826.2, 14873.0, 8849.5, 291.68, 311.15 & V & 12.60 \\
\noalign{\smallskip}
 2020.09.13(A)   & $B$     & 14  & 1.29, 2.23, 1.76  & 1.67, 4.97, 3.32, 3.90, 6.41  & 53.50, 77.75, 65.62, 27.69, 34.53  & NV & -- \\
              & $V$     & 232  & 4.63, 4.57, 4.60  & 21.41, 20.86, 21.13, 1.36, 1.50  & 11959.00, 4666.40, 8312.7, 289.92, 303.16  & V & 18.98 \\
              & $R$     & 18  & 3.57, 3.36, 3.46  &  12.71, 11.29, 12.00, 3.24, 4.92   & 215.50, 764.95, 490.22, 33.41, 40.79   & V & 13.28 \\
              & $I$     & 221  & 3.23, 2.79, 3.01  & 10.45, 7.79, 9.12, 1.37, 1.52  & 8438.00, 1750.70, 5094.35, 271.72, 290.56  & V & 15.19 \\  
 2020.09.13(B) & $R$     & 376  & 3.39, 3.44, 3.41  & 11.47, 11.83, 11.65, 1.27, 1.38  & 17804.00, 5655.20, 11729.6, 441.63, 465.36  & NV & -- \\
 2020.09.13(K) & $R$     & 233  & 3.38, 3.90, 3.64  & 11.45, 15.18, 13.32, 1.36, 1.50  & 14482.00, 3359.80, 8921.0, 285.03, 304.30 & V & 9.00 \\
\noalign{\smallskip}
 2020.09.14(A)   & $B$     & 18  & 5.91, 6.25, 6.08  & 34.90, 39.05, 36.98, 3.24, 4.92  & 1214.50, 724.54, 969.52, 33.41, 40.79  & V & 24.15 \\
              & $V$     & 224 & 9.22, 9.08, 9.15  & 85.03, 82.42, 83.73, 1.24, 1.33  & 44349.00, 16836.00, 30592.5, 275.05, 294.00  & V & 31.89 \\
              & $R$     & 18  & 8.67, 8.50, 8.59  & 75.16, 72.30, 73.73, 3.24, 4.92 & 3539.40, 1061.70, 2300.55, 33.41, 40.79   & V & 24.29 \\
              & $I$     & 223 & 7.12, 6.86, 7.02  & 51.67, 47.05, 49.36, 1.25, 1.34  & 40396.00, 9137.55, 24766.75, 513.01, 538.50  & V & 25.70 \\  
 2020.09.14(B) & $R$     & 710  & 9.69, 10.17, 9.93  & 93.87, 103.47, 98.67, 1.19, 1.26  & 192840.0, 63176.0, 128008.0, 799.53, 831.09  & V & 28.60 \\
 2020.09.14(C) & $R$     & 13  & 5.93, 5.64, 5.78  & 35.08, 31.82, 33.45, 4.16, 7.00  & 2339.10, 367.92, 1353.51, 26.22, 32.91  & V & 22.27 \\
 2020.09.14(D) & $R$     & 26  & 3.44, 3.58, 3.51  & 11.82, 12.84, 12.33, 2.60, 3.63  & 962.49, 289.91, 626.20, 44.31, 52.62  & V & 22.58 \\
\noalign{\smallskip}
\enddata
\tablecomments{Variability status is abbreviated as follows: V~-- variable, PV~-- probably variable, NV~-- non-variable.}
\end{deluxetable*}
\end{longrotatetable}

\begin{figure*}[t!]
\gridline{\fig{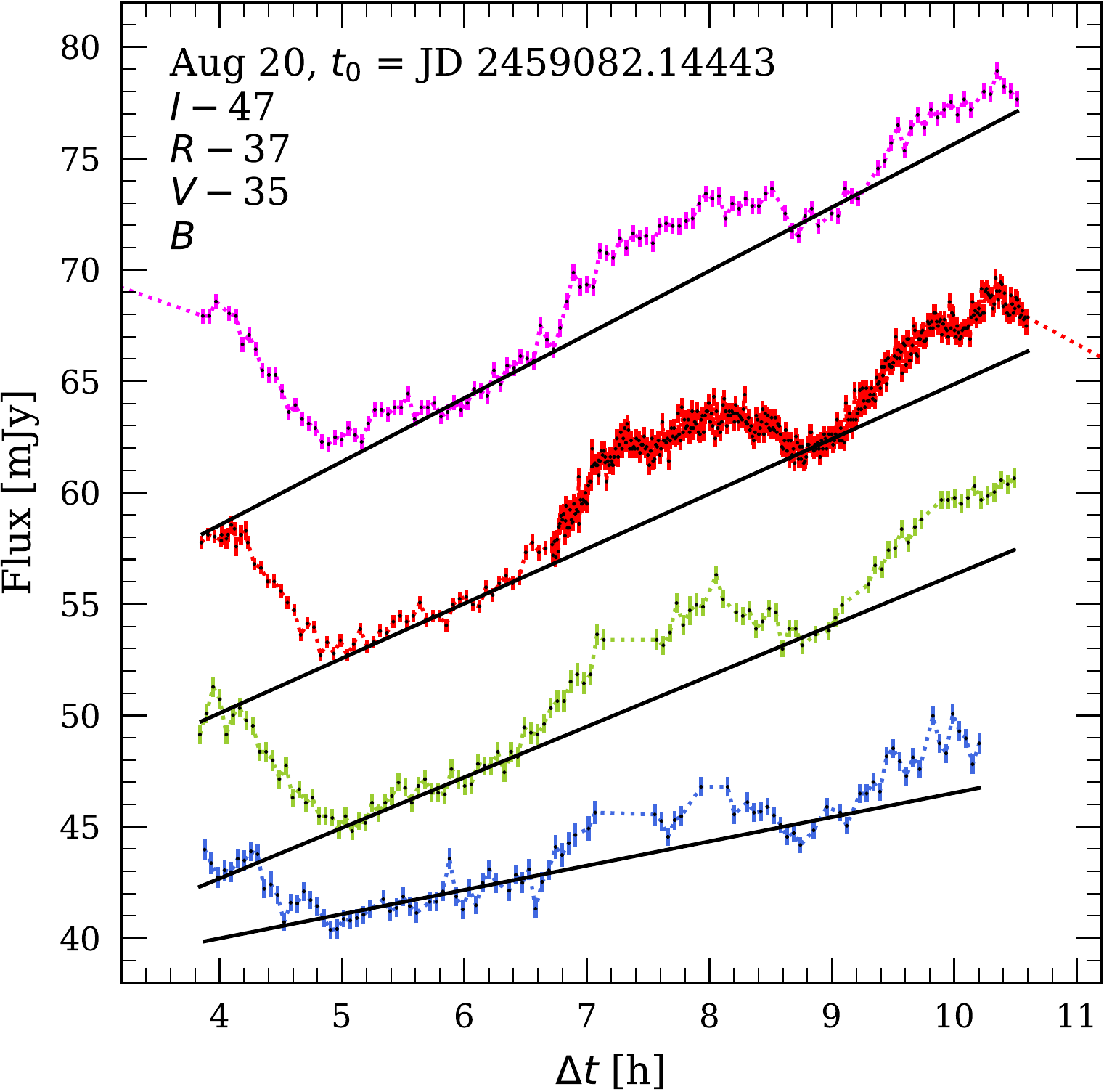}{0.315\textwidth}{}
          \fig{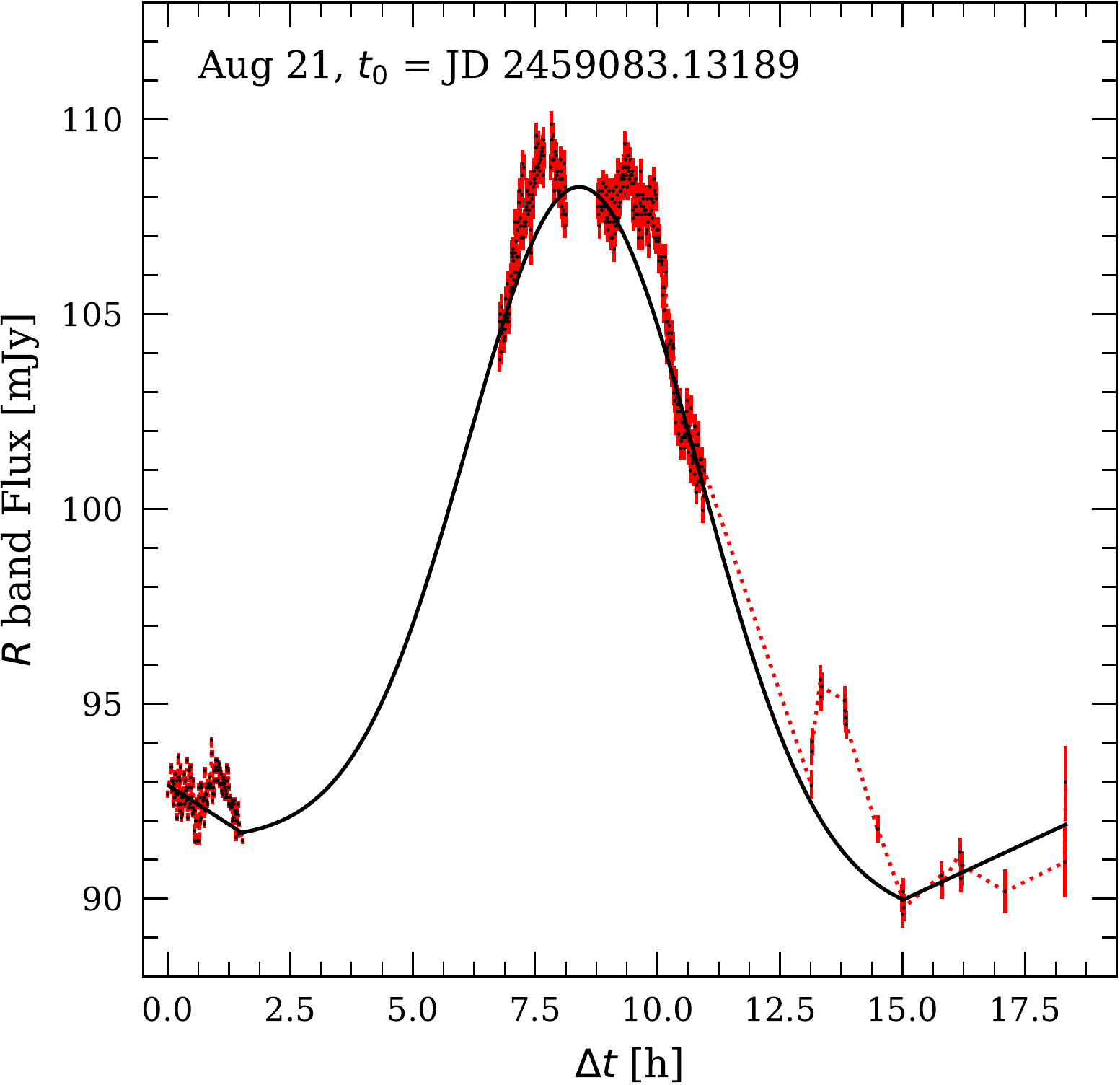}{0.315\textwidth}{}
          \fig{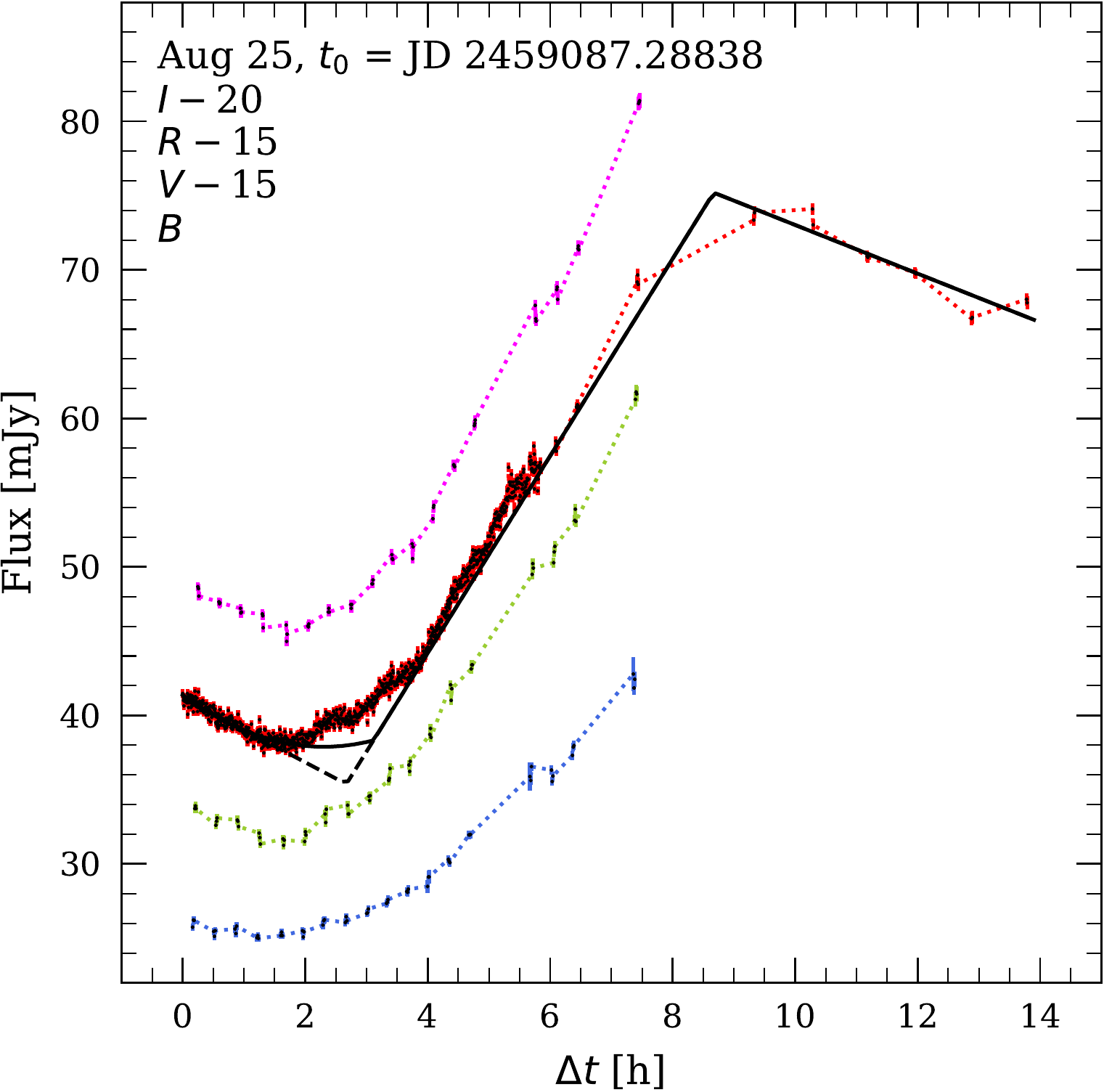}{0.315\textwidth}{}}
\gridline{\fig{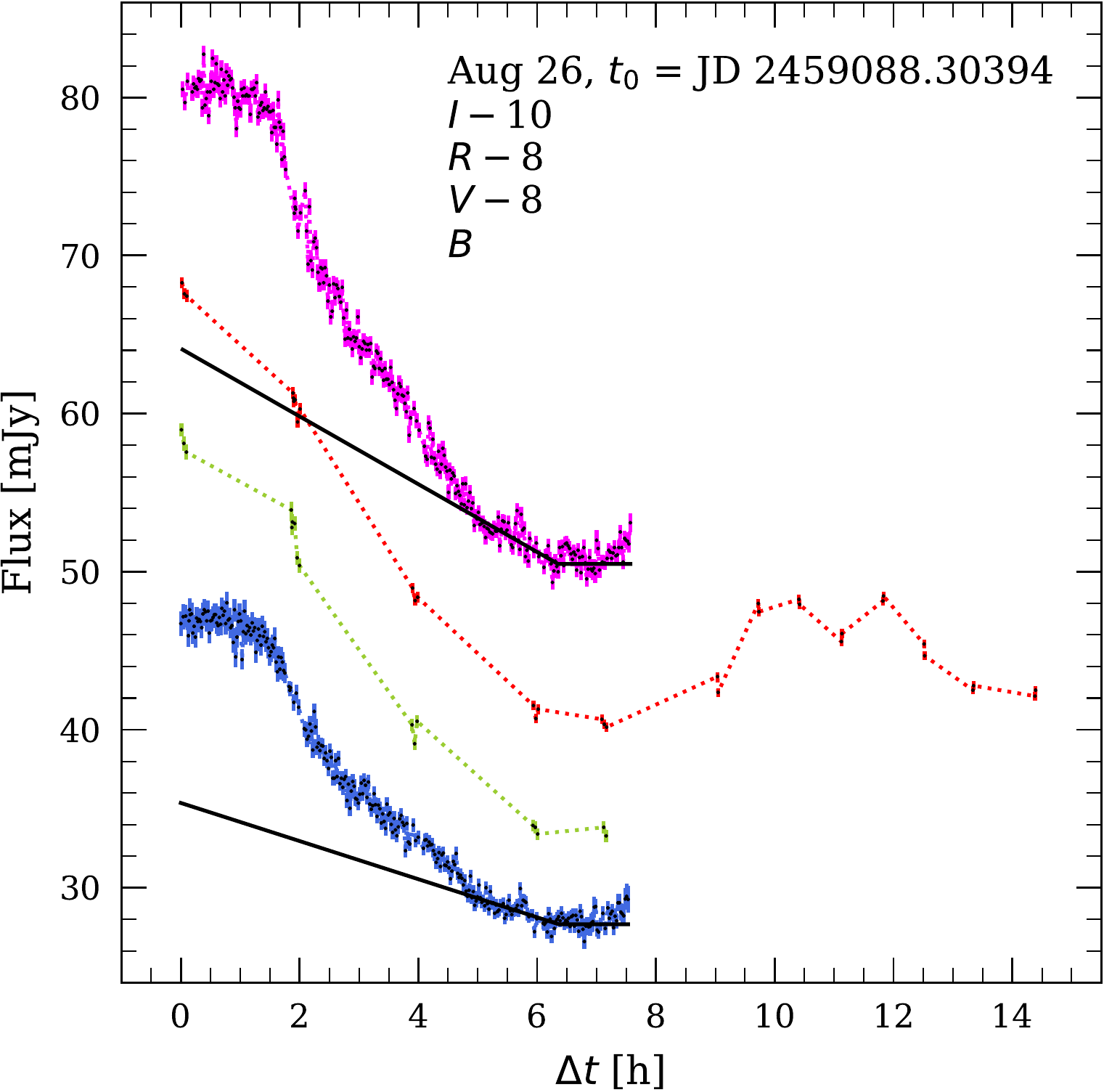}{0.315\textwidth}{}
          \fig{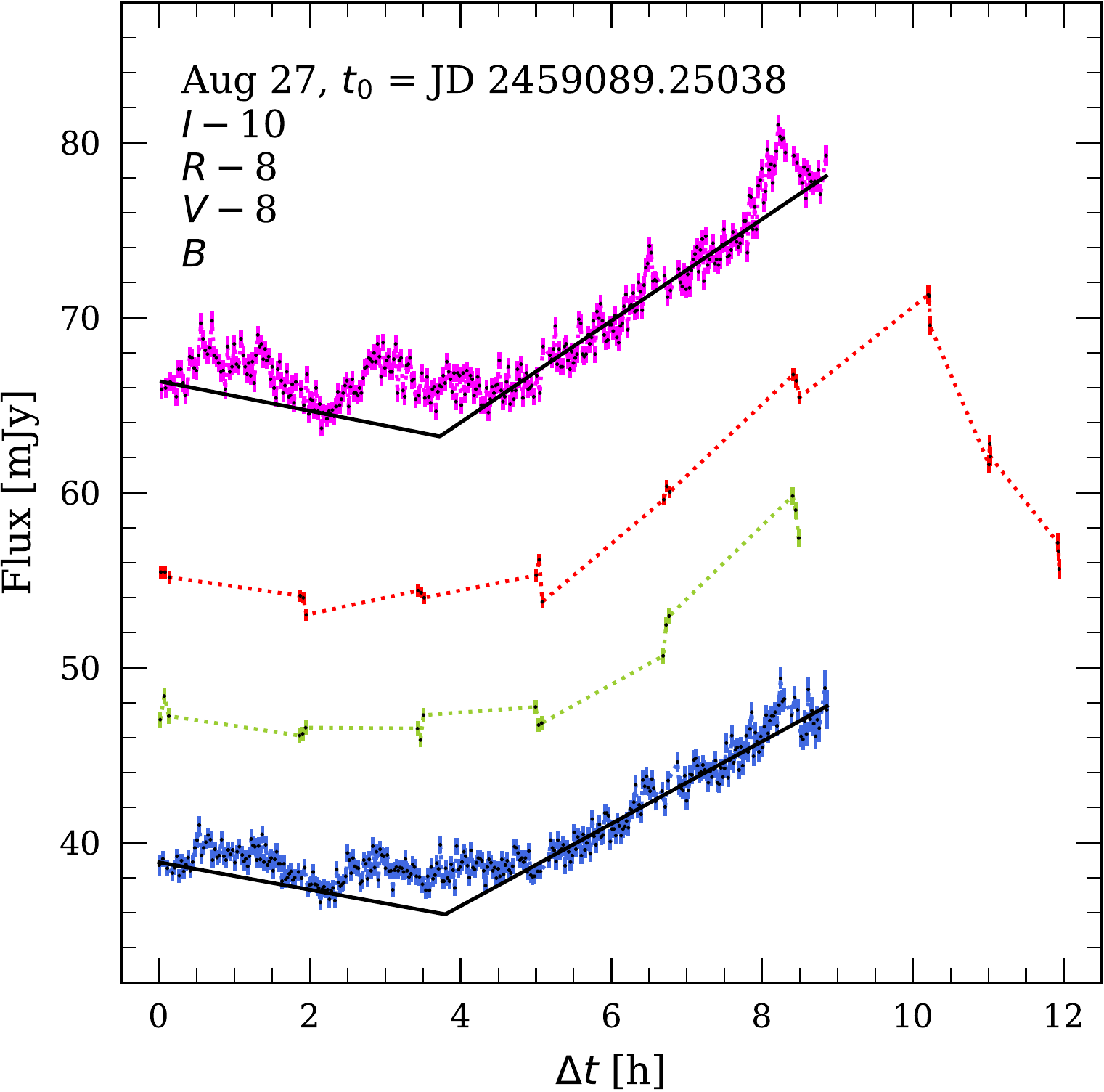}{0.315\textwidth}{}
          \fig{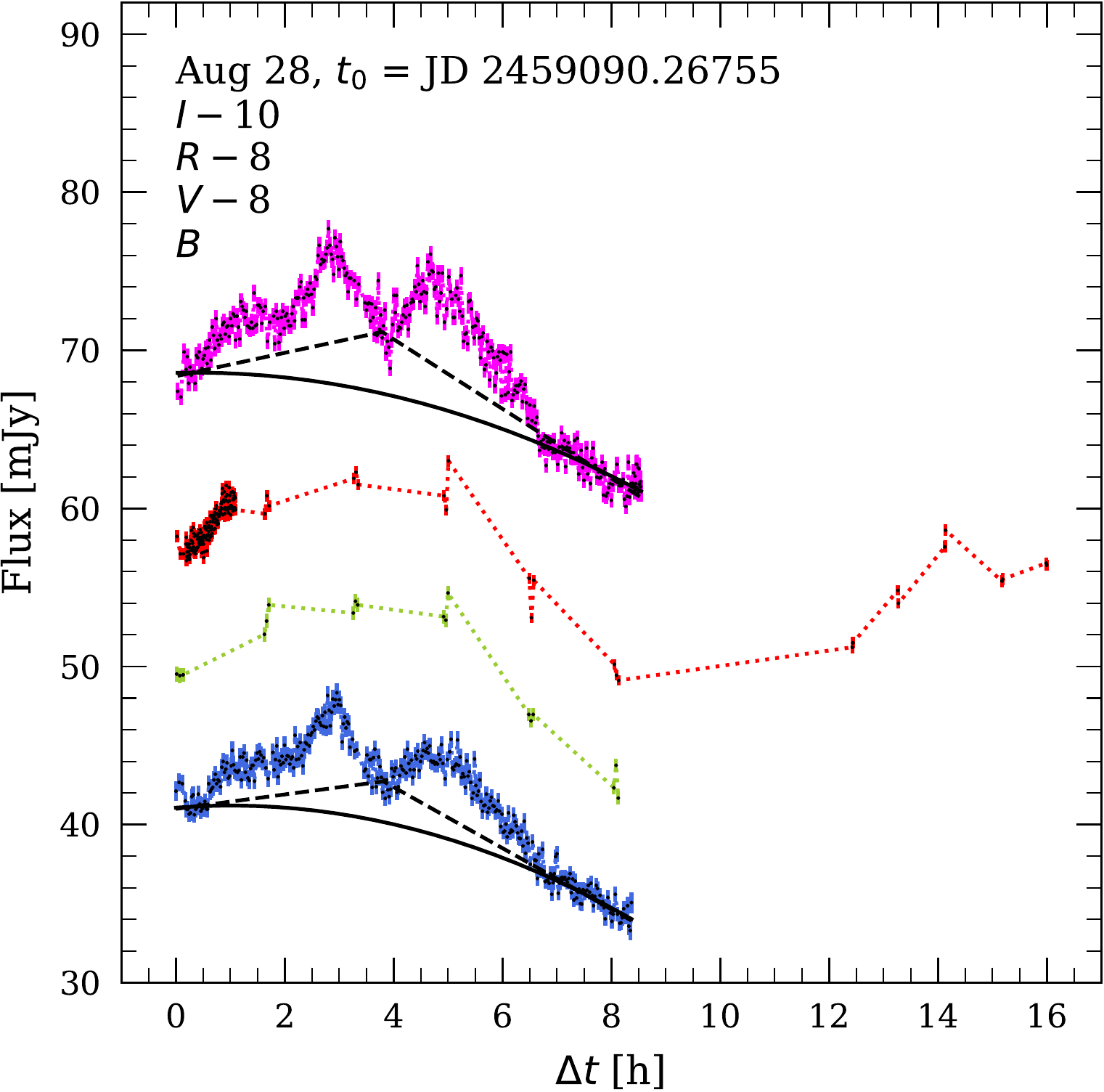}{0.315\textwidth}{}}
\gridline{\fig{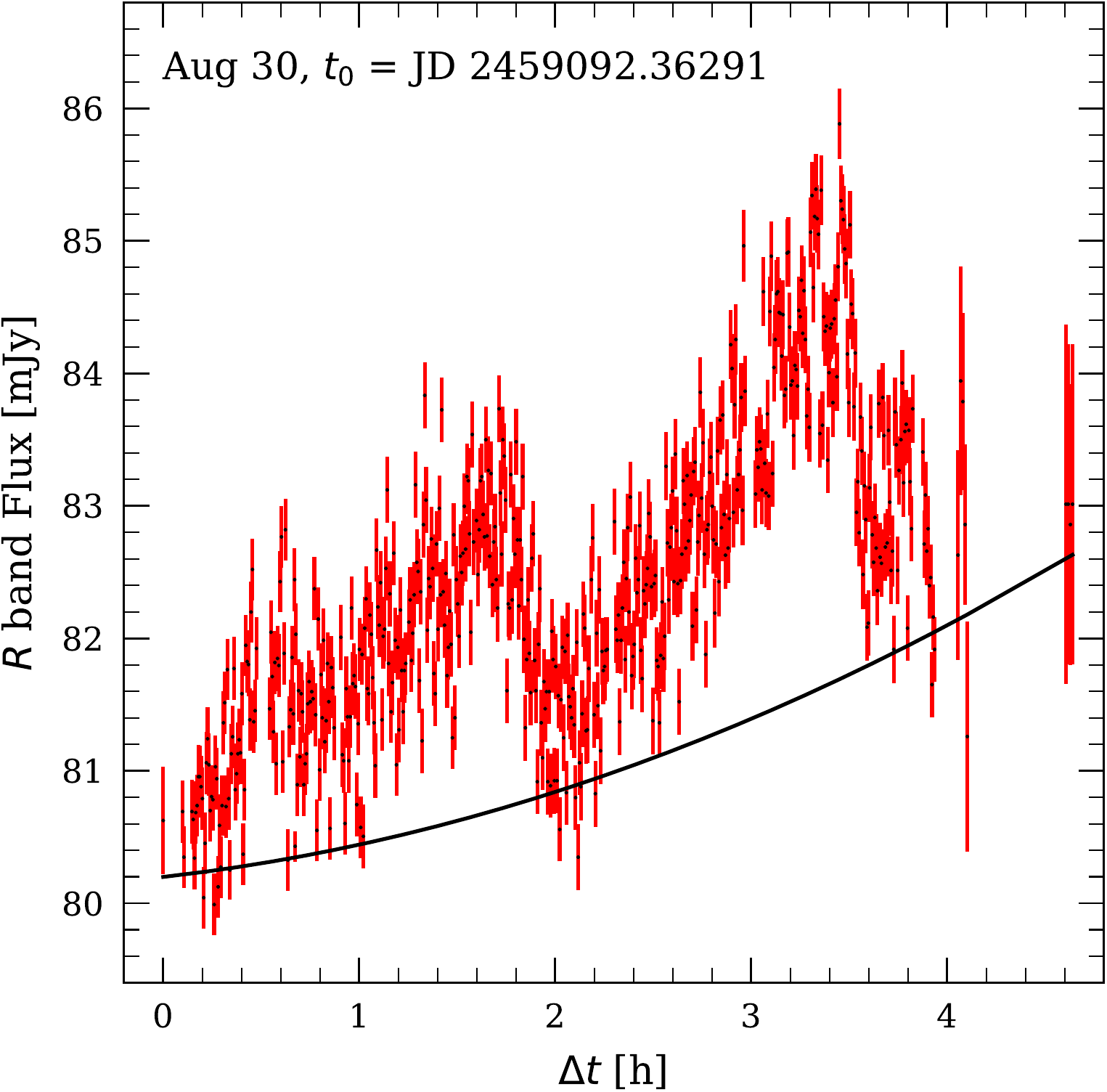}{0.315\textwidth}{}
          \fig{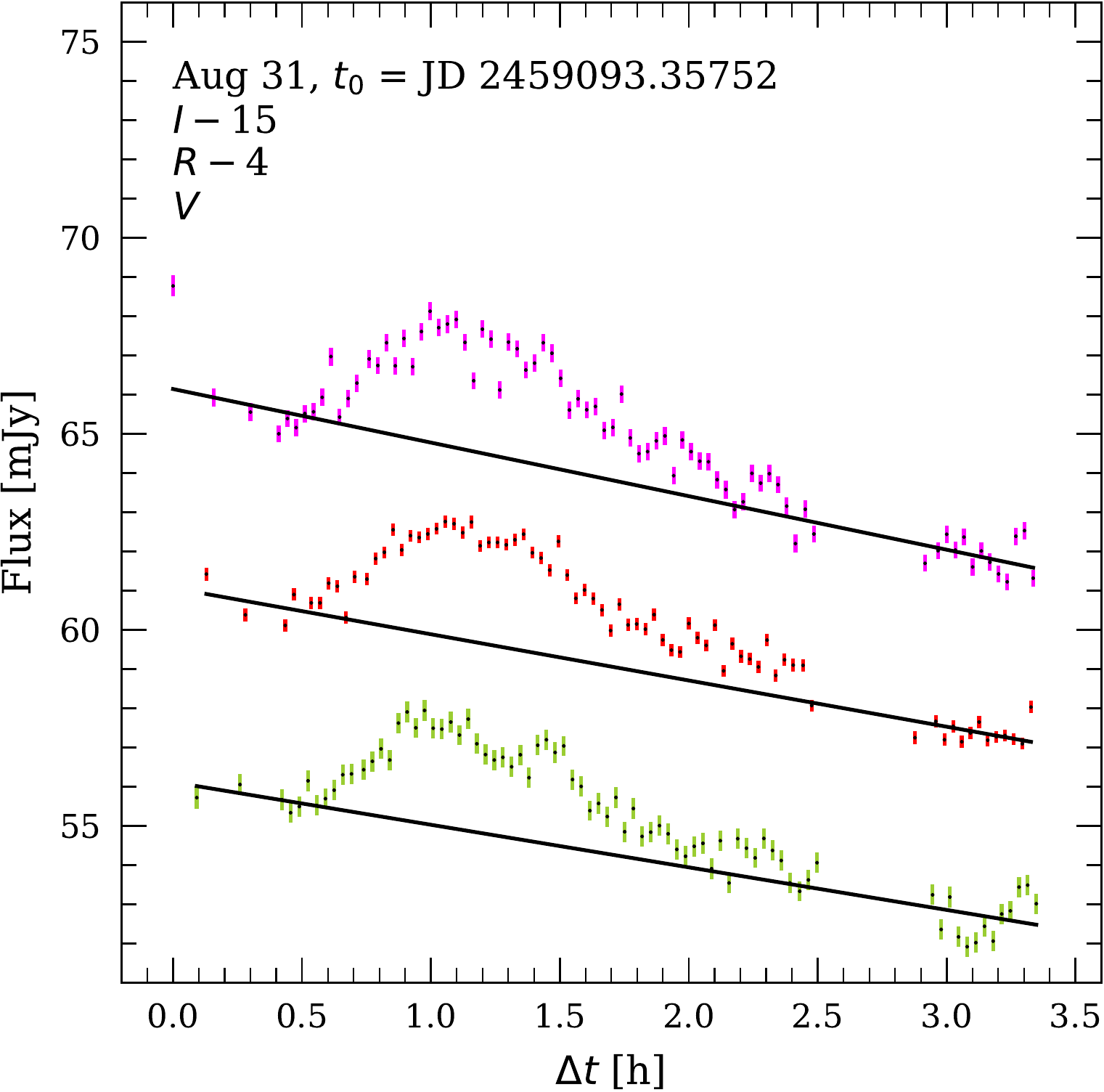}{0.315\textwidth}{}
          \fig{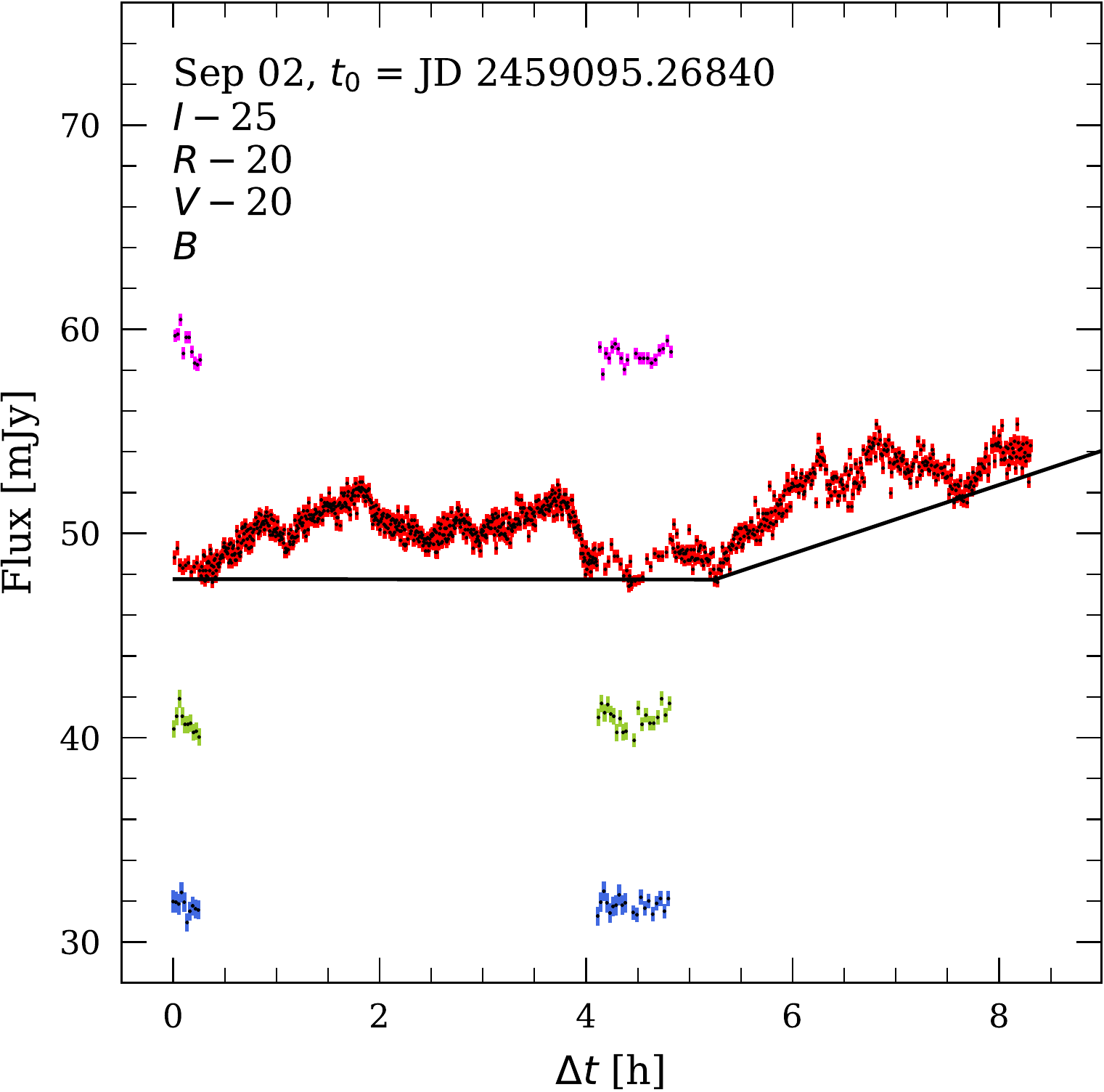}{0.315\textwidth}{}}
\caption{Combined $BVRI$ band LCs for the nights of intra-night flaring activity. In each plot, the LCs are ordered from bottom to top as follows: $B$ band~-- blue, $V$ band~-- green, $R$ band~-- red, $I$ band~-- magenta. The $(V)RI$ band LCs are shifted for display purposes downward by the corresponding offsets indicated in the plots. The (composite) fitting functions, used to detrend the corresponding LCs, are overplotted (black solid lines); the alternative fitting functions are plotted as black dashed lines (see text). For the sake of clarity, (i) if a few data points are available for a given band, then its LC is not shown and (ii) only the portion of the LC used in the decomposition is shown.}
\label{fig:inlc:comb}
\end{figure*}

\setcounter{figure}{6}
\begin{figure*}[t!]
\gridline{\fig{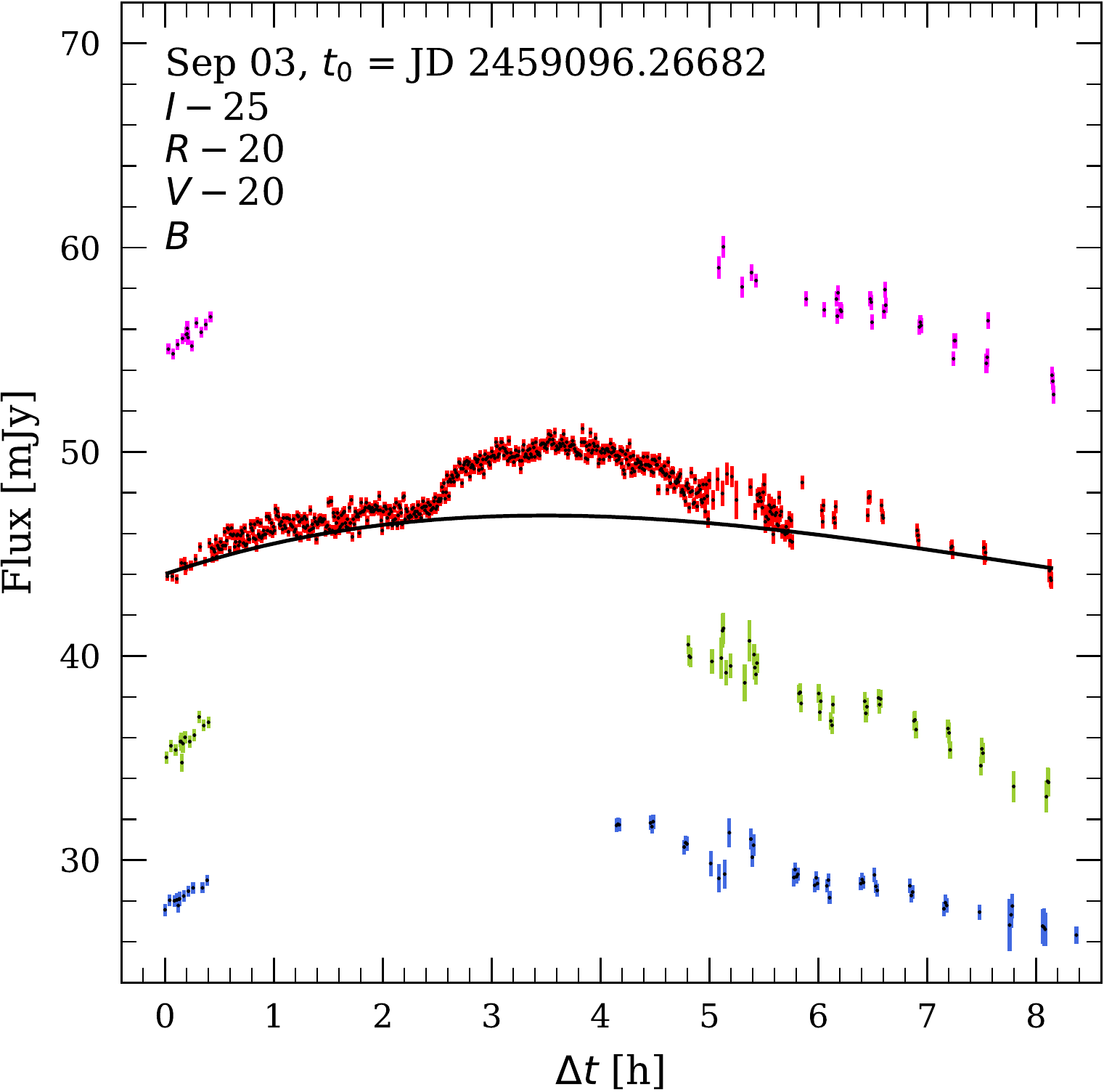}{0.315\textwidth}{}
          \fig{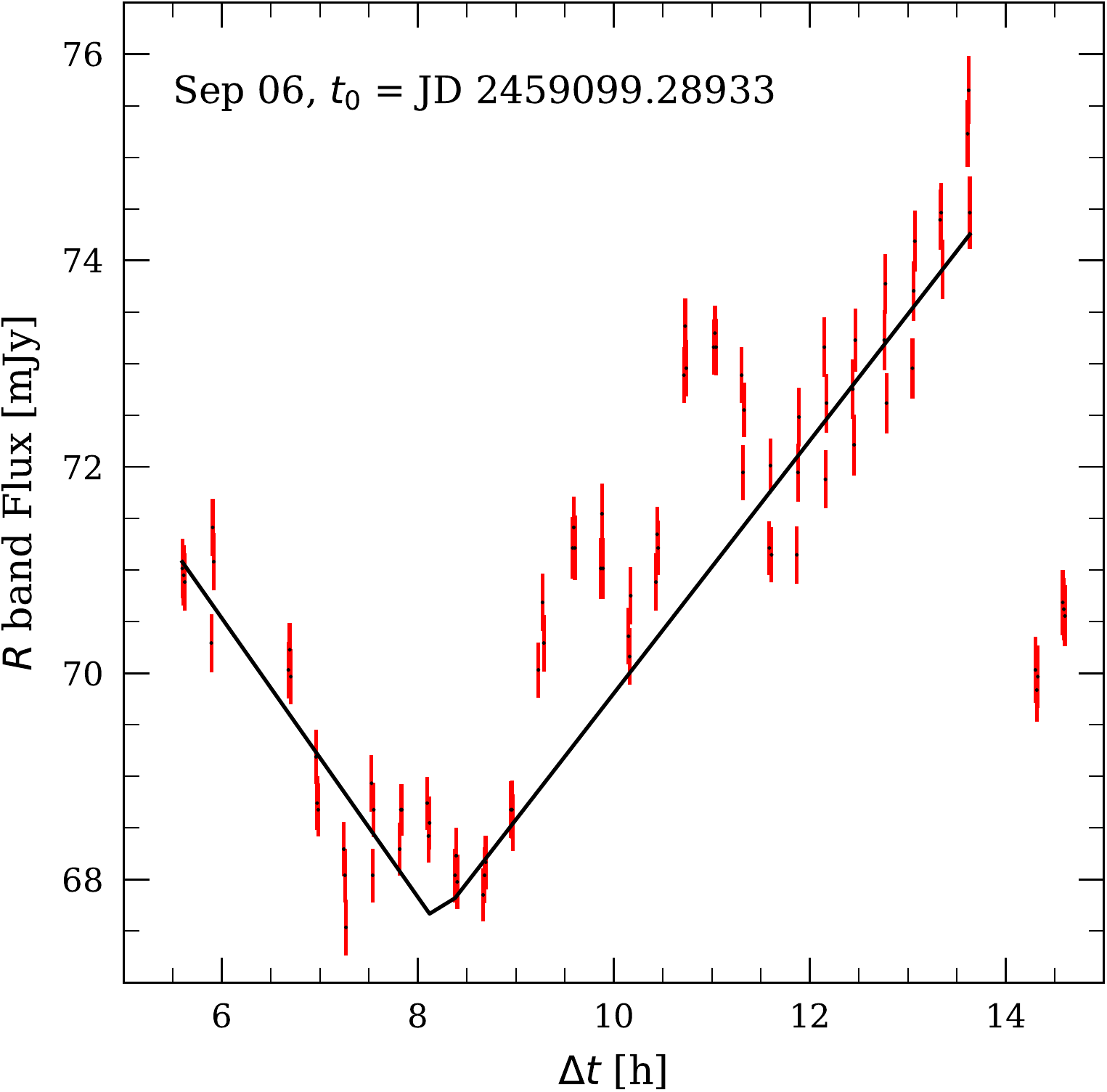}{0.315\textwidth}{}
          \fig{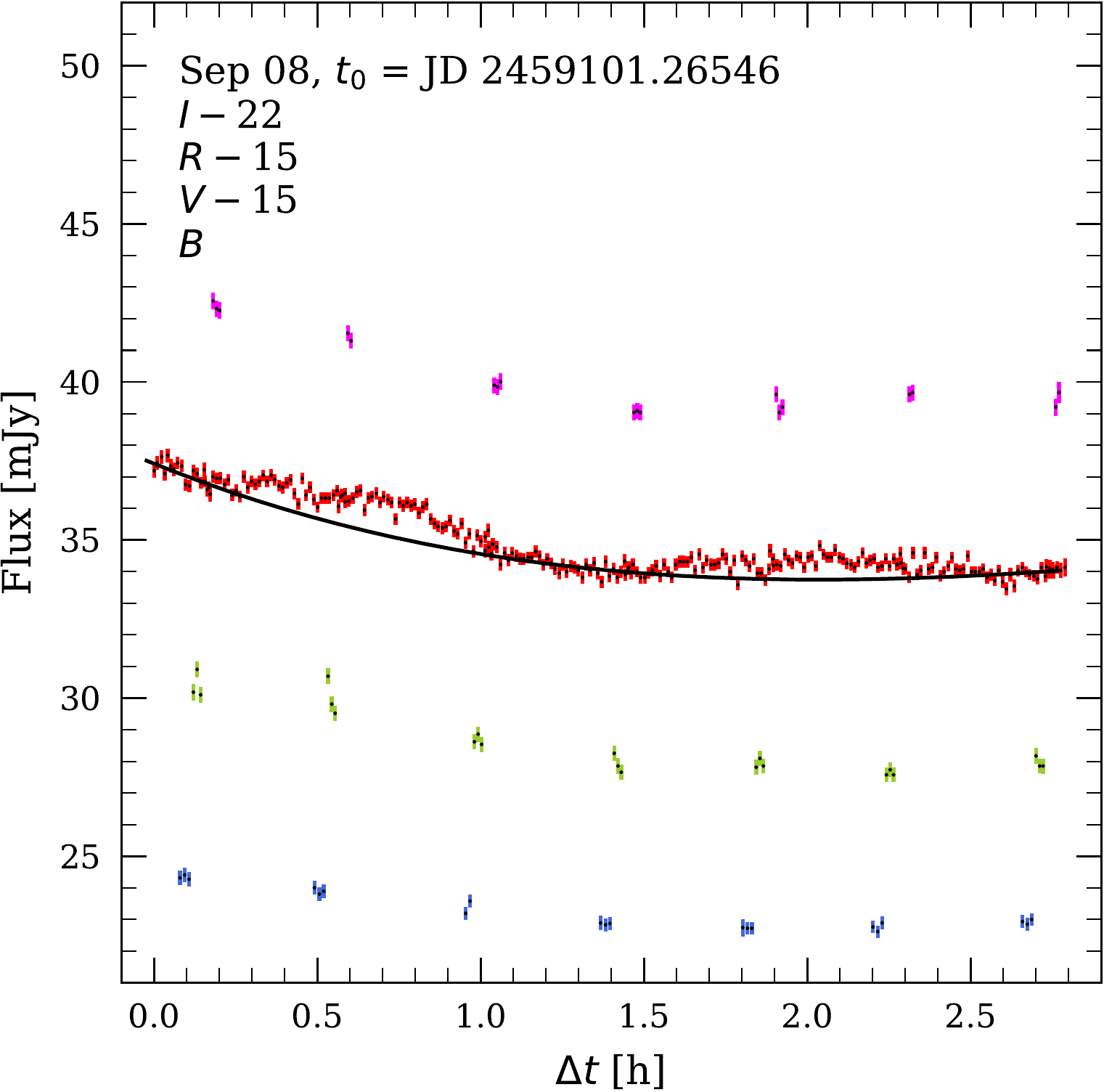}{0.315\textwidth}{}}
\gridline{\fig{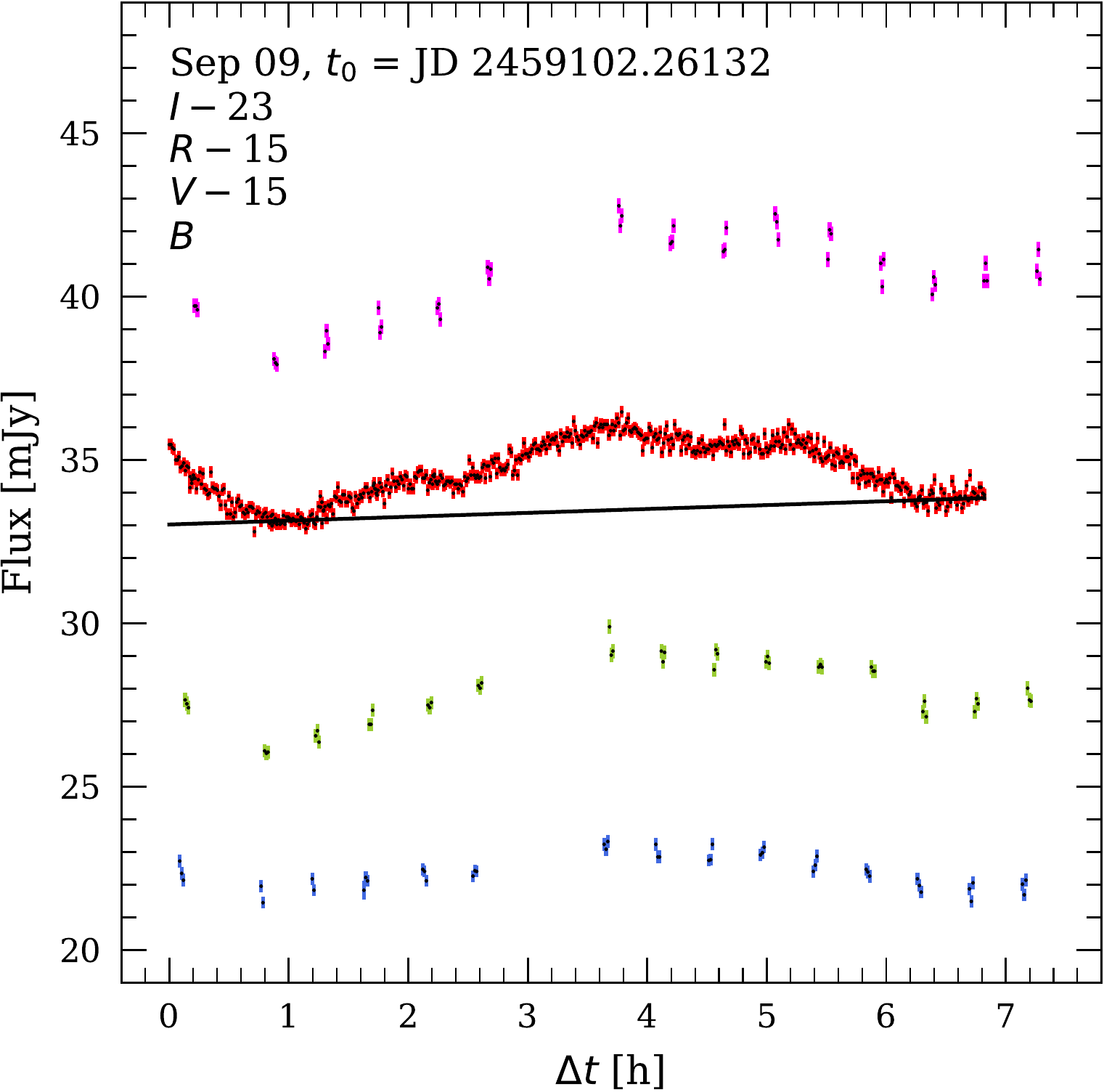}{0.315\textwidth}{}
          \fig{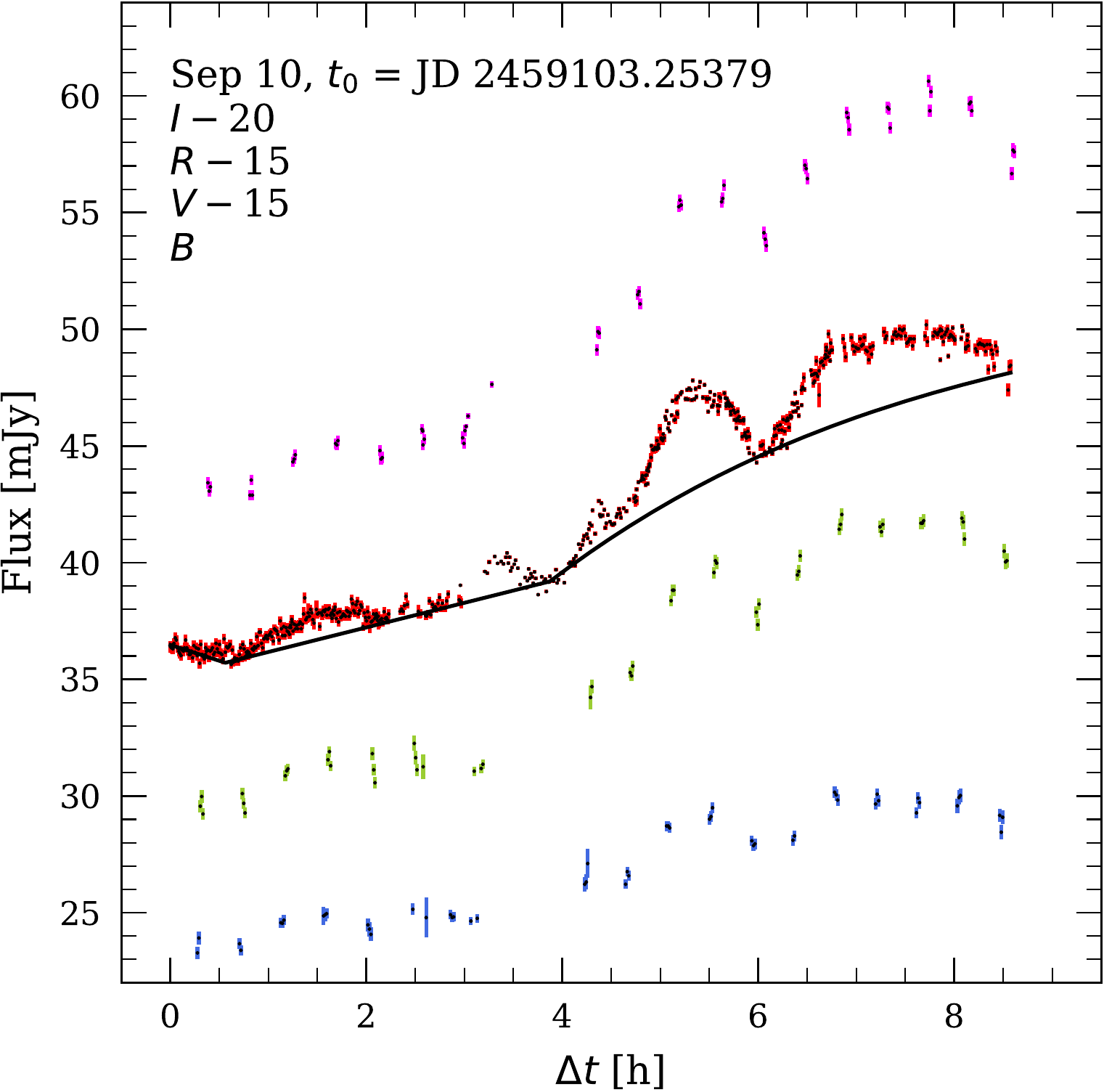}{0.315\textwidth}{}
          \fig{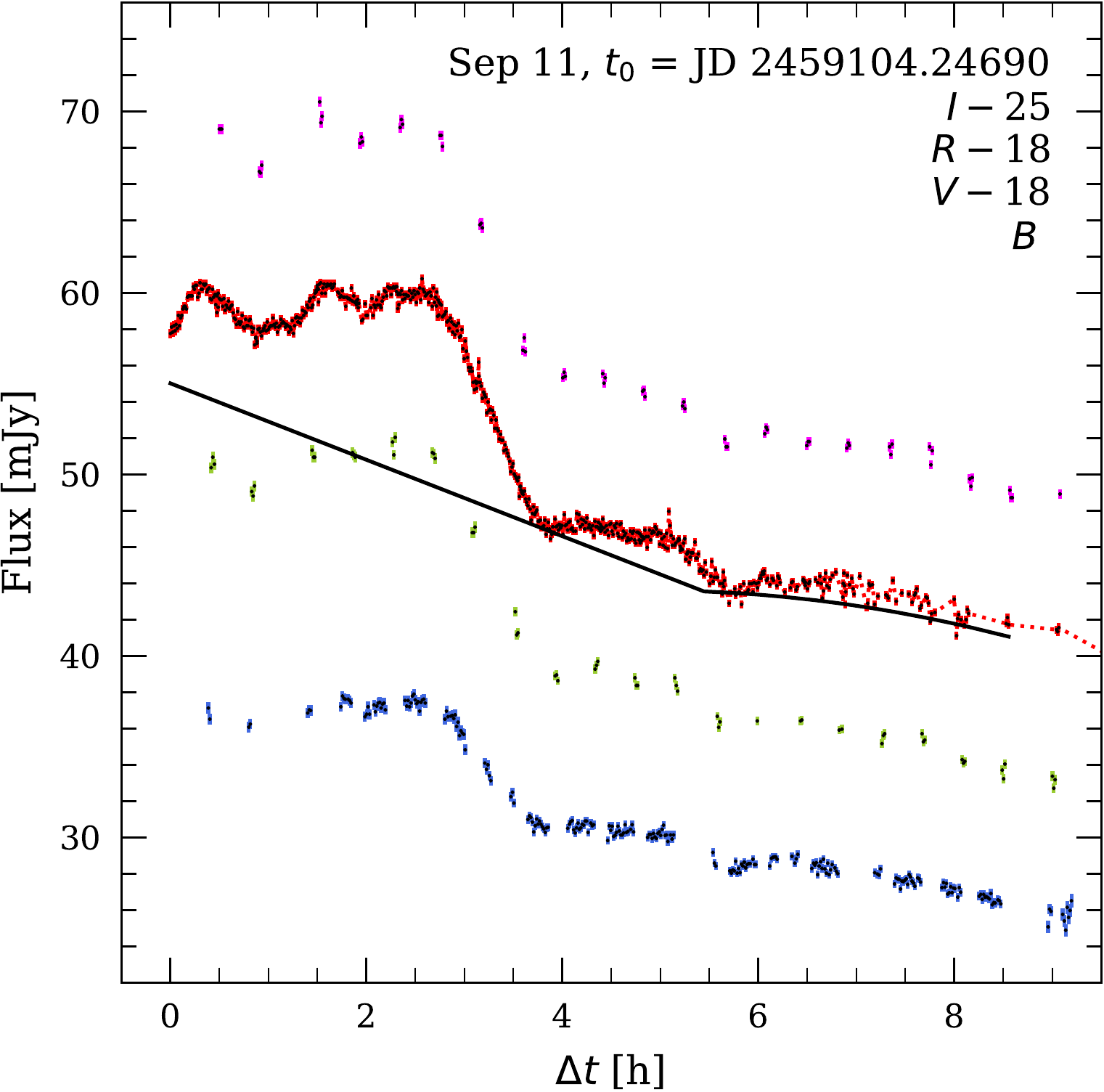}{0.315\textwidth}{}}
\gridline{\fig{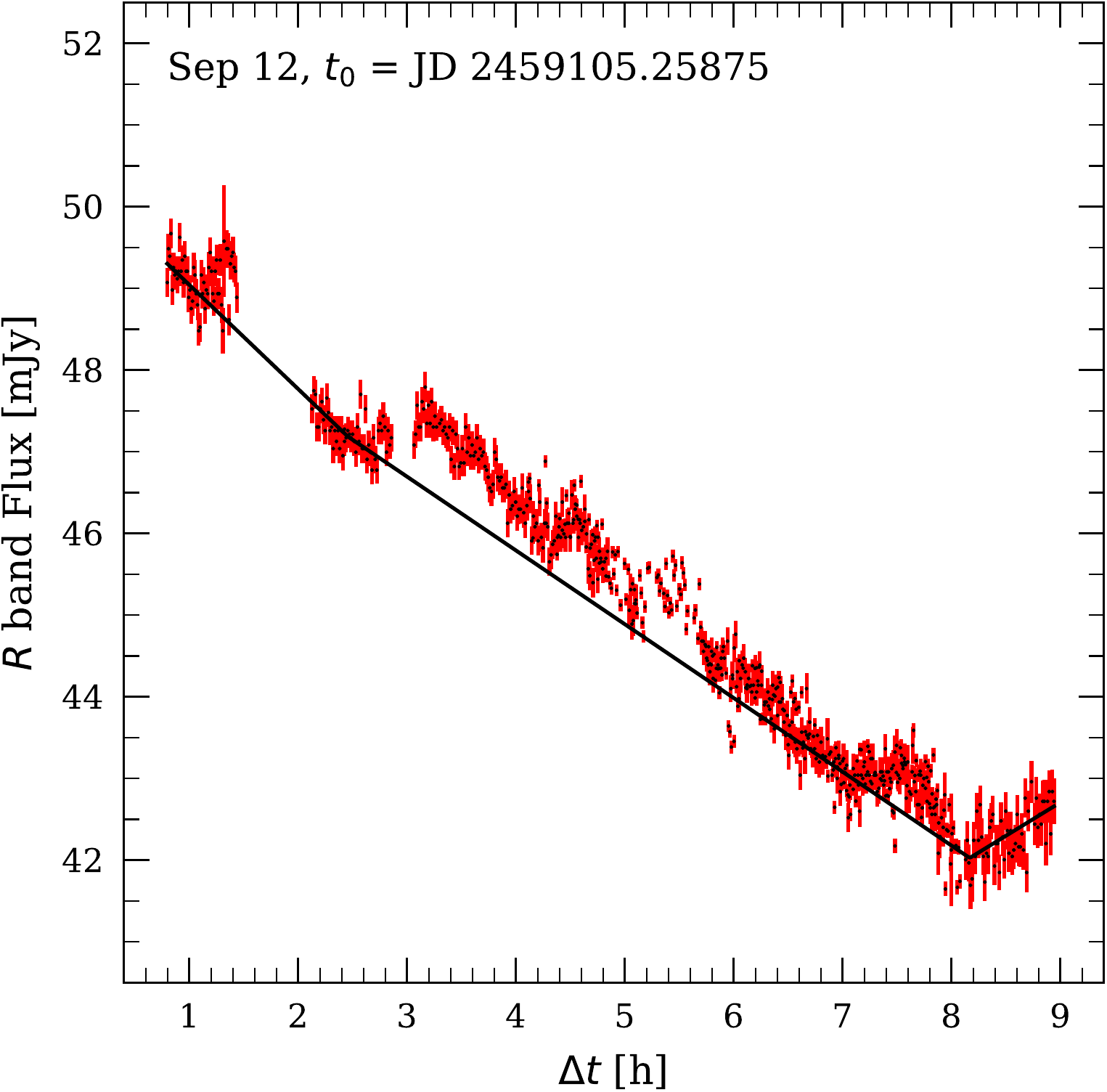}{0.315\textwidth}{}
          \fig{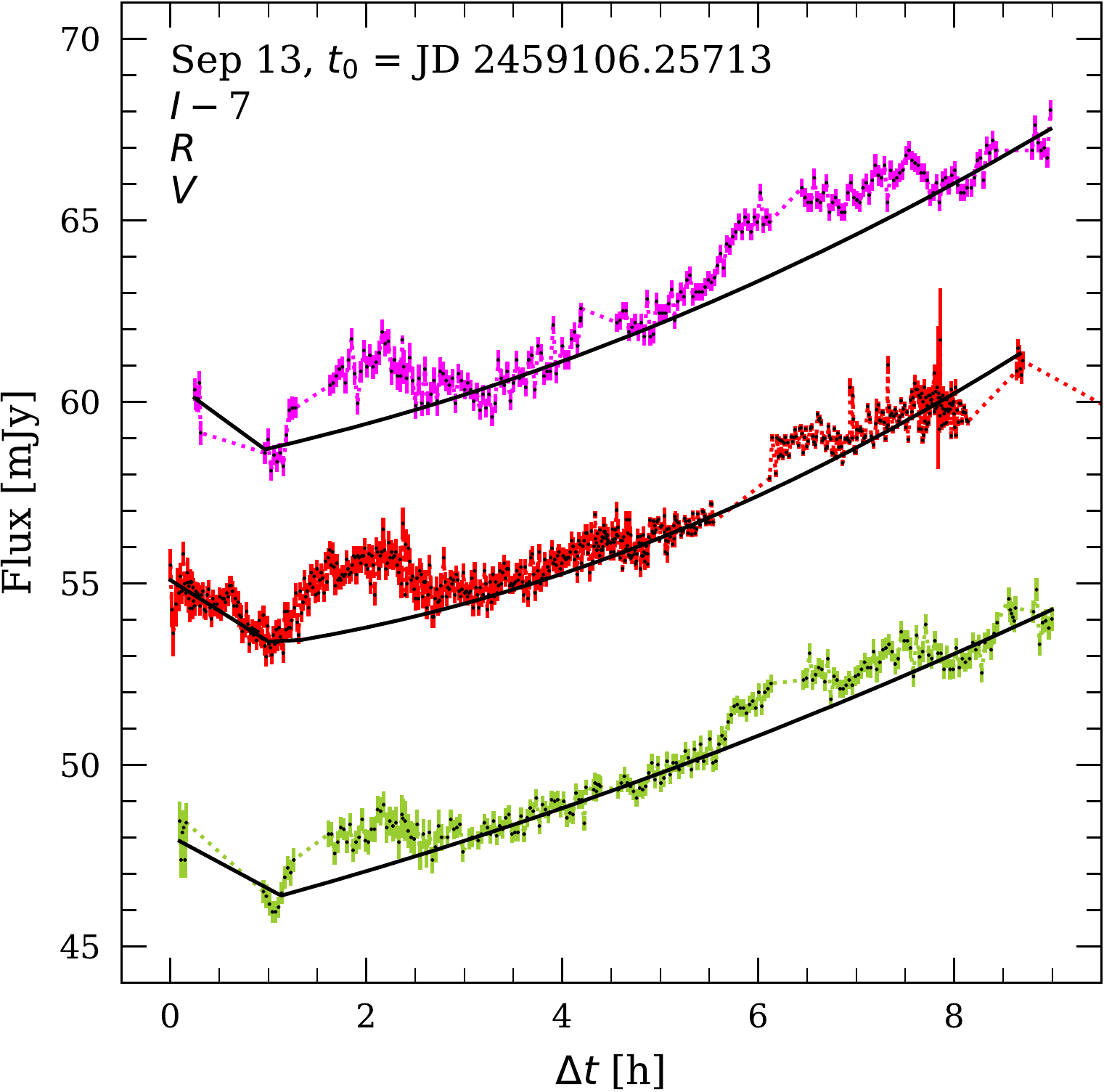}{0.315\textwidth}{}
          \fig{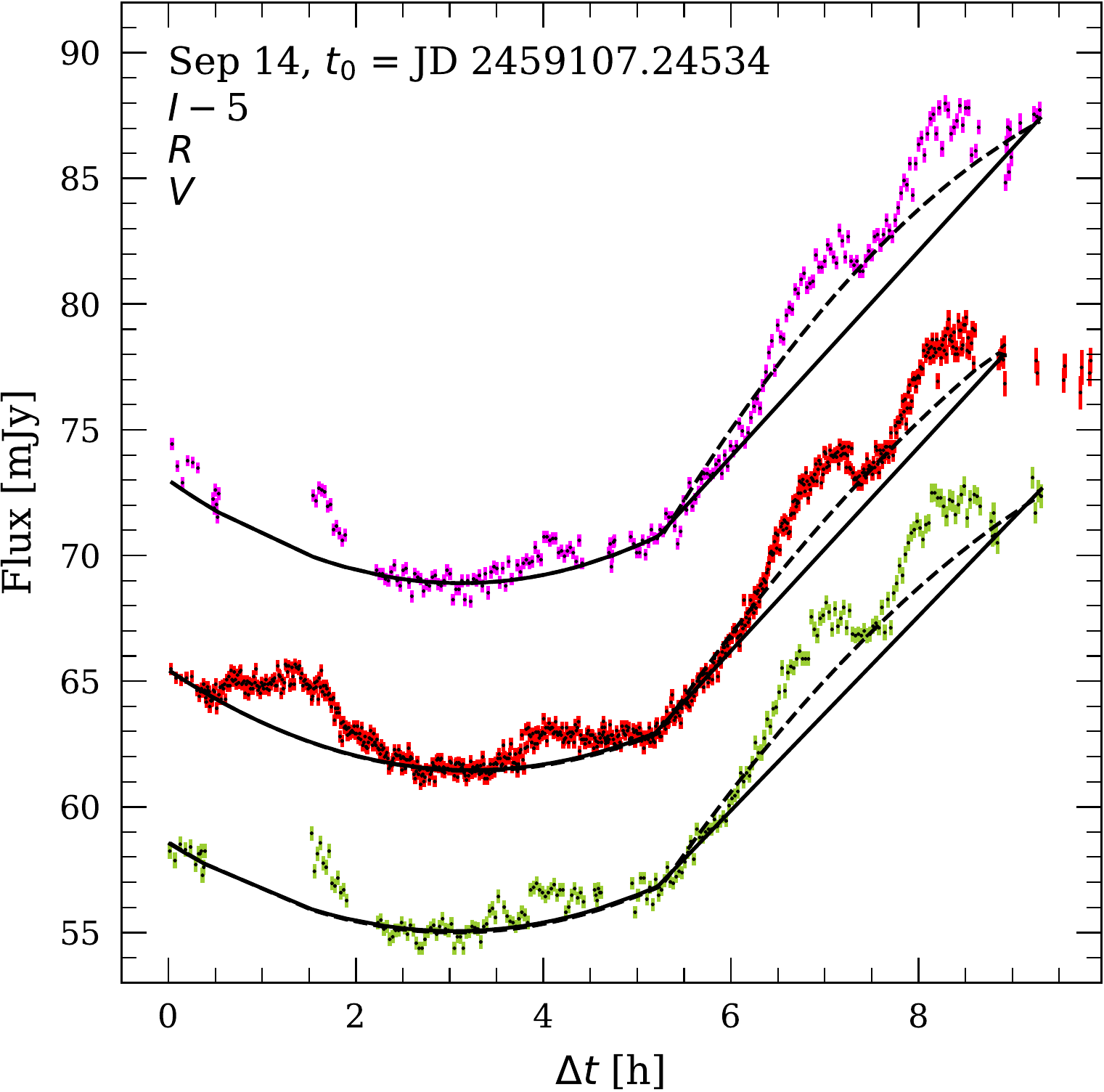}{0.315\textwidth}{}}
\caption{Continued.}
\end{figure*}

\begin{figure*}[t!]
\gridline{\fig{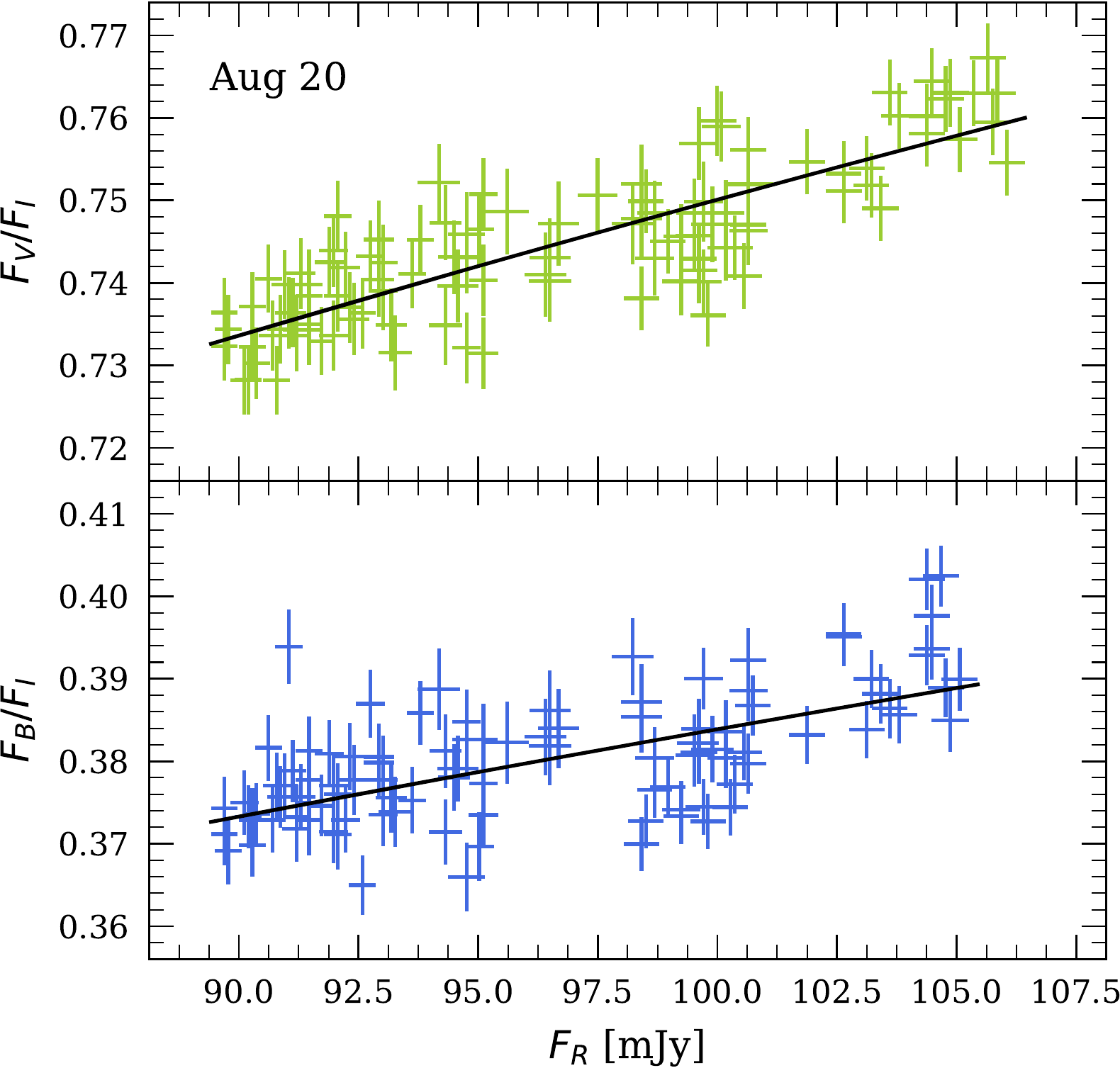}{0.31\textwidth}{}
          \fig{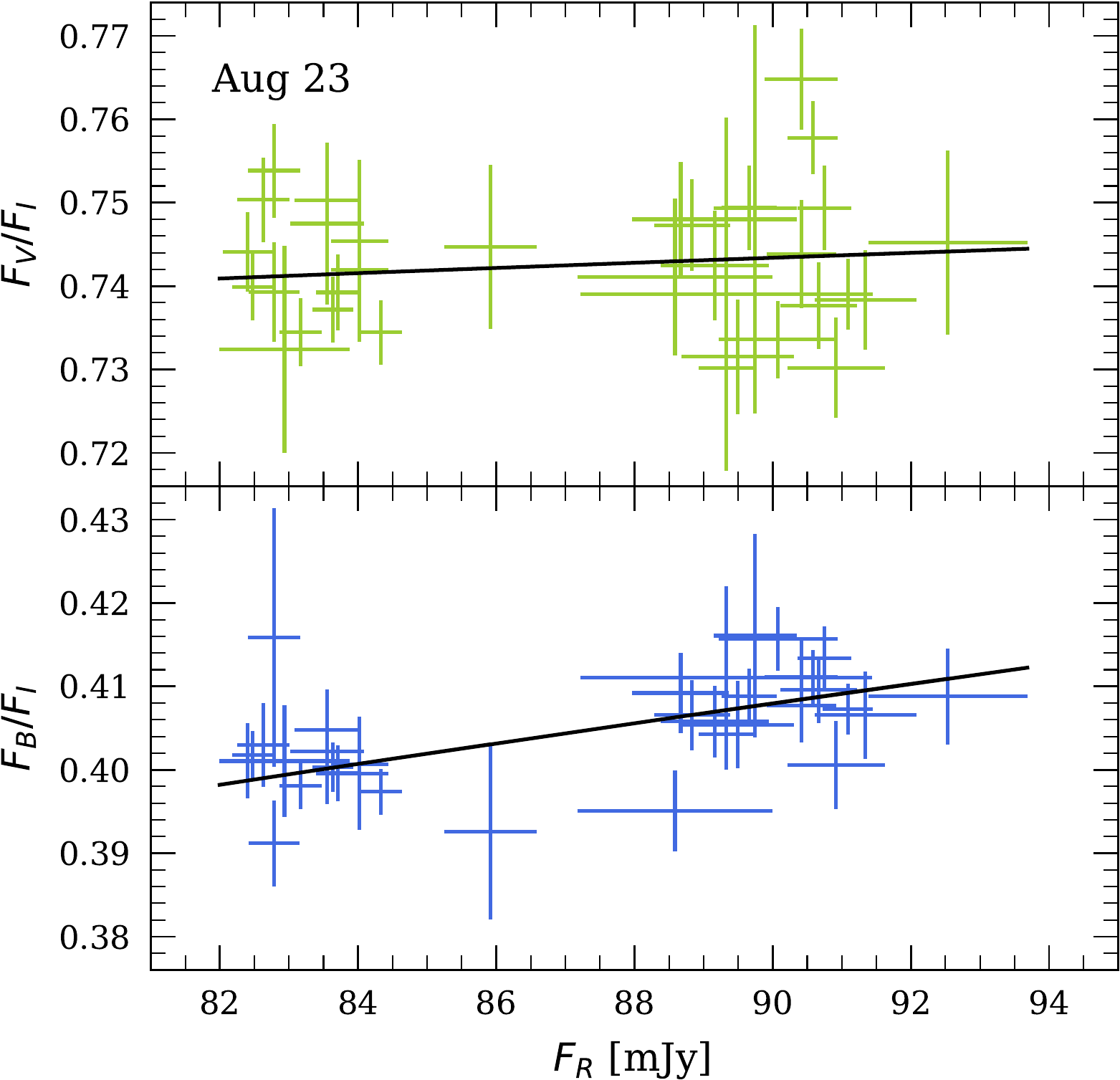}{0.31\textwidth}{}
          \fig{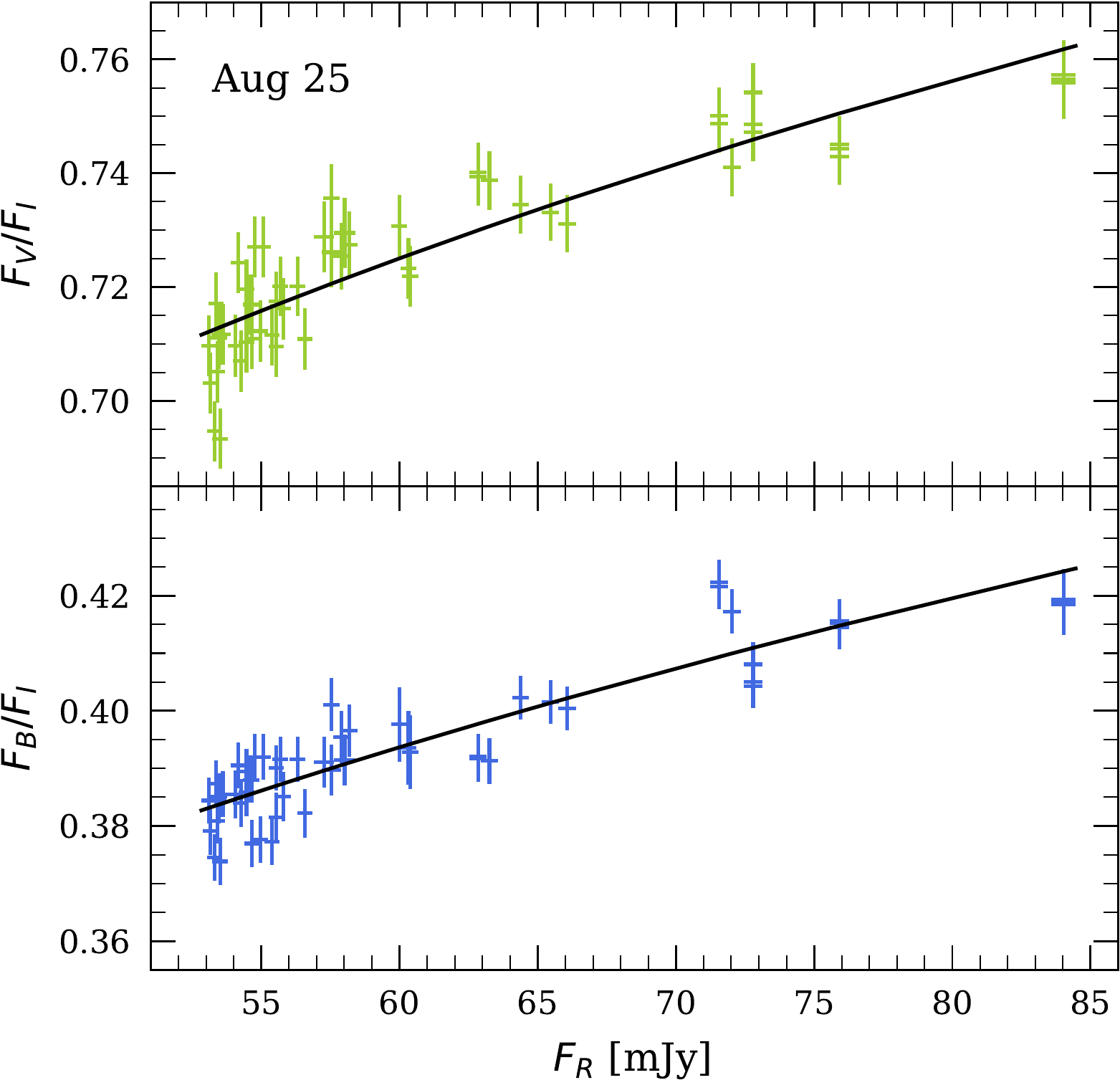}{0.31\textwidth}{}}
\gridline{\fig{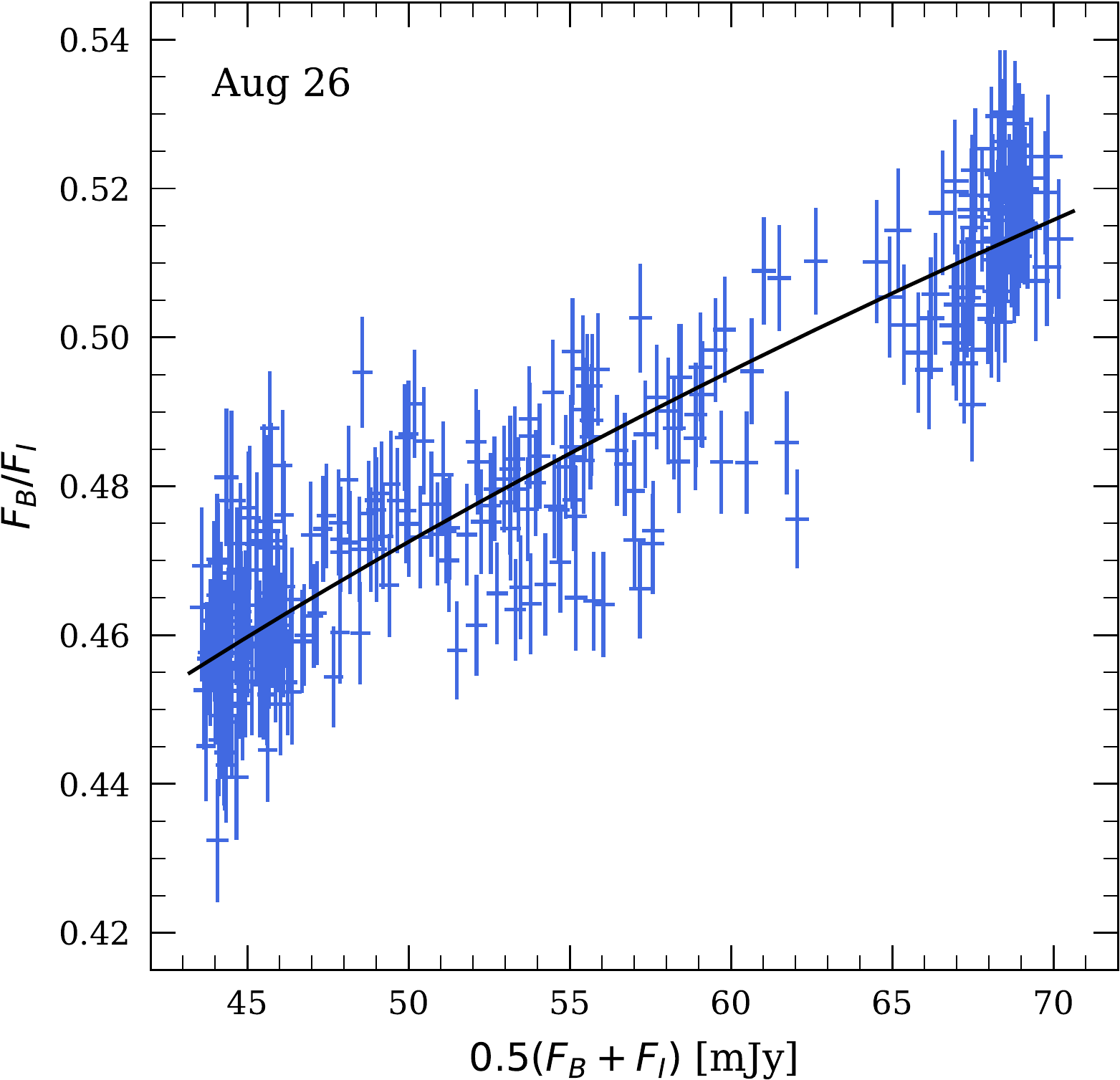}{0.31\textwidth}{}
          \fig{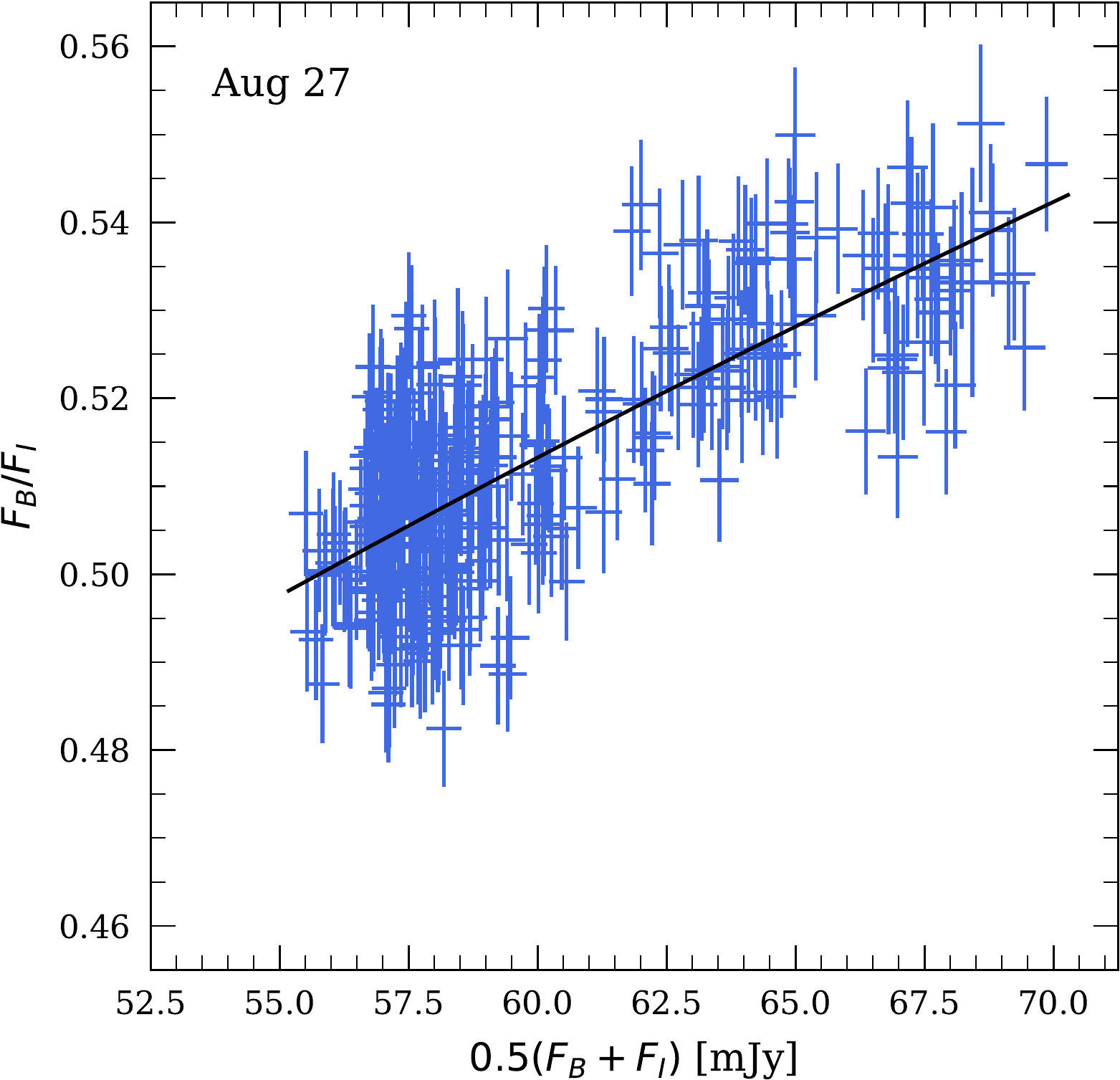}{0.31\textwidth}{}
          \fig{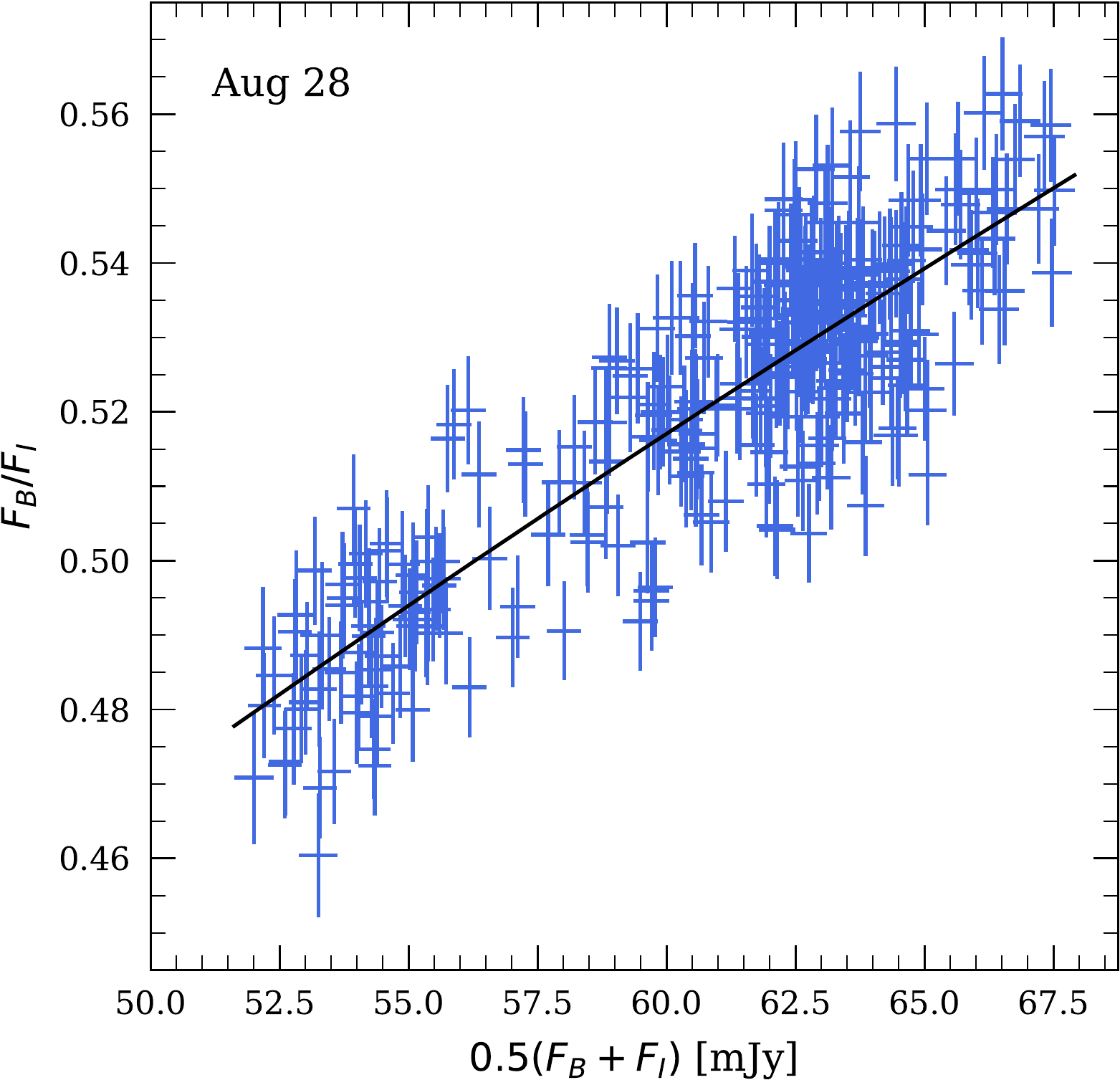}{0.31\textwidth}{}}
\gridline{\fig{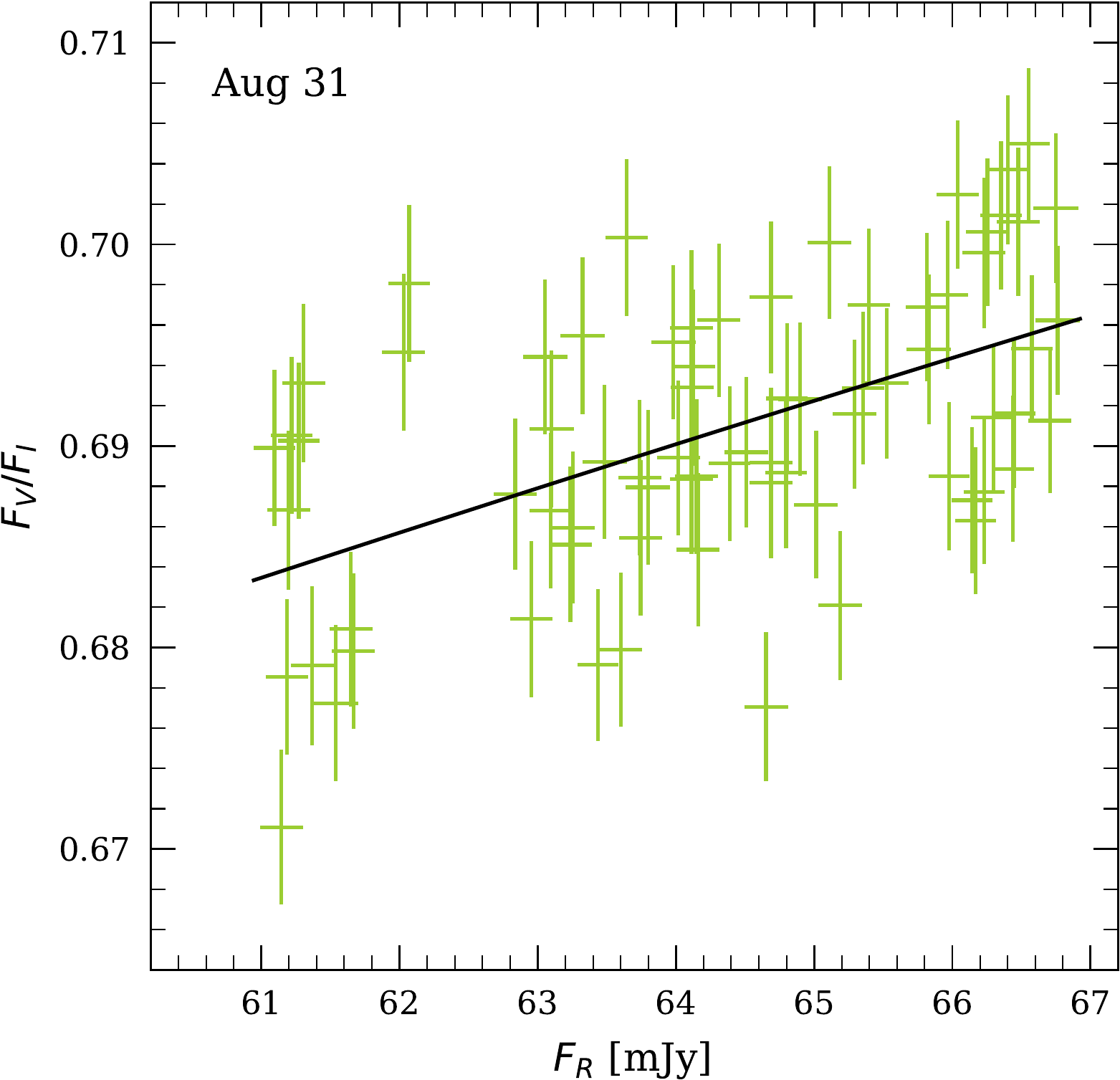}{0.31\textwidth}{}
          \fig{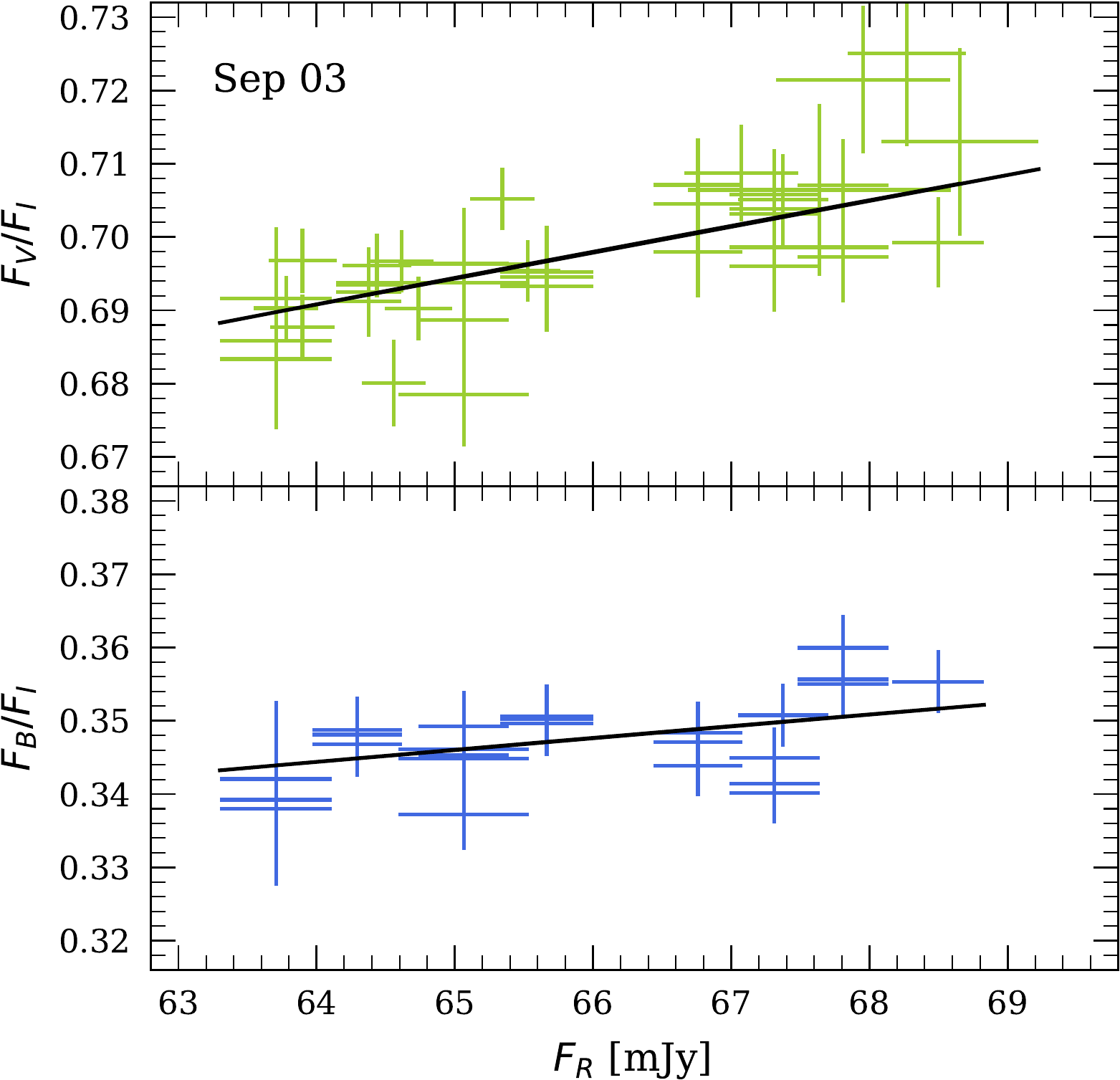}{0.31\textwidth}{}
          \fig{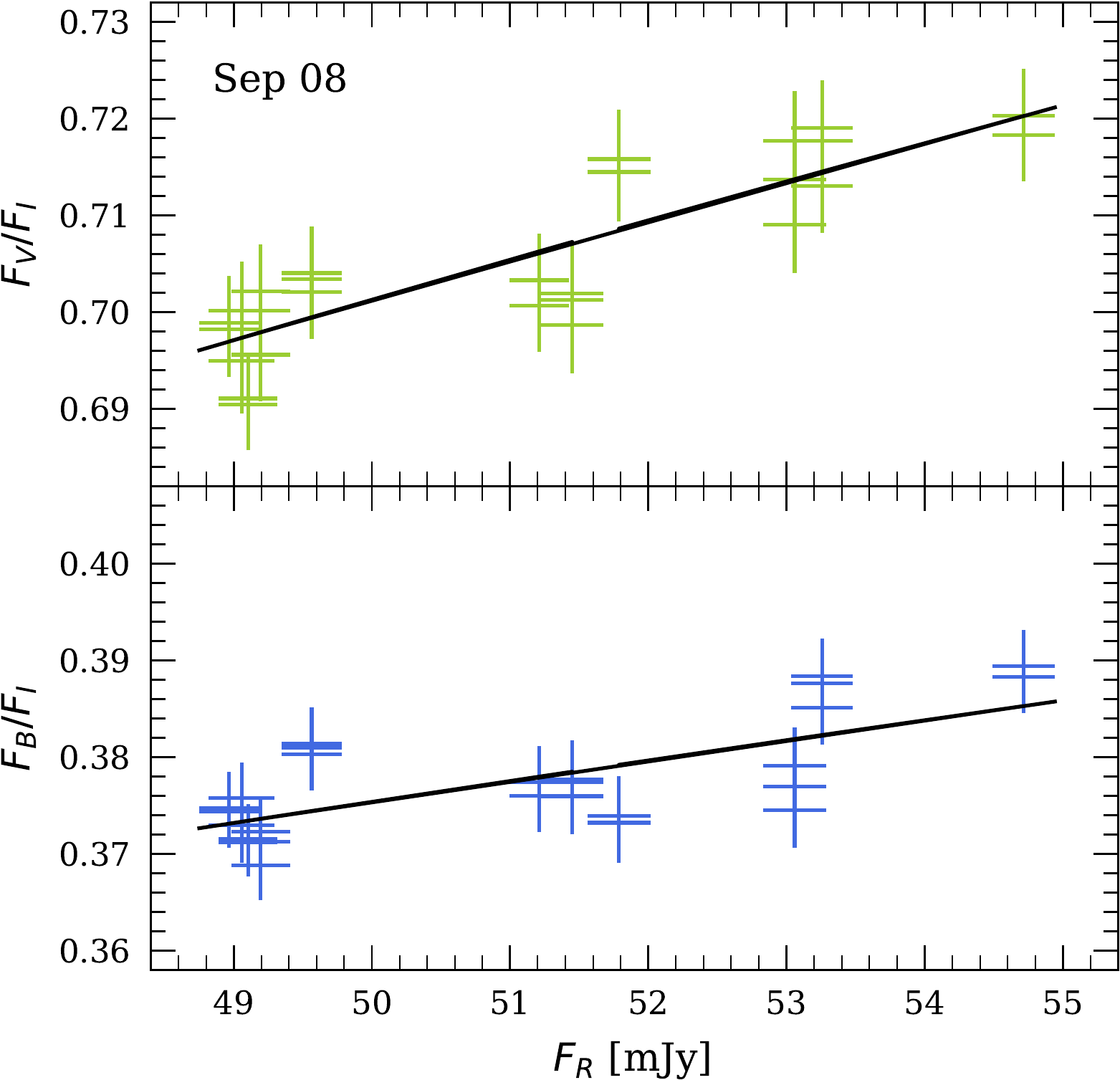}{0.31\textwidth}{}
          }
\caption{Color-magnitude diagrams built using the non-corrected LCs that show INV. The fitted power-law models are overplotted.}
\label{app:cmd:fig}
\end{figure*}

\setcounter{figure}{7}
\begin{figure*}[t!]
\gridline{\fig{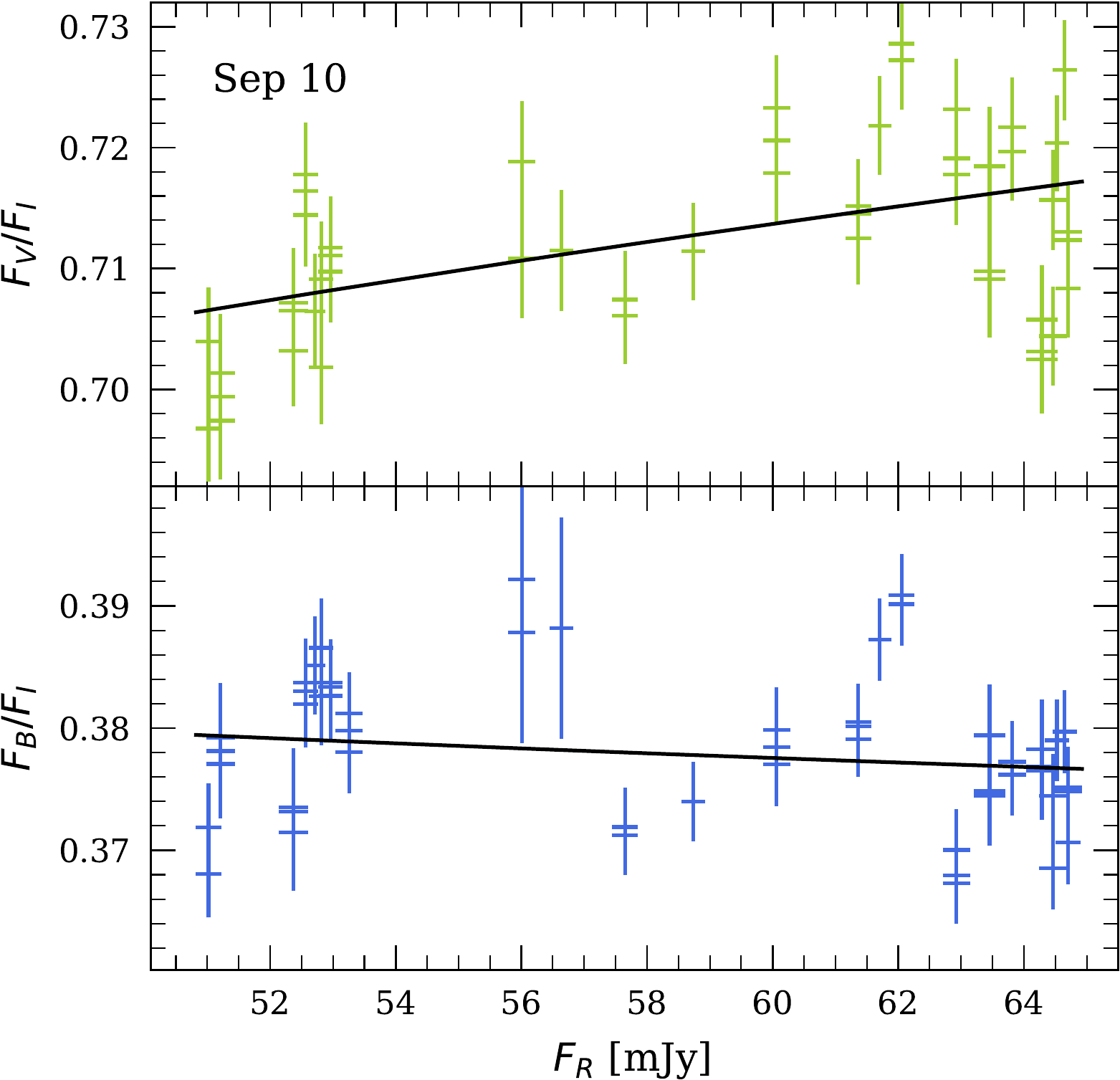}{0.31\textwidth}{}
          \fig{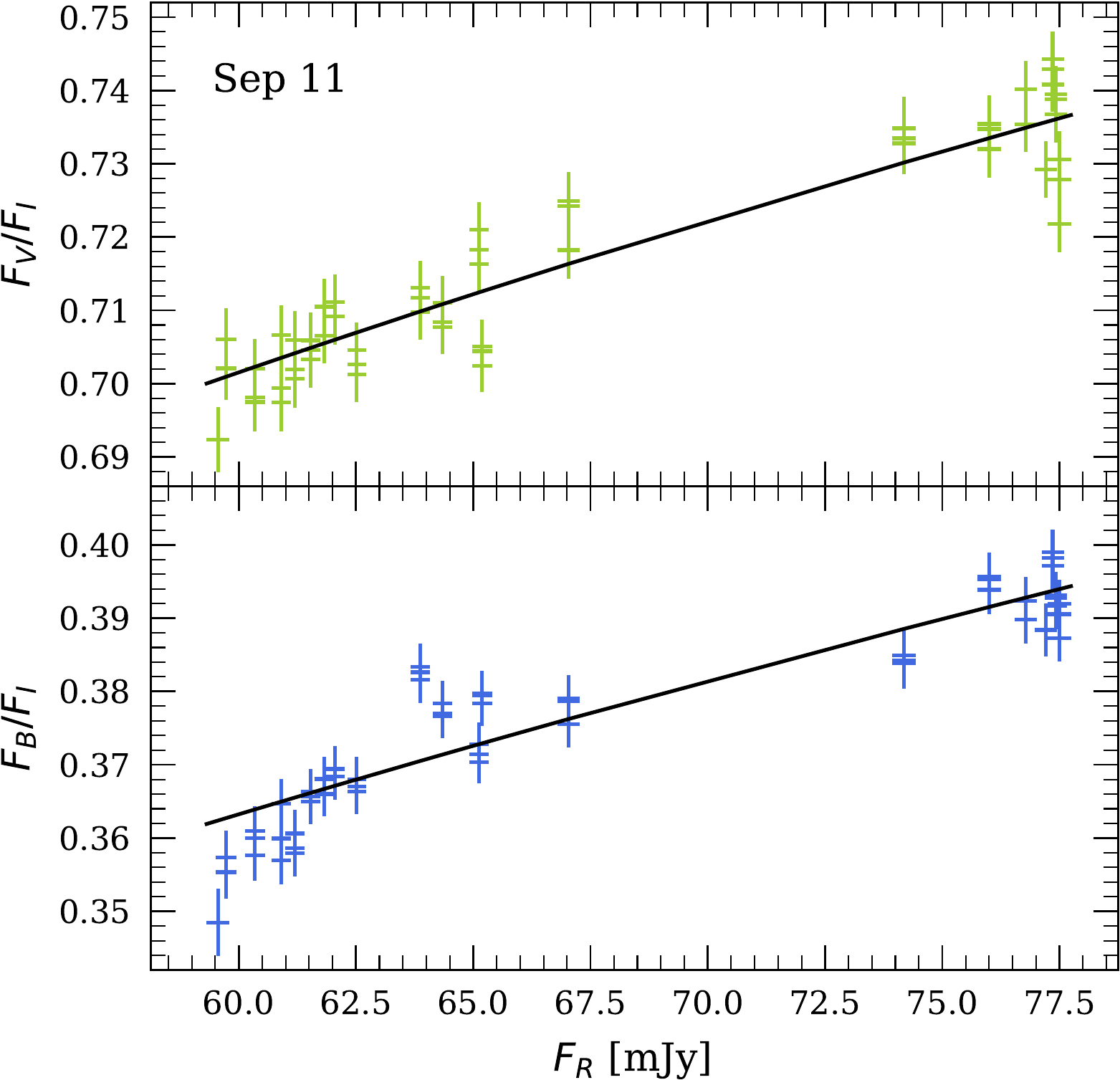}{0.31\textwidth}{}
          \fig{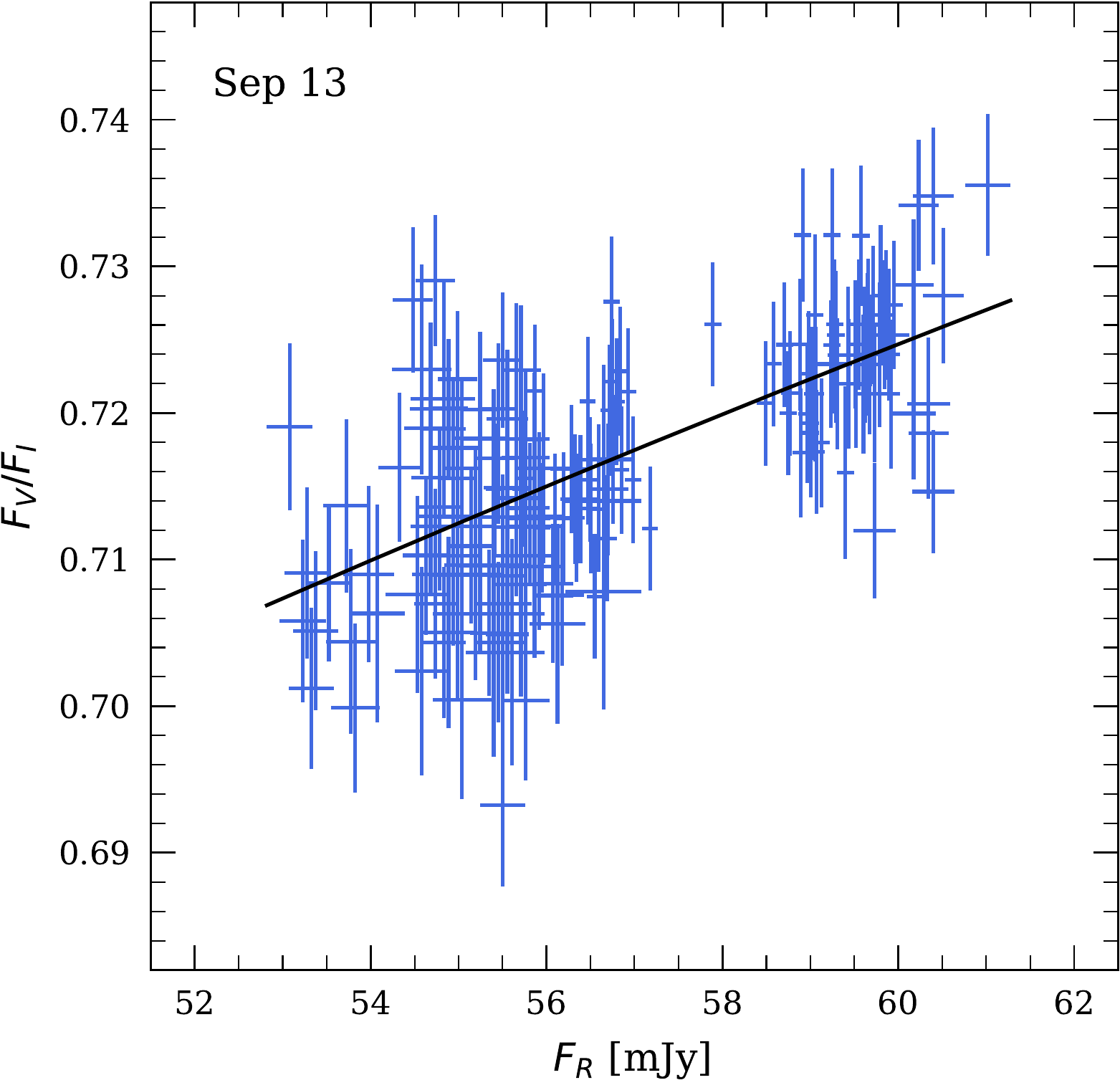}{0.31\textwidth}{}}
\gridline{\fig{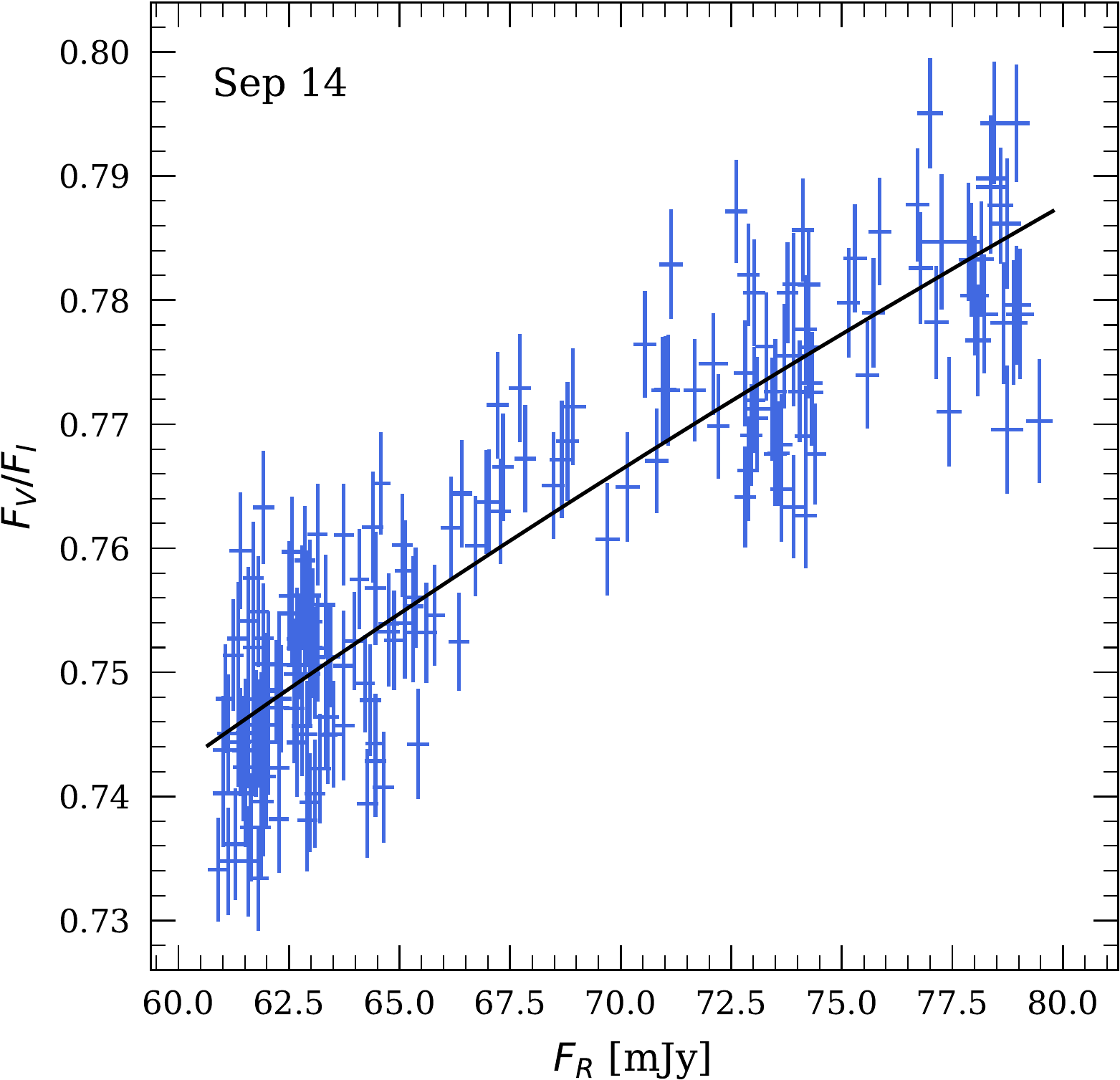}{0.31\textwidth}{}}
\caption{Continued.}
\end{figure*}

\begin{figure*}[t!]
\gridline{\fig{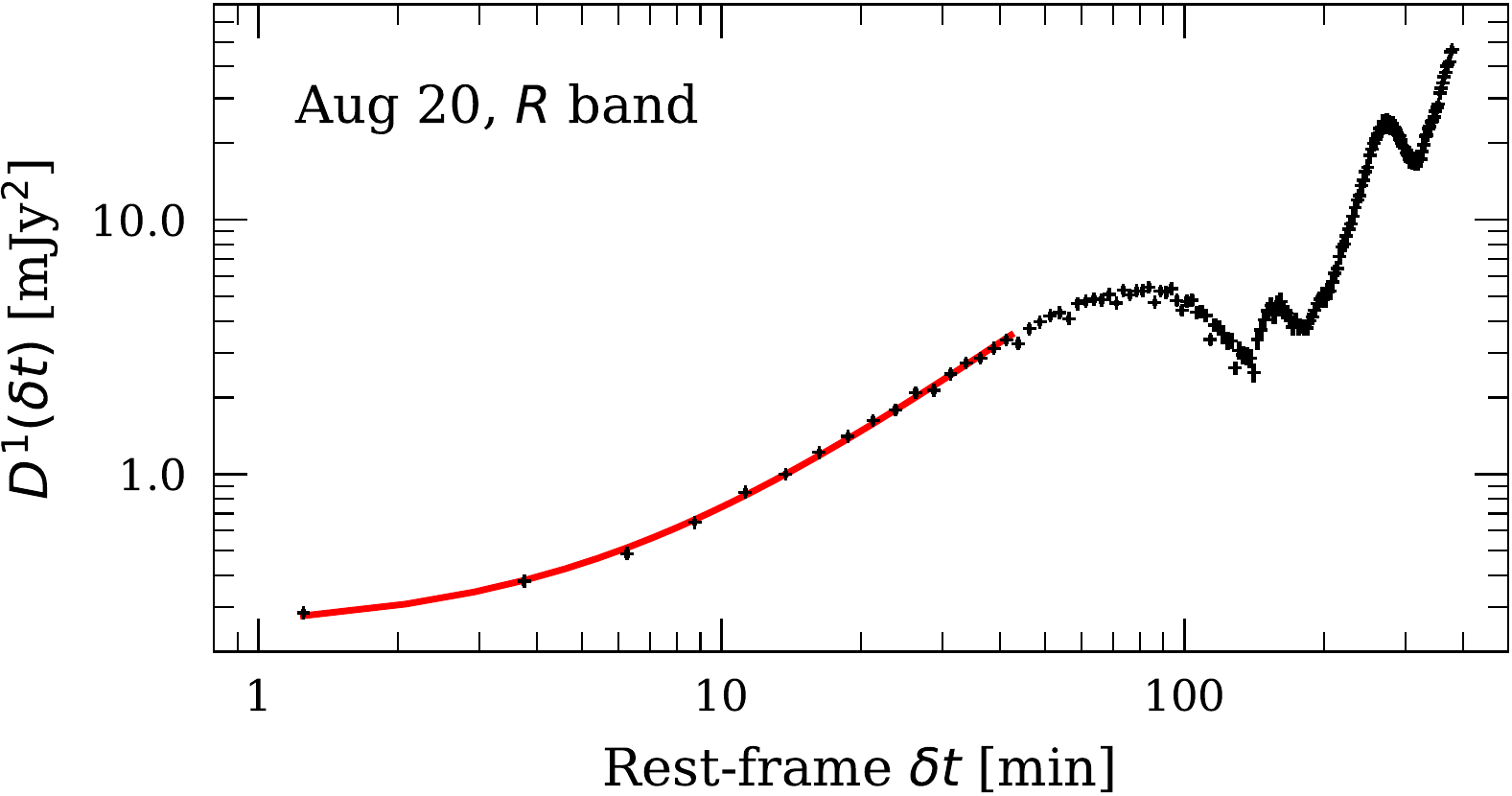}{0.31\textwidth}{}
          \fig{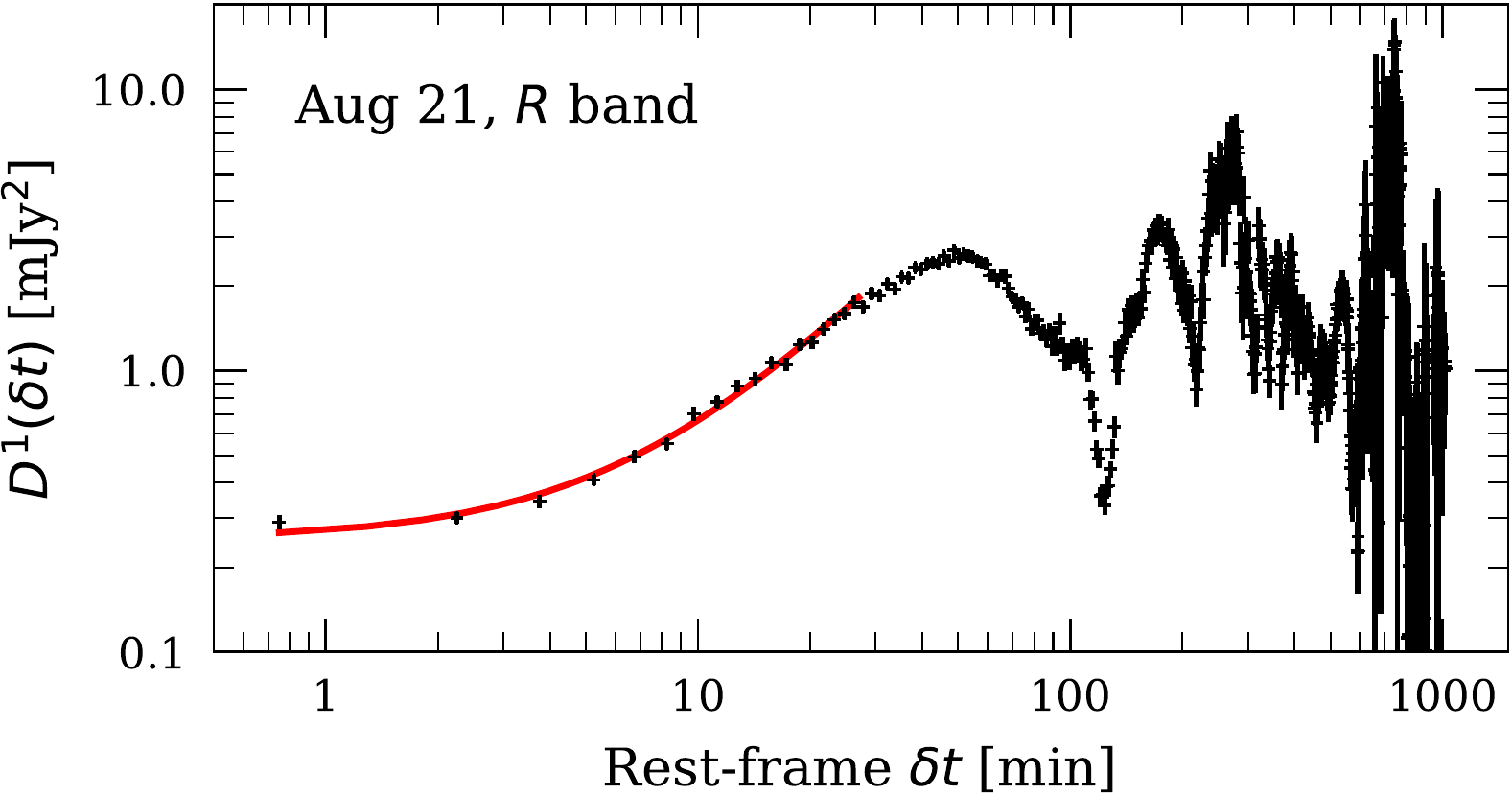}{0.31\textwidth}{}
          \fig{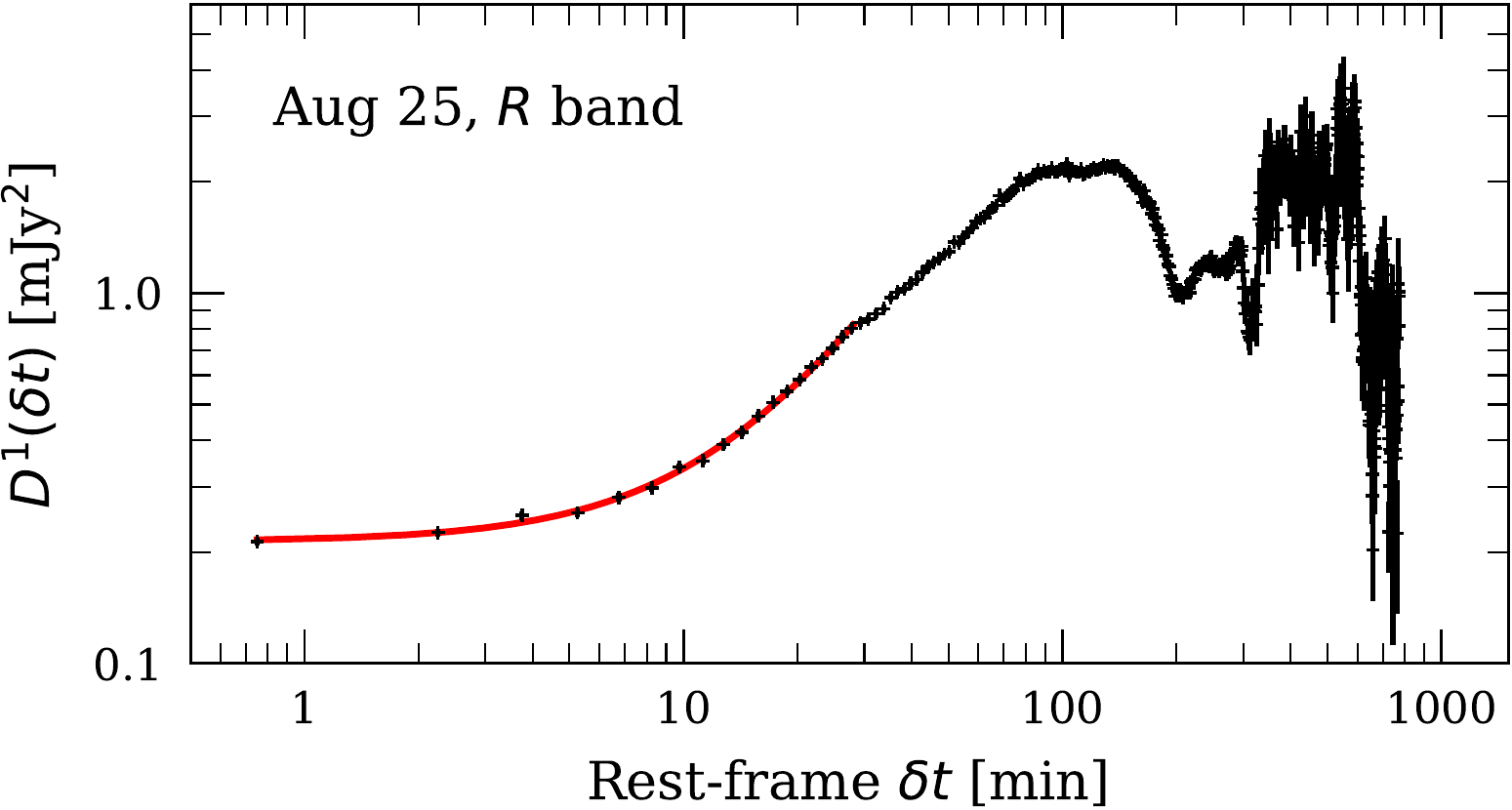}{0.31\textwidth}{}}
\gridline{\fig{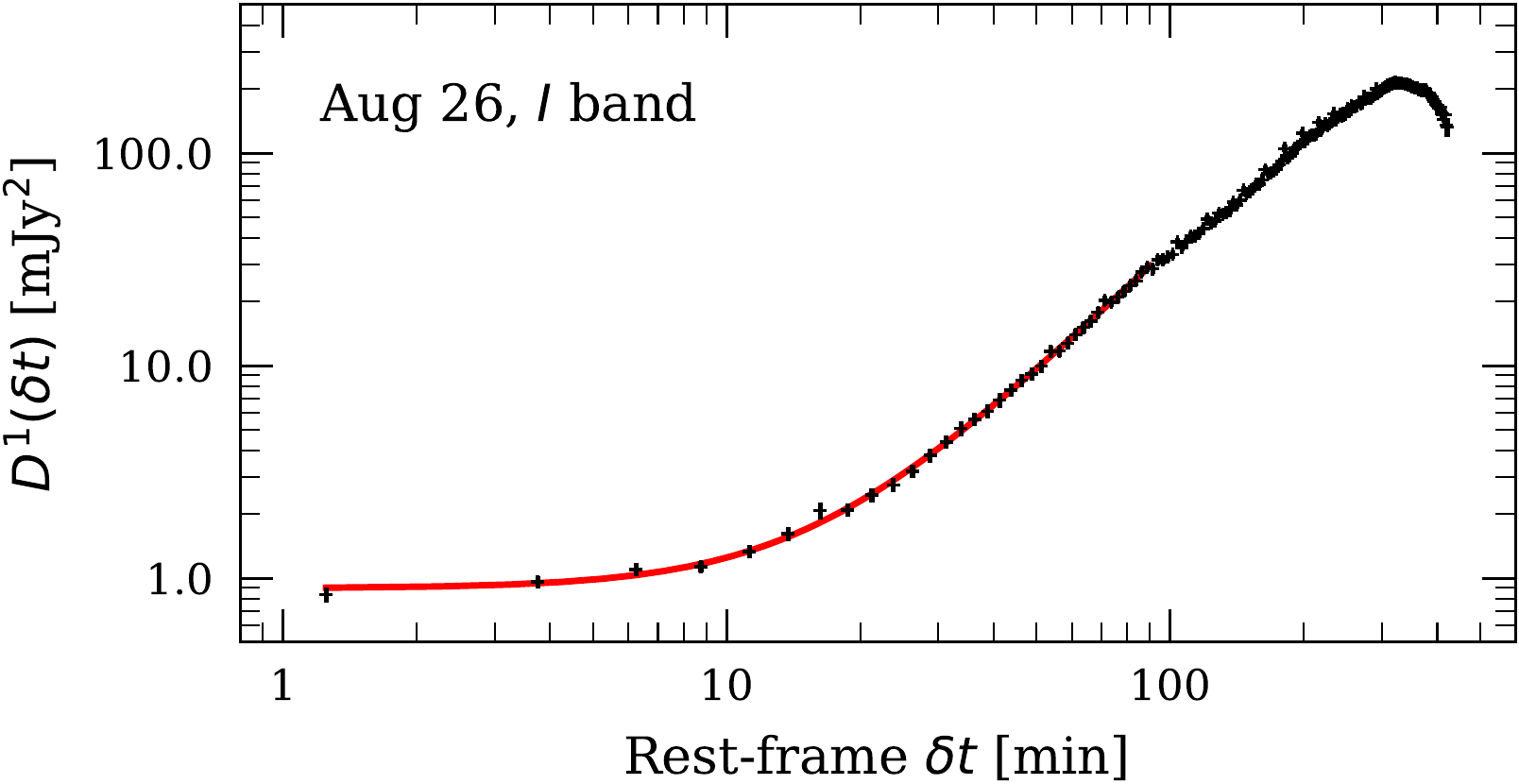}{0.31\textwidth}{}
          \fig{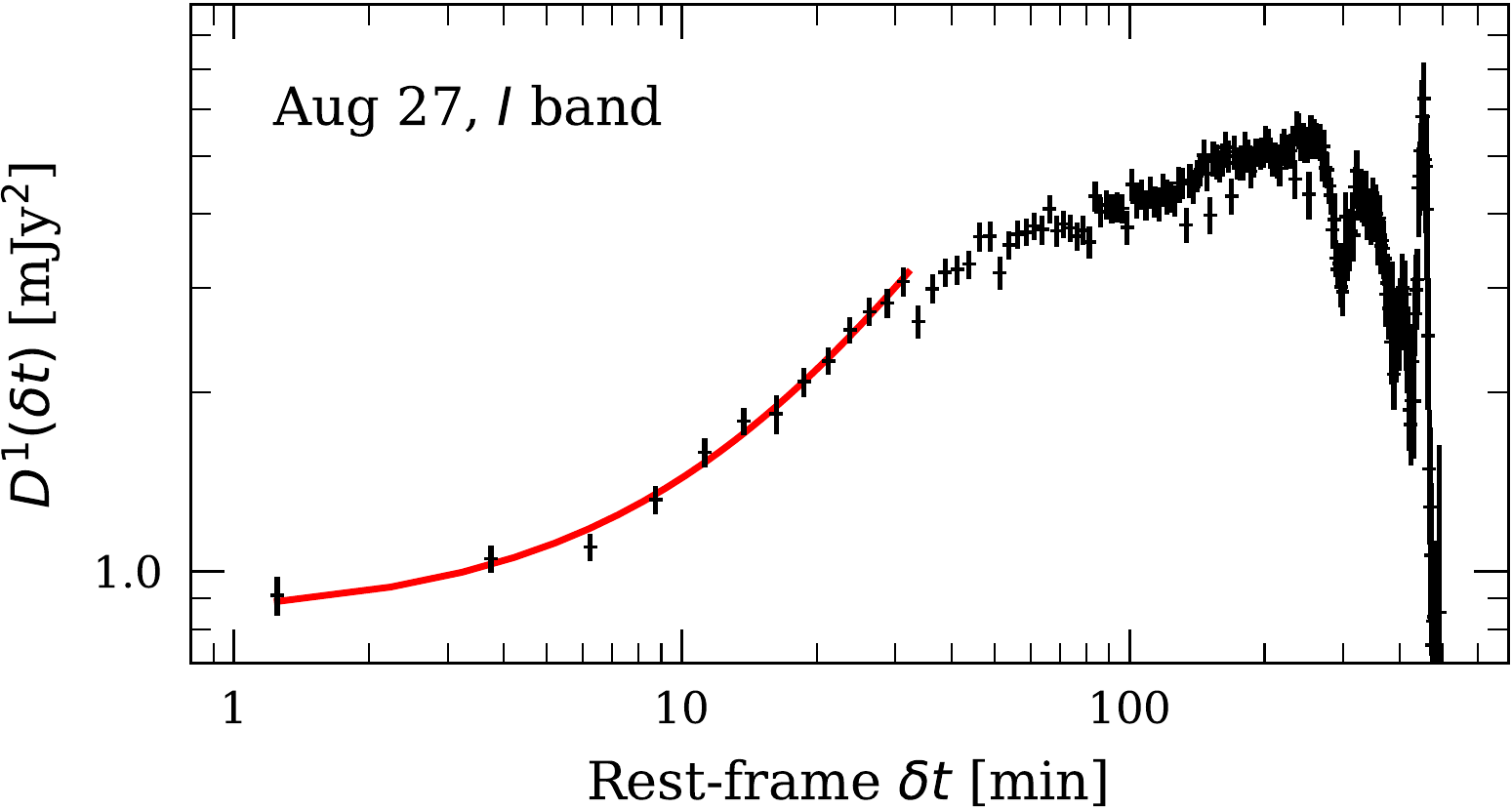}{0.31\textwidth}{}
          \fig{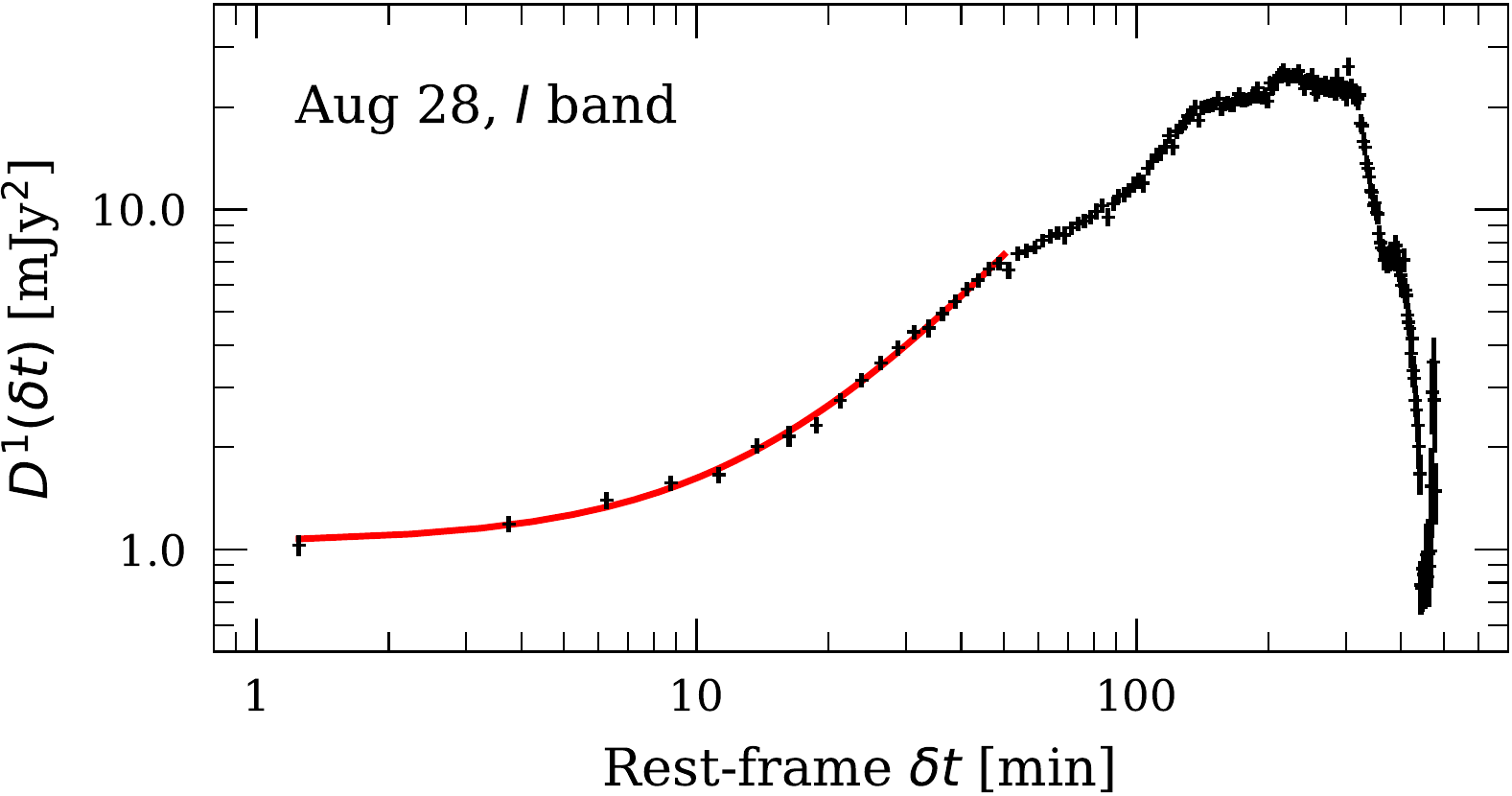}{0.31\textwidth}{}}
\gridline{\fig{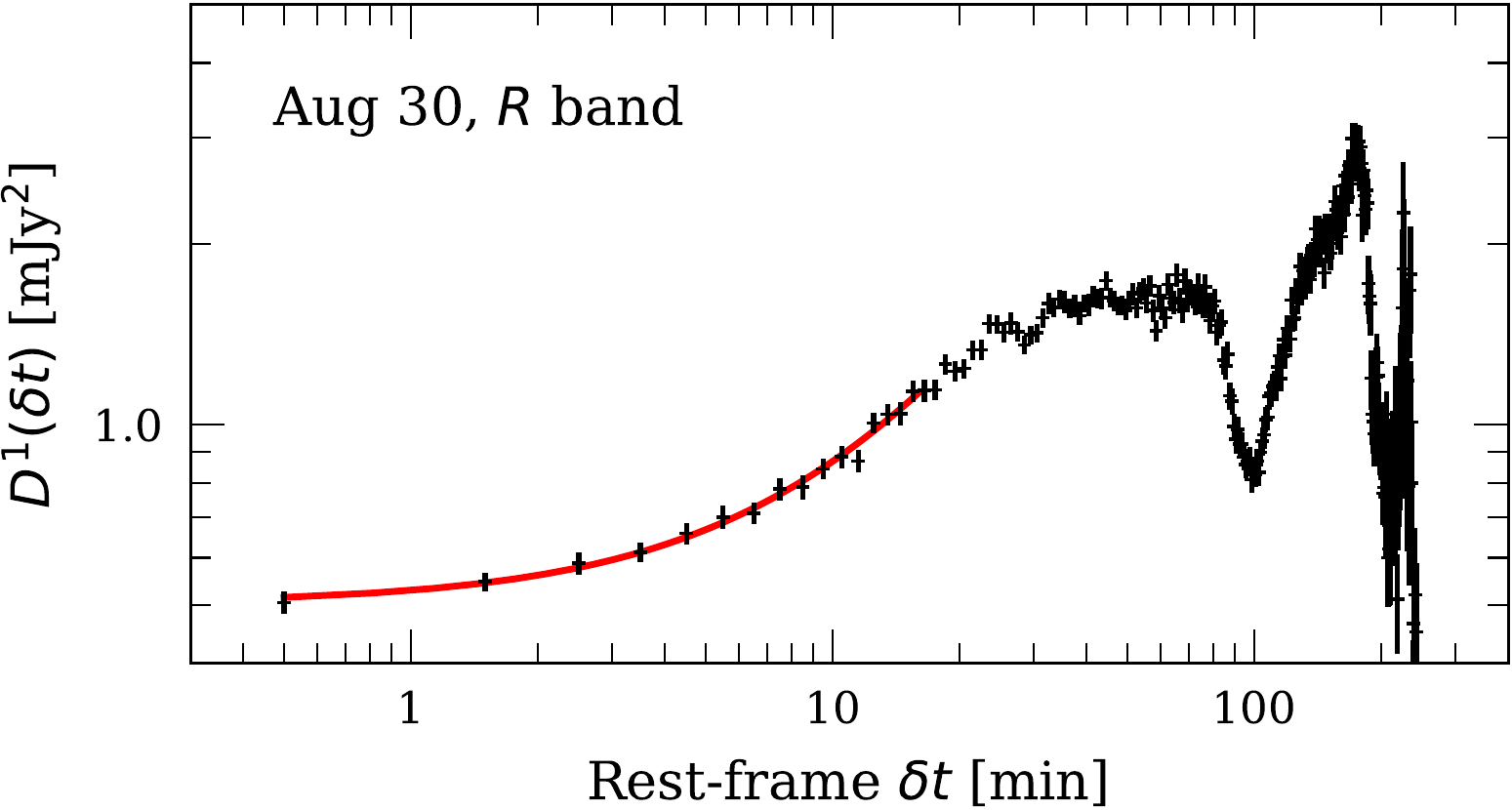}{0.31\textwidth}{}
          \fig{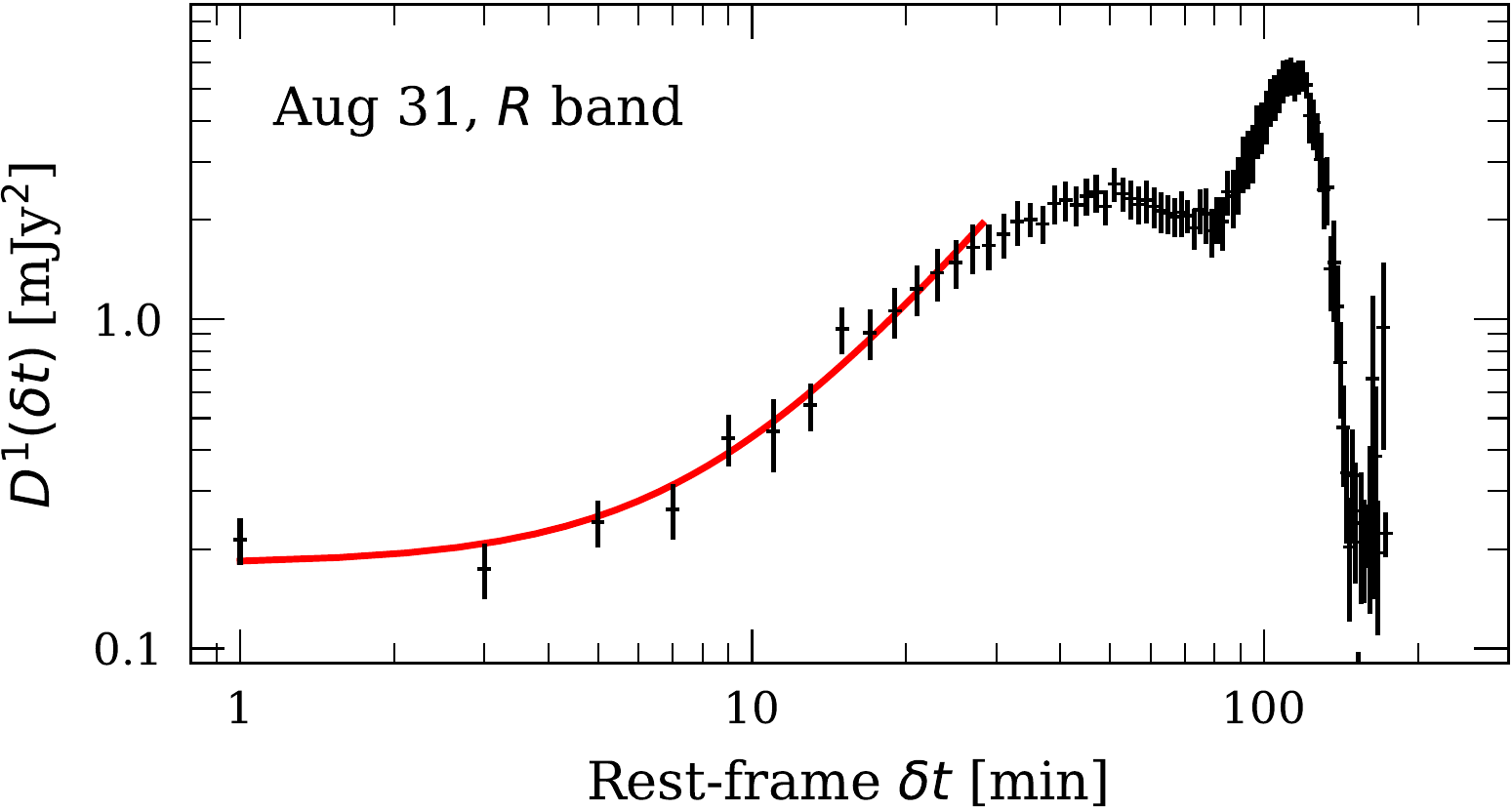}{0.31\textwidth}{}
          \fig{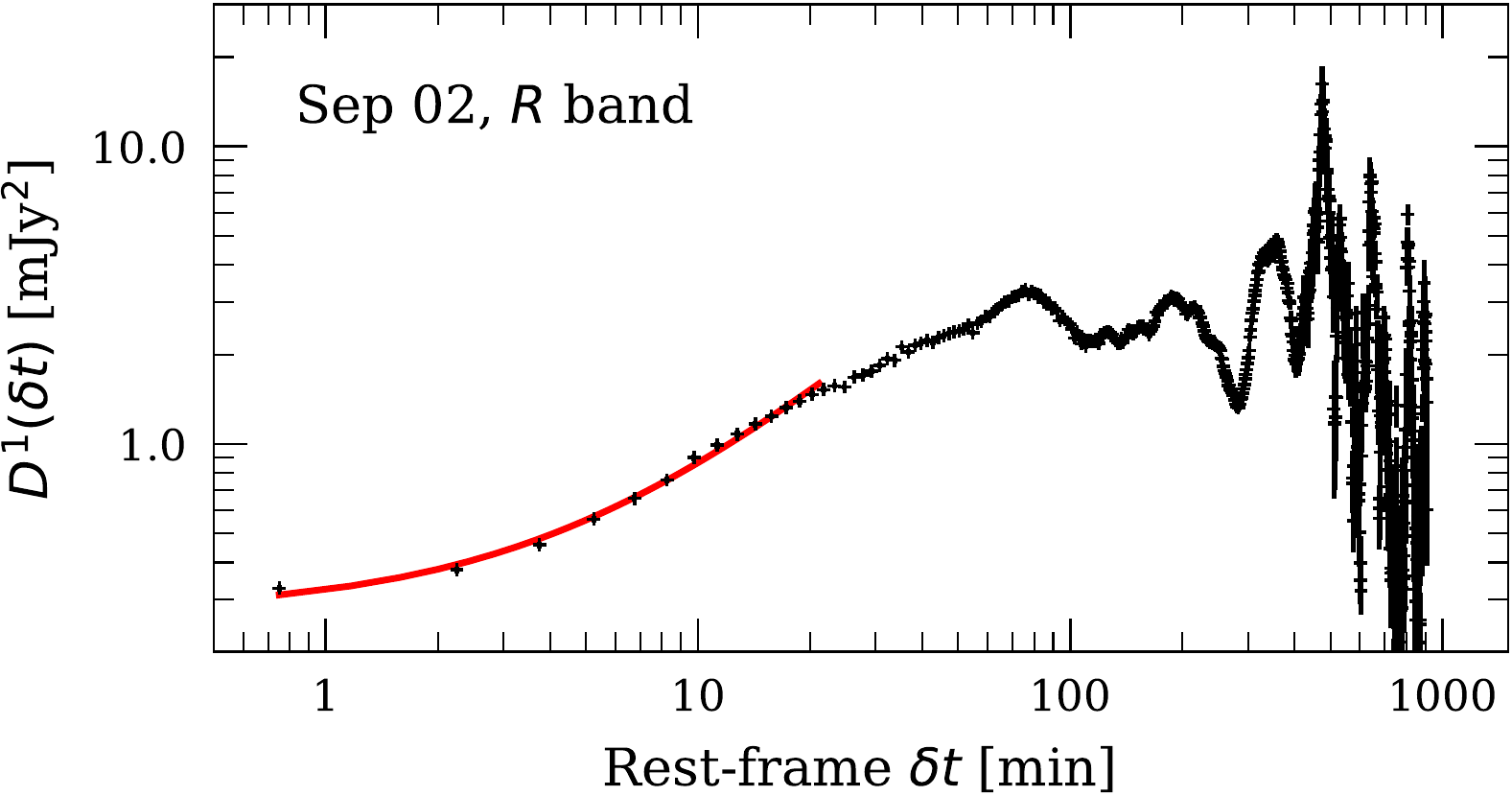}{0.31\textwidth}{}}
\gridline{\fig{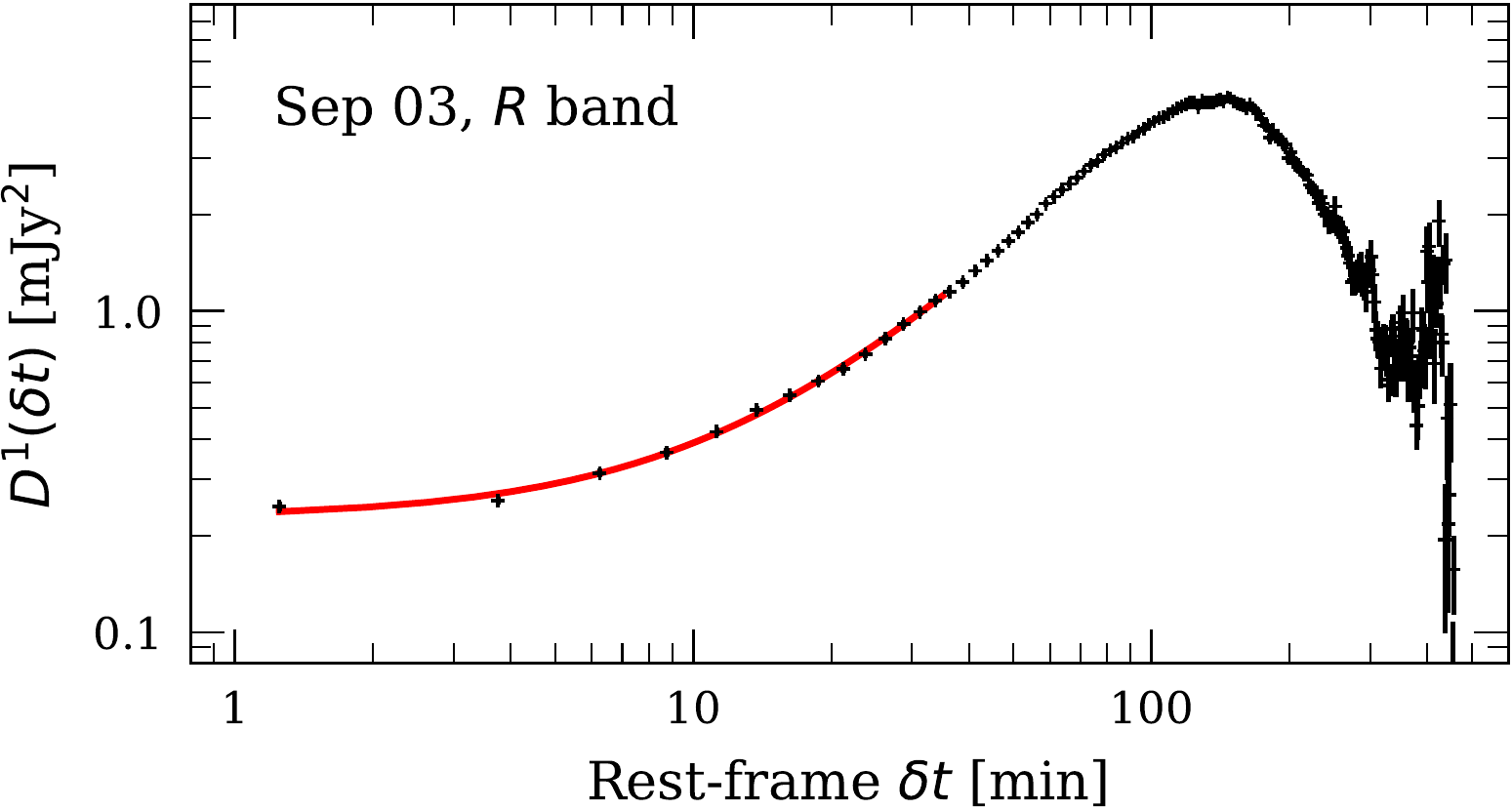}{0.31\textwidth}{}
          \fig{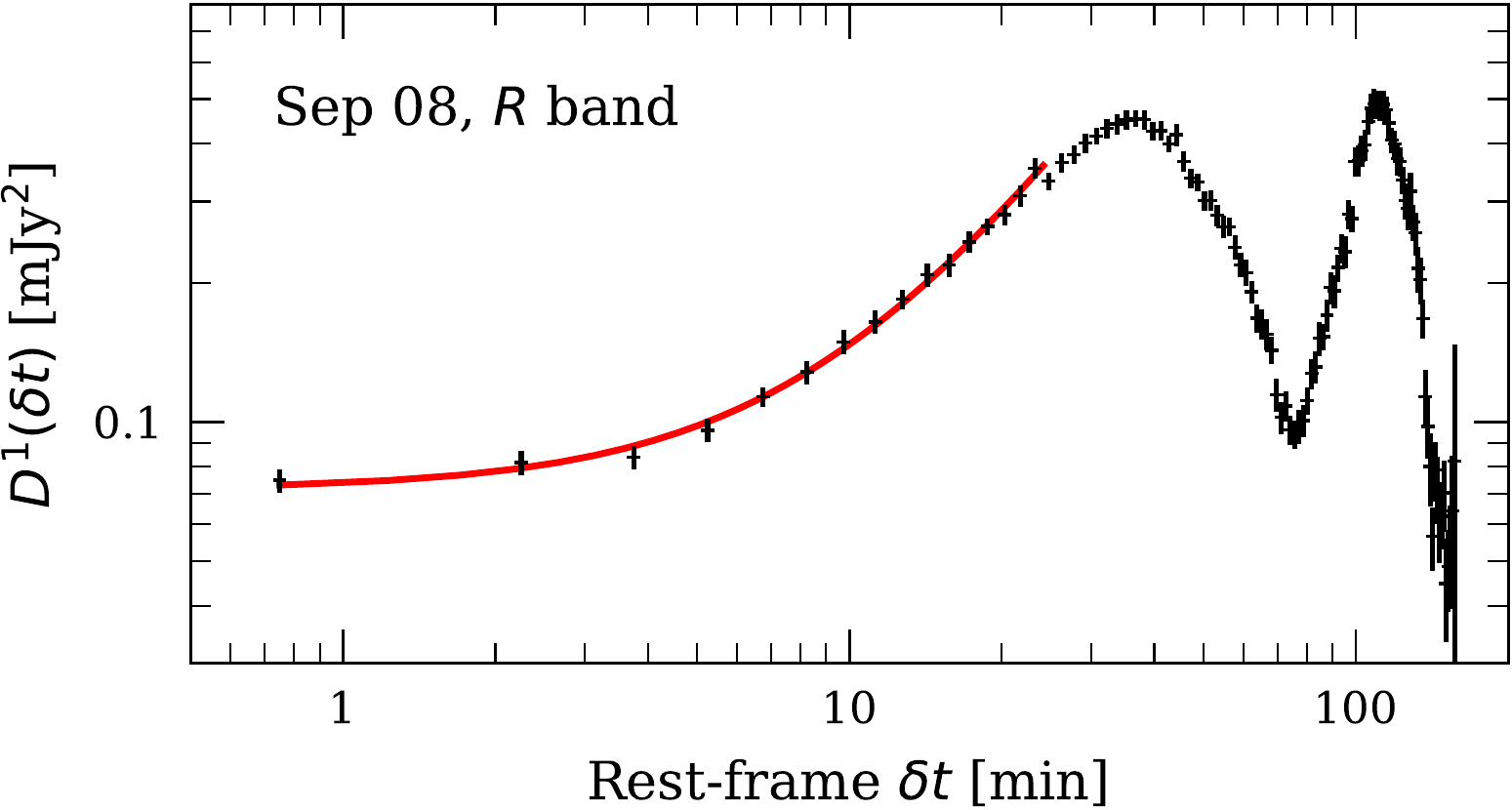}{0.31\textwidth}{}
          \fig{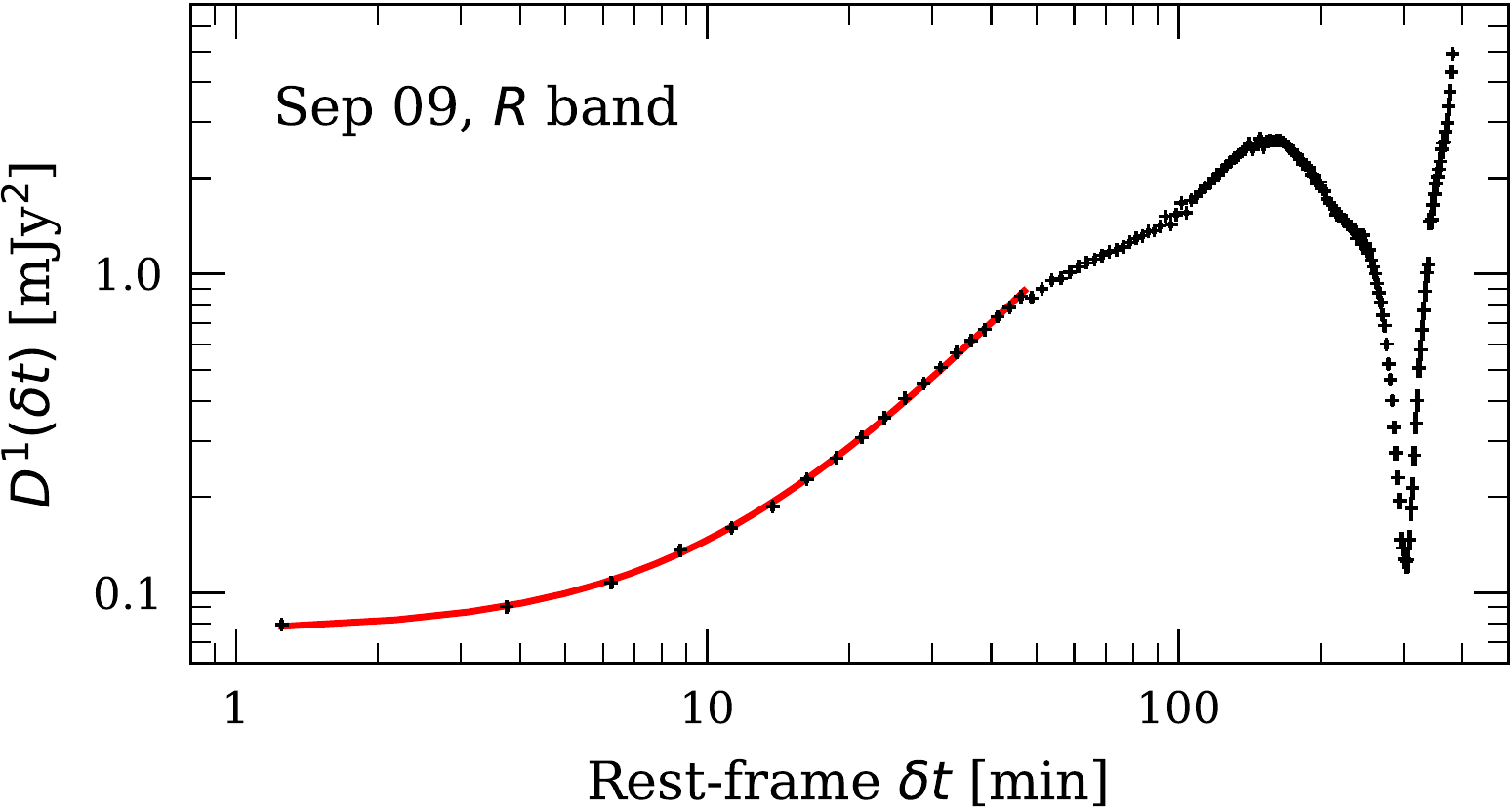}{0.31\textwidth}{}}
\gridline{\fig{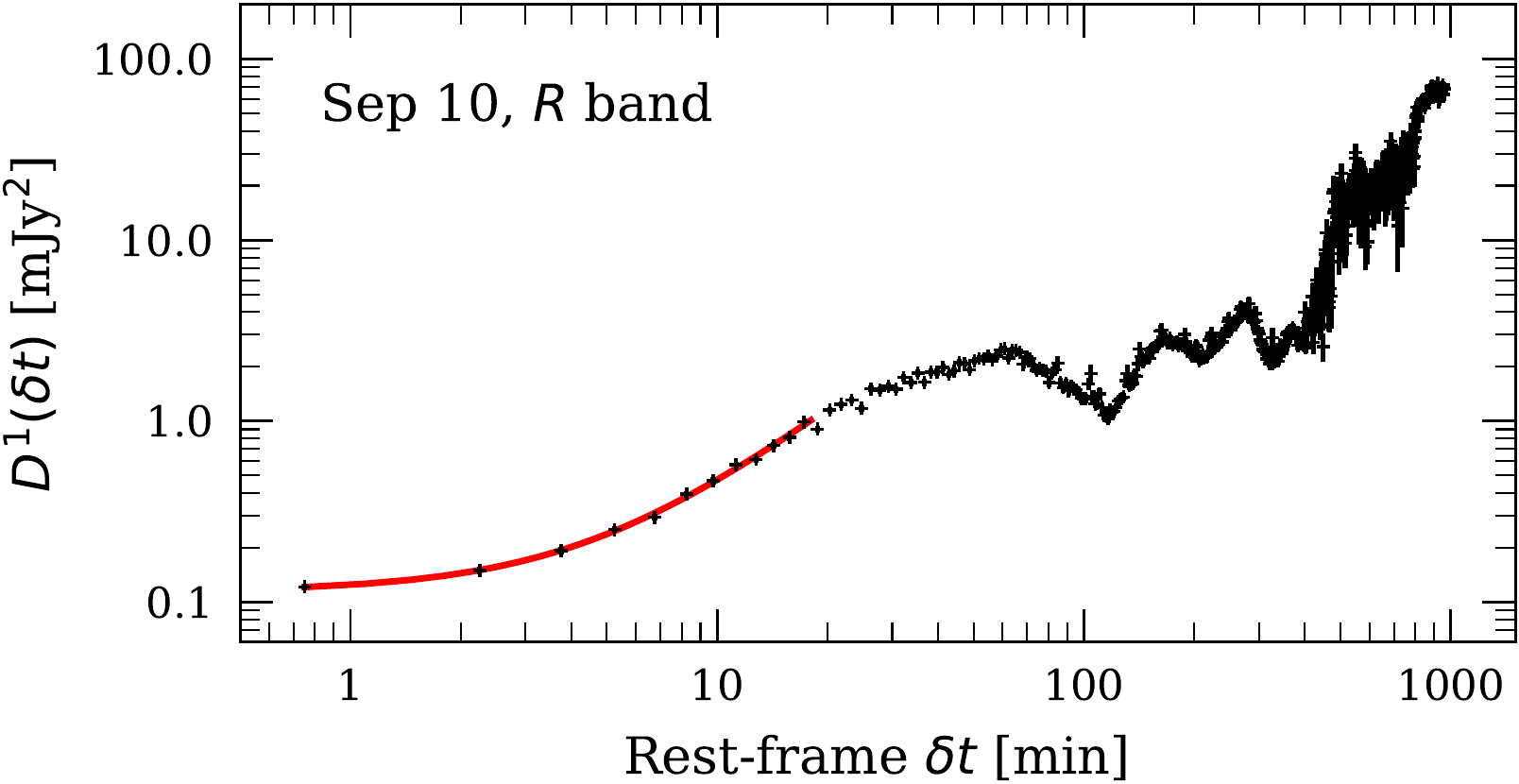}{0.31\textwidth}{}
          \fig{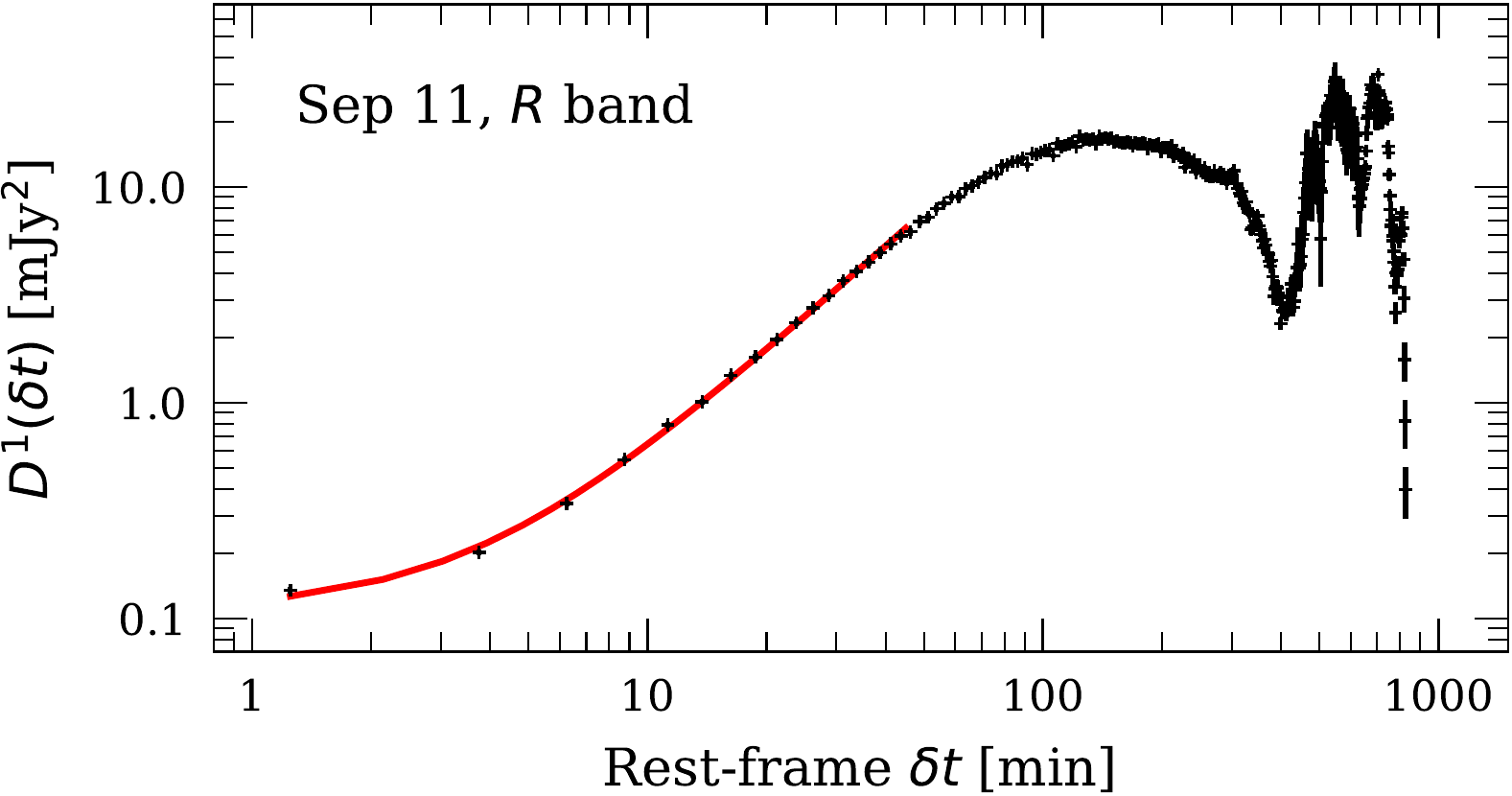}{0.31\textwidth}{}
          \fig{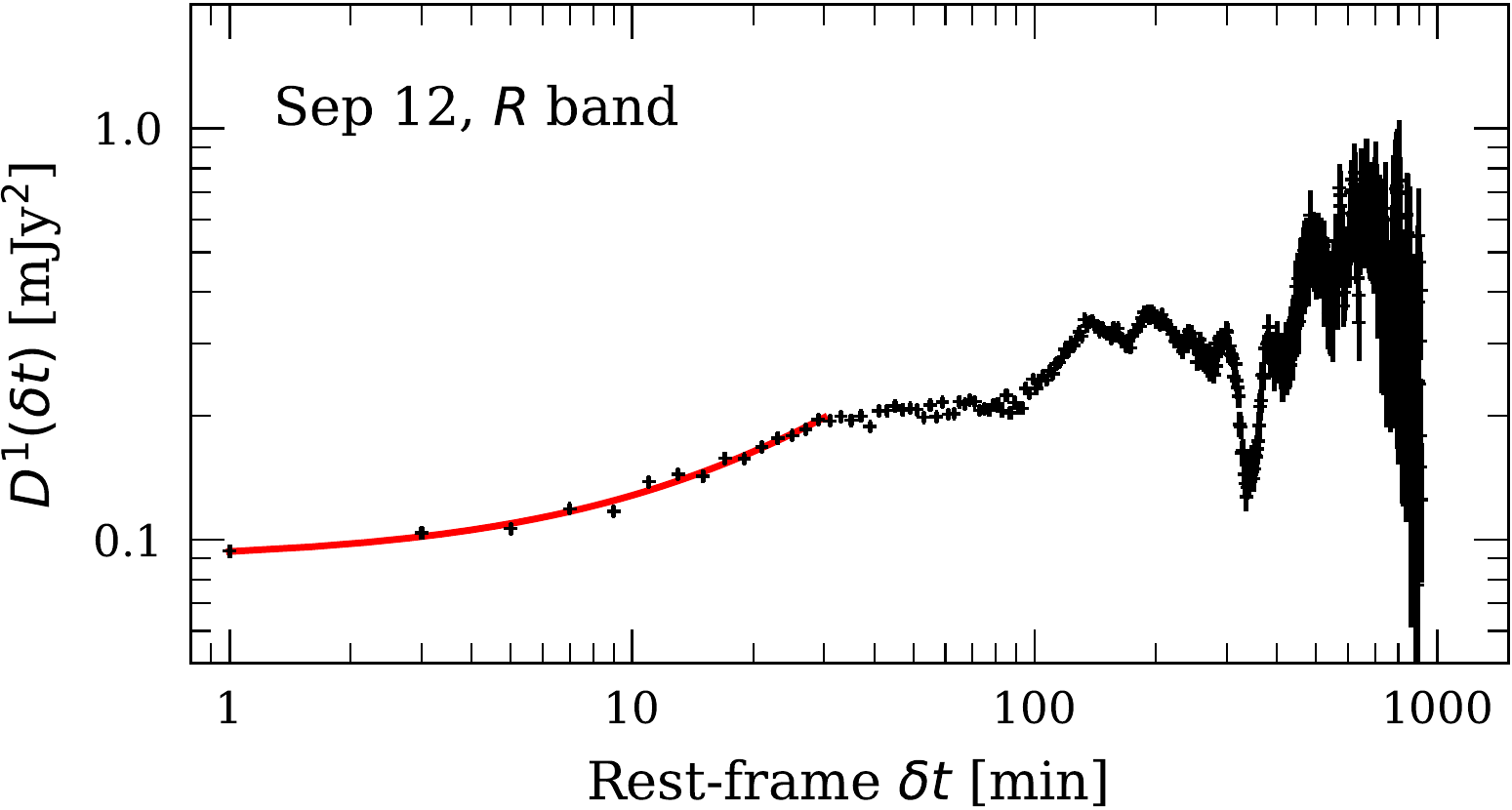}{0.31\textwidth}{}}
\gridline{\fig{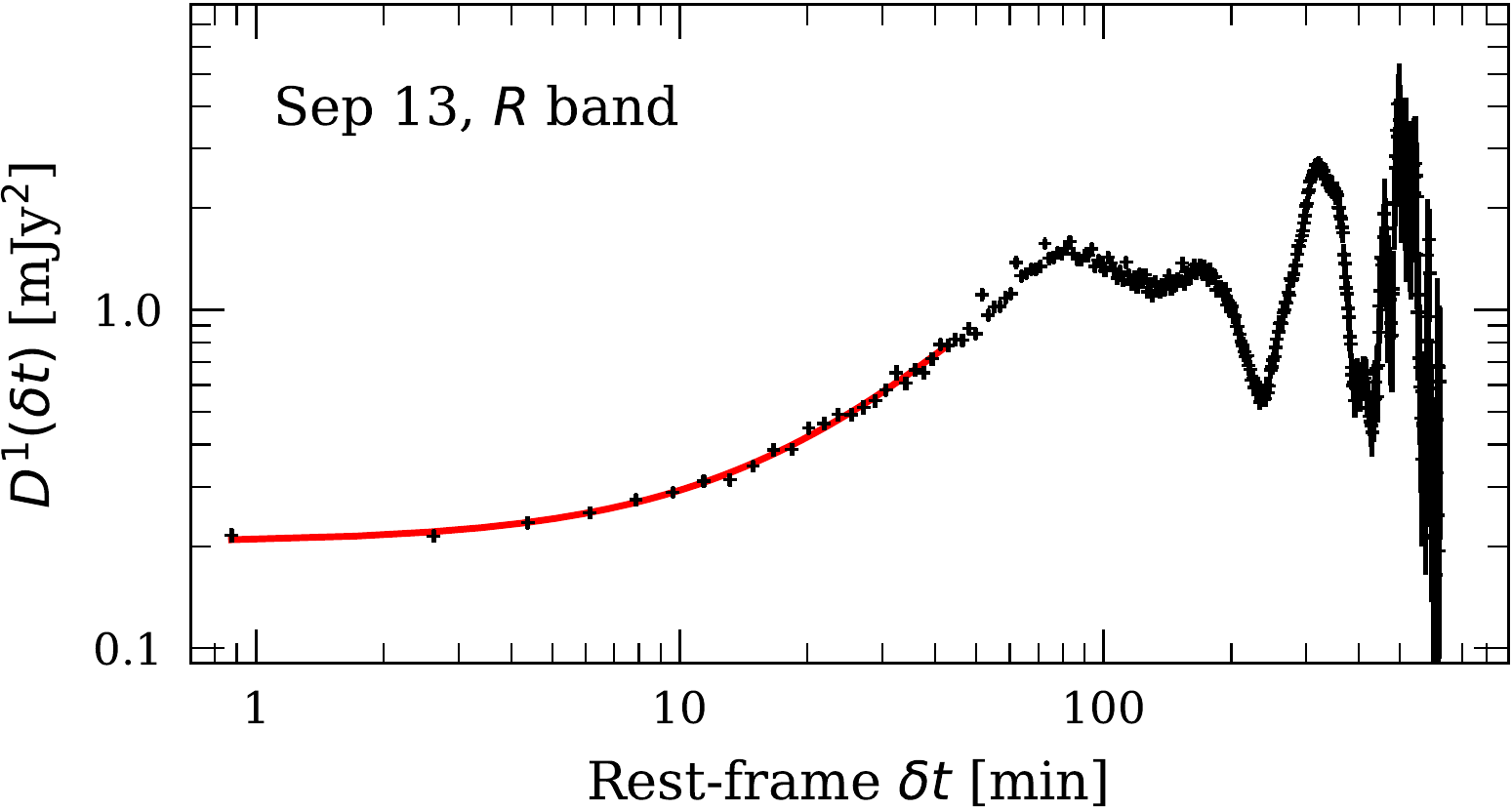}{0.31\textwidth}{}
          \fig{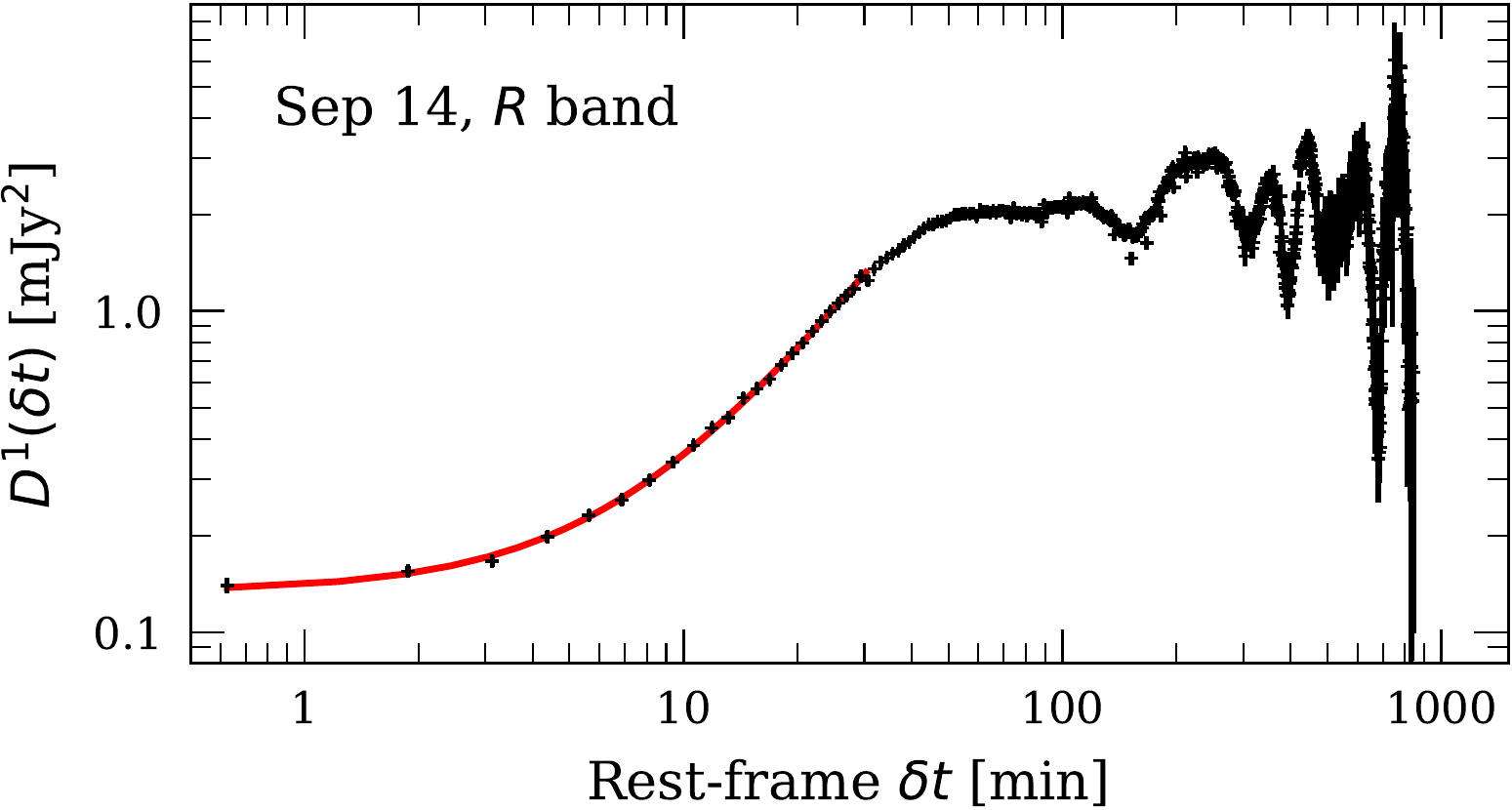}{0.31\textwidth}{}}
\caption{Structure functions built using the corrected LCs. For MWL LCs, only $R$ (or $I$) band SFs are shown. The SPL function fits are overplotted with a red line.}
\label{app:sf:fig}
\end{figure*}

\begin{figure*}[t!]
\gridline{\fig{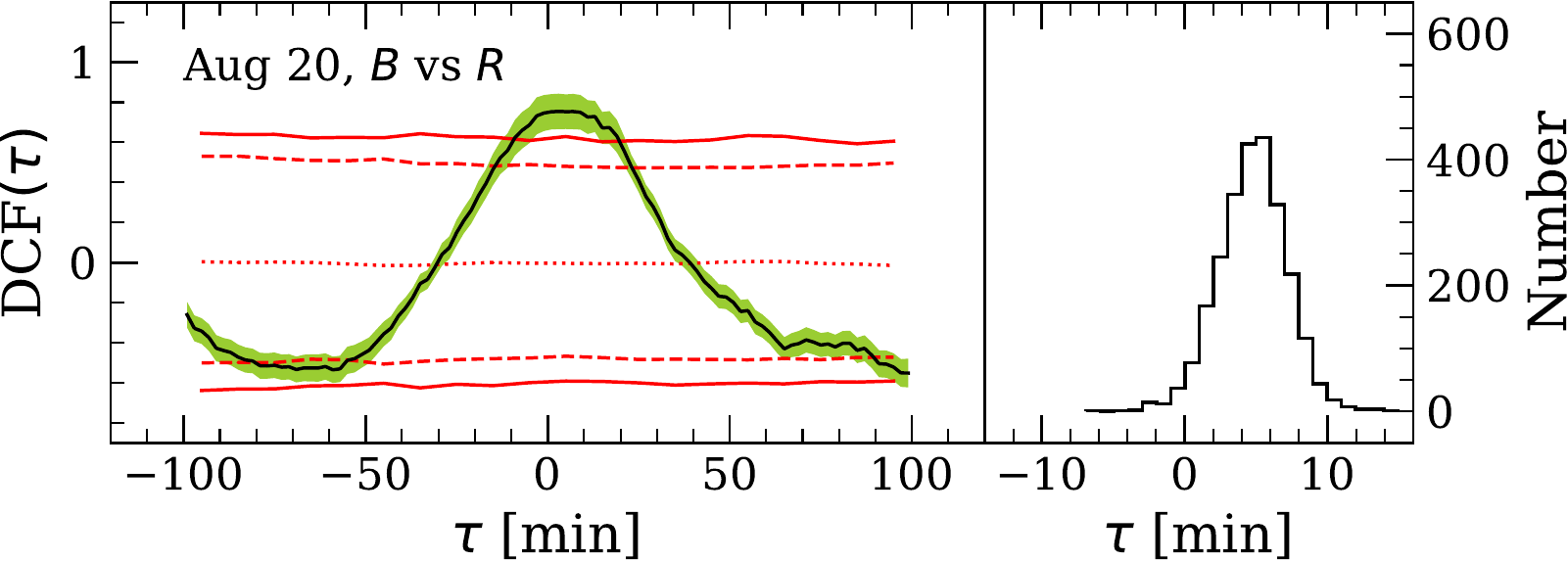}{0.325\textwidth}{}
          \fig{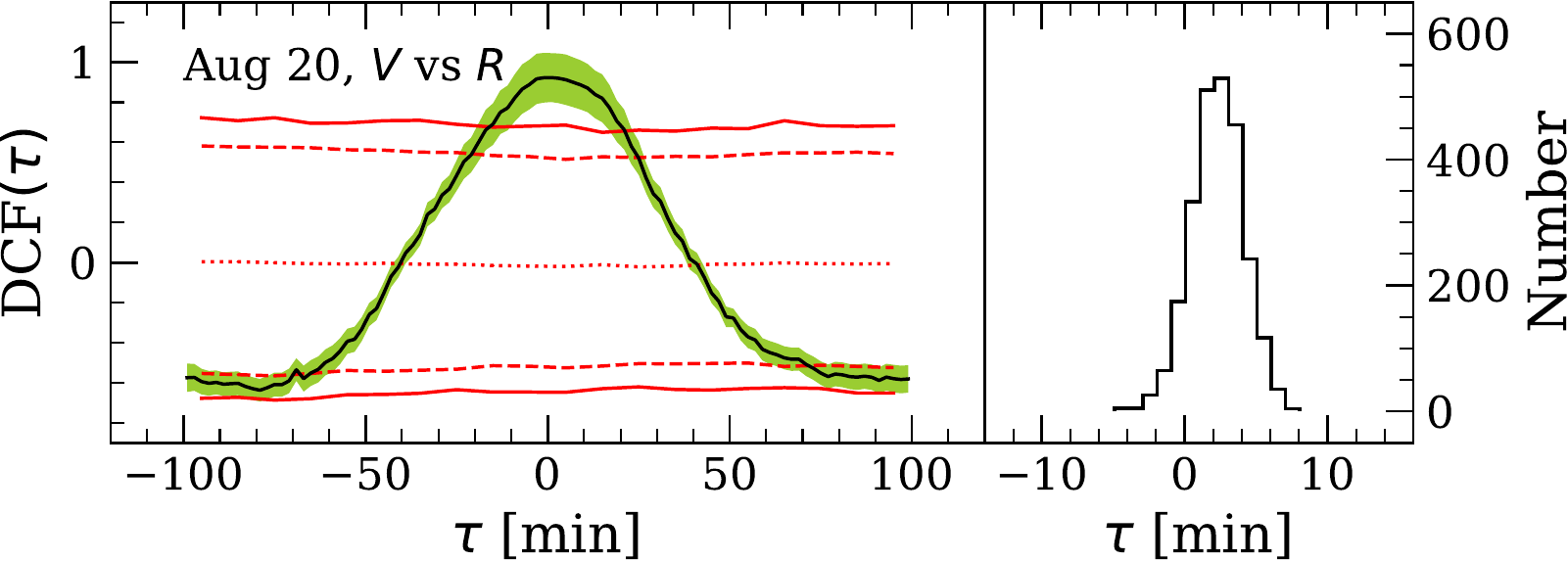}{0.325\textwidth}{}
          \fig{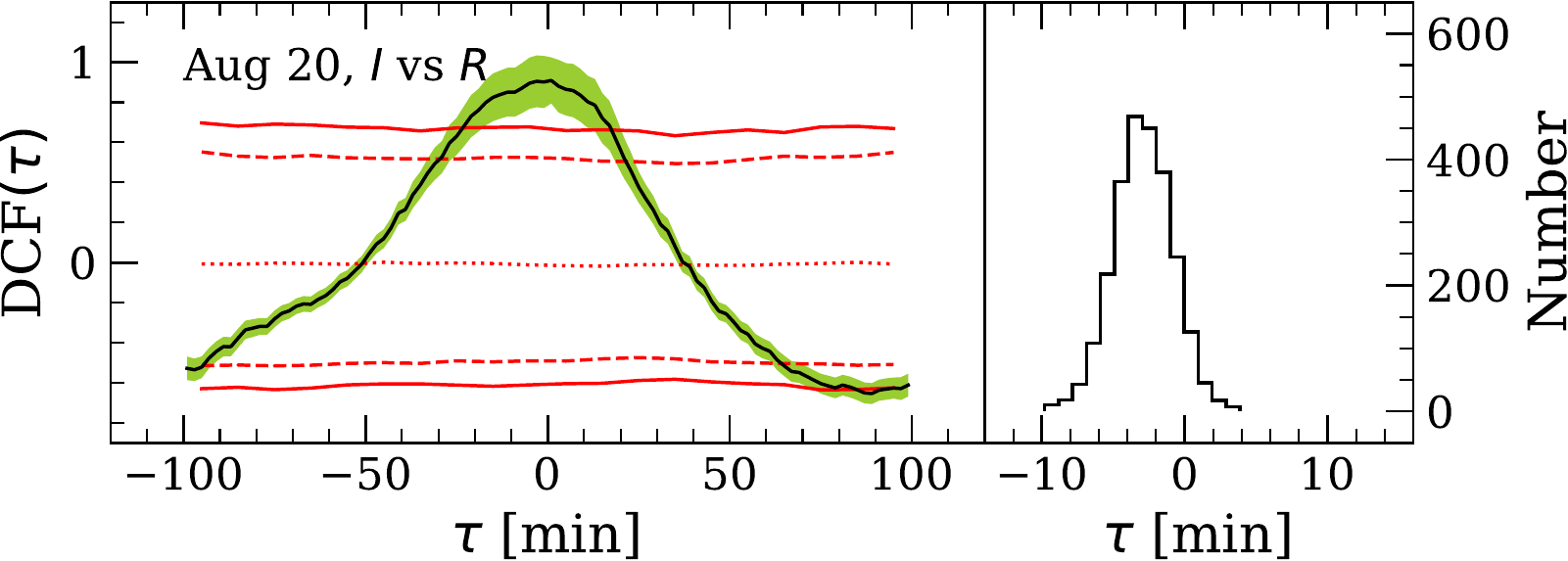}{0.325\textwidth}{}}
\gridline{\fig{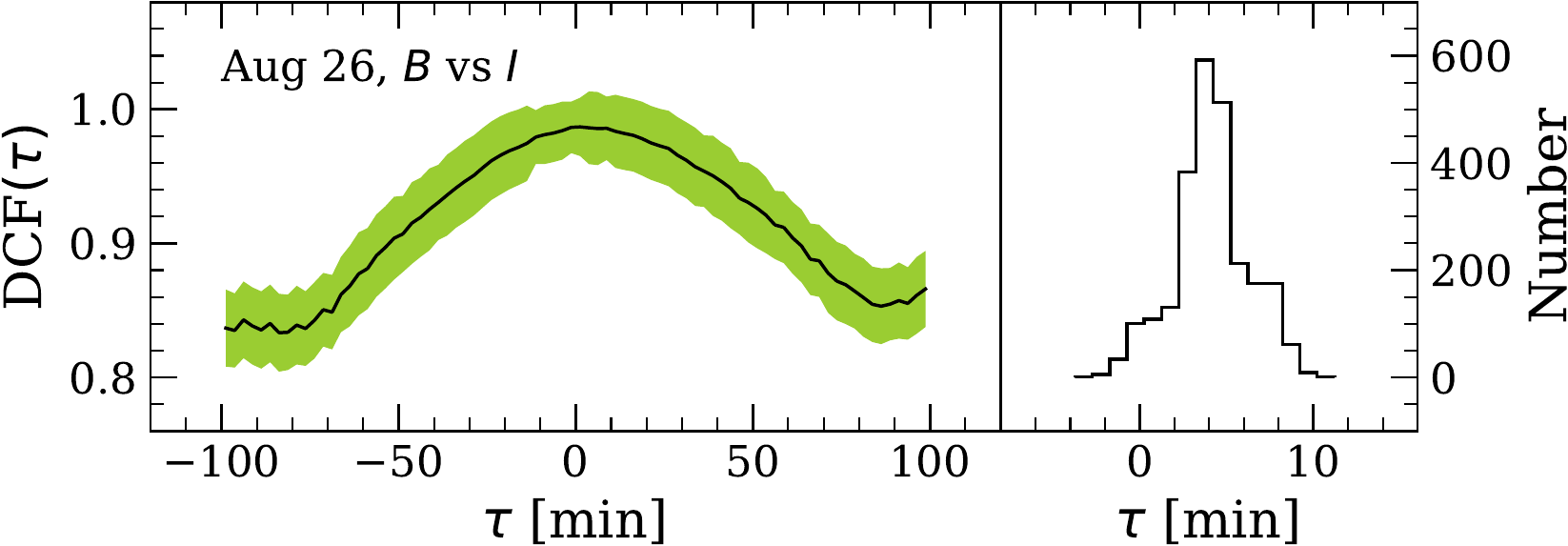}{0.325\textwidth}{}
          \fig{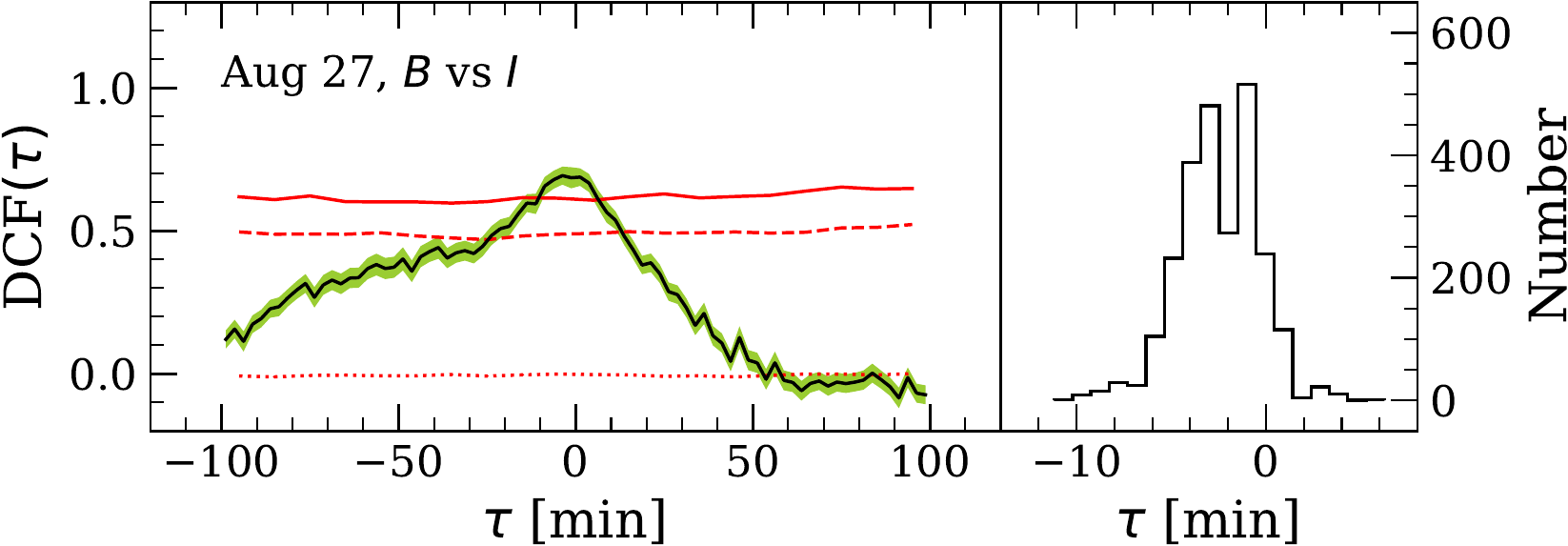}{0.325\textwidth}{}
          \fig{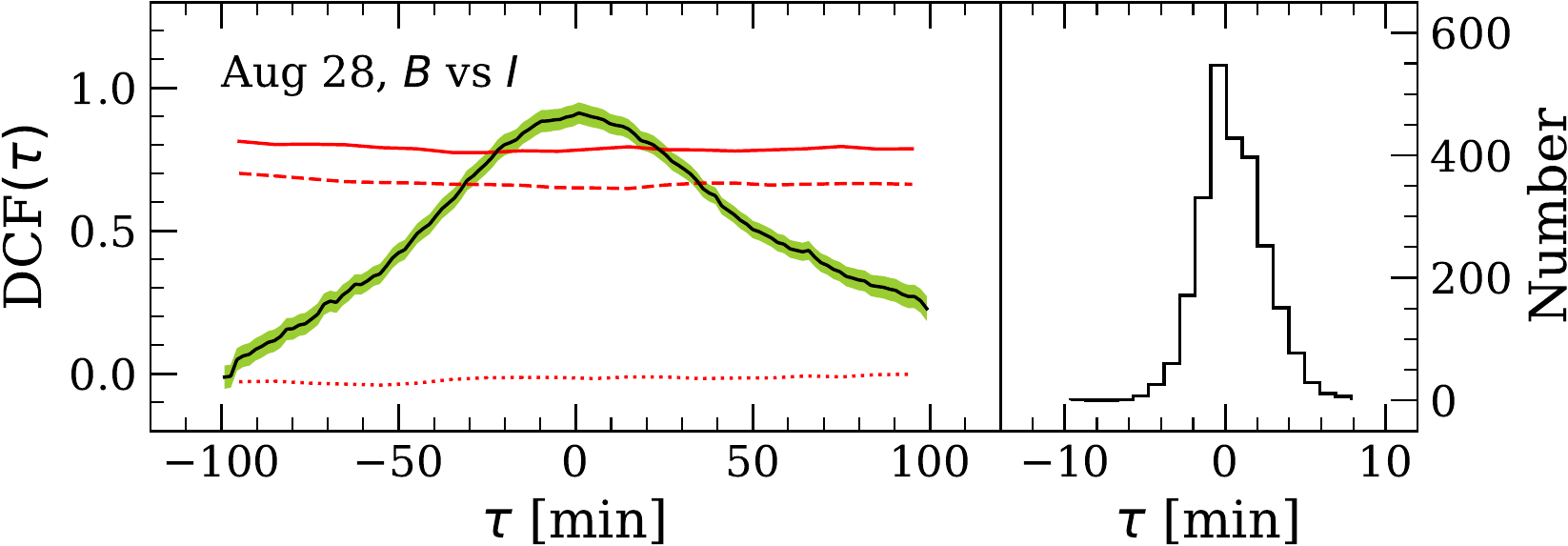}{0.325\textwidth}{}}
\gridline{\fig{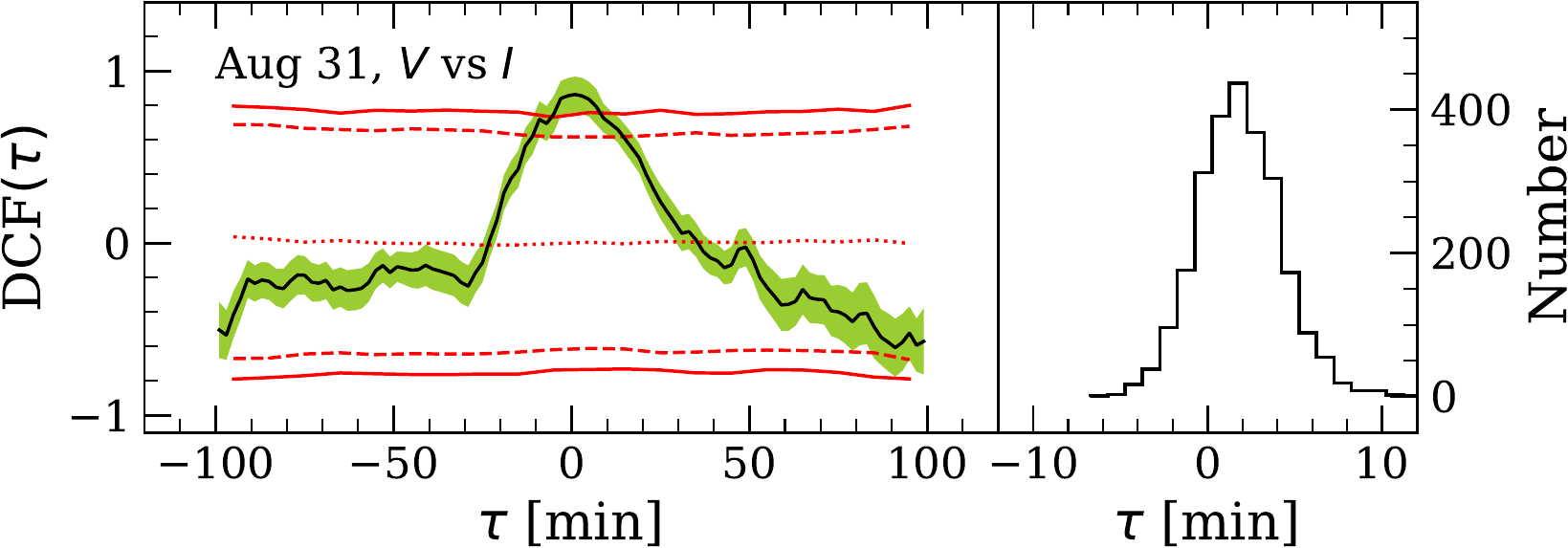}{0.325\textwidth}{}
          \fig{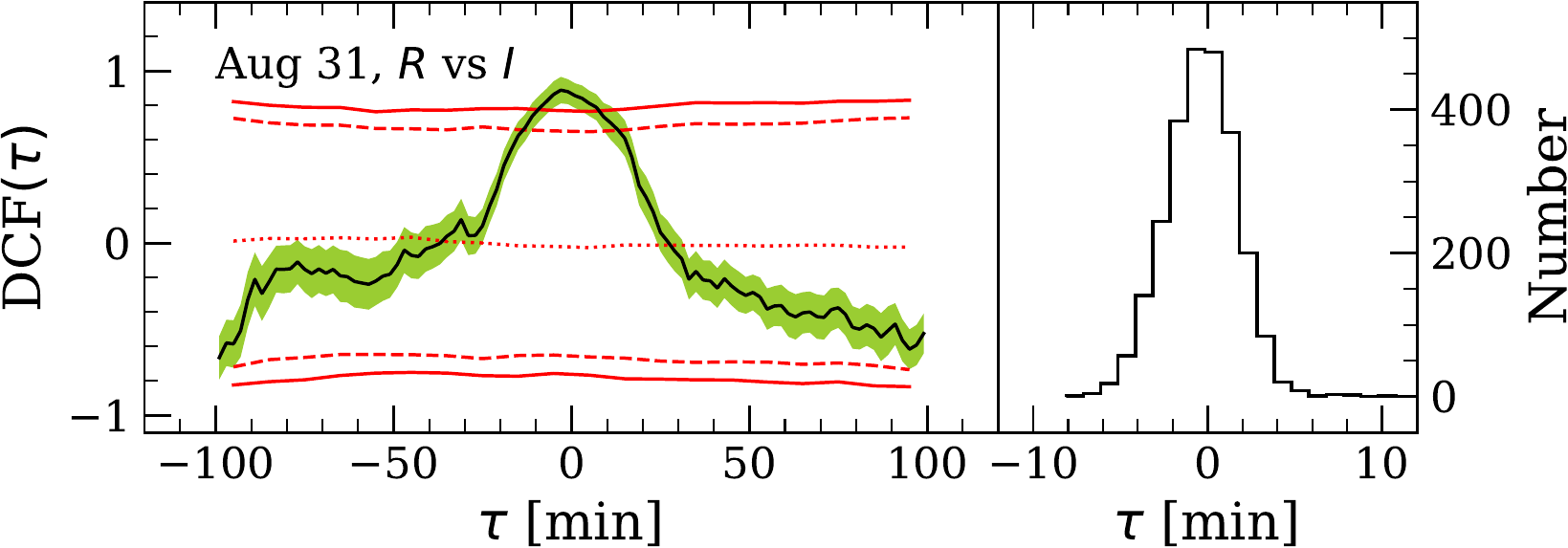}{0.325\textwidth}{}
          \fig{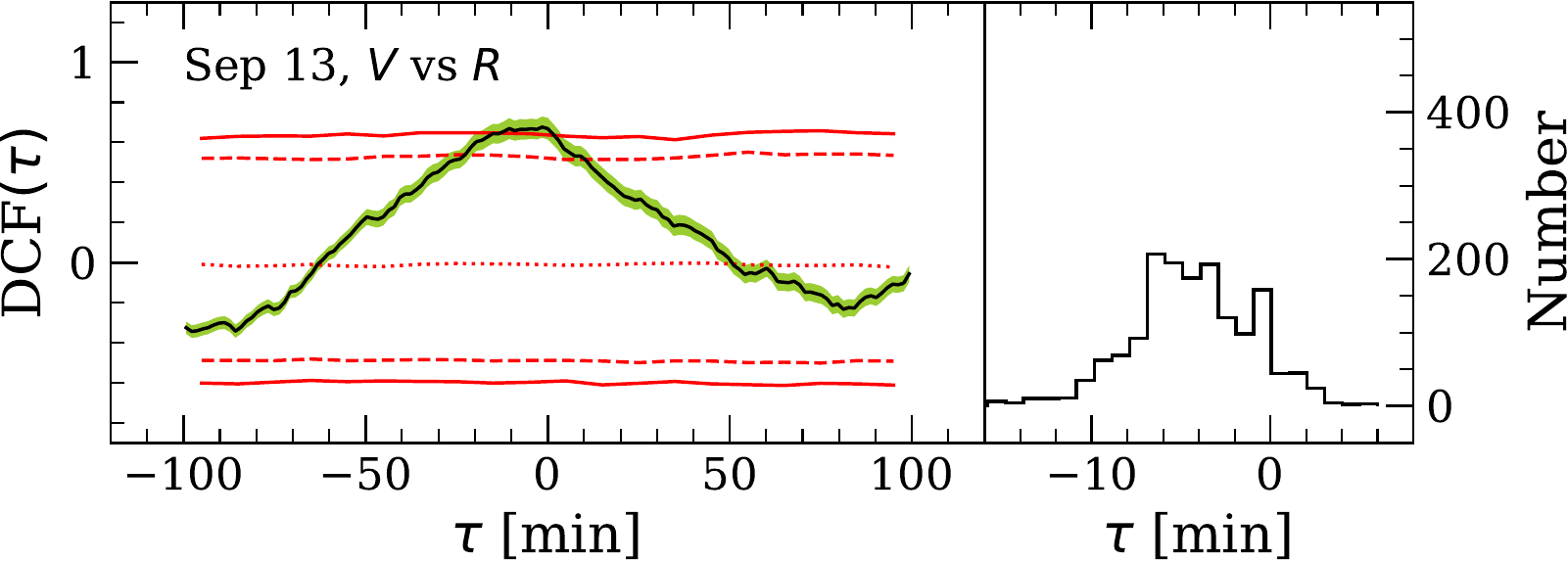}{0.325\textwidth}{}}
\gridline{\fig{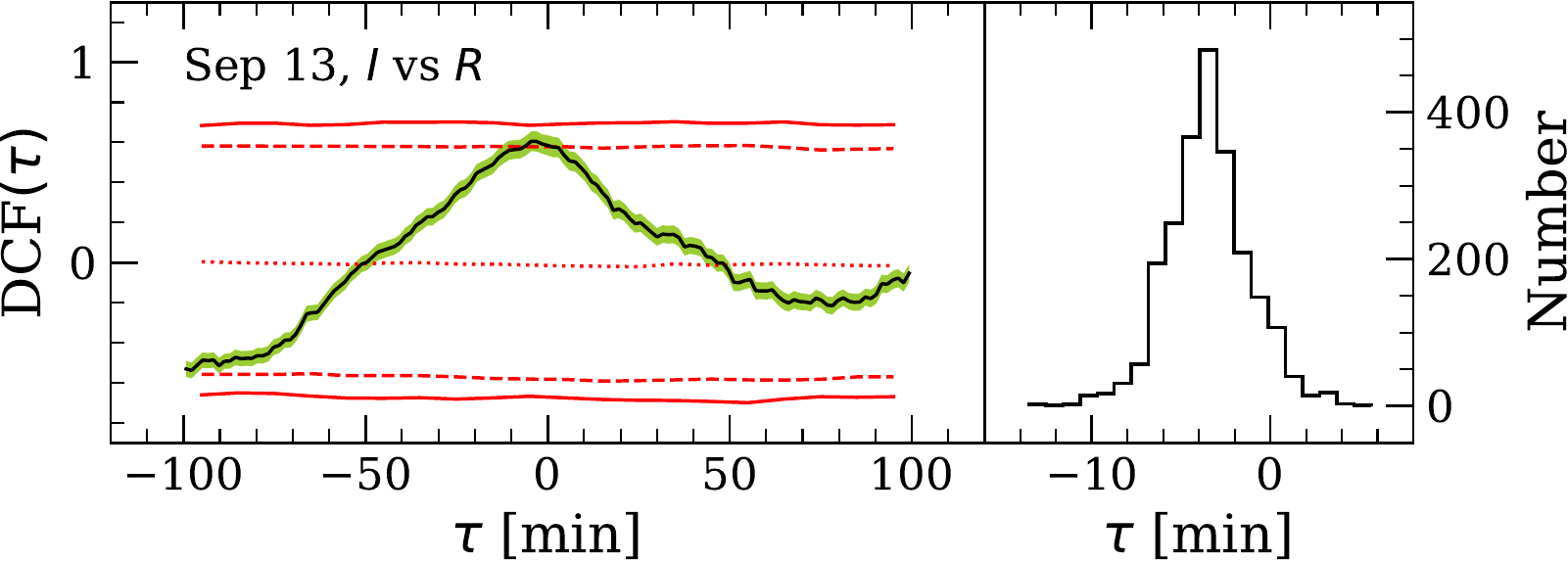}{0.325\textwidth}{}
          \fig{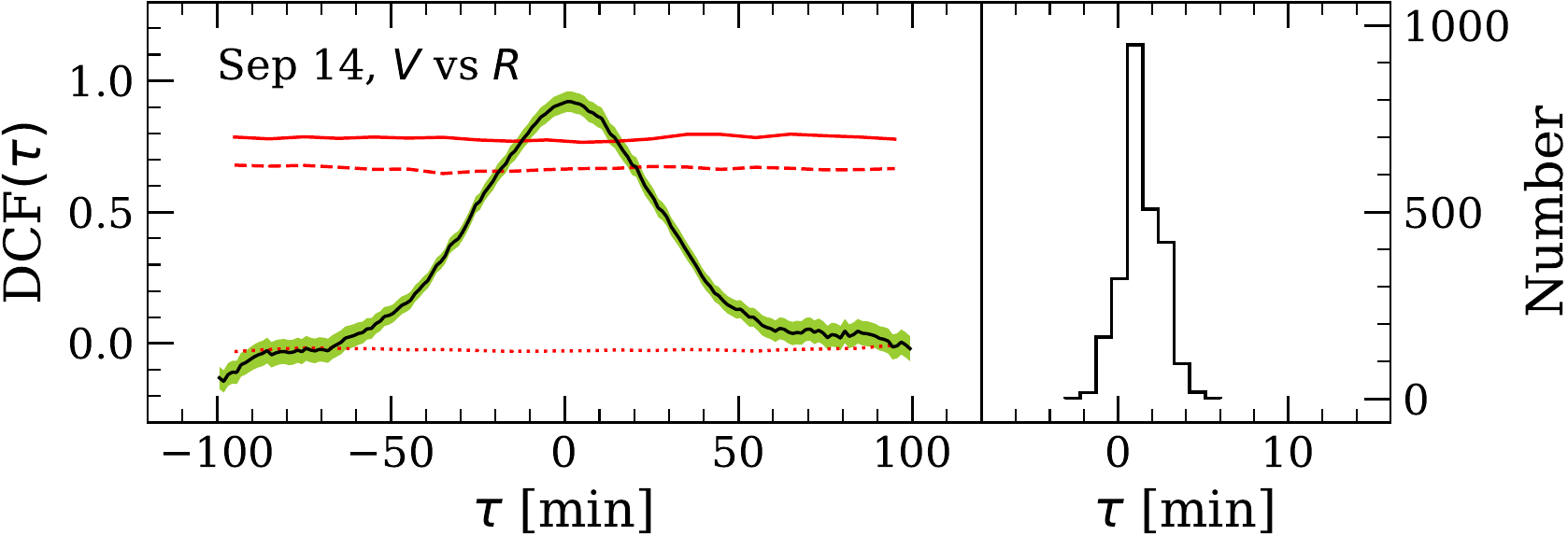}{0.325\textwidth}{}
          \fig{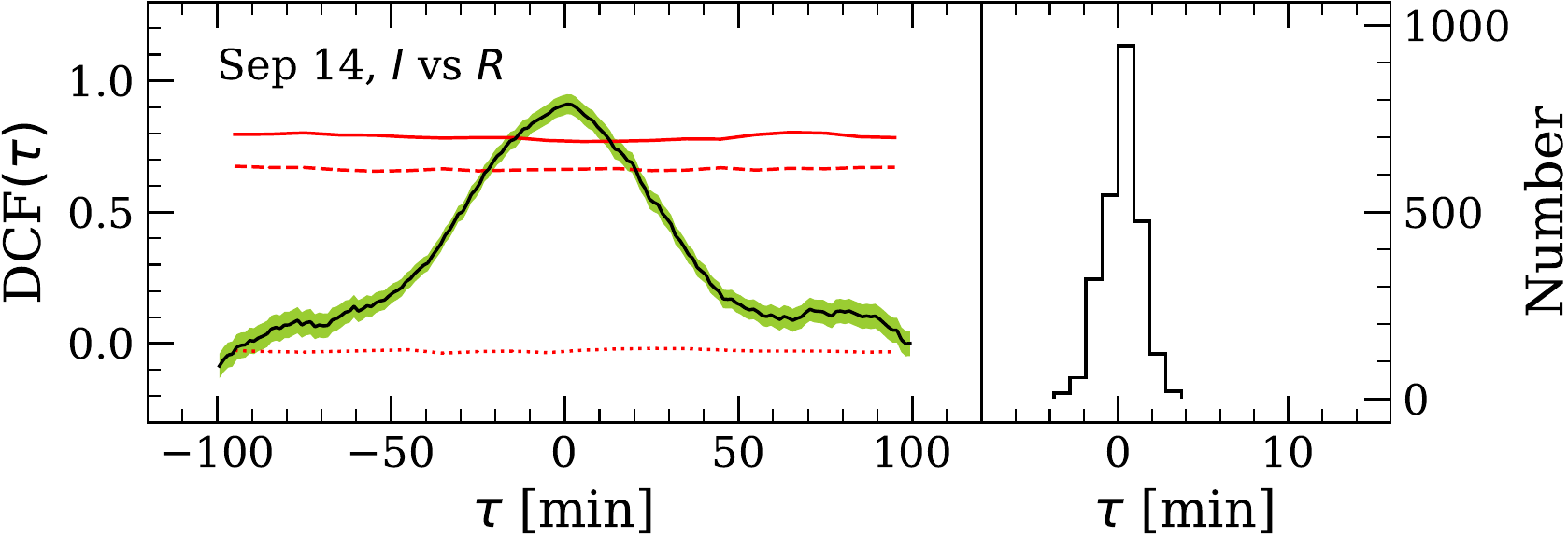}{0.325\textwidth}{}}
\caption{Results from the cross-correlation analysis of the corrected MWL LCs. In each plot the left panel shows the DCF (black lines) and its uncertainties (green shaded area). The red solid and dashed lines indicate the significance levels of 99\% and 95\%, respectively, while the red dotted line indicates the zero correlation. The right panel in each plot shows the corresponding CCCD.}
\label{app:dcf:fig}
\end{figure*}

\begin{figure*}[t!]
\gridline{\fig{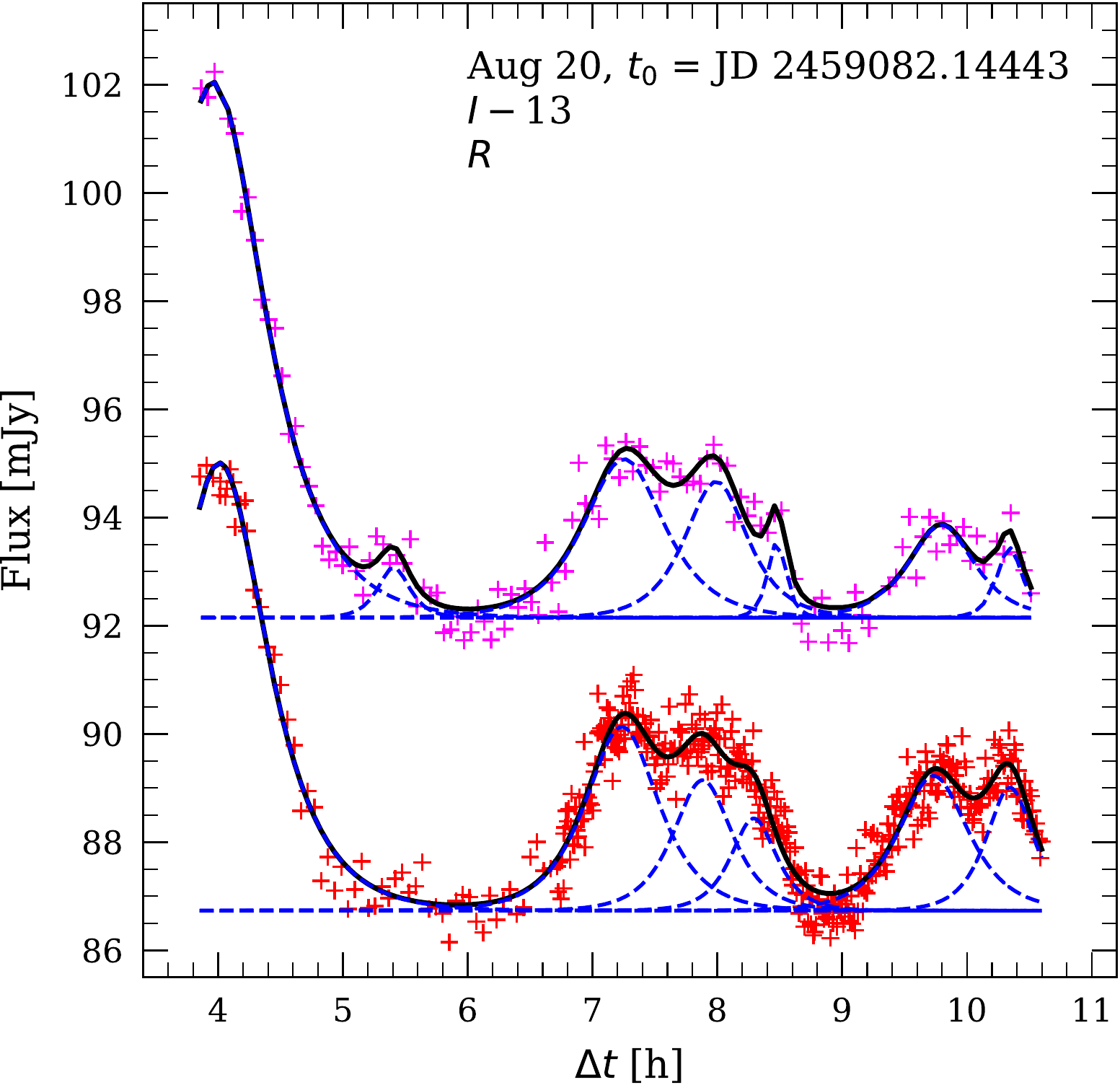}{0.31\textwidth}{}
          \fig{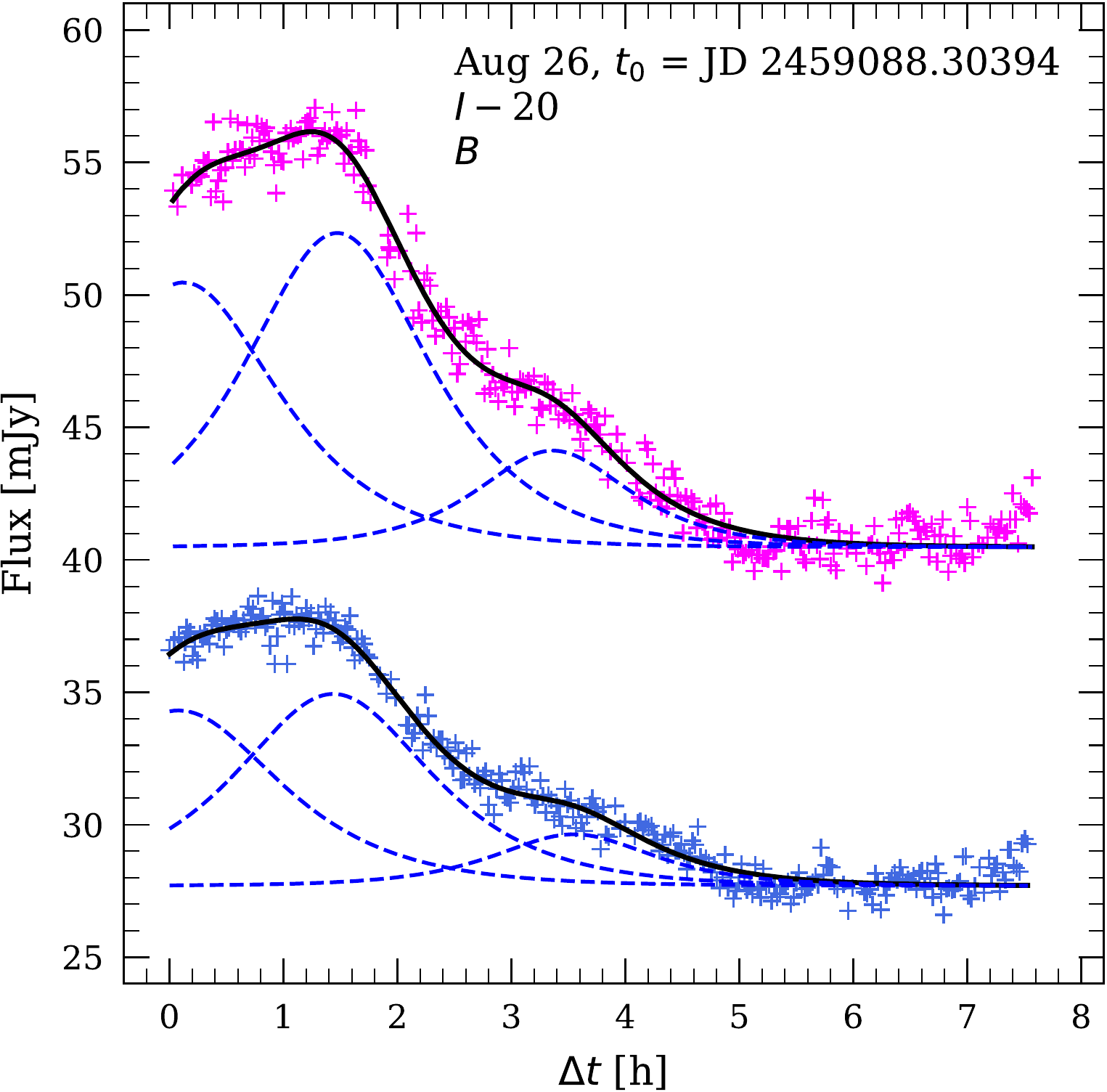}{0.31\textwidth}{}
          \fig{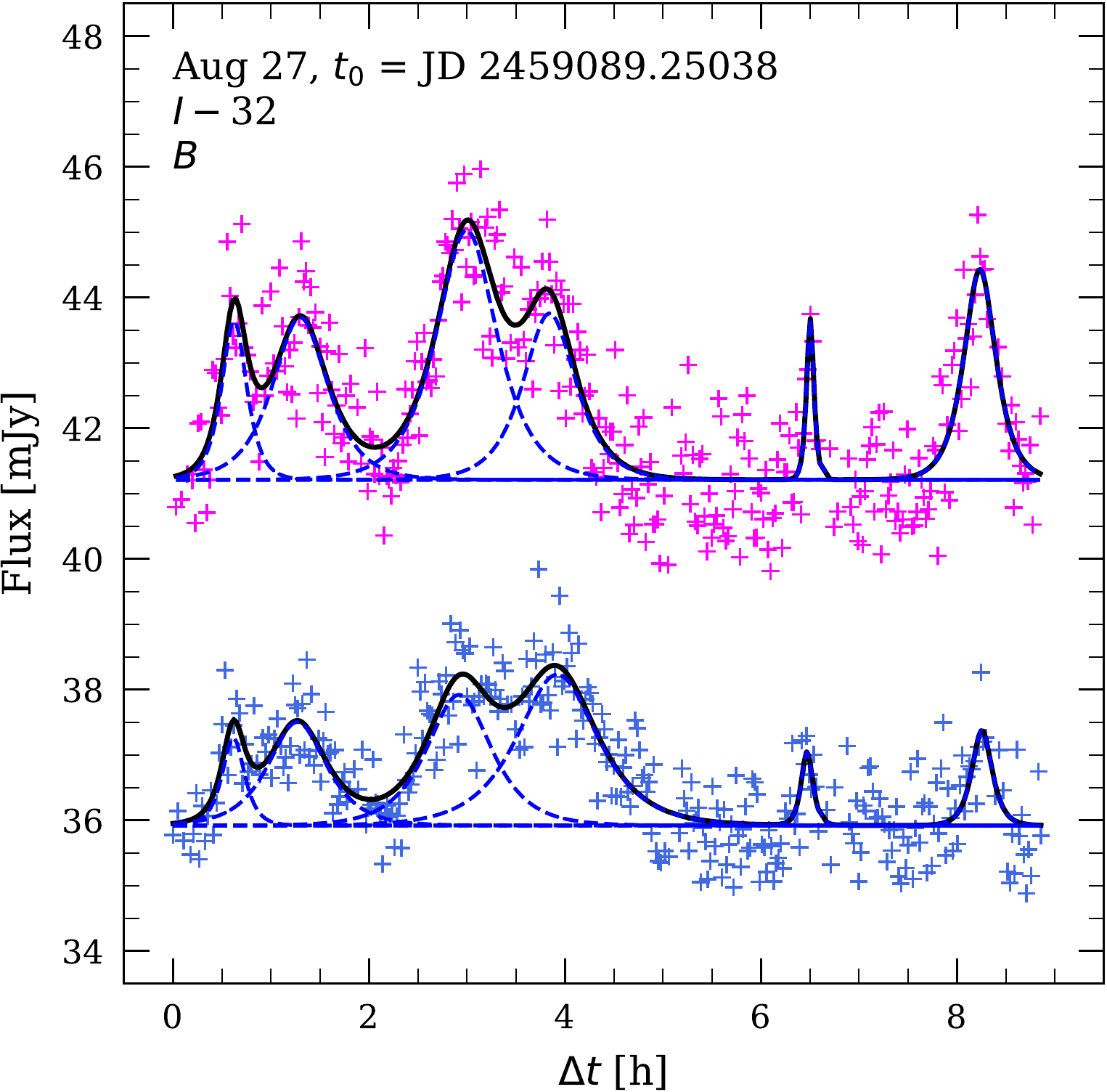}{0.31\textwidth}{}}
\gridline{\fig{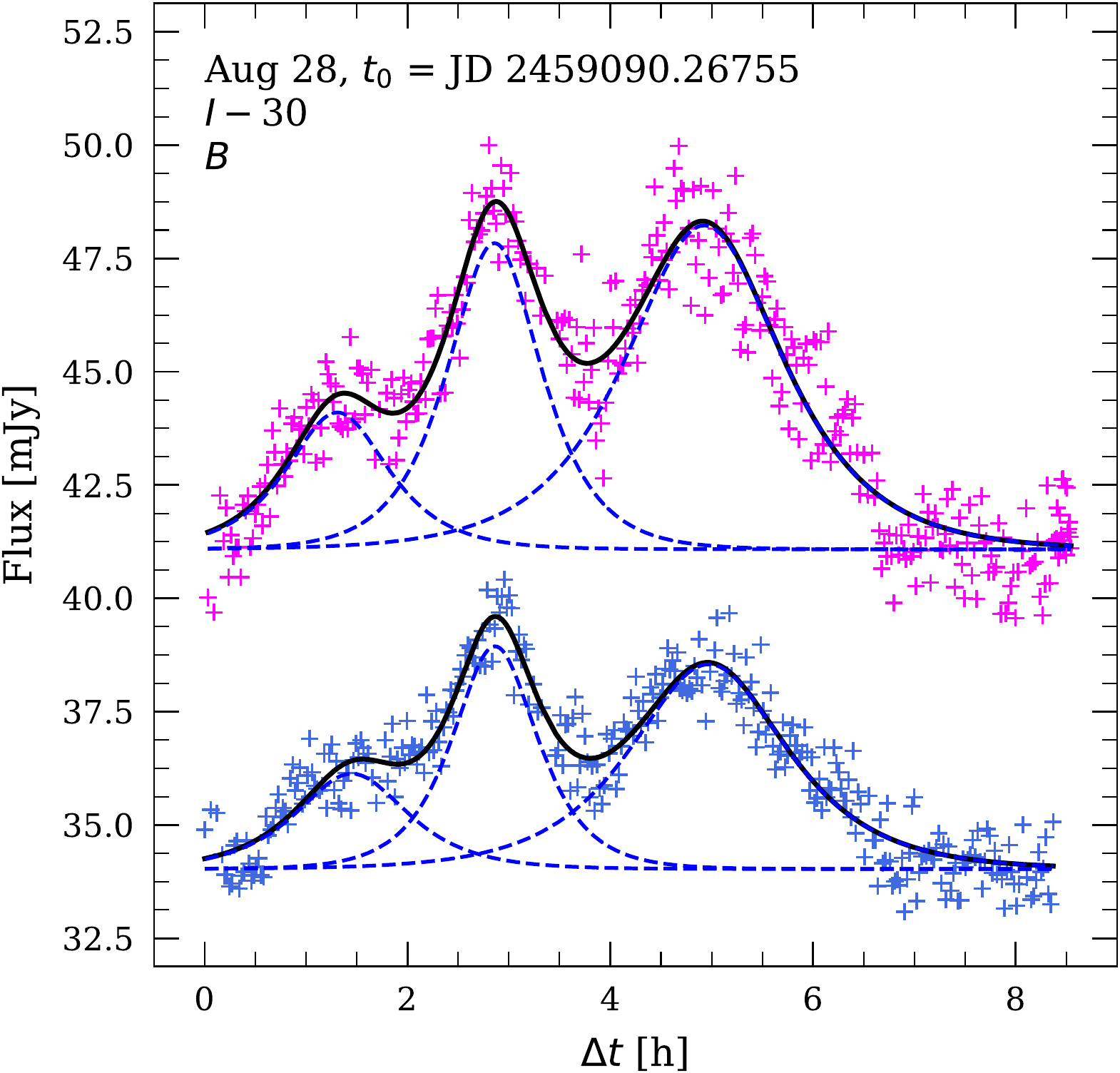}{0.31\textwidth}{}
          \fig{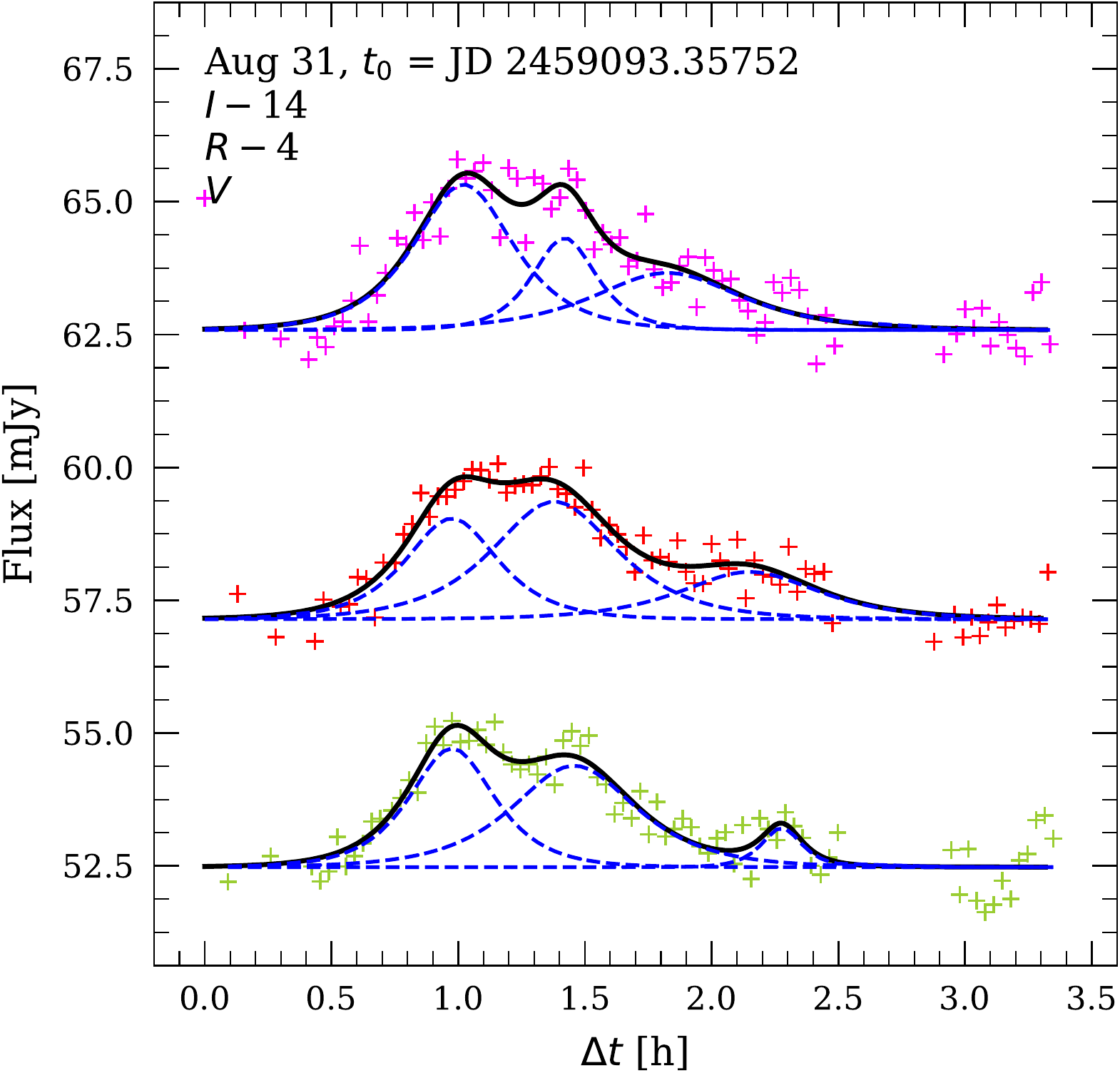}{0.31\textwidth}{}
          \fig{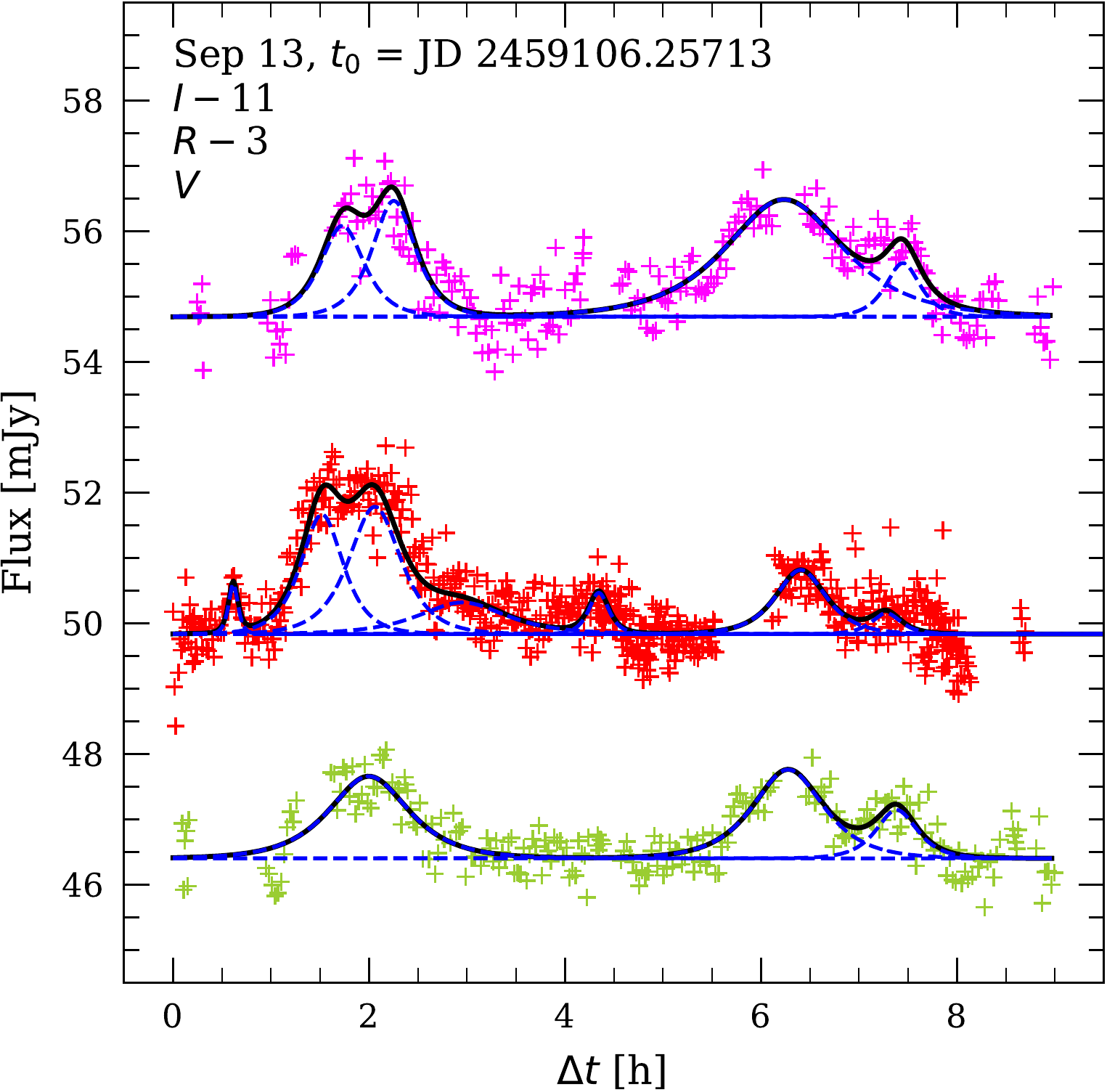}{0.31\textwidth}{}}
\gridline{\fig{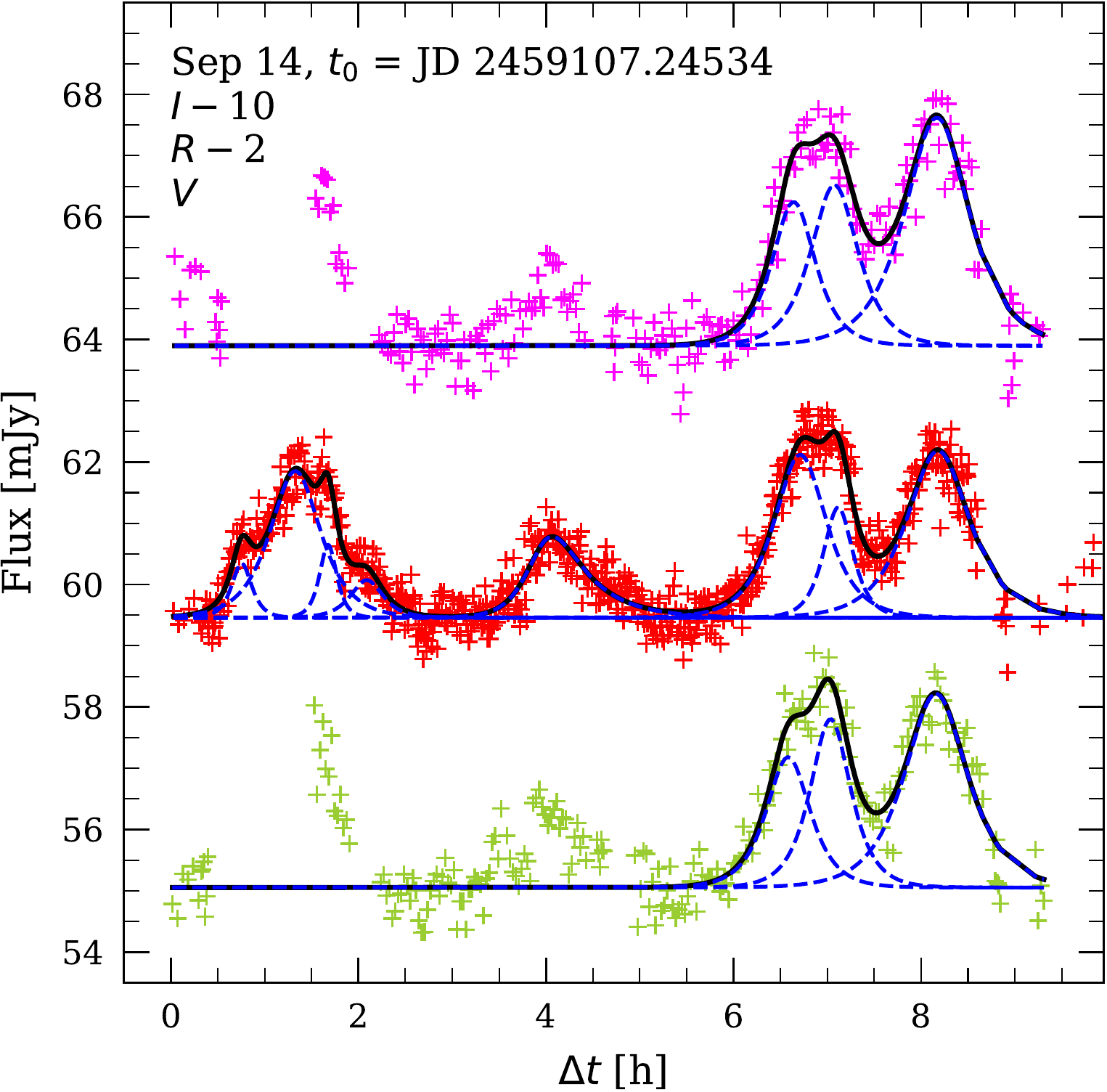}{0.31\textwidth}{}}
\caption{Decomposition of the corrected MWL LCs. In each plot, we indicate the evening date, the value of $t_0$, the bands plotted, and the corresponding offsets used for display purposes. The bands are coded as follows: $B$~-- blue, $V$~-- green, $R$~-- red, $I$~-- magenta. The blue dashed lines are the individual flares to which the LC is decomposed, while the black solid line is the model LC. The error bars are not shown for the sake of clarity.}
\label{app:decompo:fig:mwl}
\end{figure*}

\begin{figure*}[t!]
\gridline{\fig{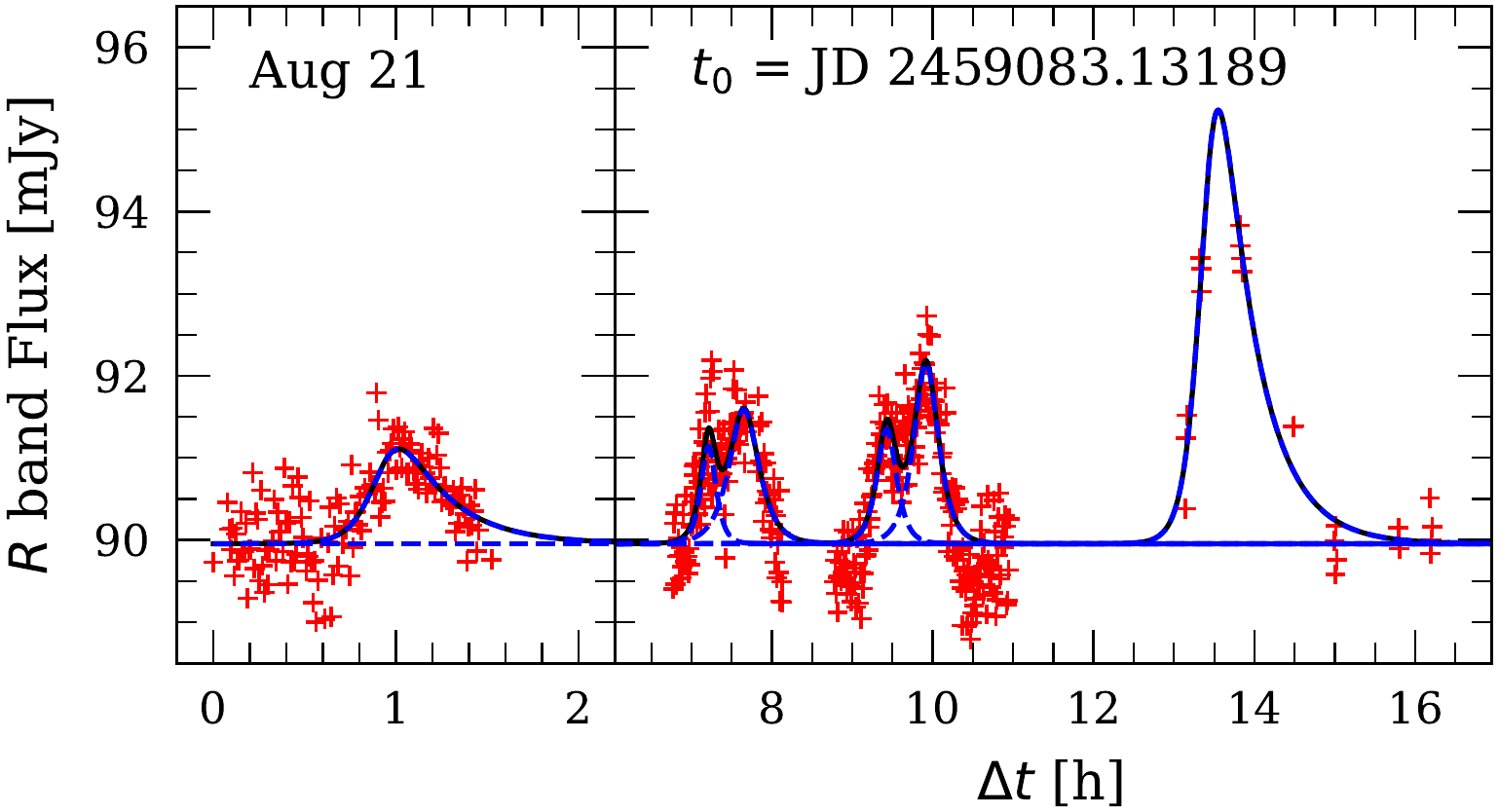}{0.31\textwidth}{}
          \fig{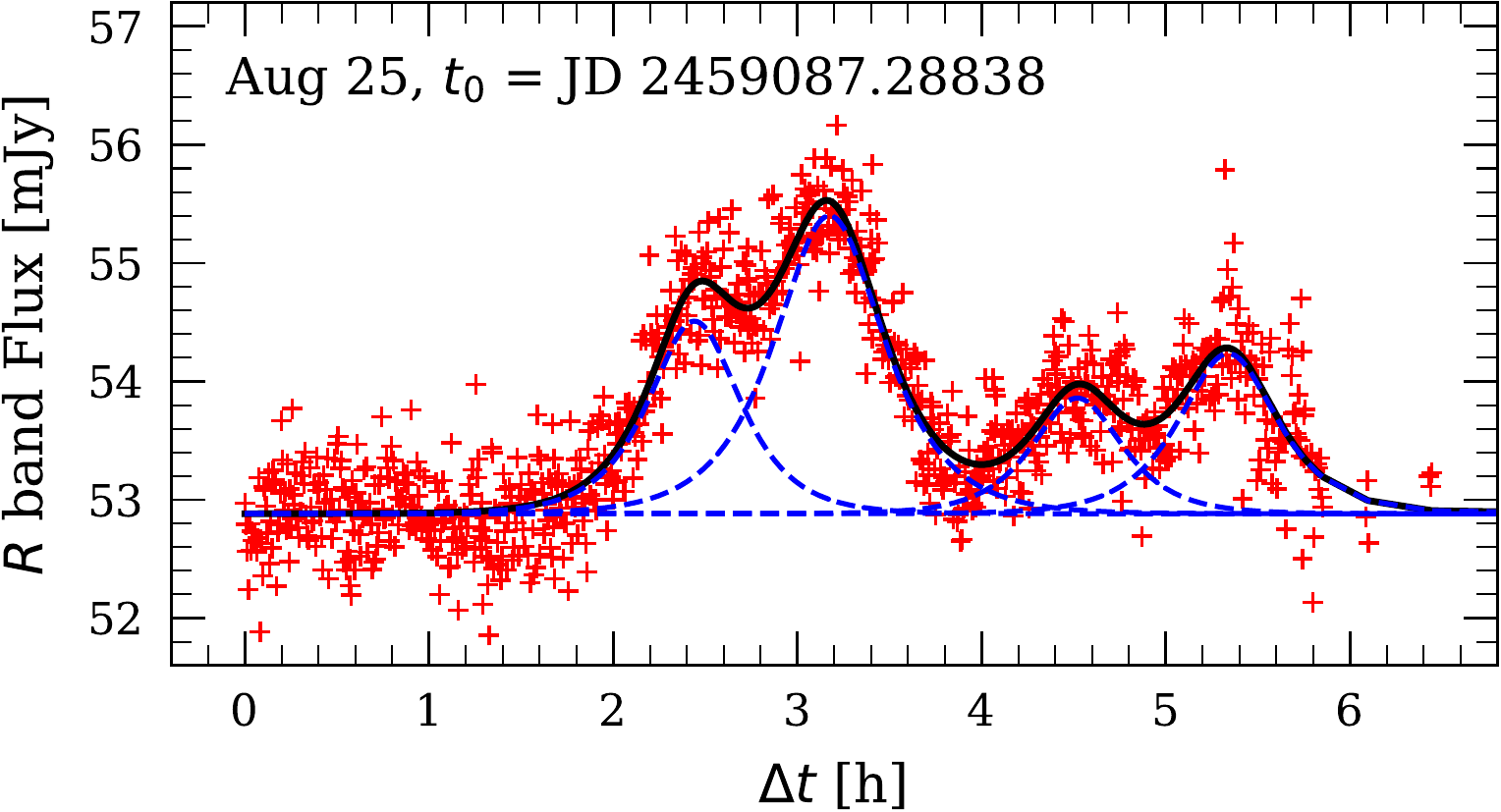}{0.31\textwidth}{}
          \fig{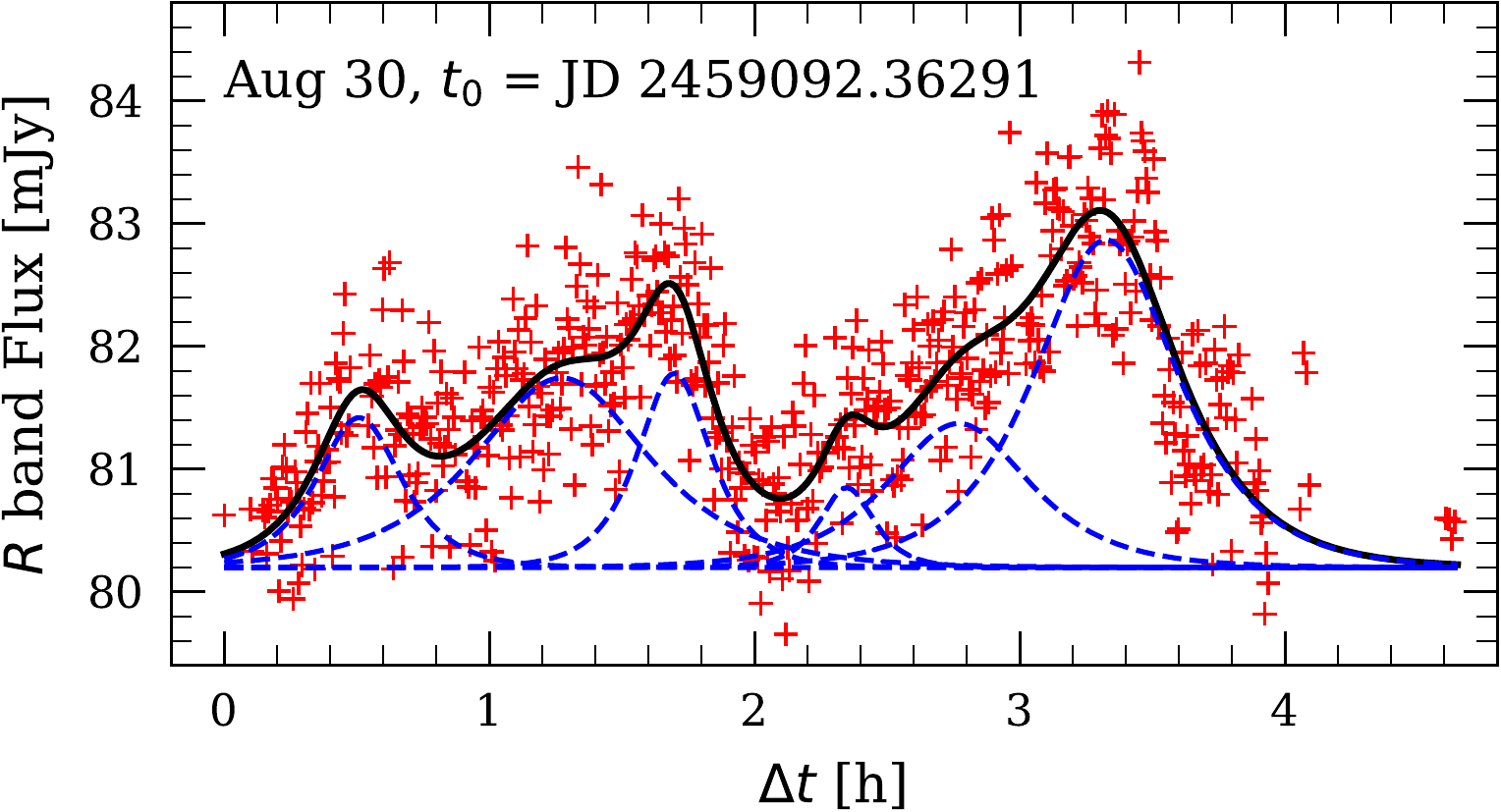}{0.31\textwidth}{}}
\gridline{\fig{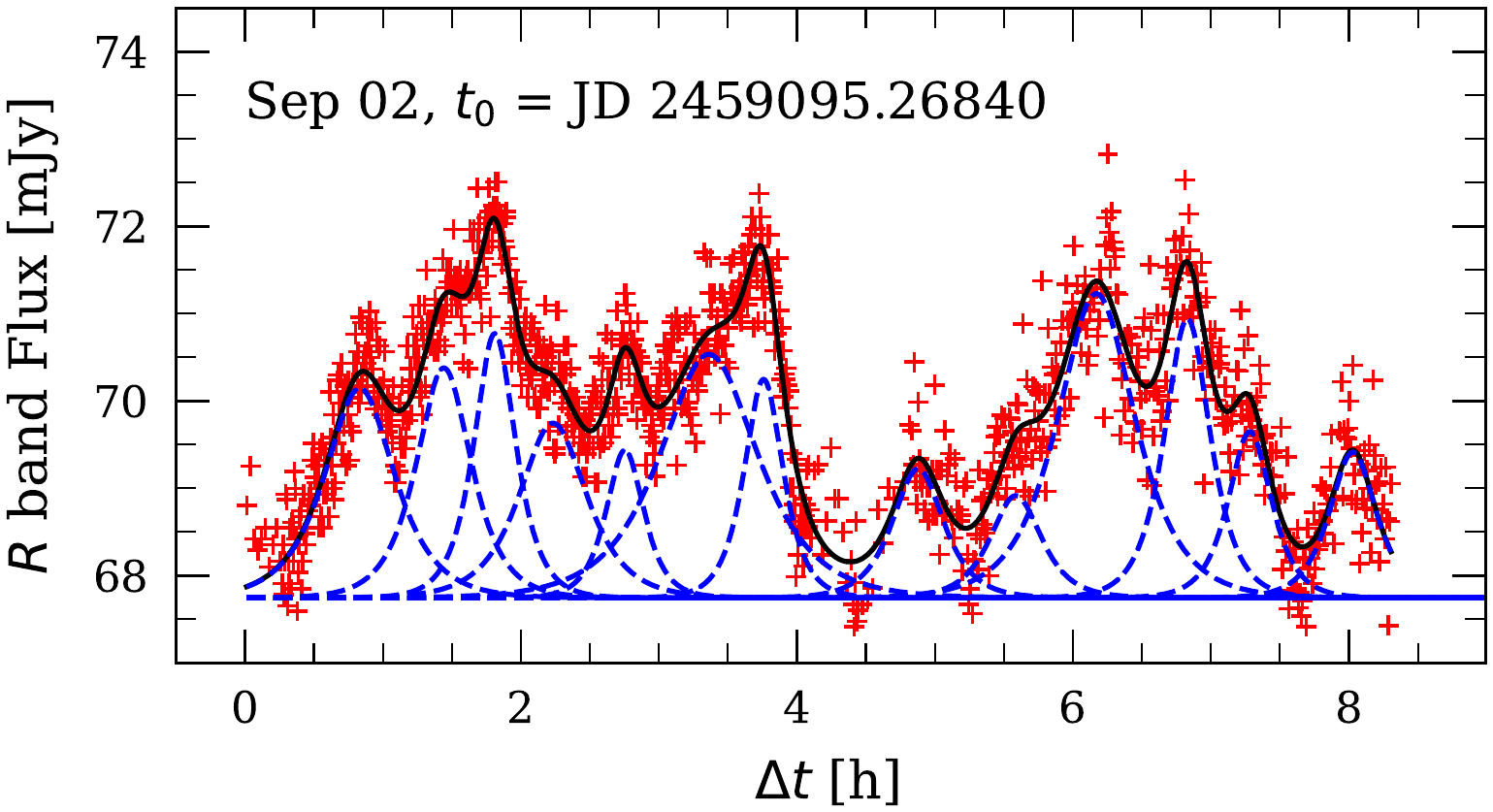}{0.31\textwidth}{}
          \fig{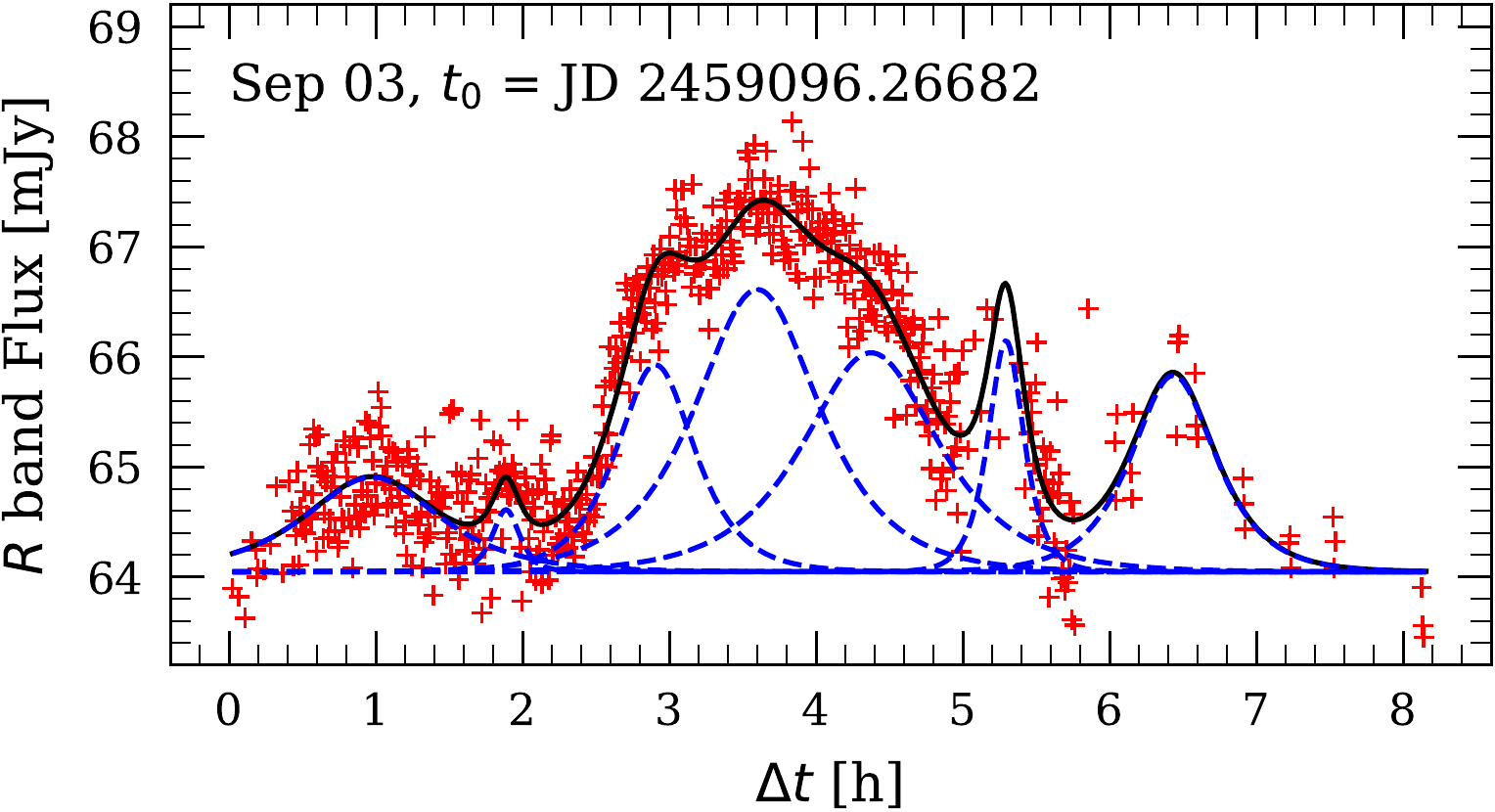}{0.31\textwidth}{}
          \fig{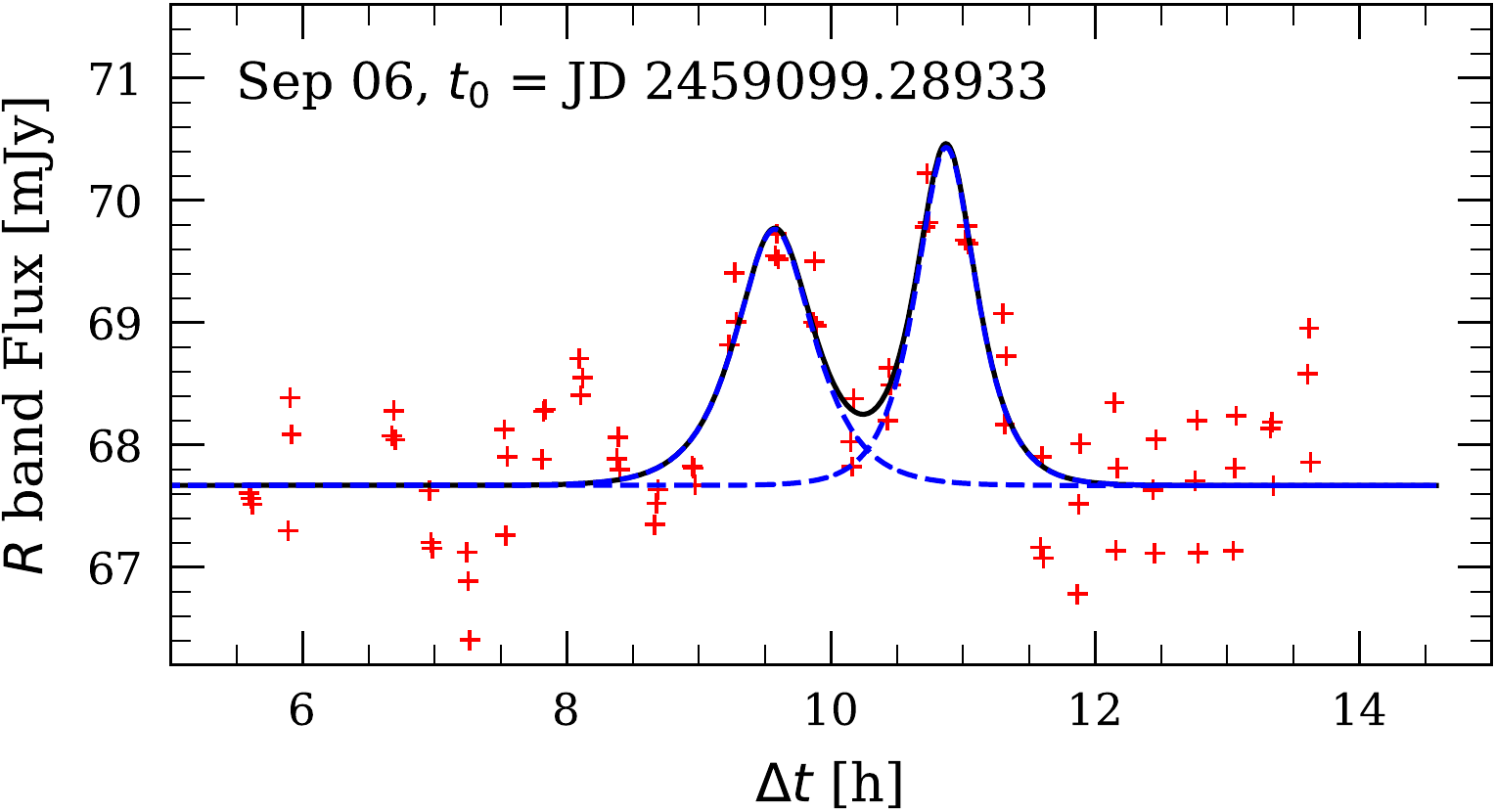}{0.31\textwidth}{}}
\gridline{\fig{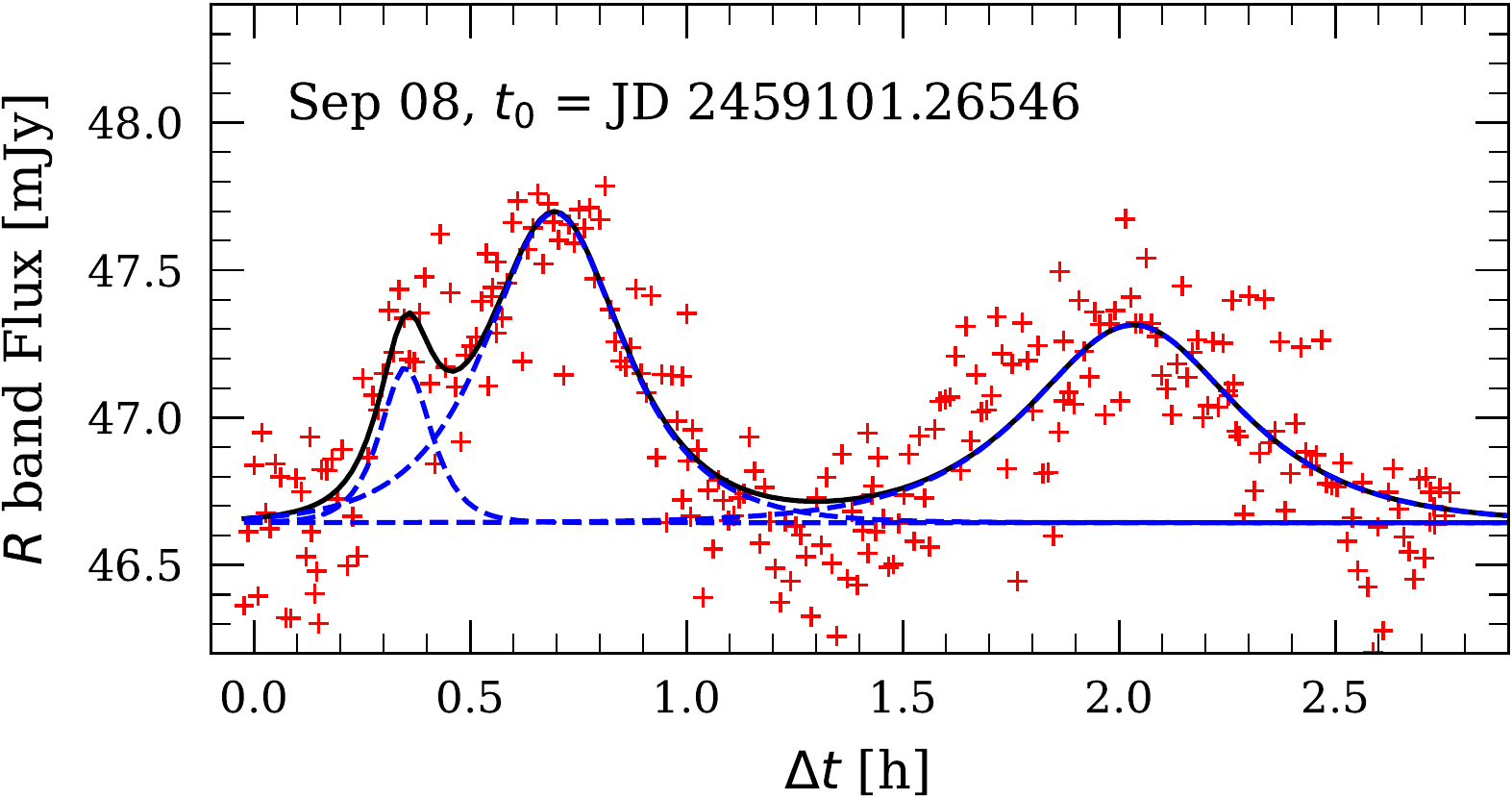}{0.31\textwidth}{}
          \fig{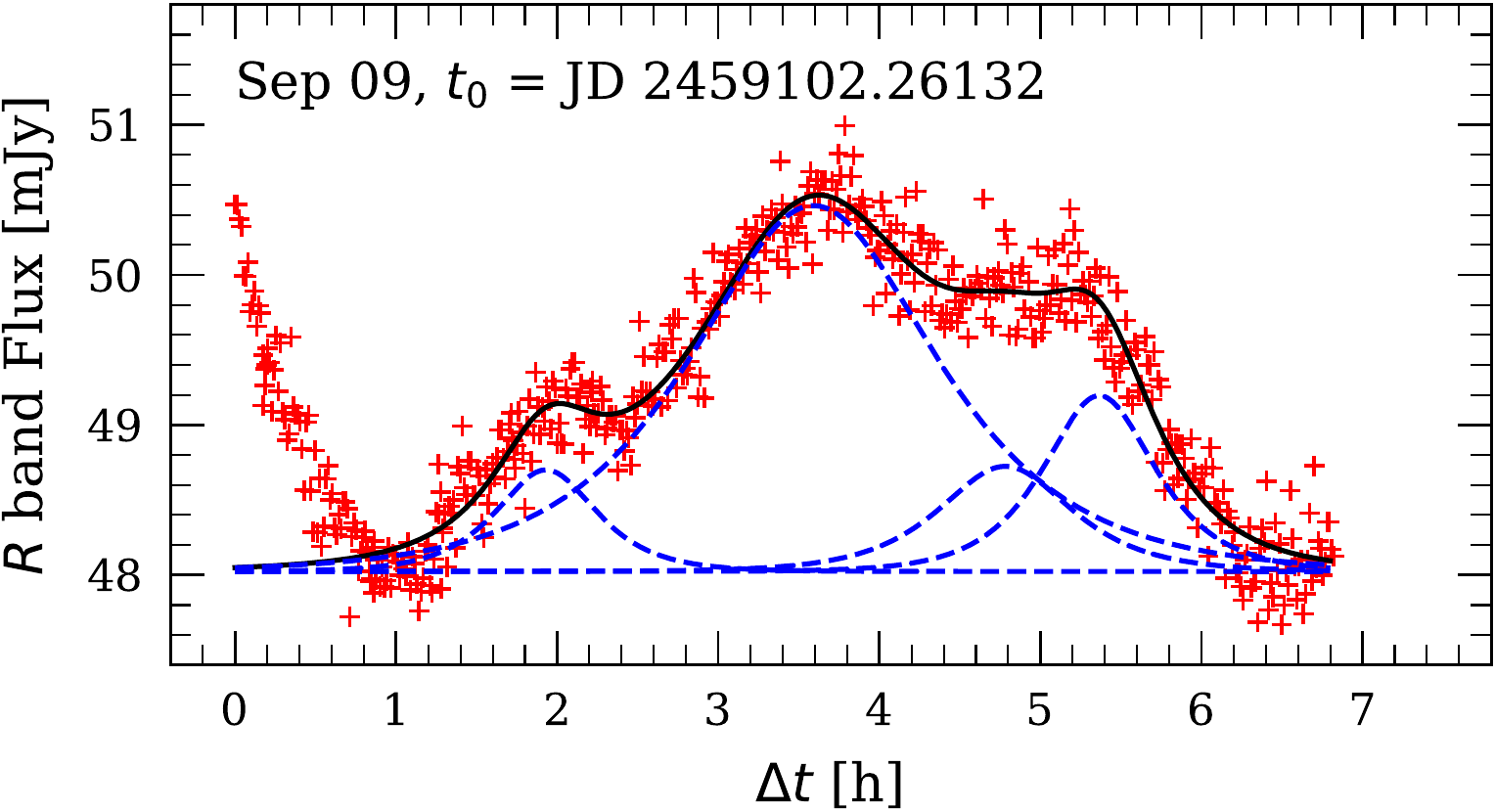}{0.31\textwidth}{}
          \fig{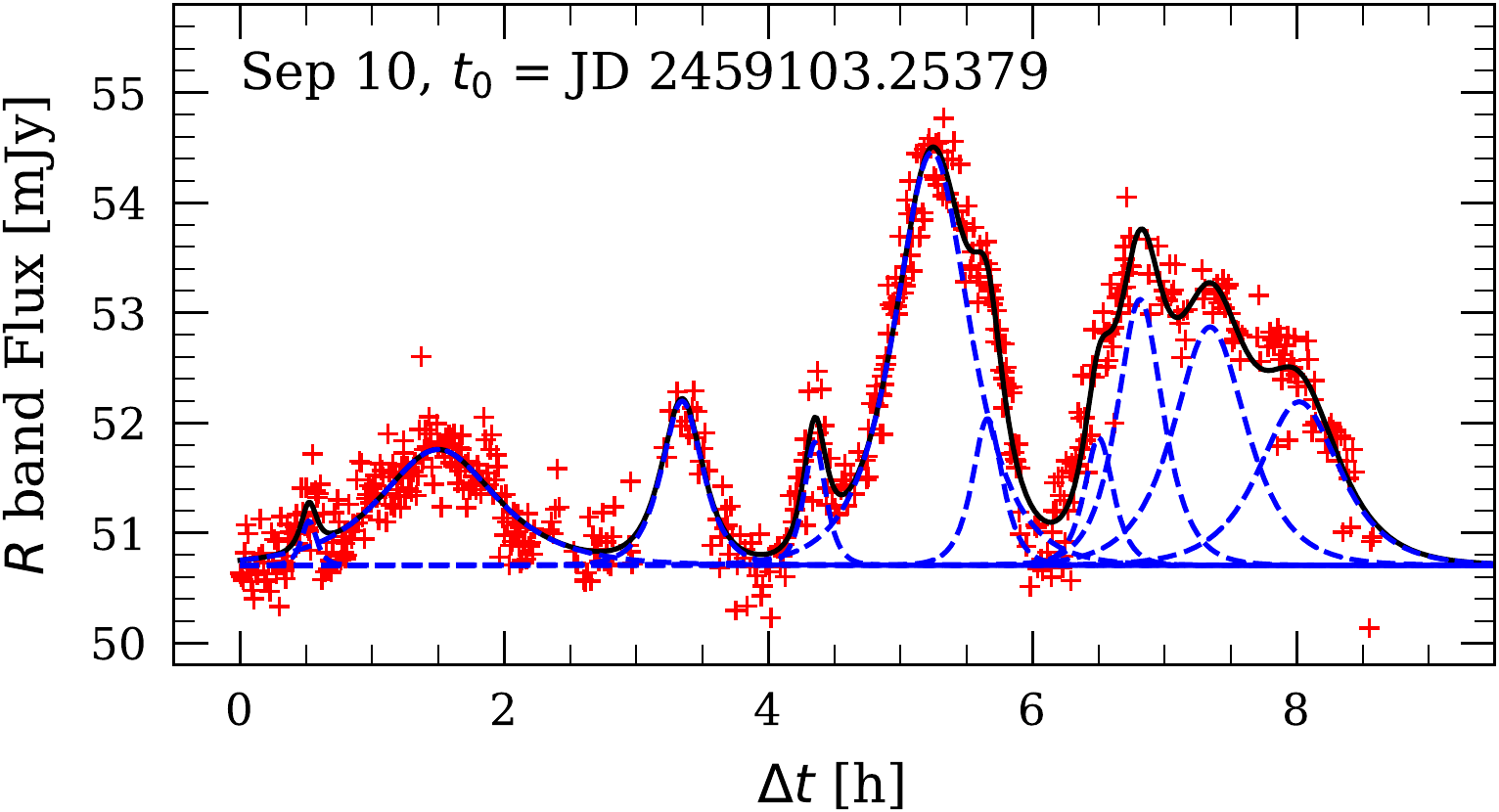}{0.31\textwidth}{}}
\gridline{\fig{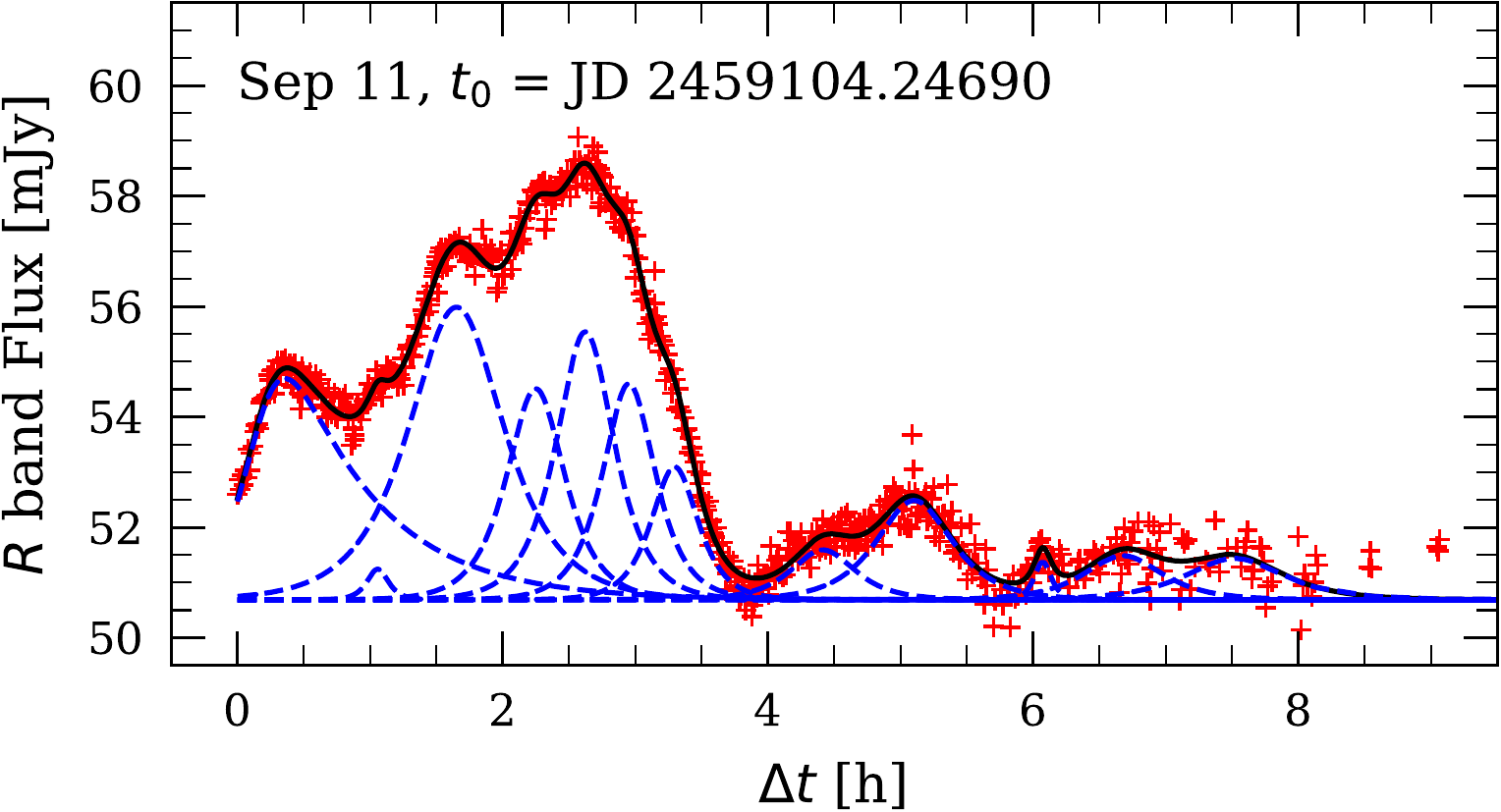}{0.31\textwidth}{}
          \fig{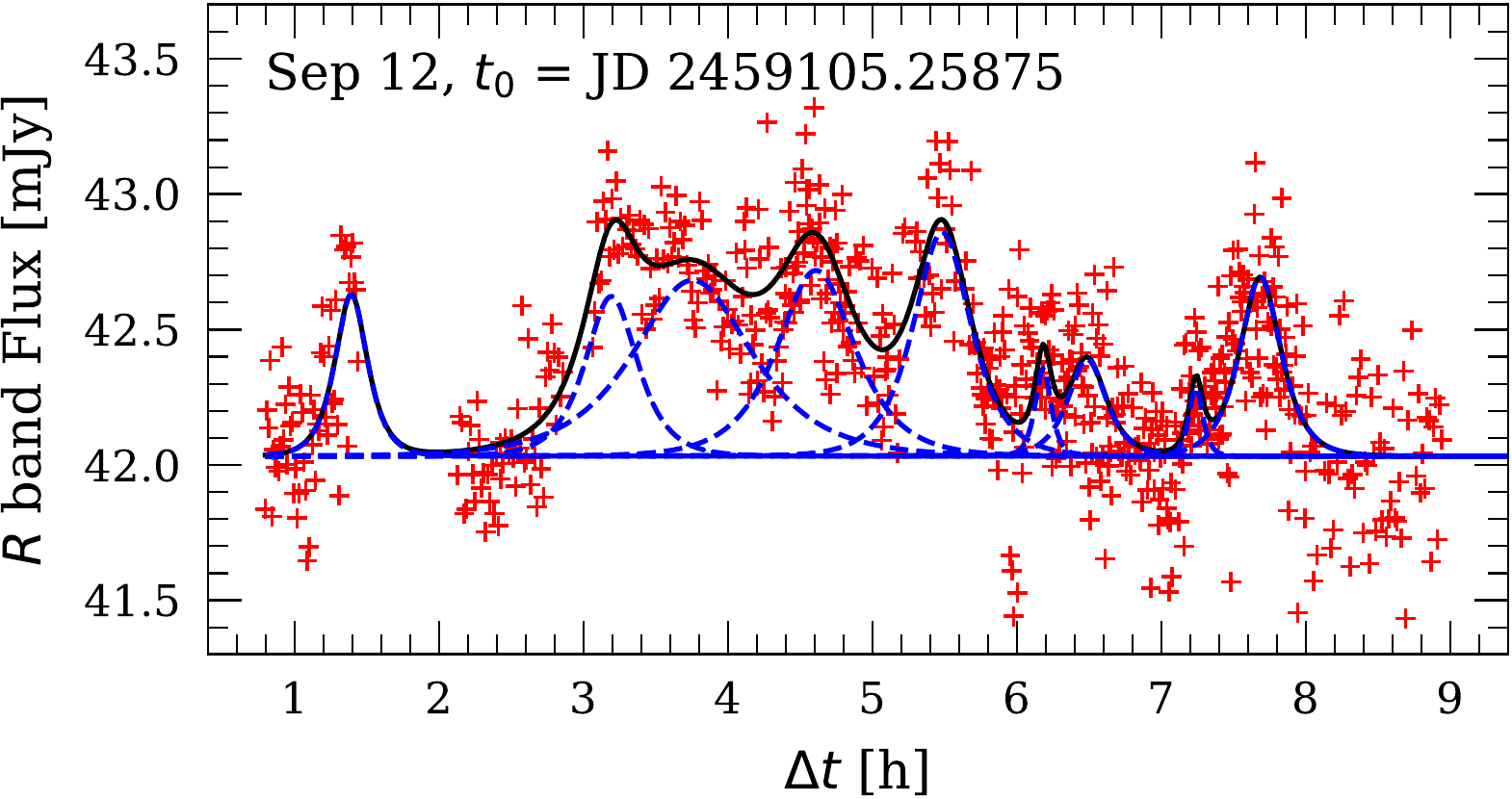}{0.31\textwidth}{}}
\caption{Same as in Figure~\ref{app:decompo:fig:mwl}, but for the $R$-band-only LCs.}
\label{app:decompo:fig}
\end{figure*}

\subsection{Intra-night Variability}
\label{sec:res:inv}

To study the INV of \bl, we included those nights that have more than two hours of monitoring.
In this way, we got a total of 48 INLCs. They are shown in Figure~\ref{fig:inlc:mag} and the results from the INV tests are summarized in Table~\ref{tab:tests}. 

We tested for variability in the INLCs of each telescope individually for a total of 25 nights. For 22 of them, \bl\ was found to show variable status, for two of them, probably variable status, and for one of them, non-variable status. If we define\footnote{A discussion about the duty cycle definition could be found in \citet{2021Galax...9..114W}.} the duty cycle as the number of nights the blazar shows INV over the total number of nights the blazar being monitored, then we found a duty cycle of 96\% (the probably variable cases considered variable) or 88\% (the probably variable cases considered non-variable).

After the magnitudes were transformed into fluxes, the multi-telescope data for the given night and band were combined. In what follows, we shall use the combined LCs unless otherwise specified. 
After the combination, we selected a total of 18 nights of intra-night monitoring suitable to perform an analysis of the INV of \bl; the corresponding LCs are of good sampling and show flaring activity (Figure~\ref{fig:inlc:comb}).

The so-combined LCs were then detrended~-- the (composite) fitting functions used are shown in Figure~\ref{fig:inlc:comb} along with the LCs. The detrending of the Aug 26 $BI$ and Sep 11 $R$ band LCs deserves special attention. For these LCs, we were not able to derive the shape of the smooth components that are to be fitted because of the shape of the LCs themselves (Figure~\ref{fig:inlc:comb}).
So, we had to take into account the data for the preceding night to get an idea of what the smooth component looks like. 
According to Figure~\ref{fig:stv:lc}, the Aug 26 $R$ band flux variations are superimposed onto a linearly decaying flux trend marked by a blue dashed line. We used that fit to determine what regions to fit for the $BI$ bands.
For Sep 11, we also assumed a linear trend, but it is obvious that alternative functional forms are also possible (Figure~\ref{fig:stv:lc}).

The above considerations show that the main source of uncertainty in the detrending process is the unknown shape of the underlying, smooth variable component. In general, the shape, assumed by us for each night, should be considered as an approximate one; however, the determination of the accurate shape of the smooth component is beyond the scope of the presented paper. To test the influence of that shape on the LC decomposition, a few LCs were detrended using alternative fitting functions (these functions are denoted in Figure~\ref{fig:inlc:comb} with dashed lines). Another source of uncertainty is the choice of regions free of flares. However, the choice of these regions is dependent to some extent on the assumed shape of the underlying component, and so we shall consider it as an uncertainty source of lower importance.

Generally, the presence of enough data points on the LC that could be attributed to the smooth component is of utmost importance to estimate its shape accurately. This requires dense sampling and the large duration of the LCs that could be achieved performing ``around-the-world'' observations \citep[e.g.][]{2013A&A...558A..92B}.

\subsubsection{Color Behaviour}
\label{sec:cmd:res}

The CMDs of \bl\ are shown in Figure~\ref{app:cmd:fig} and the fitting results are listed in Table~\ref{tab:cmd}; CMDs for the nights at which the MWL LCs are probably variable or non-variable according to Table~\ref{tab:tests} were not analyzed.
Most of the non-corrected CMDs show significant BWB trends on intra-night timescales, already observed by other authors \citep[e.g.][]{2003A&A...397..565P}. We found no loops in the CMDs.

\subsubsection{Structure Function}
\label{sec:res:sf}

The SFs built using the corrected LCs are presented in Figure~\ref{app:sf:fig}, and the results from the SPL fits are listed in Table~\ref{tab:sf}. 
We found no dependence of the SF slopes on the bands, and so we weight-averaged all slopes together~-- their mean value is $\langle \varrho \rangle_{\rm wt}=1.624 \pm 0.007$ (a weighted standard deviation of 0.275). Regarding the turnover point, its median value (in the observer's frame) over all nights and bands is $\langle \delta t_{\rm to} \rangle_{\rm med}=36.1 \pm 3.7$\,min (a standard deviation of 19.8\,min).

\subsubsection{Cross-correlation Analysis}
\label{sec:dcf:res}

For each night of MWL LCs of good sampling, we calculated DCFs using the original and detrended LCs and ICFs using the detrended LCs.
For our further analysis, we shall consider only the time lags obtained using the DCF, based on the detrended LCs, while the results from the other two cross-correlation functions will serve as a check: the consistency among the various values for a given night and bands supports the reliability of the lag obtained.
The DCFs of \bl\ are shown in Figure~\ref{app:dcf:fig}, and the resulting lags are listed in Table~\ref{tab:dcf}.

We have a total of seven nights suitable for cross-correlation analysis. To consider a given time lag real, we require the lag under consideration to be larger than (i) the modal sampling of the LCs, (ii) the bin size used to build the DCF, and (iii) the lag uncertainties obtained by the FR/RSS method; in addition, the DCF should exceed the 99\% confidence limit, and there should be consistency among the different cross-correlation functions used (see above). From Figure~\ref{app:dcf:fig} and Table~\ref{tab:dcf} one can see that the lags satisfying the above conditions are those for Aug 20 and Aug 26. In both cases, the variability at shorter wavelengths is leading; that is, we have soft lags. The lag values themselves are consistent with the previous lag estimates for \bl.

For Aug 20, the $VI$ band LCs sampling is larger, while the $R$ band LC sampling is smaller than the lag found (Table~\ref{tab:dcf}). To check the reliability of the lags obtained using such LCs, we performed the following test. We shifted the detrended $R$ band LC with the measured $V$ vs $R$ time lag (2.2\,min); we choose $V$ band LC for this test because it is of worse sampling compared to the $I$ band one (Figure~\ref{fig:inlc:comb}). Then, the shifted $R$ band LC was interpolated onto the $V$ band JDs. Finally, the $V$ band LC uncertainties were assigned to the transformed $R$ band LC. The so-generated fake $V$ band LC was cross-correlated with the original $R$ band LC~-- the time lag found is $2.9^{+6.0}_{-4.8}\,\rm min$; that is, it is consistent with the lag found using the original detrended $V$ band LC. Hence, we can conclude that the lags obtained for Aug 20 are reliable and could be used for further analysis.
Regarding Aug 26, we were not able to estimate the significance levels because of the specific LC shape (Figure~\ref{fig:inlc:comb}).

The LCs used for the cross-correlation analysis are a combination of various numbers of flares, and so the measured time lags are a kind of weight-averaged lags over the individual flares \citep{2019ApJ...884...92X}. The attempts to measure the lags using the individual flares, forming the INLCs, lead to inaccurate results either because of the flare overlapping (mainly) or because of the bad flare sampling. 

\begin{figure}[t!]
\centering
\includegraphics[width=\linewidth,clip=true]{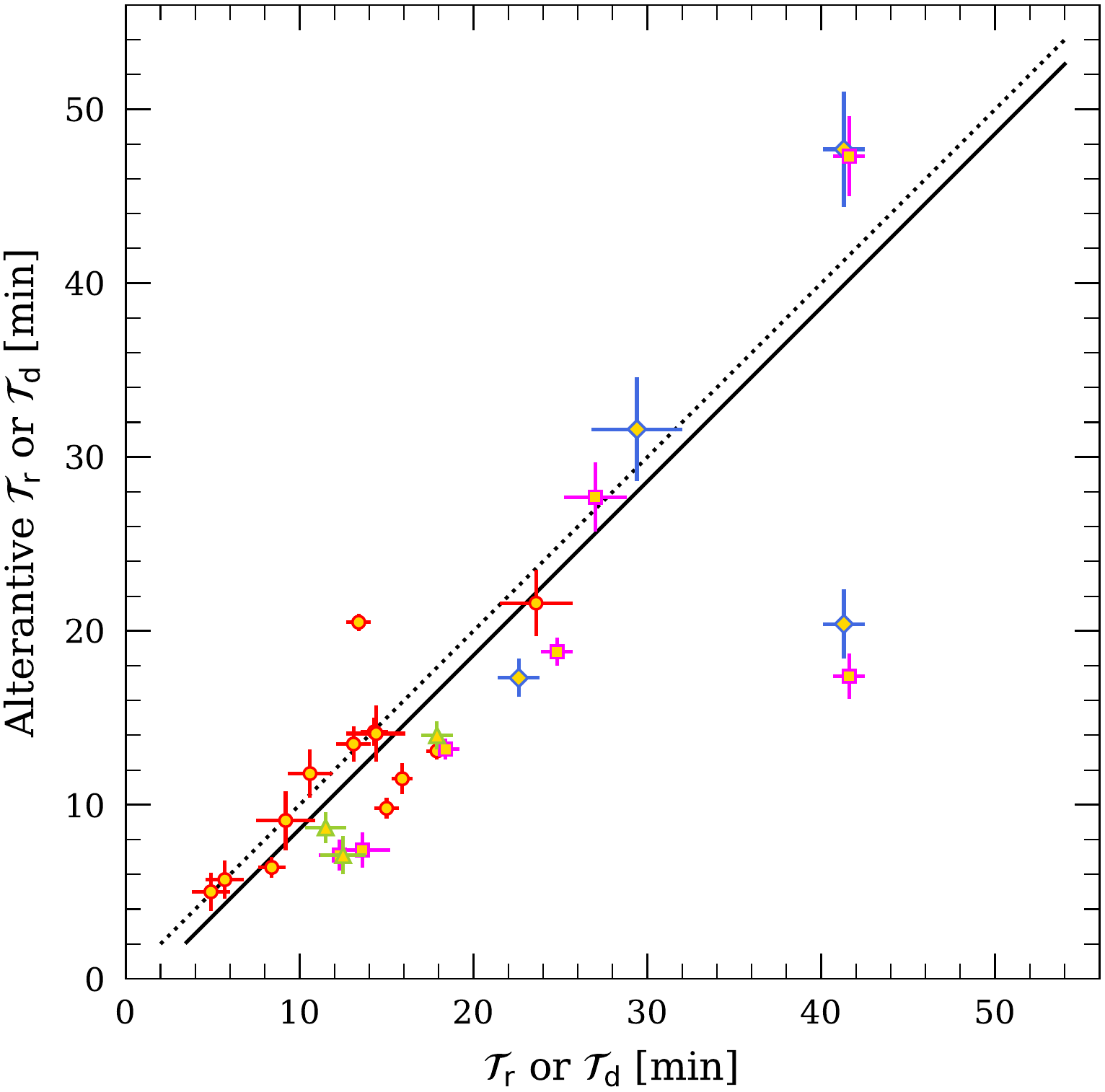} 
\caption{Comparison of the timescales obtained after the decomposition of the LCs detrended using two alternative fitting functions. The symbols denote the bands as follows: $B$~-- blue diamonds, $V$~-- green triangles, $R$~-- red circles, $I$~-- magenta squares. The timescales along the $x$-axis are those adopted by us for the further analysis. The dotted line is the line of exact correspondence. The solid line is the line corresponding to the clipped mean difference between the timescales of 1.4\,min.}
\label{fig:decompo:compar}
\end{figure}

\begin{figure}[t!]
\includegraphics[width=\linewidth,clip=true]{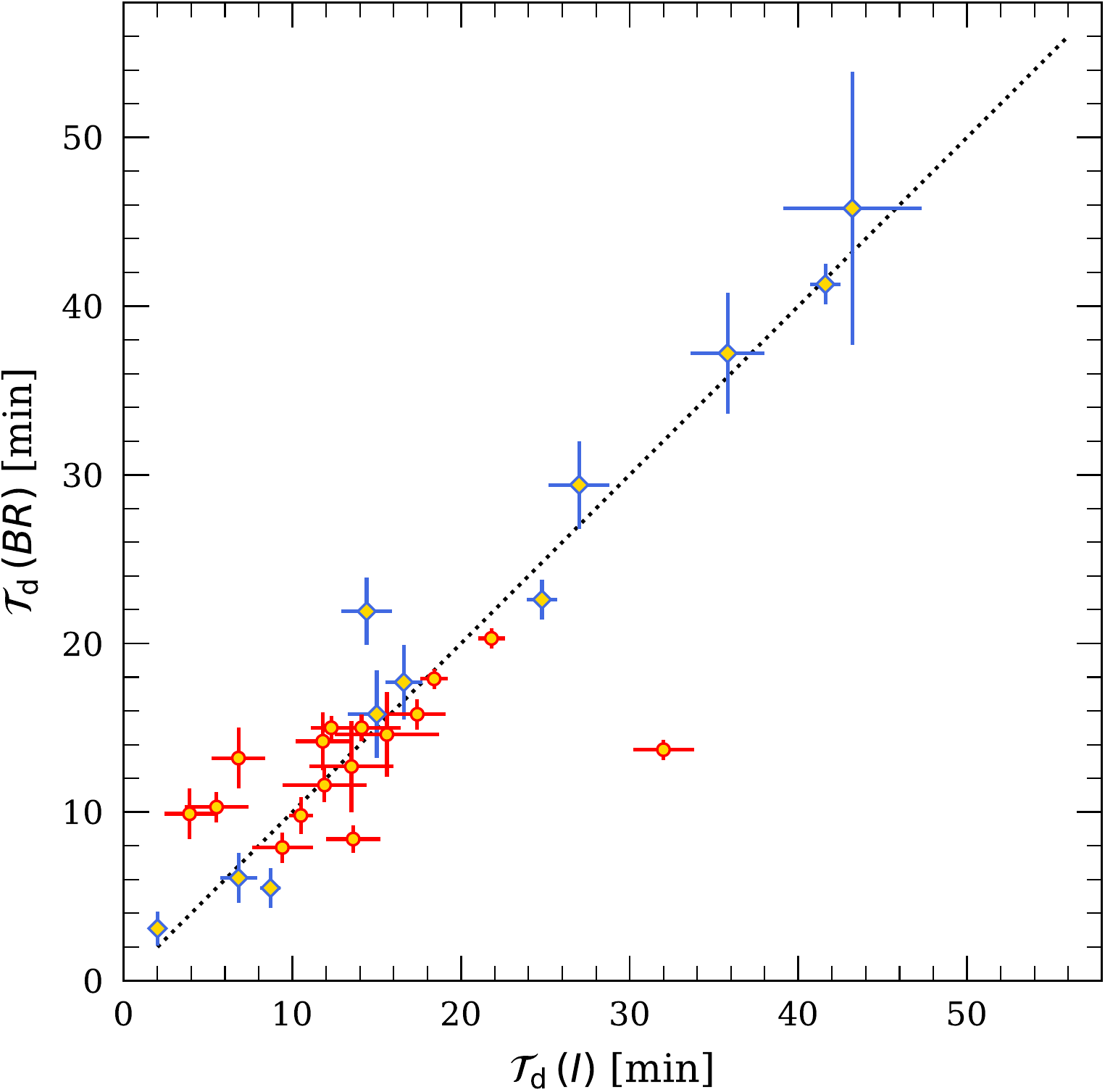} 
\caption{Plot of the $I$ band decay timescales against the $B$ band (blue diamonds) and $R$ band (red circles) ones. The dotted line is the line of exact correspondence.}
\label{fig:decompo:times}
\end{figure}

\begin{figure}[t!]
\includegraphics[width=\linewidth,clip=true]{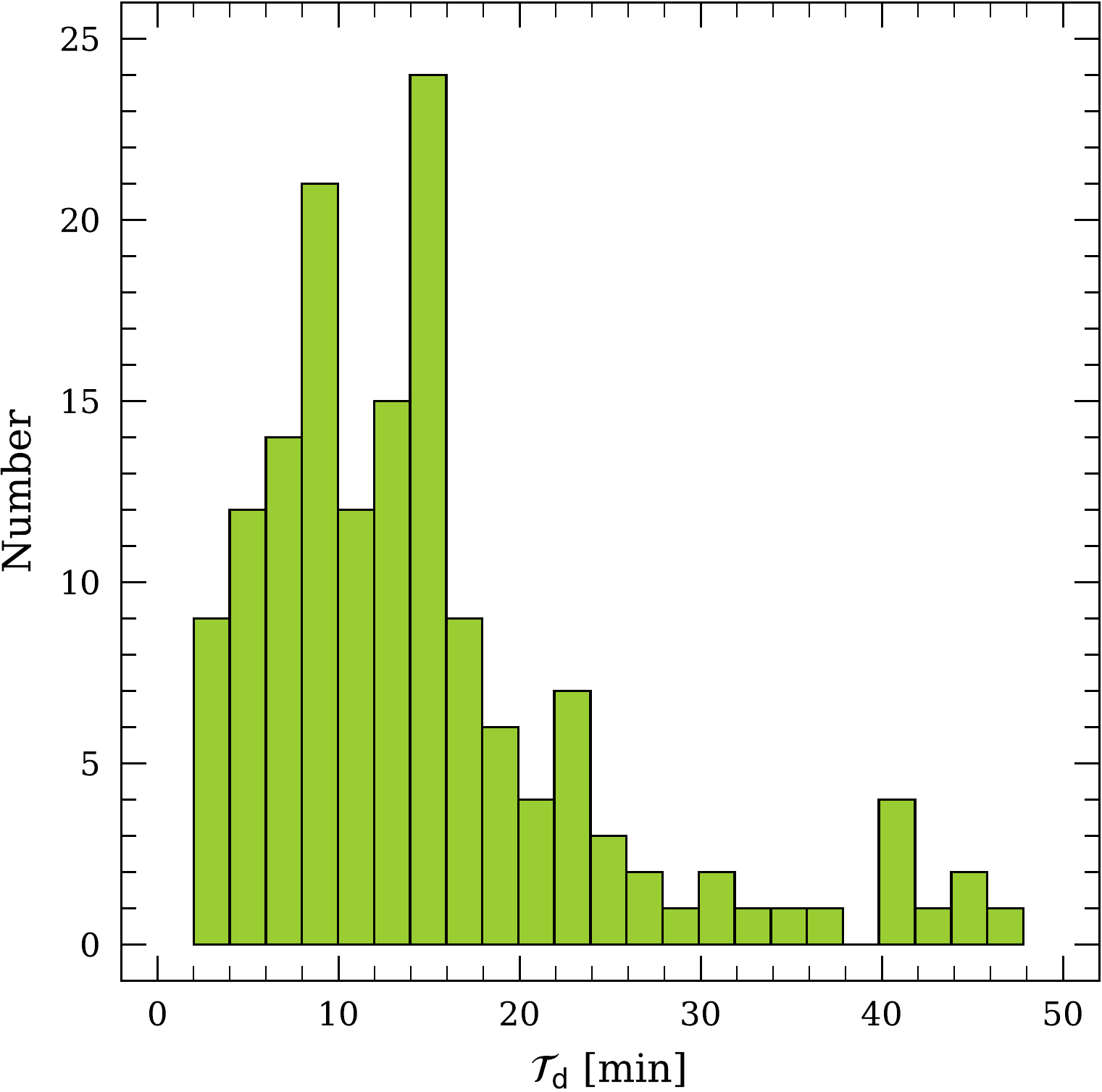} 
\caption{Distribution of the decay timescales jointly for all bands.}
\label{fig:decompo:times:histo}
\end{figure}

\begin{figure}[t!]
\includegraphics[width=\linewidth,clip=true]{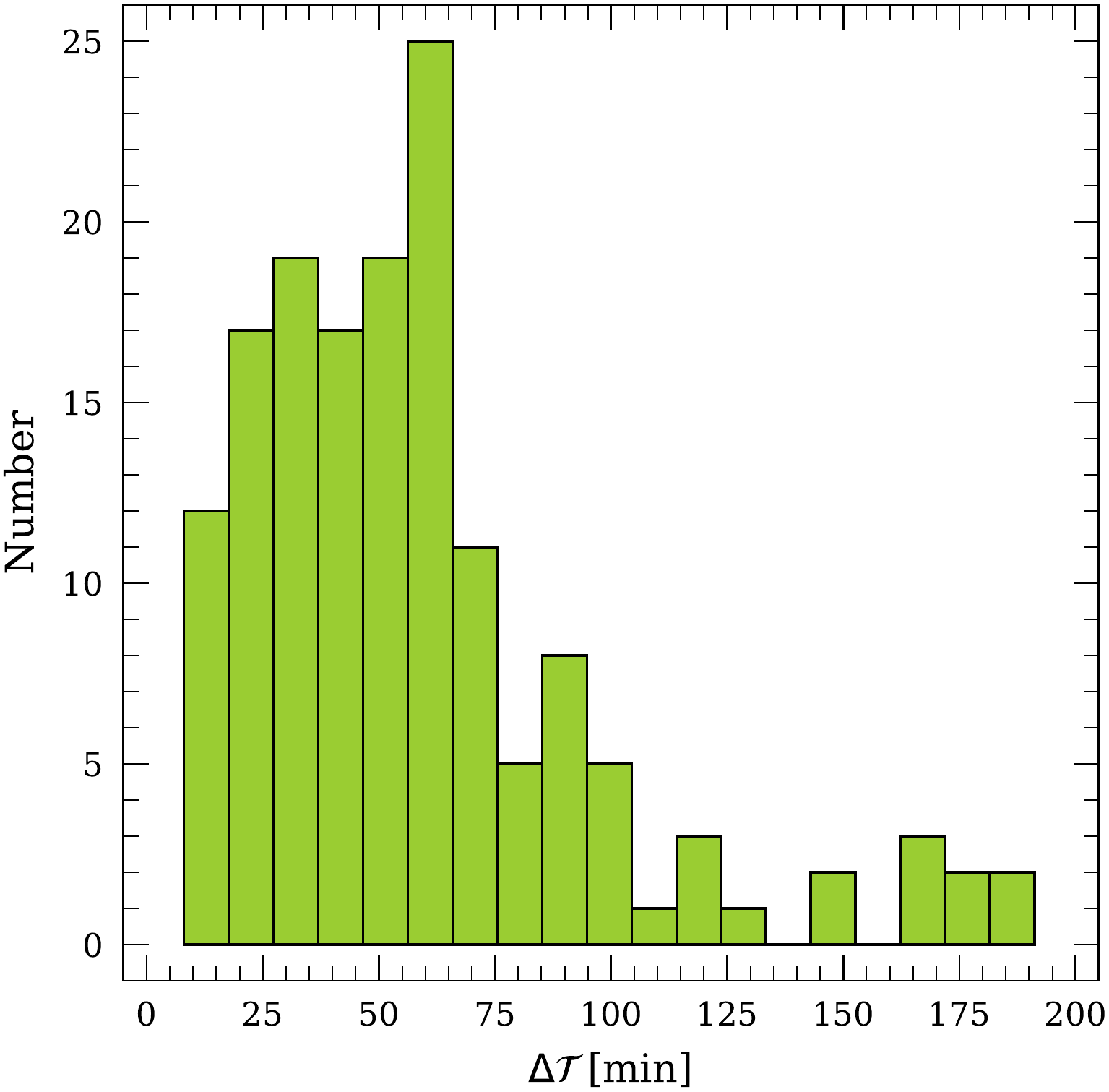} 
\caption{Distribution of the flare durations jointly for all bands.}
\label{fig:decompo:durat:histo}
\end{figure}

\subsubsection{Decomposition of the INLCs}
\label{sec:decompo:res}

The decomposition of the detrended LCs was done employing a non-linear least-squares technique implemented into the {\tt MPFIT} fitter \citep{2009ASPC..411..251M}.
If (i) a flare is not fully recorded, (ii) a flare is of low amplitude, or (iii) flares overlap to a great extent, then we used a symmetric DE function for fitting. In addition, if, for a flare, the fitted uncertainties are comparable to or larger than the fitted values after a general DE fit, then we have redone the decomposition using the symmetric DE function.

Once we have the flare model at hand, we need to estimate how many flares to fit. For most of the LCs the number of flares to be fitted, $N_{\rm fla}$, could easily be obtained; for complex or noisy LCs, however, that tack could be difficult. Hence, to avoid the overfitting, we used the Bayesian Information Criterion \citep[BIC,][]{1978AnSta...6..461S} to get the final estimate of $N_{\rm fla}$. The BIC penalizes the $\chi^2$ of the fit for the newly added parameters as follows: 
\begin{equation}
{\rm BIC}=\chi^2+N_{\rm pars}\,\ln (N_{\rm data}),
\end{equation}
where $N_{\rm pars}$ is the number of model free parameters and $N_{\rm data}$ the number of the data points of the fitted LC. Using BIC, we could identify the number of flares beyond which the addition of a new flare does not significantly improve the fit. To accept the addition of a new flare, we required BIC to decrease by ten or larger: $\Delta{\rm BIC}={\rm BIC({\it N_{\rm fla}}+1)}-{\rm BIC({\it N_{\rm fla}})}\ge10$.
The decompositions are shown in Figures~\ref{app:decompo:fig:mwl} and \ref{app:decompo:fig}; the fitted parameters are listed in Table~\ref{tab:fit}.

As we mentioned in Section~\ref{sec:res:inv}, the unknown shape of the smooth variability component is the main source of systematic uncertainties in the timescales. To make a crude estimate of these uncertainties, we compare in Figure~\ref{fig:decompo:compar} the timescales obtained using two alternative fitting functions to detrend the original LC (see Figure~\ref{fig:inlc:comb}). 
The mean difference between the timescales was found to be 1.4\,min with a standard deviation of 3.9\,min; these values were obtained after the most deviant data points were clipped out. 
These results give a crude estimate of the systematic uncertainty of the timescales due to the unknown shape of the underlying smooth component.
The difference, however, is within the scatter of individual data points, and so we shall neglect it in our further considerations.

Next, we searched for the dependence of the derived decay timescales on the band. We plot in Figure~\ref{fig:decompo:times} the $I$ band timescales against the $BR$ band ones: one can see the lack of significant dependence of ${\mathcal T}_{\rm d}$ on the band; the same applies for the rise timescales as long as all of the flare fits are done using symmetric DE functions (we have four exceptions of this). Hence, we plot the distribution of the decay timescales jointly for all bands (Figure~\ref{fig:decompo:times:histo})~-- the clipped modal value is $\langle {\mathcal T}_{\rm d} \rangle_{\rm mode}=11.6^{+10.5}_{-5.1}$\,min.
The lack of dependence on the band was found for the flare duration as well, and so we plot in Figure~\ref{fig:decompo:durat:histo} the distribution of the flare duration altogether for all bands~-- the clipped modal value is $\langle \Delta {\mathcal T} \rangle_{\rm mode}=46.6^{+41.0}_{-20.6}$\,min.
The parameter uncertainties listed above represent the 16-th and 84-th percentiles of the corresponding distributions.
Finally, using the four asymmetric flares, we calculated a weighted mean asymmetry parameter $\langle \xi \rangle_{\rm wt}=0.49 \pm 0.10$.

\begin{deluxetable*}{ccr@{ $\pm$ }lr@{.}lcc}
\tablecaption{Results from the power-law fits to the non-corrected CMDs \label{tab:cmd}}
\tablewidth{0pt}
\tablehead{
\colhead{Date, 2020} & \colhead{CMD} & \multicolumn{2}{c}{$\varpi$} & \multicolumn{2}{c}{$r$} & \colhead{$p$} & \colhead{Trend}}
\tabletypesize{\small}
\startdata
Aug 20 & $F_B/F_I$ vs $F_R$                 &    0.265 & 0.020 &    0 & 636 & $<\!10^{-5}$ & BWB \\
       & $F_V/F_I$ vs $F_R$                 &    0.211 & 0.011 &    0 & 836 & $<\!10^{-5}$ &     \\
\noalign{\smallskip}
Aug 23 & $F_B/F_I$ vs $F_R$                 &    0.260 & 0.045 &    0 & 555 & $<\!10^{-3}$ & \ldots \\
       & $F_V/F_I$ vs $F_R$                 &    0.036 & 0.032 &    0 & 058 &        0.739 &     \\
\noalign{\smallskip}
Aug 25 & $F_B/F_I$ vs $F_R$                 &    0.222 & 0.010 &    0 & 909 & $<\!10^{-5}$ & BWB \\
       & $F_V/F_I$ vs $F_R$                 &    0.147 & 0.007 &    0 & 898 & $<\!10^{-5}$ &     \\
\noalign{\smallskip}
Aug 26 & $F_B/F_I$ vs $\frac{F_B+F_I}{2}$ &    0.260 & 0.006 &    0 & 916 & $<\!10^{-5}$ & BWB \\
\noalign{\smallskip}
Aug 27 & $F_B/F_I$ vs $\frac{F_B+F_I}{2}$ &    0.357 & 0.014 &    0 & 757 & $<\!10^{-5}$ & BWB \\
\noalign{\smallskip}
Aug 28 & $F_B/F_I$ vs $\frac{F_B+F_I}{2}$ &    0.525 & 0.012 &    0 & 879 & $<\!10^{-5}$ & BWB \\
\noalign{\smallskip}
Aug 31 & $F_V/F_I$ vs $F_R$                 &    0.201 & 0.024 &    0 & 533 & $<\!10^{-5}$ & BWB \\
\noalign{\smallskip}
Sep 3 & $F_B/F_I$ vs $F_R$                 &    0.307 & 0.132 &    0 & 548 &        0.004 & BWB \\
       & $F_V/F_I$ vs $F_R$                 &    0.336 & 0.060 &    0 & 751 & $<\!10^{-5}$ &     \\
\noalign{\smallskip}
Sep 8 & $F_B/F_I$ vs $F_R$                 &    0.289 & 0.051 &    0 & 688 & $<\!10^{-5}$ & BWB \\
       & $F_V/F_I$ vs $F_R$                 &    0.296 & 0.035 &    0 & 866 & $<\!10^{-5}$ &     \\
\noalign{\smallskip}
Sep 10 & $F_B/F_I$ vs $F_R$                 & $-$0.030 & 0.015 & $-$0 & 173 &        0.210 & \ldots \\
       & $F_V/F_I$ vs $F_R$                 &    0.062 & 0.010 &    0 & 512 & $<\!10^{-3}$ &     \\
\noalign{\smallskip}
Sep 11 & $F_B/F_I$ vs $F_R$                 &    0.317 & 0.011 &    0 & 923 & $<\!10^{-5}$ & BWB \\
       & $F_V/F_I$ vs $F_R$                 &    0.189 & 0.007 &    0 & 936 & $<\!10^{-5}$ &     \\
\noalign{\smallskip}
Sep 13 & $F_V/F_I$ vs $F_R$                 &    0.195 & 0.014 &    0 & 666 & $<\!10^{-5}$ & BWB \\
\noalign{\smallskip}
Sep 14 & $F_V/F_I$ vs $F_R$                 &    0.206 & 0.005 &    0 & 901 & $<\!10^{-5}$ & BWB \\
\noalign{\smallskip}
\enddata
\tablecomments{
To derive the values of $\varpi$, $r$, and $p$, the CMDs were fitted in a ``log-log'' form.}
\end{deluxetable*}

\begin{deluxetable}{ccccc}
\tablecaption{Results from the SF fits \label{tab:sf}}
\tablewidth{0pt}
\tablehead{
\colhead{Date, 2020} & \colhead{Band} & \colhead{Bin Size} & \colhead{$\varrho$} & \colhead{$\delta t_{\rm to}$} \\
\colhead{} & \colhead{} & \colhead{(min)} & \colhead{} & \colhead{(min)}
}
\decimalcolnumbers
\tabletypesize{\small}
\startdata
    Aug 20     & $B$ & 2.50 & $0.86 \pm 0.11$ & 46.8 \\
               & $V$ & 2.50 & $1.72 \pm 0.08$ & 44.1 \\
               & $R$ & 2.50 & $1.31 \pm 0.03$ & 44.1 \\
               & $I$ & 2.50 & $1.61 \pm 0.06$ & 44.1 \\
\noalign{\smallskip}                                   
    Aug 21     & $R$ & 1.50 & $1.34 \pm 0.06$ & 28.1 \\
\noalign{\smallskip}                                   
    Aug 25     & $R$ & 1.50 & $1.55 \pm 0.03$ & 29.7 \\
\noalign{\smallskip}                                   
    Aug 26     & $B$ & 2.50 & $2.00 \pm 0.02$ & 94.8 \\
               & $I$ & 2.50 & $2.00 \pm 0.02$ & 94.8 \\
\noalign{\smallskip}                                   
    Aug 27     & $B$ & 2.50 & $0.89 \pm 0.08$ & 52.1 \\
               & $I$ & 2.50 & $1.17 \pm 0.09$ & 33.4 \\
\noalign{\smallskip}                                   
    Aug 28     & $B$ & 2.50 & $1.62 \pm 0.04$ & 49.4 \\
               & $I$ & 2.50 & $1.49 \pm 0.04$ & 52.1 \\
\noalign{\smallskip}                                   
    Aug 30     & $R$ & 1.00 & $1.15 \pm 0.09$ & 16.6 \\
\noalign{\smallskip}                                   
    Aug 31     & $V$ & 2.00 & $1.30 \pm 0.10$ & 26.7 \\
               & $R$ & 2.00 & $1.86 \pm 0.18$ & 28.9 \\
               & $I$ & 2.00 & $1.79 \pm 0.20$ & 28.9 \\
\noalign{\smallskip}                                   
    Sep 2     & $R$ & 1.50 & $1.07 \pm 0.06$ & 21.6 \\
\noalign{\smallskip}                                   
    Sep 3     & $R$ & 2.50 & $1.36 \pm 0.05$ & 36.1 \\
\noalign{\smallskip}                                   
    Sep 8     & $R$ & 1.50 & $1.51 \pm 0.05$ & 24.8 \\
\noalign{\smallskip}                                   
    Sep 9     & $R$ & 2.50 & $1.58 \pm 0.02$ & 49.4 \\
\noalign{\smallskip}                                   
    Sep 10     & $R$ & 1.50 & $1.55 \pm 0.05$ & 18.4 \\
\noalign{\smallskip}                                   
    Sep 11     & $R$ & 2.50 & $1.64 \pm 0.02$ & 46.8 \\
\noalign{\smallskip}                                   
    Sep 12     & $R$ & 2.00 & $0.93 \pm 0.10$ & 31.0 \\
\noalign{\smallskip}                                   
    Sep 13     & $V$ & 2.00 & $1.16 \pm 0.08$ & 65.2 \\
               & $R$ & 1.75 & $1.30 \pm 0.06$ & 43.9 \\
               & $I$ & 2.00 & $1.36 \pm 0.08$ & 52.4 \\
\noalign{\smallskip}                                   
    Sep 14     & $V$ & 1.75 & $1.57 \pm 0.17$ & 23.4 \\
               & $R$ & 1.25 & $1.50 \pm 0.02$ & 31.4 \\
               & $I$ & 1.75 & $1.97 \pm 0.26$ & 29.0 \\
\noalign{\smallskip}
\enddata
\tablecomments{Column 3: Bin sizes used to build the SFs. Column 5: Position of the SF turn-off point in the observer's frame; the SPL is fitted up to this point.}
\end{deluxetable}

\begin{deluxetable*}{cccccc}
\tablecaption{Results from the cross-correlation analysis of the LCs \label{tab:dcf}}
\tablewidth{0pt}
\tablehead{
\colhead{Date, 2020} & \colhead{DCF} & \colhead{Sampling} & \colhead{$\tau$} & \colhead{Bin Size} & \colhead{Detrended?} \\
\colhead{} & \colhead{} & \colhead{(min)} & \colhead{(min)} & \colhead{(min)} & \colhead{}}
\decimalcolnumbers
\tabletypesize{\small}
\startdata
    Aug 20 & $B$ vs $R$ & 3.24, 0.63 &$+4.9^{+2.1}_{-2.5}$ & 2.00 & Yes \\
           &            &            &$+4.4^{+4.1}_{-2.9}$ &   -- & Yes \\
           &            &            &$+4.3^{+2.7}_{-2.3}$ & 2.00 &  No \\
\noalign{\smallskip}
           & $V$ vs $R$ & 3.24, 0.63 &$+2.2^{+1.9}_{-1.8}$ & 2.00 & Yes \\
           &            &            &$+1.5^{+1.8}_{-2.9}$ &   -- & Yes \\
           &            &            &$+1.0^{+2.0}_{-2.0}$ & 2.00 &  No \\
\noalign{\smallskip}
           & $I$ vs $R$ & 3.24, 0.63 &$-2.9^{+2.0}_{-2.0}$ & 2.00 & Yes \\
           &            &            &$-2.6^{+2.3}_{-1.2}$ &   -- & Yes \\
           &            &            &$-1.0^{+2.0}_{-2.0}$ & 2.00 &  No \\
\noalign{\smallskip}
    Aug 26 & $B$ vs $I$ & 1.40, 1.41 &$+3.8^{+2.5}_{-1.3}$ & 2.50 & Yes \\
           &            &            &$+3.4^{+2.8}_{-2.8}$ &   -- & Yes \\
           &            &            &$+6.2^{+2.5}_{-2.5}$ & 2.50 &  No \\
\noalign{\smallskip}
    Aug 27 & $B$ vs $I$ & 1.44, 1.44 &$-2.5^{+2.2}_{-2.1}$ & 2.50 & Yes \\
           &            &            &$-0.8^{+2.8}_{-2.8}$ &   -- & Yes \\
           &            &            &$+1.0^{+2.0}_{-2.1}$ & 2.00 &  No \\
\noalign{\smallskip}
    Aug 28 & $B$ vs $I$ & 1.44, 1.43 &$+0.4^{+2.2}_{-2.2}$ & 1.75 & Yes \\
           &            &            &$+0.6^{+1.4}_{-0.0}$ &   -- & Yes \\
           &            &            &$+4.4^{+2.3}_{-1.8}$ & 1.75 &  No \\    
\noalign{\smallskip}
    Aug 31 & $V$ vs $I$ & 2.01, 2.01 &$+1.8^{+2.3}_{-2.1}$ & 2.00 & Yes \\
           &            &            &$+1.6^{+1.6}_{-1.6}$ &   -- & Yes \\
           &            &            &$+0.0^{+2.7}_{-2.0}$ & 2.00 &  No \\
\noalign{\smallskip}
           & $R$ vs $I$ & 2.01, 2.01 &$-0.3^{+2.0}_{-2.4}$ & 2.00 & Yes \\
           &            &            &$-2.3^{+2.3}_{-0.0}$ &   -- & Yes \\
           &            &            &$+0.0^{+2.0}_{-2.1}$ & 2.00 &  No \\
\noalign{\smallskip}
    Sep 13 & $V$ vs $R$ & 1.86, 0.72 &$-4.5^{+4.0}_{-3.0}$ & 1.50 & Yes \\
           &            &            &$+0.7^{+2.7}_{-3.3}$ &   -- & Yes \\
           &            &            &$+3.1^{+3.6}_{-2.3}$ & 1.50 &  No \\
\noalign{\smallskip}
           & $I$ vs $R$ & 1.87, 0.72 &$-3.6^{+2.4}_{-1.9}$ & 1.50 & Yes \\
           &            &            &$-1.3^{+3.3}_{-3.3}$ &   -- & Yes \\
           &            &            &$+4.5^{+2.3}_{-2.1}$ & 1.50 &  No \\
\noalign{\smallskip}
    Sep 14 & $V$ vs $R$ & 1.86, 0.68 &$+1.3^{+1.2}_{-1.2}$ & 1.25 & Yes \\
           &            &            &$+1.3^{+1.3}_{-1.3}$ &   -- & Yes \\
           &            &            &$+1.5^{+1.5}_{-0.8}$ & 1.50 &  No \\
\noalign{\smallskip}
           & $I$ vs $R$ & 1.87, 0.68 &$+0.0^{+1.2}_{-0.9}$ & 1.25 & Yes \\
           &            &            &$+0.7^{+1.3}_{-2.0}$ &   -- & Yes \\
           &            &            &$+0.0^{+1.5}_{-1.5}$ & 1.50 &  No \\
\noalign{\smallskip}
\enddata
\tablecomments{Time lags are in the observer’s frame. In our DCF notation, namely ``band1'' vs ``band2'', the positive lag means that the variability at ``band1'' is the leading one (see also Section~\ref{sec:disc:pars}). Column 2: Cross-correlated LCs. Column 3: Modal sampling of the cross-correlated LCs. Column 4: Time lag and its lower and upper uncertainties. Zero lower uncertainties are due to the strongly asymmetric shape of the lag distribution. Column 5: Bin size used to build the DCF. The lags with no bin size specified are obtained by means of the ICF. Column 6: Indication whether the used LCs are detrended or not.}
\end{deluxetable*}

\clearpage
\startlongtable
\begin{deluxetable*}{ccr@{ $\pm$ }lr@{ $\pm$ }lr@{ $\pm$ }lr@{ $\pm$ }lr@{ $\pm$ }lc}
\tablecaption{Results from the corrected LC decompositions \label{tab:fit}}
\tablewidth{0pt}
\tablehead{
\colhead{Date, 2020} & \colhead{Band} & \multicolumn{2}{c}{$F_0$} & \multicolumn{2}{c}{$\Delta t_0$} & \multicolumn{2}{c}{${\mathcal T}_{\rm r}$} & \multicolumn{2}{c}{${\mathcal T}_{\rm d}$} & \multicolumn{2}{c}{$\Delta{\mathcal T}$} & \colhead{$\sigma_{\rm fit}$} \\
\colhead{} & \colhead{} & \multicolumn{2}{c}{(mJy)}        & \multicolumn{2}{c}{(min)}        & \multicolumn{2}{c}{(min)}                  & \multicolumn{2}{c}{(min)}                  & \multicolumn{2}{c}{(min)}                 & \colhead{(mJy)} \\
\colhead{(1)} & \colhead{(2)} & \multicolumn{2}{c}{(3)}        & \multicolumn{2}{c}{(4)}        & \multicolumn{2}{c}{(5)}                  & \multicolumn{2}{c}{(6)}                  & \multicolumn{2}{c}{(7)} & \colhead{(8)}               
}
\tabletypesize{\small}
\startdata
    Aug 20 & $R$ & 16.6 &   0.2 &  240.6 &   0.7 &   20.3 &   0.6 &   20.3 &   0.6 &   81.2 &   1.7 & 0.41 \\
           &     &  6.8 &   0.3 &  434.1 &   1.2 &   15.8 &   0.9 &   15.8 &   0.9 &   63.2 &   2.5 &      \\
           &     &  4.8 &   0.6 &  472.9 &   1.8 &   12.7 &   2.7 &   12.7 &   2.7 &   50.8 &   7.6 &      \\
           &     &  3.4 &   0.8 &  497.5 &   1.8 &    9.9 &   1.5 &    9.9 &   1.5 &   39.6 &   4.2 &      \\
           &     &  5.0 &   0.2 &  583.9 &   0.9 &   15.0 &   0.8 &   15.0 &   0.8 &   60.0 &   2.3 &      \\
           &     &  4.5 &   0.2 &  620.6 &   0.6 &   10.3 &   0.9 &   10.3 &   0.9 &   41.2 &   2.5 &      \\
\noalign{\smallskip}
           & $I$ & 19.8 &   0.3 &  237.7 &   1.1 &   21.8 &   0.8 &   21.8 &   0.8 &   87.2 &   2.3 & 0.36 \\
           &     &  1.9 &   0.4 &  324.4 &   2.0 &    6.8 &   2.0 &    6.8 &   2.0 &   27.2 &   5.7 &      \\
           &     &  5.9 &   0.4 &  435.0 &   2.1 &   17.4 &   1.7 &   17.4 &   1.7 &   69.6 &   4.8 &      \\
           &     &  5.0 &   0.4 &  479.4 &   1.7 &   13.5 &   2.5 &   13.5 &   2.5 &   54.0 &   7.1 &      \\
           &     &  2.8 &   0.6 &  508.5 &   1.1 &    3.9 &   1.5 &    3.9 &   1.5 &   15.6 &   4.2 &      \\
           &     &  3.4 &   0.3 &  587.6 &   2.0 &   14.1 &   2.3 &   14.1 &   2.3 &   56.4 &   6.5 &      \\
           &     &  2.6 &   0.5 &  620.5 &   1.4 &    5.5 &   1.9 &    5.5 &   1.9 &   22.0 &   5.4 &      \\
\noalign{\smallskip}
    Aug 21 & $R$ &  2.0 &   0.1 &   56.4 &   0.7 &    5.3 &   0.4 &   16.5 &   1.1 &   43.6 &   2.3 & 0.41 \\
           &     &  2.4 &   0.2 &  432.6 &   0.6 &    5.0 &   0.7 &    5.0 &   0.7 &   20.0 &   2.0 &      \\
           &     &  3.3 &   0.2 &  459.2 &   0.8 &    9.8 &   0.8 &    9.8 &   0.8 &   39.2 &   2.3 &      \\
           &     &  2.8 &   0.2 &  565.6 &   0.6 &    6.5 &   0.6 &    6.5 &   0.6 &   26.0 &   1.7 &      \\
           &     &  4.4 &   0.1 &  595.0 &   0.4 &    8.1 &   0.4 &    8.1 &   0.4 &   32.4 &   1.1 &      \\
           &     &  9.0 &   0.9 &  805.3 &   2.0 &    8.0 &   1.3 &   27.5 &   4.2 &   71.0 &   8.8 &      \\   
\noalign{\smallskip}
    Aug 25 & $R$ &  3.3 &   0.1 &  146.4 &   0.7 &   13.4 &   0.7 &   13.4 &   0.7 &   53.6 &   2.0 & 0.35 \\
           &     &  5.0 &   0.1 &  190.4 &   0.5 &   15.9 &   0.6 &   15.9 &   0.6 &   63.6 &   1.7 &      \\   
           &     &  2.0 &   0.1 &  271.1 &   0.9 &   13.1 &   1.0 &   13.1 &   1.0 &   52.4 &   2.8 &      \\
           &     &  2.7 &   0.1 &  320.3 &   0.7 &   14.3 &   0.8 &   14.3 &   0.8 &   57.2 &   2.3 &      \\
\noalign{\smallskip}
    Aug 26 & $B$ & 13.2 &   5.0 &    4.8 &   8.7 &   47.8 &  34.5 &   47.8 &  34.5 &  191.2 &  97.6 & 0.54 \\
           &     & 14.5 &   7.5 &   86.2 &  10.7 &   45.8 &   8.1 &   45.8 &   8.1 &  183.2 &  22.9 &      \\
           &     &  3.9 &   0.4 &  212.9 &   4.3 &   37.2 &   3.6 &   37.2 &   3.6 &  148.8 &  10.2 &      \\
\noalign{\smallskip}
           & $R$ & 15.3 &   0.5 &  604.8 &   1.2 &   30.4 &   1.0 &   30.4 &   1.0 &  121.6 &   2.8 & 0.58 \\
           &     & 13.4 &   0.5 &  716.0 &   1.5 &   30.4 &   1.0 &   30.4 &   1.0 &  121.6 &   2.8 &      \\
\noalign{\smallskip}
           & $I$ & 20.0 &   3.3 &    7.5 &   4.0 &   44.3 &  14.2 &   44.3 &  14.2 &  177.2 &  40.2 & 0.76 \\
           &     & 23.7 &   4.8 &   88.3 &   4.7 &   43.2 &   4.1 &   43.2 &   4.1 &  172.8 &  11.6 &      \\
           &     &  7.3 &   0.5 &  201.8 &   2.6 &   35.8 &   2.2 &   35.8 &   2.2 &  143.2 &   6.2 &      \\
\noalign{\smallskip}
    Aug 27 & $B$ &  2.7 &   0.4 &   37.0 &   1.3 &    6.1 &   1.5 &    6.1 &   1.5 &   24.4 &   4.2 & 0.53 \\
           &     &  3.2 &   0.3 &   76.1 &   2.2 &   15.8 &   2.6 &   15.8 &   2.6 &   63.2 &   7.4 &      \\
           &     &  4.0 &   0.3 &  175.2 &   2.1 &   17.7 &   2.2 &   17.7 &   2.2 &   70.8 &   6.2 &      \\
           &     &  4.6 &   0.3 &  235.3 &   2.3 &   21.9 &   2.0 &   21.9 &   2.0 &   87.6 &   5.7 &      \\
           &     &  2.3 &   0.6 &  388.2 &   1.0 &    3.1 &   1.0 &    3.1 &   1.0 &   12.4 &   2.8 &      \\
           &     &  2.9 &   0.5 &  495.3 &   1.5 &    5.5 &   1.2 &    5.5 &   1.2 &   22.0 &   3.4 &      \\
\noalign{\smallskip}
           & $I$ &  4.9 &   0.5 &   37.5 &   1.0 &    6.8 &   1.1 &    6.8 &   1.1 &   27.2 &   3.1 & 0.68 \\
           &     &  5.0 &   0.3 &   77.9 &   1.5 &   15.0 &   1.7 &   15.0 &   1.7 &   60.0 &   4.8 &      \\
           &     &  7.6 &   0.3 &  179.8 &   1.1 &   16.6 &   1.1 &   16.6 &   1.1 &   66.4 &   3.1 &      \\
           &     &  5.1 &   0.3 &  230.5 &   1.5 &   14.4 &   1.5 &   14.4 &   1.5 &   57.6 &   4.2 &      \\
           &     &  4.9 &   0.7 &  390.2 &   0.4 &    2.0 &   0.4 &    2.0 &   0.4 &    8.0 &   1.1 &      \\
           &     &  6.4 &   0.4 &  494.0 &   0.8 &    8.7 &   0.6 &    8.7 &   0.6 &   34.8 &   1.7 &      \\
\noalign{\smallskip}
    Aug 28 & $B$ &  4.2 &   0.2 &   87.0 &   2.8 &   29.4 &   2.6 &   29.4 &   2.6 &  117.6 &   7.4 & 0.57 \\
           &     &  9.8 &   0.3 &  171.8 &   0.9 &   22.6 &   1.2 &   22.6 &   1.2 &   90.4 &   3.4 &      \\
           &     &  9.0 &   0.2 &  298.3 &   1.2 &   41.3 &   1.2 &   41.3 &   1.2 &  165.2 &   3.4 &      \\
\noalign{\smallskip}
           & $I$ &  6.0 &   0.2 &   78.5 &   1.7 &   27.0 &   1.8 &   27.0 &   1.8 &  108.0 &   5.1 & 0.85 \\
           &     & 13.5 &   0.2 &  171.5 &   0.7 &   24.8 &   0.9 &   24.8 &   0.9 &   99.2 &   2.5 &      \\
           &     & 14.3 &   0.2 &  295.6 &   0.8 &   41.6 &   0.9 &   41.6 &   0.9 &  166.4 &   2.5 &      \\
\noalign{\smallskip}
    Aug 30 & $R$ &  2.4 &   0.2 &   30.5 &   0.5 &    8.6 &   0.8 &    8.6 &   0.8 &   34.4 &   2.3 & 0.54 \\
           &     &  3.1 &   0.2 &   76.3 &   2.4 &   17.5 &   2.5 &   17.5 &   2.5 &   70.0 &   7.1 &      \\
           &     &  3.2 &   0.4 &  101.8 &   0.5 &    7.4 &   1.0 &    7.4 &   1.0 &   29.6 &   2.8 &      \\
           &     &  1.3 &   0.3 &  140.8 &   1.0 &    5.5 &   1.7 &    5.5 &   1.7 &   22.0 &   4.8 &      \\
           &     &  2.3 &   0.2 &  166.1 &   1.8 &   14.3 &   3.7 &   14.3 &   3.7 &   57.2 &  10.5 &      \\
           &     &  5.3 &   0.3 &  199.5 &   1.0 &   14.5 &   0.7 &   14.5 &   0.7 &   58.0 &   2.0 &      \\
\noalign{\smallskip}
    Aug 31 & $V$ &  4.5 &   0.4 &   58.5 &   1.0 &    9.0 &   1.0 &    9.0 &   1.0 &   36.0 &   2.8 & 0.36 \\
           &     &  3.8 &   0.2 &   87.4 &   1.8 &   13.1 &   1.6 &   13.1 &   1.6 &   52.4 &   4.5 &      \\
           &     &  1.5 &   0.3 &  136.6 &   1.3 &    4.2 &   1.3 &    4.2 &   1.3 &   16.8 &   3.7 &      \\
\noalign{\smallskip}
           & $R$ &  3.8 &   0.6 &   58.5 &   1.2 &    9.8 &   1.1 &    9.8 &   1.1 &   39.2 &   3.1 & 0.30 \\
           &     &  4.4 &   0.4 &   82.8 &   1.7 &   13.2 &   1.8 &   13.2 &   1.8 &   52.8 &   5.1 &      \\
           &     &  1.8 &   0.2 &  129.0 &   2.1 &   14.6 &   2.5 &   14.6 &   2.5 &   58.4 &   7.1 &      \\
\noalign{\smallskip}
           & $I$ &  5.5 &   0.2 &   61.3 &   0.9 &   10.5 &   0.7 &   10.5 &   0.7 &   42.0 &   2.0 & 0.50 \\
           &     &  3.5 &   0.6 &   85.1 &   0.9 &    6.8 &   1.6 &    6.8 &   1.6 &   27.2 &   4.5 &      \\
           &     &  2.2 &   0.3 &  109.5 &   4.7 &   15.6 &   3.1 &   15.6 &   3.1 &   62.4 &   8.8 &      \\
\noalign{\smallskip}
    Sep 2 & $R$ &  4.8 &   0.2 &   49.9 &   0.8 &   13.5 &   0.6 &   13.5 &   0.6 &   54.0 &   1.7 & 0.48 \\
           &     &  5.3 &   0.4 &   86.5 &   1.0 &   10.6 &   1.1 &   10.6 &   1.1 &   42.4 &   3.1 &      \\
           &     &  6.1 &   0.7 &  108.7 &   0.6 &    8.1 &   1.1 &    8.1 &   1.1 &   32.4 &   3.1 &      \\
           &     &  4.0 &   0.3 &  134.0 &   1.6 &   13.1 &   2.3 &   13.1 &   2.3 &   52.4 &   6.5 &      \\
           &     &  3.4 &   0.3 &  165.2 &   0.6 &    7.3 &   1.0 &    7.3 &   1.0 &   29.2 &   2.8 &      \\
           &     &  5.6 &   0.3 &  201.7 &   2.0 &   19.0 &   1.8 &   19.0 &   1.8 &   76.0 &   5.1 &      \\
           &     &  5.0 &   0.6 &  225.5 &   0.4 &    7.5 &   0.9 &    7.5 &   0.9 &   30.0 &   2.5 &      \\
           &     &  3.0 &   0.1 &  292.7 &   0.9 &    9.9 &   1.0 &    9.9 &   1.0 &   39.6 &   2.8 &      \\
           &     &  2.3 &   0.2 &  335.2 &   1.1 &    9.9 &   1.5 &    9.9 &   1.5 &   39.6 &   4.2 &      \\
           &     &  7.0 &   0.1 &  370.2 &   0.6 &   15.1 &   1.0 &   15.1 &   1.0 &   60.4 &   2.8 &      \\
           &     &  6.4 &   0.2 &  409.8 &   0.4 &    8.8 &   0.6 &    8.8 &   0.6 &   35.2 &   1.7 &      \\
           &     &  3.8 &   0.2 &  436.9 &   0.6 &    8.4 &   0.7 &    8.4 &   0.7 &   33.6 &   2.0 &      \\
           &     &  3.3 &   0.1 &  481.7 &   0.6 &    9.1 &   0.7 &    9.1 &   0.7 &   36.4 &   2.0 &      \\
\noalign{\smallskip}
    Sep 3 & $R$ &  1.7 &   0.1 &   58.3 &   1.5 &   24.5 &   2.0 &   24.5 &   2.0 &   98.0 &   5.7 & 0.37 \\
           &     &  1.1 &   0.1 &  113.1 &   0.9 &    5.0 &   1.0 &    5.0 &   1.0 &   20.0 &   2.8 &      \\
           &     &  3.8 &   0.6 &  174.3 &   1.3 &   14.4 &   1.7 &   14.4 &   1.7 &   57.6 &   4.8 &      \\
           &     &  5.1 &   0.8 &  215.9 &   1.9 &   22.6 &   4.6 &   22.6 &   4.6 &   90.4 &  13.0 &      \\
           &     &  4.0 &   0.8 &  262.4 &   4.2 &   24.7 &   3.7 &   24.7 &   3.7 &   98.8 &  10.5 &      \\
           &     &  4.2 &   0.9 &  317.6 &   1.0 &    6.5 &   1.4 &    6.5 &   1.4 &   26.0 &   4.0 &      \\
           &     &  3.6 &   0.3 &  386.0 &   2.1 &   15.5 &   1.9 &   15.5 &   1.9 &   62.0 &   5.4 &      \\
\noalign{\smallskip}
    Sep 6 & $R$ &  4.2 &   0.3 &  574.4 &   1.7 &   15.8 &   1.5 &   15.8 &   1.5 &   63.2 &   4.2 & 0.47 \\
           &     &  5.5 &   0.4 &  652.3 &   0.9 &   12.2 &   1.2 &   12.2 &   1.2 &   48.8 &   3.4 &      \\
\noalign{\smallskip}
    Sep 8 & $R$ &  1.1 &   0.2 &   21.0 &   0.6 &    3.2 &   0.7 &    3.2 &   0.7 &   12.8 &   2.0 & 0.20 \\
           &     &  2.1 &   0.1 &   41.7 &   0.6 &    8.4 &   0.6 &    8.4 &   0.6 &   33.6 &   1.7 &      \\
           &     &  1.3 &   0.1 &  122.0 &   0.9 &   12.9 &   1.0 &   12.9 &   1.0 &   51.6 &   2.8 &      \\
\noalign{\smallskip}
    Sep 9 & $R$ &  1.4 &   0.1 &  115.9 &   1.3 &   16.8 &   1.9 &   16.8 &   1.9 &   67.2 &   5.4 & 0.21 \\
           &     &  4.9 &   0.2 &  215.5 &   2.7 &   40.6 &   2.4 &   40.6 &   2.4 &  162.4 &   6.8 &      \\
           &     &  1.4 &   0.7 &  286.9 &   8.4 &   22.1 &   9.8 &   22.1 &   9.8 &   88.4 &  27.7 &      \\
           &     &  2.3 &   0.7 &  321.9 &   3.6 &   18.7 &   2.3 &   18.7 &   2.3 &   74.8 &   6.5 &      \\              
\noalign{\smallskip}
    Sep 10 & $R$ &  0.8 &   0.2 &   31.3 &   0.9 &    3.1 &   0.9 &    3.1 &   0.9 &   12.4 &   2.5 & 0.26 \\
           &     &  2.1 &   0.1 &   90.1 &   0.9 &   23.4 &   1.1 &   23.4 &   1.1 &   93.6 &   3.1 &      \\
           &     &  3.0 &   0.1 &  200.8 &   0.3 &    8.3 &   0.3 &    8.3 &   0.3 &   33.2 &   0.8 &      \\
           &     &  2.2 &   0.1 &  261.1 &   0.2 &    4.4 &   0.2 &    4.4 &   0.2 &   17.6 &   0.6 &      \\
           &     &  7.5 &   0.0 &  314.4 &   0.3 &   15.2 &   0.2 &   15.2 &   0.2 &   60.8 &   0.6 &      \\
           &     &  2.7 &   0.1 &  339.5 &   0.2 &    6.1 &   0.3 &    6.1 &   0.3 &   24.4 &   0.8 &      \\
           &     &  2.3 &   0.3 &  390.1 &   0.6 &    5.9 &   0.7 &    5.9 &   0.7 &   23.6 &   2.0 &      \\
           &     &  4.8 &   0.3 &  408.8 &   0.6 &    9.2 &   1.4 &    9.2 &   1.4 &   36.8 &   4.0 &      \\
           &     &  4.3 &   0.3 &  440.6 &   1.2 &   14.7 &   2.0 &   14.7 &   2.0 &   58.8 &   5.7 &      \\
           &     &  3.0 &   0.2 &  481.1 &   1.5 &   16.3 &   1.4 &   16.3 &   1.4 &   65.2 &   4.0 &      \\
\noalign{\smallskip}
    Sep 11 & $R$ &  6.5 &   0.3 &   10.3 &   1.4 &    9.7 &   0.9 &   40.4 &   5.0 &  100.2 &  10.2 & 0.26 \\
           &     &  1.1 &   0.2 &   63.4 &   0.7 &    4.3 &   1.0 &    4.3 &   1.0 &   17.2 &   2.8 &      \\
           &     & 10.6 &   0.5 &   99.0 &   0.8 &   19.2 &   1.0 &   19.2 &   1.0 &   76.8 &   2.8 &      \\
           &     &  7.6 &   1.8 &  134.9 &   1.4 &   11.7 &   1.7 &   11.7 &   1.7 &   46.8 &   4.8 &      \\
           &     &  9.7 &   2.9 &  157.3 &   1.3 &   11.7 &   2.8 &   11.7 &   2.8 &   46.8 &   7.9 &      \\
           &     &  7.8 &   2.6 &  177.0 &   1.4 &   10.6 &   2.5 &   10.6 &   2.5 &   42.4 &   7.1 &      \\
           &     &  4.8 &   1.0 &  197.8 &   1.2 &    9.5 &   0.8 &    9.5 &   0.8 &   38.0 &   2.3 &      \\
           &     &  1.8 &   0.1 &  264.8 &   1.1 &   12.9 &   1.1 &   12.9 &   1.1 &   51.6 &   3.1 &      \\
           &     &  3.6 &   0.1 &  306.1 &   0.6 &   15.3 &   0.7 &   15.3 &   0.7 &   61.2 &   2.0 &      \\
           &     &  1.4 &   0.2 &  364.3 &   0.5 &    3.4 &   0.7 &    3.4 &   0.7 &   13.6 &   2.0 &      \\
           &     &  1.6 &   0.1 &  400.7 &   2.0 &   15.8 &   2.7 &   15.8 &   2.7 &   63.2 &   7.6 &      \\
           &     &  1.5 &   0.1 &  451.1 &   3.3 &   19.7 &   3.4 &   19.7 &   3.4 &   78.8 &   9.6 &      \\
\noalign{\smallskip}
    Sep 12 & $R$ &  1.2 &   0.1 &   83.5 &   1.8 &    6.0 &   1.3 &    6.0 &   1.3 &   24.0 &   3.7 & 0.22 \\
           &     &  1.2 &   0.3 &  191.5 &   1.5 &    9.7 &   2.9 &    9.7 &   2.9 &   38.8 &   8.2 &      \\
           &     &  1.3 &   0.1 &  225.2 &   4.6 &   22.0 &   4.5 &   22.0 &   4.5 &   88.0 &  12.7 &      \\
           &     &  1.4 &   0.1 &  276.5 &   0.9 &   14.4 &   1.2 &   14.4 &   1.2 &   57.6 &   3.4 &      \\
           &     &  1.7 &   0.1 &  328.9 &   0.4 &   10.7 &   0.4 &   10.7 &   0.4 &   42.8 &   1.1 &      \\
           &     &  0.6 &   0.1 &  370.9 &   0.5 &    2.9 &   0.6 &    2.9 &   0.6 &   11.6 &   1.7 &      \\
           &     &  0.7 &   0.1 &  389.1 &   0.8 &    7.1 &   0.8 &    7.1 &   0.8 &   28.4 &   2.3 &      \\
           &     &  0.5 &   0.1 &  434.3 &   0.6 &    2.5 &   0.6 &    2.5 &   0.6 &   10.0 &   1.7 &      \\
           &     &  1.3 &   0.1 &  461.1 &   0.4 &    7.7 &   0.4 &    7.7 &   0.4 &   30.8 &   1.1 &      \\
\noalign{\smallskip}
    Sep 13 & $V$ &  2.5 &   0.2 &  119.9 &   2.5 &   21.7 &   2.3 &   21.7 &   2.3 &   86.8 &   6.5 & 0.30 \\
           &     &  2.7 &   0.2 &  376.6 &   1.5 &   19.3 &   1.9 &   19.3 &   1.9 &   77.2 &   5.4 &      \\
           &     &  1.5 &   0.2 &  443.0 &   2.1 &   11.2 &   2.1 &   11.2 &   2.1 &   44.8 &   5.9 &      \\              
\noalign{\smallskip}
           & $R$ &  1.5 &   0.3 &   36.9 &   0.7 &    2.6 &   0.7 &    2.6 &   0.7 &   10.4 &   2.0 & 0.35 \\
           &     &  3.7 &   0.3 &   91.2 &   1.1 &   11.6 &   1.0 &   11.6 &   1.0 &   46.4 &   2.8 &      \\              
           &     &  3.9 &   0.3 &  123.6 &   1.4 &   14.2 &   1.7 &   14.2 &   1.7 &   56.8 &   4.8 &      \\
           &     &  1.0 &   0.1 &  177.2 &   6.8 &   23.9 &   4.7 &   23.9 &   4.7 &   95.6 &  13.3 &      \\             
           &     &  1.3 &   0.1 &  260.7 &   0.6 &    5.5 &   0.5 &    5.5 &   0.5 &   22.0 &   1.4 &      \\
           &     &  2.0 &   0.0 &  384.3 &   0.4 &   13.7 &   0.6 &   13.7 &   0.6 &   54.8 &   1.7 &      \\              
           &     &  0.6 &   0.1 &  436.7 &   0.9 &    7.9 &   0.9 &    7.9 &   0.9 &   31.6 &   2.5 &      \\
\noalign{\smallskip}
           & $I$ &  2.8 &   0.3 &  103.7 &   2.7 &   11.9 &   2.5 &   11.9 &   2.5 &   47.6 &   7.1 & 0.38 \\
           &     &  3.6 &   0.4 &  135.2 &   1.8 &   11.8 &   1.6 &   11.8 &   1.6 &   47.2 &   4.5 &      \\
           &     &  3.6 &   0.1 &  374.1 &   1.5 &   32.0 &   1.8 &   32.0 &   1.8 &  128.0 &   5.1 &      \\            
           &     &  1.6 &   0.2 &  446.9 &   1.6 &    9.4 &   1.8 &    9.4 &   1.8 &   37.6 &   5.1 &      \\
\noalign{\smallskip}
    Sep 14 & $V$ &  4.3 &   0.5 &  395.0 &   2.0 &   12.5 &   1.3 &   12.5 &   1.3 &   50.0 &   3.7 & 0.36 \\
           &     &  5.5 &   0.4 &  422.0 &   1.3 &   11.5 &   1.2 &   11.5 &   1.2 &   46.0 &   3.4 &      \\
           &     &  6.3 &   0.2 &  489.5 &   0.7 &   17.9 &   0.9 &   17.9 &   0.9 &   71.6 &   2.5 &      \\   
\noalign{\smallskip}
           & $R$ &  1.8 &   0.2 &   45.3 &   0.7 &    5.7 &   1.1 &    5.7 &   1.1 &   22.8 &   3.1 & 0.31 \\
           &     &  4.8 &   0.1 &   79.6 &   1.3 &   14.4 &   1.7 &   14.4 &   1.7 &   57.6 &   4.8 &      \\
           &     &  2.4 &   0.4 &  100.6 &   0.6 &    4.9 &   1.1 &    4.9 &   1.1 &   19.6 &   3.1 &      \\  
           &     &  1.2 &   0.2 &  125.3 &   1.7 &    9.2 &   1.7 &    9.2 &   1.7 &   36.8 &   4.8 &      \\
           &     &  2.5 &   0.1 &  238.2 &   2.0 &   10.6 &   1.3 &   23.6 &   2.1 &   68.4 &   4.9 &      \\
           &     &  5.3 &   0.2 &  402.7 &   1.2 &   15.0 &   0.7 &   15.0 &   0.7 &   60.0 &   2.0 &      \\
           &     &  3.6 &   0.3 &  426.9 &   0.7 &    8.4 &   0.8 &    8.4 &   0.8 &   33.6 &   2.3 &      \\
           &     &  5.4 &   0.1 &  490.8 &   0.5 &   17.9 &   0.6 &   17.9 &   0.6 &   71.6 &   1.7 &      \\
\noalign{\smallskip}
           & $I$ &  4.7 &   0.7 &  398.4 &   1.8 &   12.3 &   1.2 &   12.3 &   1.2 &   49.2 &   3.4 & 0.42 \\
           &     &  5.2 &   0.5 &  424.6 &   1.9 &   13.6 &   1.6 &   13.6 &   1.6 &   54.4 &   4.5 &      \\
           &     &  7.4 &   0.2 &  490.1 &   0.6 &   18.4 &   0.8 &   18.4 &   0.8 &   73.6 &   2.3 &      \\   
\noalign{\smallskip}
\enddata
\tablecomments{Timescales are in the observer's frame. Column 3: Twice the flare amplitude. Column 4: Approximate position of the flare maximum (the actual position is equal to $\Delta t_0$ only for symmetric flares). Column 5: $e$-folding rise timescale. Column 6: $e$-folding decay timescale. Column 7: Approximate duration of the flare. Column 8: Standard deviation about the fitted sum of DE functions.}
\end{deluxetable*}

\begin{deluxetable}{ccccc}
\tablecaption{Characteristics of the emitting regions \label{tab:pars}}
\tablewidth{0pt}
\tablehead{
\colhead{Parameter} & \colhead{Min} & \colhead{Max} & \colhead{Median} & \colhead{Mode}}
\tabletypesize{\small}
\startdata
\noalign{\smallskip}
${\widetilde {\mathcal B}}_{\rm min}$ & 7.5  & 76.3 & $20.5^{+13.3}_{-6.4}$ & $17.7^{+16.1}_{-3.6}$ \\
${\mathcal B}_{\rm min}(11.0)$        & 3.4  & 34.3 & $9.2^{+6.0}_{-2.9}$   & $8.0^{+7.2}_{-1.6}$   \\
\noalign{\smallskip}
${\widetilde \gamma}_{\rm e,max}$     & 1189 & 5110 & $2540^{+584}_{-548}$  & $2470^{+655}_{-478}$  \\
$\gamma_{\rm e,max}(11.0)$            & 535  & 2298 & $1142^{+263}_{-246}$  & $1111^{+295}_{-215}$  \\
\noalign{\smallskip}
${\widetilde {\mathcal R}}_{\rm max}$ & 0.2  &  5.4 &  $1.4^{+1.0}_{-0.8}$  & $1.2^{+1.2}_{-0.6}$   \\
${\mathcal R}_{\rm max}(11.0)$        & 2.2  & 59.4 & $15.4^{+11.0}_{-8.8}$ & $13.2^{+13.2}_{-6.6}$ \\
\noalign{\smallskip}
\enddata
\tablecomments{Magnetic field strength is in units of Gauss and the radius is in astronomical units (AU). The mode is calculated using a clipping technique. The uncertainties represent the 16-th and 84-th percentiles of the corresponding distributions.}
\end{deluxetable}

\begin{figure}[t!]
\centering
\includegraphics[width=\linewidth,clip=true]{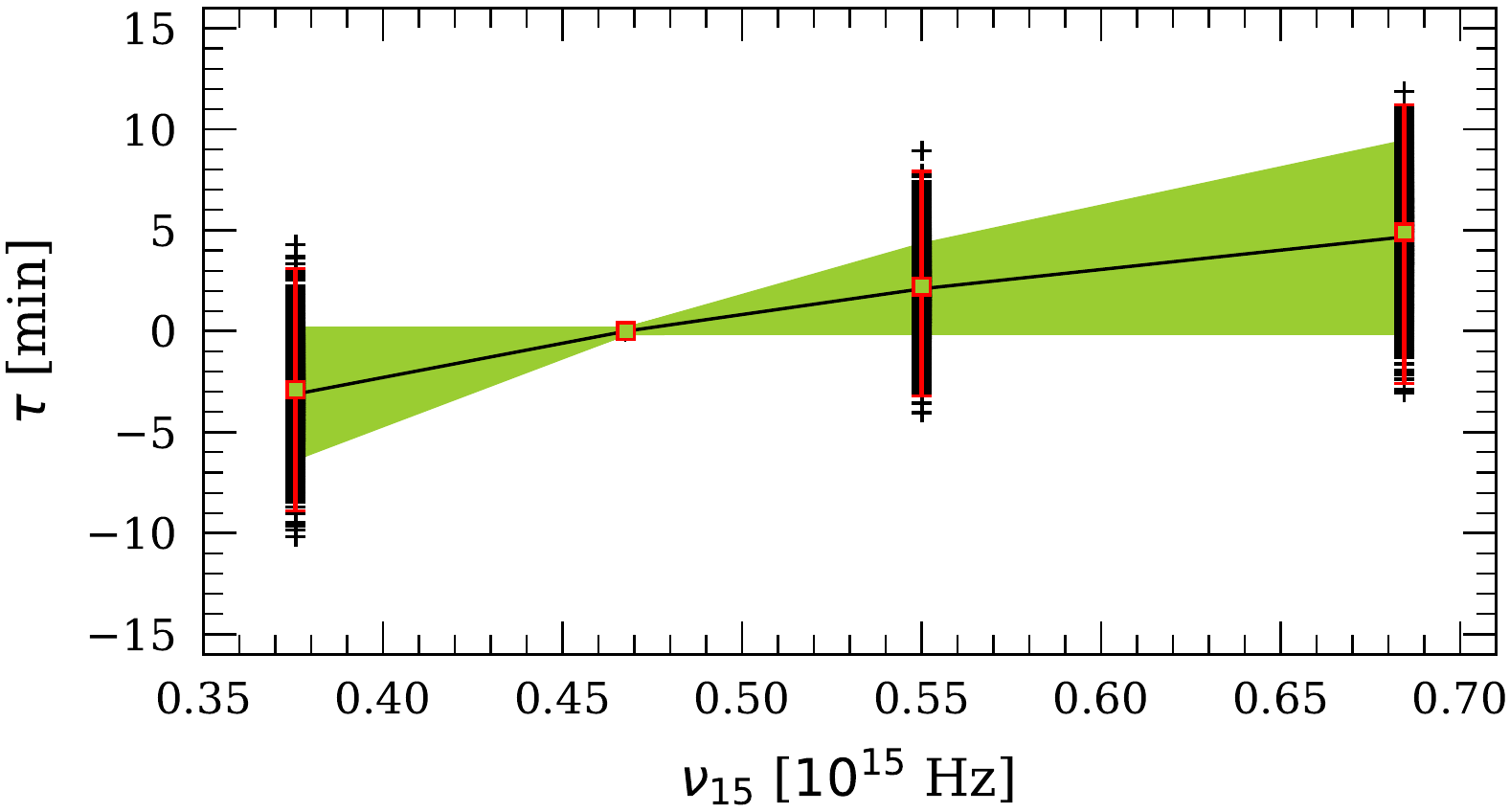} 
\caption{Time lag of the $BVI$ band variations (with respect to the $R$ band ones) against the frequency of the corresponding bands (squares). The black solid curve is fit for this frequency dependence. The black plus signs mark the randomized lag values, while the green lines are the fits to each set of randomized time lags (see text). We show the $3\sigma$ error bars for the sake of comparison with the randomized lag values.}
\label{fig:tcool}
\end{figure}

\begin{figure}[t!]
\includegraphics[width=\linewidth,clip=true]{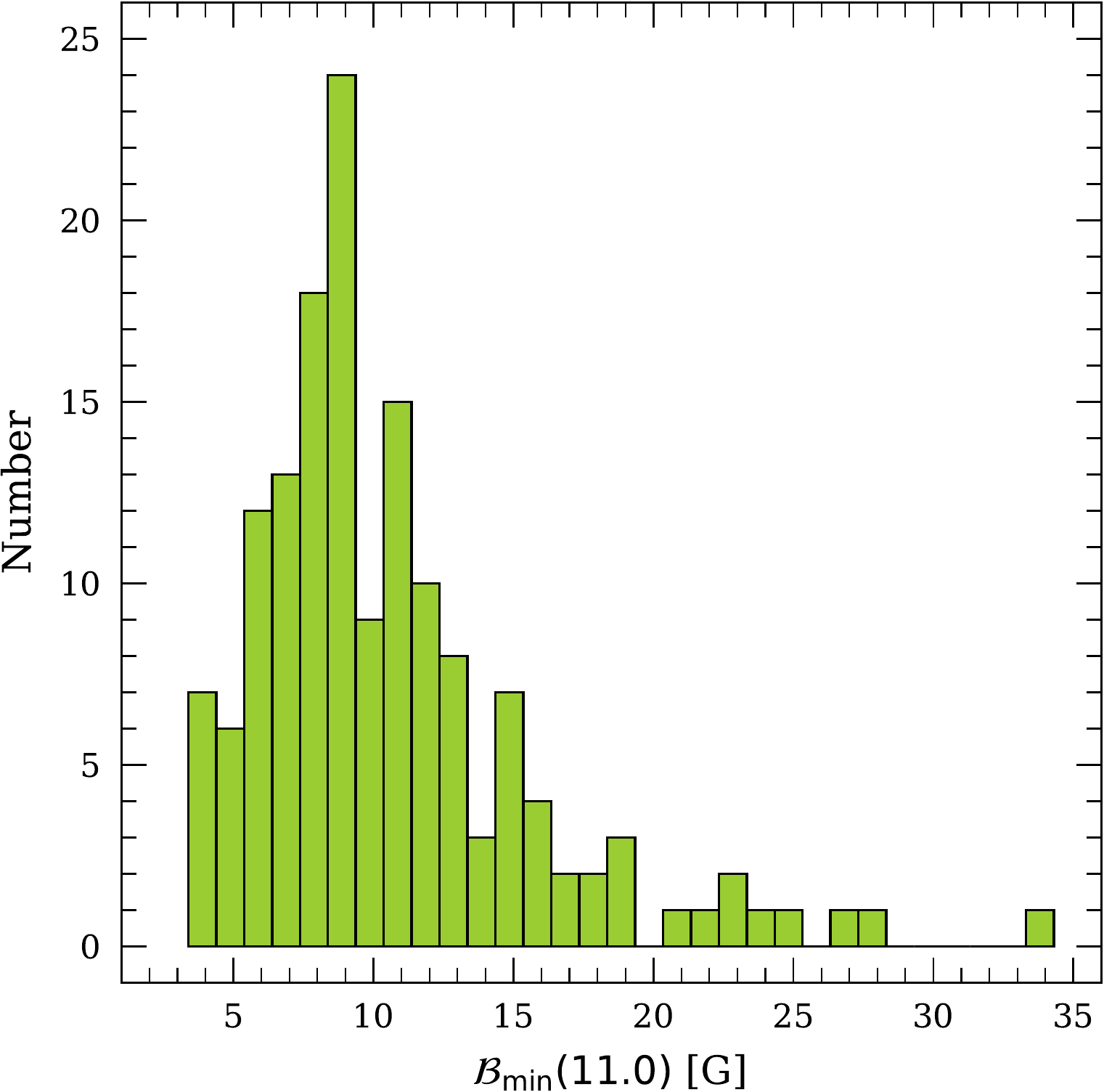}
\caption{Distribution of the minimal magnetic field strengths.}
\label{fig:magnet:histo}
\end{figure}

\begin{figure}[t!]
\includegraphics[width=\linewidth,clip=true]{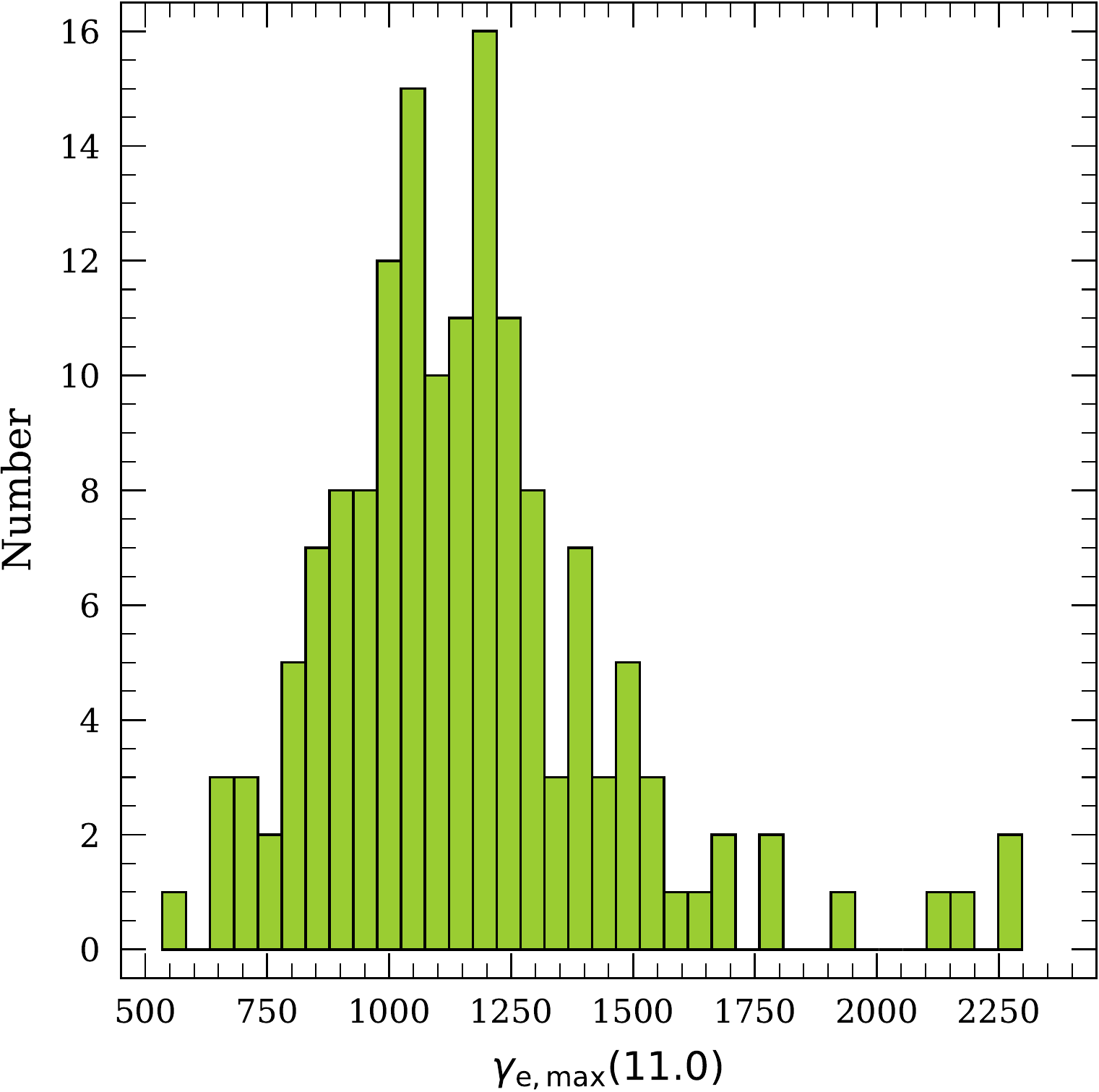} 
\caption{Distribution of the maximal electron Lorentz factors.}
\label{fig:gamma:histo}
\end{figure}

\begin{figure}[t!]
\includegraphics[width=\linewidth,clip=true]{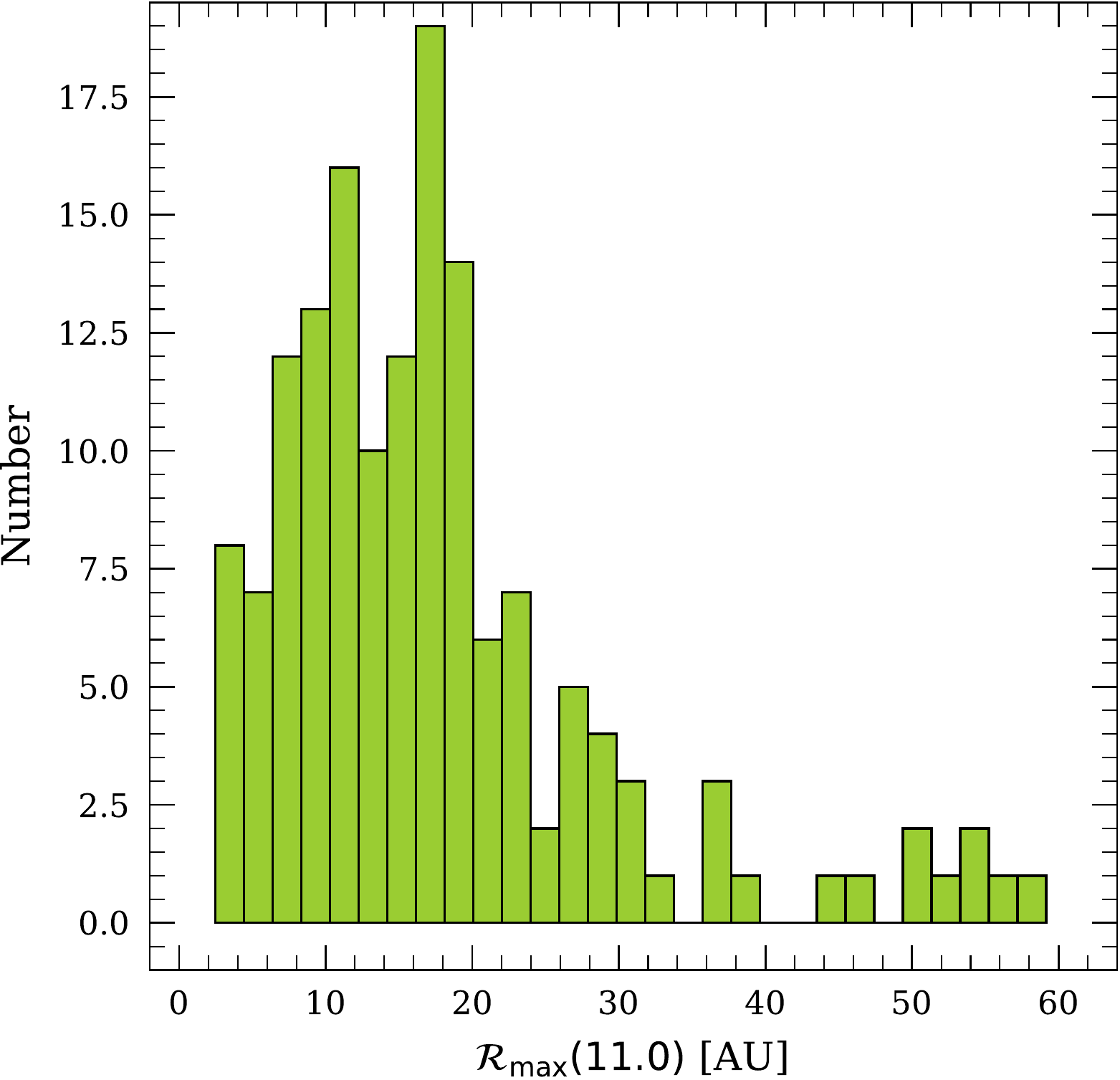} 
\caption{Distribution of the maximal radii.}
\label{fig:radius:histo}
\end{figure}

\section{Discussion}
\label{sec:disc}

In this paper, we have presented the results from the optical monitoring of the blazar \bl\ for the period Jul 11~-- Sep 14, 2020, which encompasses the August 2020 flare. During this period (more specifically, starting from the second half of August), we have performed intense intra-night monitoring of \bl. The blazar showed very high intra-night activity with a duty cycle over that period of 96\% or 88\%, depending on whether the probably variable cases are considered variable or not. We performed a thorough analysis of the INV of \bl\ during the August 2020 flare, and now we shall discuss some constraints that the results from our analysis can place on the blazar jet parameters.

\subsection{Emitting Region Parameters}
\label{sec:disc:pars}

First of all, we adopted the turbulent jet model \citep[e.g.][]{2013A&A...558A..92B} in order to interpret the INV observed. Within this model, a plane shock hits a turbulent cell and accelerates (energize) the cell electrons, which are then cooled by synchrotron emission. In this way, a flux pulse is produced, which manifests itself as a flare on the LC.
The combination of the individual pulses coming from cells of various characteristics leads to the observed INV. Within this model, the high duty cycle obtained by us means that there is well-developed turbulence within the jet \citep[e.g.][]{2021Galax...9..114W}. 
In a recent study, \citet{2023ApJ...943..135K} reported results from the \bl\ monitoring from Oct 1 to Nov 23, 2020 in the optical. According to their Table~2, the source showed INV during four nights out of ten (the probably variable cases considered non-variable); see also \citet{2023MNRAS.519.3798S} regarding the source monitoring in that period.
Therefore, the duty cycle could be estimated as $\sim$40\%, which is significantly lower than ours. The probably variable cases, however, are associated with the intra-night monitoring duration of $\lesssim$3\,h, which could affect the source variability status and, hence, the duty cycle estimate. In any case, the above-obtained value could be considered as a lower limit. If, however, we assume that the duty cycle decrease is real, and not an artifact of the insufficient monitoring duration, then, following the turbulent jet model, the turbulence within the jet subsides significantly within about two months since the August 2020 flare onset.

The details about the processes of particle acceleration taking place in the jet are not directly relevant to the present scenario, and so we assumed for simplicity a quasi-instantaneous injection within a time $t'_{\rm inj} \le {\mathcal R}/c$ of a mono-energetic population of high-energy electrons in a homogeneous region of radius $\mathcal R$ (here $c$ stands for the speed of light); here and below the primed quantities are in the rest frame. These electrons cool by synchrotron emission and lose half of their energy within the cooling time, $t_{\rm cool}(\nu)$:
\begin{equation}
t_{\rm cool}(\nu) \simeq 4.73\!\times\!10^4\,{\mathcal B}^{-3/2}\,\nu_{15}^{-1/2}\,\left(\frac{\delta}{1+z}\right)^{-1/2} ~ [\rm s],
\label{eq:tcool}
\end{equation}
where $\nu_{15}$ is the observed photon frequency (in units of $10^{15}$\,Hz, $\nu = 10^{15}\nu_{15}$\,Hz) and ${\mathcal B}$ the magnetic field strength (in units of Gauss). Here we neglected the cooling by the inverse Compton processes; that is, a zero Compton dominance parameter was assumed. This assumption is justified because \citet{2010ApJ...716...30A} reported a Compton dominance parameter of 0.2 for \bl.

In the framework of this scenario, the low-energy electrons result from initially more energetic ones after their synchrotron cooling, thereby leading to the soft time lag
\citep[e.g.][]{1997ApJ...486..799U,1998ApJ...509..608T}: the time lag between two bands corresponding to frequencies $\nu_1$ and $\nu_2$ ($\nu_1>\nu_2$) is equal to $\tau(\nu_2,\nu_1) = t_{\rm cool}(\nu_2)-t_{\rm cool}(\nu_1)$.
Therefore, if we have estimated the time lags among the $BVRI$ bands, then we can derive ${\mathcal B}$ and $\delta$ simultaneously. 
This technique was applied using the Aug 20 lags (Table~\ref{tab:dcf}), and so we have $\tau(\nu_R,\nu_k) = t_{\rm cool}(\nu_R)-t_{\rm cool}(\nu_k)$, $k=B,V,I$. In this notation the lags $\tau(\nu_R,\nu_B)$ and $\tau(\nu_R,\nu_V)$ are positive, while the lag $\tau(\nu_R,\nu_I)$ is negative. The frequency dependence of the observed lags is shown in Figure~\ref{fig:tcool}.
Technically, we did randomization of the time lags within the corresponding asymmetric lag uncertainties to estimate the parameters and their uncertainties. For each set of randomly drawn lags, we estimated ${\mathcal B}$ and $\delta$ performing an unweighted fit using the Nelder-Mead fitting method; we ran a total of 2500 cycles. 
Finally, we built the parameter distributions and used them to get ${\mathcal B}=5.6^{+1.3}_{-0.8}\,\rm G$ and $\delta=11.0^{+0.3}_{-0.3}$; the weighted Nelder-Mead fit without randomization gave very similar results. The parameter uncertainties represent the 16-th and 84-th percentiles of the corresponding distributions, and the fit corresponding to the so-derived parameters is drawn in Figure~\ref{fig:tcool} with a black line.
Using the same approach and MWL time lags from $\gamma$-rays to optical, \citet{2020ApJ...900..137W} obtained a magnetic field strength of $\sim$3.0\,G for \bl.

We have only $B$ vs $I$ time lag for Aug 26, and so we can apply the following expression to derive the magnetic field strength \citep[e.g.][]{1998ApJ...509..608T,2003A&A...397..565P}:
\begin{multline}
{\mathcal B}\,\delta^{1/3} \simeq 1.31\!\times\!10^3\,\left(\frac{1+z}{\nu_{15,I}}\right)^{1/3}\times \\
\left[\frac{1-(\nu_{15,I}/\nu_{15,B})^{1/2}}{\tau(\nu_I,\nu_B)}\right]^{2/3} ~ [\rm G],
\end{multline}
where $\nu_{15,B}$ and $\nu_{15,I}$ are the frequencies corresponding to the $BI$ bands, respectively (in units of $10^{15}$\,Hz) and $\tau(\nu_I,\nu_B)$ the $B$ vs $I$ time lag (in units of seconds). Having a $B$ vs $I$ time lag of $3.8^{+2.5}_{-1.3}$\,min, we got ${\mathcal B}\,\delta^{1/3}\simeq20.3^{+6.4}_{-5.7}\,\rm G$ or ${\mathcal B}\simeq9.1^{+2.9}_{-2.6}\,\rm G$ if we assume a Doppler factor of $11.0^{+0.3}_{-0.3}$ as estimated above \citep[see also][]{2023MNRAS.519.3798S}. The uncertainties of $\mathcal B$ were derived using the lag and Doppler factor randomization. 

As we mentioned in Section~\ref{sec:dcf}, the measured time lags are INLC lags rather than individual flare lags. Therefore, we shall assume the parameters determined above to be
an average over the emitting regions, contributing to the given INLC.
In this regard, our estimate of the Doppler factor is a kind of local estimate related to the regions, contributing to the Aug 20 INLC. Nevertheless, it is consistent with the literature values of $\delta$ for \bl\ as mentioned before. We see that the various Doppler factor estimates for \bl\ are consistent with each other irrespective of the band and method used to get them. This is in contrast with the estimates of $\delta$ for the high-energy synchrotron-peaked blazars, for which dependence on the band and method used is observed \citep[this is termed as the ``Doppler crisis'', e.g.][]{2018mgm..conf.3074P,2021A&A...645A.137A}. An explanation of that dependence could lie in the more complex internal jet structure in these sources compared to the other kind of blazars.
Hence, our results are in support of this scenario as far as \bl\ is classified as a low-energy synchrotron-peaked blazar: the lack of discrepancy among the Doppler factor estimates could mean a simple structure of its jet.

An independent magnetic field strength estimate could be obtained using the results from the LC decompositions and considering the decay timescale, ${\mathcal T}_{\rm d}$, as an upper limit of $t_{\rm cool}$; that is, ${\mathcal T}_{\rm d} \ge t_{\rm cool}$ \citep[e.g.][]{2021RAA....21..302F}.
Thus, the lower limit (or the minimum value) of the magnetic field strength, ${\mathcal B}_{\rm min}(\delta)$, inside the emitting region could be derived by rewriting the Equation\,(\ref{eq:tcool}) as follows:
\begin{equation} \label{eq:magnet}
\begin{split}
{\widetilde {\mathcal B}}_{\rm min} & = 1.31\!\times\!10^3\,{\mathcal T}_{\rm d}^{\,-2/3}\,\nu_{15}^{-1/3}\,(1+z)^{1/3} ~ [\rm G]; \\ 
{\mathcal B}_{\rm min}(\delta) & = {\widetilde {\mathcal B}}_{\rm min}\,\delta^{-1/3}; \\
{\mathcal B} & \ge {\mathcal B}_{\rm min}(\delta),
\end{split}
\end{equation}
where ${\mathcal T}_{\rm d}$ is in units of seconds. 
In addition, the results from the LC decompositions could also be used to set limits on the electron Lorentz factor in the emitting region and on the radius of the emitting region.

The electron Lorentz factor, $\gamma_{\rm e}$, which is the electron energy in units of $m_{\rm e}c^2$ can be associated with the observed frequency of the emitted synchrotron radiation via \citep{1997A&A...327...61G}
\begin{equation}
\nu = \frac{4}{3}\,\gamma^2_{\rm e}\,\nu_{\mathcal B}\frac{\delta}{1+z},
\end{equation}
where $\nu_{\mathcal B} = 2.80 \times 10^6\mathcal B$ is the cyclotron frequency.
This equation, coupled with Equation\,(\ref{eq:tcool}), yields $\gamma_{\rm e} \propto t^{1/3}_{\rm cool}(\nu)$. Assuming again that ${\mathcal T}_{\rm d} \ge t_{\rm cool}$, we get an upper limit (or a maximal value) of the electron Lorentz factor for the corresponding frequency:
\begin{equation} \label{eq:gamma}
\begin{split}
{\widetilde \gamma_{\rm e,max}} & = 4.53\!\times\!10^2\,\nu^{2/3}_{15}\left[{\mathcal T}_{\rm d}\,(1+z)\right]^{1/3}; \\
\gamma_{\rm e,max}(\delta) & = {\widetilde \gamma_{\rm e,max}}\,\delta^{-1/3}; \\
\gamma_{\rm e} & \le \gamma_{\rm e,max}(\delta),
\end{split}
\end{equation}
where ${\mathcal T}_{\rm d}$ is in units of seconds.

Accounting for our assumption about the injection time of electrons, the rising part of the flare LC constrains the light-crossing time, $t_{\rm cros}$ (${\mathcal T}_{\rm r} \ge t_{\rm cros}$), thus setting an upper limit (or a maximal value) on the emitting region radius as follows:
\begin{equation} \label{eq:radius}
\begin{split}
{\widetilde {\mathcal R}_{\rm max}} & = \frac{c\,{\mathcal T}_{\rm r}}{1+z} ~ [\rm cm]; \\
{\mathcal R}_{\rm max}(\delta) & = {\widetilde {\mathcal R}_{\rm max}}\,\delta; \\
{\mathcal R} & \le {\mathcal R}_{\rm max}(\delta),
\end{split}
\end{equation}
where ${\mathcal T}_{\rm r}$ is in units of seconds.
The dominance of the light-crossing time means also that the rising timescale and, hence, the emitting region radius are not frequency dependent.

Taking the values of ${\mathcal T}_{\rm r}$ and ${\mathcal T}_{\rm d}$ from Table~\ref{tab:fit}, assuming $\delta=11.0$, and using Equations (\ref{eq:magnet}), (\ref{eq:gamma}), and (\ref{eq:radius}), we obtained the minimal values of the magnetic field strength, maximal values of the electron Lorentz factor, and maximal values of the radius that characterize the emitting regions. 
The distributions of ${\mathcal B}_{\rm min}(\delta=11.0) \equiv {\mathcal B}_{\rm min}(11.0)$, $\gamma_{\rm e,max}(\delta=11.0) \equiv \gamma_{\rm e,max}(11.0)$, and ${\mathcal R}_{\rm max}(\delta=11.0) \equiv {\mathcal R}_{\rm max}(11.0)$ are shown in Figures~\ref{fig:magnet:histo}, \ref{fig:gamma:histo}, and \ref{fig:radius:histo}. Some characteristics of the emitting regions are listed in Table~\ref{tab:pars}.

Using the same approach, \citet{2015A&A...578A..68C} found the following characteristics for the emitting regions of \bl\ assuming $\delta=10.0$ and a Compton dominance parameter of unity: a lower limit for the magnetic field strength of 6.0\,G and an upper limit for the radius of $3\times10^{-5}\,{\rm pc}=6.2$\,AU. In addition, \citet{2020ApJ...900..137W} obtained a magnetic field strength of $\sim$3.0\,G using a minimal timescale of $\sim$30.0\,min, derived on the basis of the \bl\ MWL variability.

\subsubsection{Turbulent Cell Sizes}

Following the turbulent jet model, the INLCs are a combination of synchrotron pulses coming from various turbulent cells within the emitting region in the jet, hit by a plane shock. The turbulence is a stochastic process, and so each INLC is a single realization of this process.
In the framework of Kolmogorov theory of turbulence \citep{1941DoSSR..30..301K} the Kolmogorov scale is the smallest spatial scale in a turbulent flow at which scale the turbulence kinetic energy dissipates. Therefore, the smallest spatial scales, found on the basis of the minimal timescales of the flux variations, directly probe the Kolmogorov scale. Following this line of reasoning, the Kolmogorov scale (i.e., the diameter of the smallest emitting region) derived by us is $\le$4.4\,AU (see Table~\ref{tab:pars}). \citet{2017MNRAS.469.3588M} fitted synchrotron pulses to the flares of the \bl\ INLCs and estimated a smallest cell size of $\sim$1.5\,AU assuming a shock speed of $0.1c$ and a Doppler factor of 7.3; the smaller cell size becomes $\sim$2.3\,AU if one assumes $\delta=11.0$.

Figure~\ref{fig:radius:histo} shows that the limits on the turbulent cell sizes are distributed continuously up to about 70\,AU; the cell size is assumed to be twice ${\mathcal R}_{\rm max}(11.0)$. This is in agreement with the results of \citet{2013A&A...558A..92B} but somewhat larger than the cell sizes obtained by \citet{2012JSARA...7...33R} and \citet{2019ApJ...884...92X}; in all three papers the authors model the INLCs by means of numerical calculation of the synchrotron pulse profiles, expected from the energized turbulent cells.

Large cell sizes are rare (Figure~\ref{fig:radius:histo}), which could mean that either (i) the large cells exist, but they are truly rare because they are unstable, or (ii) the large cells do not exist and they are actually unresolved groups of smaller cells. Our INLCs provide examples in support of both possibilities (see Figures \ref{app:decompo:fig:mwl} and \ref{app:decompo:fig}): the INLCs for Aug 28 shows two large, well-pronounced flares, while the INLC for Sep 11 shows a barely resolved group of four flares, which in the case of noisy data could be misidentified as a large, single flare. A few other INLCs show similar, barely resolved groups of flares (Sep 2, Sep 3, and Sep 14).
These groups reflect the synchrotron pulses from compact, fragmented regions within the jet hit by a shock. A crude estimate of the upper limit of the radius of such a fragmented region could be made by summation of the radii of the corresponding individual region that form it~-- the upper limits are in the interval $(39-77)$\,AU $=(5.8 - 12) \times 10^{14}\,\rm cm$ for the above-mentioned dates.

The typical jet radii used in the blazar SED modeling are $(10^{16}\!-\!10^{17})\,\rm cm$ \citep[e.g.][]{2018ApJS..235...39C}; that is, the fragmented regions we studied are smaller than the jet by a factor of at least $\sim$8.

\subsection{Power Spectral Density Slopes}

Characterization of the PSD on intra-night timescales in the optical has been performed occasionally in the past. The first systematic study on this topic, based on a sample of blazars, was published just recently by \citet{2021ApJ...909...39G}.
The PSD characteristics reflect the emission processes in blazars, and so it is important to deepen our research in that field both by enlarging the samples and by increasing the number of the single-source INLCs analyzed (as we did). Here, we used the SF to study the temporal characteristics of the detrended \bl\ INLCs.

We built and fitted a total of 29 SFs of \bl. The weighted mean (over the nights and bands) slope was found to be $1.624 \pm 0.007$. Using the approximate relation between the SF and PSD slopes, we found a mean PSD slope of $\varkappa\simeq2.6$ (a standard deviation of 0.3, see Section \ref{sec:sf}), that is our PSD slope is steeper than that of a pure random walk/red-noise process for which $\varkappa=2$. We should point out, however, that there could be an additional offset in our estimate because of the PSD slope approximation used by us~-- this should be accounted for in the discussion that follows. 
\citet[][]{2003A&A...397..565P} estimated the PSD slope for \bl\ on intra-night timescales to be $\varkappa=1.87 \pm 0.16$; the individual PSDs were averaged over nights and bands before the fitting.
\citet{2011AJ....141...49C} found the SF slopes for the blazar S5\,0716+714 to lie mostly between 1 and 2 (corresponding to the PSD slopes in the range 2--3).
Recently, \citet{2021ApJ...909...39G} found a mean PSD slope of $3.1\pm 0.3$ for a sample of seven BL\,Lacs.
Our result is consistent with that of \citet{2011AJ....141...49C} and \citet{2021ApJ...909...39G} to within the scatter quoted and steeper than the PSD slope obtained by \citet[][]{2003A&A...397..565P}.
The above groups, however, did not apply any detrending procedure, and so their results could be affected by the long-term component when present: the results will be dependent on the number of the INLCs showing a long-term component (the INLCs without such component could be considered as being detrended already).

Our assumption about the INLC generation is related to the turbulent jet model as we mentioned above. In this regard, \citet{2015JApA...36..255C} and \citet{2016ApJ...820...12P} estimated the PSD slopes expected from the turbulence within the jet flow. Their computations are based on the numerical 2D modeling of relativistic jet propagation, and both groups found the PSD slopes to average around $\varkappa=2$ for timescales from a few days to years. Our mean PSD slope is steeper, but it is derived on the intra-night timescales. However, the detailed analysis of the PSDs for our data is beyond the scope of the present paper.

\section{Summary}

The main results of the presented study could be summarized as follows:
\begin{enumerate}
\item Short-timescale flux variations displayed a total amplitude variation of $\sim$2.2\,mag in $R$ band. In addition, we found that on a short-term basis the spectral index has a weak dependence on the flux level and the variations could be mildly chromatic;
\item During the August 2020 flare, the median spectral index was calculated to be $\langle \alpha_{VRI} \rangle_{\rm med}=0.885\pm0.020$;
\item We did not find any significant periodicity;
\item The source was found to display BWB chromatism on intra-night timescales;
\item The duty cycle was estimated to be $\sim$90\% or higher;
\item The weighted mean SF slope was found to be $\langle \varrho \rangle_{\rm wt}=1.624 \pm 0.007$;
\item The cross-correlation analysis resulted in two cases of significant inter-band time lags~-- the lags were of order of a few minutes;
\item We obtained an estimate of the Doppler factor, $\delta=11.0^{+0.3}_{-0.3}$, using the inter-band time lags;
\item We derived the values or limits for the magnetic field strength in the emitting regions using the inter-band time lags or LC decomposition results, respectively. The typical values/limits for $\mathcal B$ were found to be $\sim$10.0\,G if we assume $\delta=11.0$;
\item Using the LC decomposition results, we obtained limits for the Lorentz factors of the emitting electrons and the radii of the emitting regions. In particular, the smallest upper limit on the radius is 2.2\,AU, which we related to the Kolmogorov scale of the turbulent flow;
\item The mean slope of the power spectral density on intra-night timescales, roughly estimated from the mean SF slope, is steeper than that of a pure random walk/red-noise process.
\end{enumerate}

\acknowledgments

We thank the anonymous referee for valuable comments and suggestions, which helped in improving the paper.
The work is partly supported by the NCN grant No 2018/29/B/ST9/01793.
A.A. and A.O. were supported by the Scientific and Technological Research Council of Turkey (TUBITAK), Project No. 121F427. EE was supported by the Scientific Research Project Coordination Unit of Istanbul University, Project No. FDK-2022-19145. We thank TUBITAK National Observatory for partial support in using T60 and T100 telescopes with project numbers 19BT60-1505 and 19AT100-1486, respectively.

\label{lastpage}

\bibliographystyle{aasjournal}
\bibliography{bllac2020aug}

\end{document}